\documentclass[traditabstract]{aa}

\usepackage{txfonts}
\usepackage{graphics}
\usepackage{xcolor}
\begin{document}

\title{Long-term optical monitoring of TeV emitting Blazars}

\subtitle{I. Data analysis.}

\author{
K. Nilsson\inst{1} \and
E. Lindfors\inst{1,2} \and
L. O. Takalo\inst{2} \and
R. Reinthal\inst{2} \and
A. Berdyugin\inst{2} \and
A. Sillanp\"a\"a\inst{2}
S. Ciprini\inst{3,4} \and
A. Halkola\inst{2} \and
P. Hein\"am\"aki\inst{2} \and
T. Hovatta\inst{2,5,6} \and
V. Kadenius\inst{2} \and
P. Nurmi\inst{2} \and
L. Ostorero\inst{7,8} \and
M. Pasanen\inst{2} \and
R. Rekola\inst{2} \and
J. Saarinen\inst{2} \and
J. Sainio\inst{2} \and
T. Tuominen\inst{2} \and
C. Villforth\inst{9} \and
T. Vornanen \inst{2} \and
B. Zaprudin\inst{2}
}

\institute{
Finnish Centre for Astronomy with ESO (FINCA), Quantum, Vesilinnantie 5, FI-20014, University of Turku, Finland\and
Department of Physics and Astronomy, FI-20014, University of Turku, Finland\and
Agenzia Spaziale Italiana (ASI) Science Data Center, I-00133 Roma, Italy \and
Istituto Nazionale di Fisica Nucleare, Sezione di Perugia, I-06123 Perugia, Italy \and
Aalto University Mets\"ahovi Radio Observatory, Mets\"ahovintie 114, 02540, Kylm\"al\"a, Finland \and
Aalto University Department of Radio Science and Engineering,P.O. BOX 13000, FI-00076 AALTO, Finland \and
Dipartimento di Fisica, Universit\`{a} degli Studi di Torino, Via P. Giuria 1, 10125 Torino, Italy \and
Istituto Nazionale di Fisica Nucleare (INFN), Via P. Giuria 1, 10125 Torino, Italy \and
Department of Physics, University of Bath, Claverton Down, Bath BA2 7AY, UK
}

\date{Received; accepted}

\abstract{We present 10 years of R-band monitoring data of 31 northern
  blazars which were either detected at very high energy (VHE) gamma
  rays or listed as potential VHE gamma-ray emitters. The data
  comprise 11820 photometric data points in the R-band obtained in
  2002-2012. We analyze the light curves by determining their power
  spectral density (PSD) slopes assuming a power-law dependence with a
  single slope $\beta$ and a Gaussian probability density function
  (PDF).  We use the multiple fragments variance function (MFVF)
  combined with a forward-casting approach and likelihood analysis to
  determine the slopes and perform extensive simulations to estimate
  the uncertainties of the derived slopes. We also look for periodic
  variations via Fourier analysis and quantify the false alarm
  probability through a large number of simulations. Comparing the
  obtained PSD slopes to values in the literature, we find the slopes
  in the radio band to be steeper than those in the optical and gamma
  rays. Our periodicity search yielded one target, Mrk 421, with
  a significant ($p<5\%$) period. Finding one significant period among
  31 targets is consistent with the expected false alarm rate, but the
  period found in Mrk~421 is very strong and deserves further
  consideration}.


\maketitle

\section{Introduction}

Blazars are active galactic nuclei (AGN) with a relativistic jet,
which is pointing close to our line of sight. The blazar family
consists of flat-spectrum radio quasars (FSRQs) and BL Lac
objects. Blazars are the most numerous objects in the extragalactic
gamma-ray sky.  The spectral energy distribution of blazars shows
two humps; one in the infra-red to X-ray range and the second in the
X-rays to gamma-rays.  The first hump is ascribed to synchrotron
emission and the second is typically attributed to inverse Compton
(IC) emission. The peak frequency $\nu_{peak}$ of the synchrotron peak is
commonly used to further divide the BL Lacs into low-, intermediate-
and high frequency peaked BL Lacs (LBL, IBL and HBL, respectively)
with $\log{\nu_{peak}} < 14$ defining the LBL, $14 < \log{\nu_{peak}}
< 15$ the IBL and $\log{\nu_{peak}} > 15$ the HBL classes
\citep{2010ApJ...716...30A}.

Blazars show variability in all bands from radio to Very High Energy
(VHE) gamma-rays and in time scales ranging from years to only a few
minutes. Sometimes there is correlated variability between two bands
\citep[e.g][and references therein]{2016MNRAS.456..171R}, but not
always.  The long-term variability has been most extensively studied
in the radio and optical bands
\citep[e.g.][]{2003ApJ...586...33A,2007A&A...469..899H,1988ApJ...325..628S,
  2004A&A...424..497V}, where long time series have been collected
during decades. Blazar light curves are typically characterized by a
power-law power spectral density (PSD), lacking clear and persistent
periodicities and/or breaks in the spectrum, which would signify upper
and lower limits for the variability time scales. The PSD is
notoriously difficult to determine reliably due to uneven sampling and
instrument noise \citep{1982ApJ...263..835S,1989AJ.....97..720H}. In
spite of these challenges, there have been several claims for
periodicities in both radio and optical light curves of single sources
\citep[e.g][]{{1988ApJ...325..628S},{2001A&A...377..396R},{2004A&A...424..497V},{2010AJ....139.2425N},{2013MNRAS.436L.114K}},
but e.g. \citet{2008A&A...488..897H} found no periodic changes in
large sample of radio light curves. In recent years such searches have
also become feasible in the gamma-ray band and, interestingly, for
several sources common periodicities in optical and gamma-rays have
been reported
\citep{{2014ApJ...793L...1S},{2015ApJ...813L..41A},{2016AJ....151...54S},{2016ApJ...820...20S}}.

In this paper we present a detailed analysis of optical light curves of
31 blazars extending for 10 years. The data originate from the Tuorla
Blazar monitoring program, which is introduced in Section 2, along with
the sample selection. The observations and reduction processes are
explained in Section 3, along with a detailed
analysis of the variability, in particular the intrinsic power spectral
density, and search for periodicities in the light curves in Section 4.
The entire flux data set is also published in electronic form for the first
time.

\section{Sample}

Tuorla Blazar Monitoring Program
\footnote{ http://users.utu.fi/kani/1m/index.html}
\citep{2008AIPC.1085..705T} is an optical monitoring program that was
started in autumn 2002. The monitoring program aims to support the VHE
gamma-ray observations of the MAGIC Telescopes and therefore the
original sample consisted of 24 BL Lac objects from
\cite{2002A&A...384...56C} with $\delta>+20^{\circ}$. These targets
were predicted to emit VHE gamma-rays and they are observable from
Tuorla Observatory over a large portion of the year.  The sample has
been gradually extended to include also other types of gamma-ray
emitting blazars and to the southern sky. Starting from 2004 most of
the observations have been performed with the KVA (Kungliga
Vetenskapsakademien) telescope on La Palma (see Section 3).

The sample discussed here consists of the original sample of 24
blazars from \cite{2002A&A...384...56C} along with seven additional
well-sampled blazars.  The targets are listed in Table 1 together with
their most relevant properties. The sample covers all blazar classes,
even though, due to the selection criteria, the HBLs are the most
numerous sources in the sample. The large majority of the sources have
been detected in VHE gamma-ray energies, some after triggers about
high optical state from this monitoring program (e.g. 1ES1011+496, Mrk
180, ON325, S50716+714)
\citep{2007ApJ...667L..21A,2006ApJ...648L.105A,2012A&A...544A.142A,
  2009ApJ...704L.129A}.

This paper presents photometric data of these 31 blazars from
September 2002 to September 2012. Part of these data have been
previously presented as light curves in papers reporting results of
multiwavelength campaigns of individual blazars (see complete list in
Table 1), looking for recurrent timescales and periodicities in the 
optical band \citep{2007A&A...467..465C,2010A&A...517A..63T,2016ApJ...819L..37V},
common periodicities between the optical and gamma-ray
bands \citep{{2015ApJ...813L..41A}} as well as in studies looking for
correlations between different wavebands
\citep[e.g.][]{2012ApJ...754..114H,2014ApJ...786..157A,2016MNRAS.456..171R,
2016MNRAS.462.4267J,2016A&A...593A..98L,2016A&A...593A..91A}.
However, only a small portion of the data has been published in
numerical form before \citep{2010MNRAS.402.2087V}.

\begin{table*}
  \caption{\label{bllist} Main properties of the targets and observing
    log. Targets above the dividing horizontal line belong to the
    long-term monitoring sample of 24 targets drawn from \cite{2002A&A...384...56C}.  }
  \centering
\begin{tabular}{lllllrrl}
\hline
\hline
(1)           & (2)      & (3)  & (4)   & (5)        & (6)          & (7) & (8)\\          
Target        & z        & Type & TeV   & A$_{\rm R}$ & N$_{\rm frms}$ & N$_{\rm obs}$ &ref.\\
              &          &      & det?  & [mag]      &              &        &\\ 
\hline                                                                          
1ES 0033+595  & -        & HBL  & y     & 1.911     & 1501         & 387 & 1 \\
1ES 0120+340  & 0.272    & HBL  & -     & 0.125     & 1183         & 300 & \\
RGB 0136+391  & -        & HBL  & y     & 0.168     & 1556         & 393 & \\
RGB 0214+517  & 0.049    & HBL  & -     & 0.381     & 1183         & 309 & \\
3C 66A        & 0.444    & IBL  & y     & 0.182     & 2726         & 644 & 2,3\\
1ES 0647+250  & 0.41     & HBL  & y     & 0.214     & 1134         & 303 & \\
1ES 0806+524  & 0.138    & HBL  & y     & 0.096     & 1188         & 328 & 4\\
OJ 287        & 0.306    & LBL  & y     & 0.062     & 3308         & 699 & 5,6,7,8\\
1ES 1011+496  & 0.212    & HBL  & y     & 0.027     & 1509         & 426 & 9,10,11 \\
1ES 1028+511  & 0.360    & HBL  & -     & 0.027     & 1040         & 273 & \\
Mkn 421       & 0.031    & HBL  & y     & 0.033     & 2797         & 683 & 12-20\\
RGB 1117+202  & 0.139    &  -   & -     & 0.043     &  780         & 230 & \\
Mkn 180       & 0.045    & HBL  & y     & 0.029     & 1323         & 379 & 21 \\
RGB 1136+676  & 0.135    & HBL  & y     & 0.019     &  908         & 244 & \\
ON 325        & 0.130    & IBL/HBL & y  & 0.052     & 1031         & 272 & 22 \\
1ES 1218+304  & 0.182    & HBL  & y     & 0.045     &  941         & 273 & \\
RGB 1417+257  & 0.237    & HBL  & -     & 0.041     &  907         & 246 & \\
1ES 1426+428  & 0.129    & HBL  & y     & 0.027     &  825         & 219 & \\
1ES 1544+820  & -        & HBL  & -     & 0.108     &  169         &  46 & \\
Mkn 501       & 0.034    & HBL  & y     & 0.042     & 3958         & 749 & 23-28\\
OT 546        & 0.055    & HBL  & y     & 0.064     & 1496         & 401 & 29\\
1ES 1959+650  & 0.047    & HBL  & y     & 0.384     & 2784         & 734 & 30,31\\
BL Lac        & 0.069    & LBL  & y     & 0.714     & 3122         & 771 & 32-34\\
1ES 2344+514  & 0.044    & HBL  & y     & 0.468     & 1584         & 451 & 35,36\\
\hline                                                                             
S5 0716+714   & 0.31     & LBL  & y     & 0.067     & 2789         & 511 & 37-41\\
ON 231        & 0.102    & IBL  & y     & 0.049     &  757         & 196 & 42\\
3C 279        & 0.536    &FSRQ  & y     & 0.062     & 1198         & 316 & 43-50\\
PG 1424+240   & 0.604    & IBL/HBL  & y     & 0.127     &  408     & 141 & 51\\
PKS 1510-089  & 0.360    &FSRQ  & y     & 0.209     &  994         & 272 & 52-54\\
PG 1553+113   & -        & HBL  & y     & 0.113     & 1610         & 444 & 55-60\\
PKS 2155-304  & 0.116    & HBL  & y     & 0.047     & 1097         & 190 & \\
\hline
\end{tabular}
\tablefoot{
Columns:
(1) Target name, (2) redshift, (3) broadband type (FSRQ/BL division from
the Roma-BZCAT \citep[5th Edition,][]{2015Ap&SS.357...75M}, LBL/IBL/HBL classification
form this work),  (4) is the target
detected at TeV energies, (5) galactic extinction in the R-band,
obtained from the NED,(6) number of CCD frames, (7) number of data
points and (8) references to papers where parts of these data have been
used before.
}
\tablebib{
(1)  \citet{2015MNRAS.446..217A};
(2)  \citet{2009ApJ...692L..29A};
(3)  \citet{2009ApJ...694..174B};
(4)  \citet{2015MNRAS.451..739A};
(5)  \citet{2010MNRAS.402.2087V};
(6)  \citet{2009ApJ...698..781V};
(7)  \citet{2013ApJ...764....5P};
(8)  \citet{2008Natur.452..851V};
(9)  \citet{2007ApJ...667L..21A};
(10) \citet{2016MNRAS.459.2286A};
(11) \citet{2016A&A...591A..10A};
(12) \citet{2008A&A...486..721L};
(13) \citet{2009ApJ...691L..13D};
(14) \citet{2009ApJ...703..169A};
(15) \citet{2010A&A...519A..32A};
(16) \citet{2011ApJ...736..131A};
(17) \citet{2011ApJ...738...25A};
(18) \citet{2012A&A...542A.100A};
(19) \citet{2015A&A...576A.126A};
(20) \citet{2015A&A...578A..22A};
(21) \citet{2006ApJ...648L.105A};
(22) \citet{2012A&A...544A.142A};
(23) \citet{2015A&A...573A..50A};
(24) \citet{2011ApJ...729....2A};
(25) \citet{2011ApJ...727..129A};
(26) \citet{2010A&A...524A..77A};
(27) \citet{2009ApJ...705.1624A};
(28) \citet{2007ApJ...669..862A};
(29) \citet{2014A&A...563A..90A};
(30) \citet{2006ApJ...639..761A};
(31) \citet{2008ApJ...679.1029T};
(32) \citet{2007ApJ...666L..17A};
(33) \citet{2009A&A...501..455V};
(34) \citet{2013MNRAS.436.1530R};
(35) \citet{2007ApJ...662..892A};
(36) \citet{2013A&A...556A..67A};
(37) \citet{2005A&A...429..427P};
(38) \citet{2006A&A...451..797O};
(39) \citet{2009ApJ...704L.129A};
(40) \citet{2008A&A...481L..79V};
(41) \citet{2013A&A...558A..92B};
(42) \citet{2009ApJ...707..612A};
(43) \citet{2007ApJ...670..968B};
(44) \citet{2008Sci...320.1752M};
(45) \citet{2008A&A...492..389L};
(46) \citet{2010Natur.463..919A};
(47) \citet{2011A&A...530A...4A};
(48) \citet{2012ApJ...754..114H};
(49) \citet{2014A&A...567A..41A};
(50) \citet{2016A&A...590A..10K};
(51) \citet{2014A&A...567A.135A};
(52) \citet{2009A&A...508..181D};
(53) \citet{2011A&A...529A.145D};
(54) \citet{2014A&A...569A..46A};
(55) \citet{2007ApJ...654L.119A};
(56) \citet{2009A&A...493..467A};
(57) \citet{2010A&A...515A..76A};
(58) \citet{2012ApJ...748...46A};
(59) \citet{2015MNRAS.450.4399A};
(60) \citet{2015ApJ...813L..41A};
}
\end{table*}

\section{Observations and data reduction}

The observations were made at two different telescopes using three
different CCD cameras, whose details are given in Table
\ref{ccdcameras}. The Tuorla 1.03m Dall-Kirkham telescope is located
at Tuorla Observatory, Piikki\"o, Finland at 53 m altitude from the
sea level. The focal length of the telescope is 8.45 m, which results
in a field of view (FOV) of 10 $\times$ 10 arcmin with the ST-1001E
chip. Typical seeing at the telescope is 3-6 arcsec and hence the CCD
was binned by 2$\times$2 pixels to obtain the pixel scale in Table
\ref{ccdcameras}. Depending on target brightness, 3 to 8 exposures of
60 s were obtained through the R-band filter. In addition to the
science frames, five bias, dark and dome flats were obtained. The CCD
frames were reduced by first subtracting bias and dark and then dividing
by the flat-field.

\begin{table*}
\caption{\label{ccdcameras}Telescopes and CCD cameras used in the
monitoring. The last column gives the number of CCD frames
obtained with each instrument.}
\begin{tabular}{llllllcr}
\hline
\hline
Telescope & Camera & Pixel format & Pixel scale   & Gain        & Readout noise &  Color term & N$_{\rm frms}$\\
          &        &              & [arcsec/pix.] & [e$^-$/ADU] & [e$^-$]       &  $\zeta$    &            \\
\hline
Tuorla 1.03 m & SBIG-ST1001E    & 1024$\times$1024\tablefootmark{a} & 1.17\tablefootmark{b} & 2.3 & 17 &  -0.05 & 7941 \\
KVA 35 cm     & SBIG-ST8        & 1530$\times$1020\tablefootmark{a} & 0.94\tablefootmark{b} & 2.3 & 14 & \ 0.11 & 35268 \\
KVA 35 cm     & Apogee Alta U47 & 1024$\times$1024                  & 0.68                  & 1.6 & 10 & \ 0.01 & 4597 \\
\hline
\end{tabular}
\tablefoot{
\tablefoottext{a}{Binned by 2$\times$2 pixels during the observations.}
\tablefoottext{b}{When binned.}
}
\end{table*}

\begin{table*}
  \caption{\label{cstars} Comparison and control stars used in this work. Column (1) Target name, (2)
    Comparison star name in the corresponding reference. Stars C1-C4 refer to stars
    calibrated by us, (3) R-band magnitude of the comparison star, (4) V-R color of
    the comparison star, (5) control star name, (6) reference to the comparison and control star
    magnitudes, (7) the aperture radius in arcsec used to measure the comparison star, control
  star and the target, (8) host galaxy flux within the aperture in mJy.}
\centering
\begin{tabular}{lllllccc}
\hline
\hline
(1)           & (2)   &  (3)    & (4)   & (5)     & (6) & (7) & (8)\\                 
Target        & Comp. &  R-band & V - R & Control & Ref. & r$_{ap}$ (Tuorla / KVA) & Host flux\\
              & star  &  mag    &       & star    &     & [arcsec] & [mJy]\\
\hline                                                                                                                         
1ES 0033+595  &  D & 13.66 $\pm$ 0.03 & 1.46 $\pm$ 0.04 & F  & 1 & 5.0 / 5.0 & 0.22  $\pm$ 0.03  \\
1ES 0120+340  &  C & 13.12 $\pm$ 0.03 & 0.38 $\pm$ 0.05 & G  & 1 & 4.0 / 4.0 & 0.17  $\pm$ 0.01  \\
RGB 0136+391  &  B & 13.82 $\pm$ 0.02 & 0.42 $\pm$ 0.04 & A  & 1 & 7.5 / 5.0 & -                 \\
RGB 0214+517  &  A & 13.85 $\pm$ 0.05 & 0.51 $\pm$ 0.06 & B  & 1 & 7.5 / 7.5 & 2.83  $\pm$ 0.09  \\
3C 66A        &  A & 13.38 $\pm$ 0.04 & 0.22 $\pm$ 0.06 & B  & 2 & 7.5 / 5.0 & 0.08  $\pm$ 0.01  \\
1ES 0647+250  &  E & 13.03 $\pm$ 0.04 & 0.59 $\pm$ 0.05 & B  & 1 & 7.5 / 5.0 & 0.033 $\pm$ 0.005 \\
1ES 0806+524  & C2 & 14.22 $\pm$ 0.04 & 0.39 $\pm$ 0.07 & C4 & 3 & 7.5 / 7.5 & 0.69  $\pm$ 0.04  \\
OJ 287        & 4  & 13.74 $\pm$ 0.04 & 0.44 $\pm$ 0.06 & 10 & 2 & 7.5 / 7.5 & 0.077 $\pm$ 0.013 \\
1ES 1011+496  &  E & 14.04 $\pm$ 0.03 & 0.39 $\pm$ 0.03 & B  & 1 & 7.5 / 7.5 & 0.49  $\pm$ 0.02  \\
1ES 1028+511  &  1 & 12.93 $\pm$ 0.03 & 0.27 $\pm$ 0.04 & 5  & 5 & 7.5 / 7.5 & 0.10  $\pm$ 0.02  \\
Mkn 421       &  1 & 14.04 $\pm$ 0.02 & 0.32 $\pm$ 0.03 & 2  & 5 & 7.5 / 7.5 & 8.1   $\pm$ 0.4   \\
RGB 1117+202  &  E & 13.56 $\pm$ 0.04 & 0.42 $\pm$ 0.04 & F  & 1 & 7.5 / 7.5 & 0.66  $\pm$ 0.04  \\
Mkn 180       &  1 & 13.73 $\pm$ 0.02 & 0.25 $\pm$ 0.03 & 2  & 5 & 5.0 / 5.0 & 3.2   $\pm$ 0.2   \\
RGB 1136+676  &  D & 14.58 $\pm$ 0.04 & 0.46 $\pm$ 0.05 & E  & 1 & 7.5 / 7.5 & 0.85  $\pm$ 0.04  \\
ON 325        &  B & 14.59 $\pm$ 0.04 & 0.37 $\pm$ 0.06 & C1 & 2 & 7.5 / 7.5 & 1.0   $\pm$ 0.1   \\
1ES 1218+304  &  B & 13.61 $\pm$ 0.01 & 0.40 $\pm$ 0.02 & C \tablefootmark{a} & 4 & 7.5 / 7.5 & 0.40  $\pm$ 0.02 \\
RGB 1417+257  &  A & 13.78 $\pm$ 0.04 & 0.57 $\pm$ 0.06 & C2\tablefootmark{b} & 3 & 7.5 / 7.5 & 0.52  $\pm$ 0.06 \\
1ES 1426+428  &  A & 13.23 $\pm$ 0.02 & 0.93 $\pm$ 0.03 & B  & 4 & 7.5 / 7.5 & 0.89  $\pm$ 0.03  \\
1ES 1544+820  &  A & 14.59 $\pm$ 0.03 & 0.37 $\pm$ 0.04 & B  & 1 & 7.5 / 7.5 & 0.21  $\pm$ 0.01  \\
Mkn 501       &  4 & 14.96 $\pm$ 0.02 & 0.34 $\pm$ 0.03 & 1  & 5 & 7.5 / 7.5 & 12.0  $\pm$ 0.3   \\
              &  6 & 14.99 $\pm$ 0.04 & 0.68 $\pm$ 0.06 &    & 5 &  &\\                                  
OT 546        &  B & 12.81 $\pm$ 0.06 & 0.33 $\pm$ 0.09 & H  & 2 & 7.5 / 7.5 & 1.25  $\pm$ 0.06  \\
1ES 1959+650  &  4 & 14.08 $\pm$ 0.03 & 0.45 $\pm$ 0.05 & 7  & 5 & 7.5 / 7.5 & 1.70  $\pm$ 0.04  \\         
              &  6 & 14.78 $\pm$ 0.03 & 0.42 $\pm$ 0.05 &    & 5 &\\                                       
BL Lac        &  C & 13.79 $\pm$ 0.05 & 0.47 $\pm$ 0.08 & H  & 2 & 7.5 / 7.5 & 1.38  $\pm$ 0.03  \\
1ES 2344+514  & C1 & 12.25 $\pm$ 0.04 & 0.36 $\pm$ 0.06 & C3 & 3 & 7.5 / 7.5 & 3.71  $\pm$ 0.05  \\        
\hline                                                                                                                         
S5 0716+714   &  5 & 13.18 $\pm$ 0.01 & 0.37 $\pm$ 0.03 & 6  & 5 & 7.5 / 5.0 & 0.10  $\pm$ 0.05  \\          
ON 231        &  D & 13.86 $\pm$ 0.04 & 0.95 $\pm$ 0.06 & C1 & 2 & 7.5 / 7.5 & 0.58  $\pm$ 0.08  \\        
3C 279        &  5 & 15.47 $\pm$ 0.04 & 0.51 $\pm$ 0.03 & 4  & 6 & 7.5 / 7.5 & 0.033 $\pm$ 0.0017\\          
PG 1424+240   & C1 & 13.20 $\pm$ 0.04 & 0.39 $\pm$ 0.06 & C2 & 2 & 7.5 / 7.5 & -                 \\         
PKS 1510-089  &  A & 14.25 $\pm$ 0.05 & 0.37 $\pm$ 0.08 & B  & 7 & 5.0 / 5.0 & -                 \\          
PG 1553+113   &  1 & 13.2  $\pm$ 0.3  & 0.5  $\pm$ 0.3\tablefootmark{c} & 4 & 8 & 7.5 / 7.5 & -  \\
PKS 2155-304  &  2 & 11.67 $\pm$ 0.01 & 0.38 $\pm$ 0.02 & 3 & 9 & 7.5 / 7.5 & 1.17  $\pm$ 0.12   \\
\hline
\end{tabular}
\tablefoot{
\tablefoottext{a}{A star 45 arcsec SSE of the target.}
\tablefoottext{b}{A star 1\farcm7 West of the target.}
\tablefoottext{c}{Assumed value, V-band photometry not available.}
}
\tablebib{
(1) \citet{2007A&A...475..199N};
(2) \citet{1996A&AS..116..403F};
(3) \citet{1998PASP..110..105F};
(4) \citet{1991ApJS...77...67S};
(5) \citet{1998A&AS..130..305V};
(6) \citet{1998A&AS..130..495R};
(7) \citet{1997A&AS..121..119V};
(8) http://www.chara.gsu.edu/PEGA/charts/?1553.113
(9) \citet{1989AJ.....97..720H};
}
\end{table*}

The KVA (Kungliga Vetenskapsakademien) telescope is located on
Observatorio del Roque de los Muchachos (ORM) on La Palma, Spain at
2396 m above the sea level. The KVA system consists of two
telescopes, a 60 cm telescope on a fork mount and a 35 cm Celestron-14
telescope bolted to the underbelly of the 60 cm telescope. All ``KVA''
data in this paper were obtained with the latter telescope, remotely
operated from Finland.  The 3.91
m focal length of the 35 cm telescope gave a FOV of 12$\times$8 arcmin
with the ST-8 chip and 11.6$\times$11.6 arcmin with the U47
chip. Typical seeing during the observations was 1.5 - 3.5 arcsec,
which required binning of the ST-8 chip by 2$\times$2 pixels.  Typical
exposure times were 3-8$\times$180s, depending on object
brightness. Calibration and image reduction was similar to the Tuorla
data, except that the flat-fields were obtained from twilight sky.

\subsection{Photometry}

Photometry of the targets was made in differential mode, i.e. by
comparing the object brightness to the brightness of calibrated
comparison stars near the target.  Using multiple comparison stars
improves the signal to noise (S/N) of the photometry, but in a
long-term project it is not guaranteed that all comparison stars are
always within the FOV. Since the tabulated comparison star magnitudes
always have errors, the derived zero point of the image depends on the
stars chosen to calibrate the image. This effect is likely to be small
since the above errors are usually small, a few percent, but
nevertheless we uses only one comparison star, sufficiently bright in
order to obtain good S/N. The observers were then instructed to always
include this star within the FOV. Exceptions to this rule are Mkn 501
and 1ES 1959+650, for which only relatively weak calibrated comparison
stars are available close to the target. For these targets two
comparison stars were used.  In addition to the comparison star, each
field has a ``control star'', whose photometry is performed
identically to the target and which is used to identify possible
problems during image reduction. Table \ref{cstars} lists the
comparison and control stars and their properties.

Photometry was performed with semiautomatic {\sc Diffphot} software
developed at Tuorla Observatory. In short, {\sc Diffphot} reduces each
image in turn as described above, displays the image on the screen and
waits for the user to point the target. Then the software finds the
comparison and control stars on the image using an internal database
and computes accurate positions of the targets by computing the center
of gravity of the light distribution. Aperture photometry is then
performed at these positions. We used aperture radii r$_{ap}$ between
4.0 and 7.5 arcsec depending on the object brightness (Table
\ref{cstars}). To facilitate accurate host galaxy subtraction, the
aperture was held constant for each target, except when the host
galaxy contributed less than 3\% to the total flux, in which case we
used a smaller aperture for the KVA to take advantage of the better
seeing.  The chosen aperture sizes correspond roughly to the the
optimal aperture r$_{ap} \approx 1-1.5$ FWHM
\citep{1989PASP..101..616H}, except during the best seeing conditions
at the KVA. However, this telescope suffered sometimes of bad
tracking, resulting in elongated stars and the larger than optimal
aperture size helped in compensating this.

The sky background was determined from a circular annulus, sufficiently
far from the target in order not to contaminate the sky region with
target flux and devoid of any bright background/foreground targets.
The sky pixel distribution was first sigma-cleaned and the mode of the
distribution was computed from the formula
\begin{equation}
mode = 2.5 * median - 1.5 * mean\ .
\end{equation}
Using both sigma clipping and mode for sky estimation improve
immunity against sky annulus contamination by background/foreground
targets. The sky level was subtracted from the pixel values inside the
aperture and the net counts $N$ inside the aperture were computed,
taking into account that some pixels are only partially inside the
aperture. During this and aperture centering phase we also checked and
eliminated highly deviant pixels inside the aperture by comparing the
pixel value to the median of the six adjacent pixels. This check was
inhibited within two pixels from the stellar core in order to not
wrongly correct the central pixel when good seeing prevailed.

To calibrate the photometry we computed the scaling factor $c$ from ADUs to
Flux (Jy s ADU$^{-1}$) for each image. The comparison star magnitude $R_{comp}$
was first transformed into flux $F_{comp}$ via
\begin{equation}
\label{fkaava}
F_{comp} = F_0\ 10^{-0.4*R_{comp}}
\end{equation}
with $F_0 = 3080.0$ Jy and then $c$ was computed from
\begin{equation}
\label{scalekaava}
c  = \frac{F_{comp}\ T_{exp}}{N_{comp}}\ 10^{-0.4*\zeta*(V-R)_{comp}} 
\end{equation}
where $N_{comp}$ are the comparison star net counts in ADUs,
$\zeta$ is the color term listed in Table
\ref{ccdcameras} and  $T_{exp}$ is the exposure time.
The R-band fluxes of the target and
the control star, $F$ and $F_{ctrl}$ respectively, were then computed from
\begin{equation}
\label{fluxkaava2}
F = \frac{c\ N}{T_{exp}}\ 10^{0.4*\zeta*(V-R)} 
\end{equation}
and 
\begin{equation}
\label{fluxkaava3}
F_{ctrl} = \frac{c\ N_{ctrl}}{T_{exp}}\ 10^{0.4*\zeta*(V-R)_{ctrl}}\ .
\end{equation}
For the BL Lac nuclei we used $V-R = 0.5$, which
corresponds to a power-law index $\alpha = 1.78$ ($F_\nu \propto
\nu^{-\alpha}$). 

Finally, the data were averaged into one hour bins to improve the
signal to noise (formulae given below). These averaged fluxes $F_a$
were then converted into R-band magnitudes via Eq. \ref{fkaava}.

\subsection{Error analysis}

The averaged fluxes $F_a$ derived above are affected by (i) statistical
noise arising from photon, dark and readout noise and image processing
and (ii) systematic errors arising from assumptions of target and detector
properties. The latter produce a systematic shift of the whole light
curve, but do not change the flux differences between the data points
and thus they are not included in the error bars.  Below we discuss
these errors in the order they appear in the error analysis.

Statistical variations of the fluxes in Eqs. \ref{fluxkaava2} and
\ref{fluxkaava3} arise from the noise in observed counts $N$ and the
statistical noise in the scale factor $c$, the latter of which
originates from the statistical noise in $N_{comp}$ via
Eq. \ref{scalekaava}.  The statistical errors of $c$, $F$ and
$F_{ctrl}$ were determined by first computing the statistical errors
of the corresponding observed counts $N_{comp}$, $N$ and $N_{ctrl}$
from
\begin{equation}
\label{staterrkaava}
\sigma_{N} = 
\frac{\sqrt{G N + G^2  n_{ap}\ \sigma_{sky}^2(1+\frac{n_{ap}}{n_{sky}})}}{G}\ ,
\end{equation}
where $G$ is the gain factor ($e^-$/ADU), $\sigma_{sky}$ is the
standard deviation of sky pixels, $n_{ap}$ is the number of pixels in
the aperture and $n_{sky}$ is the number of pixels in the sky annulus.
Note that $\sigma_{sky}$ is empirically measured from the image, so it
includes the photon noise of the sky, dark noise, readout noise and
any residual noise from image processing.
The statistical errors of target fluxes $F$ are then obtained from
\begin{equation}
\label{ferr}
\sigma_{F} = F \sqrt{ \left(\frac{\sigma_{N_{comp}}}{N_{comp}}\right)^2
+ \left(\frac{\sigma_N}{N}\right)^2 }\ .
\end{equation}
These errors were then used to compute the weighted average of the
one hour bin $F_{a}$ and its error $\sigma_{a}$ from
\begin{equation}
\label{weightavg}
F_{a} = \sum_i \frac{F_i}{\sigma_F^2(i)} \Biggm/ \sum_i \frac{1}{\sigma_F^2(i)}
\end{equation}
and
\begin{equation}
\label{weighterr}
\sigma_{a} = \sqrt { \frac{1}{ \sum_i 1 / \sigma_F^2(i)}}\ .
\end{equation}

Systematic flux errors arise in many ways from the color
correction term $10^{0.4*\zeta*(V-R)}$ in Eqs. \ref{fluxkaava2} and
\ref{fluxkaava3}. Firstly, since $\zeta$ varies from one instrument
to another, small offsets between the three instruments are expected.
We checked this by extracting
the light curves of 31 control stars and measuring the systematic
offsets between data obtained by different cameras.  We found offsets
between -0.051 and 0.050 mag, with 67\% of the
offsets between -0.011 and 0.019 mag.  The target and control star
data obtained by the KVA were shifted to the Tuorla data using
these offsets, thereby suppressing the systematic differences between
the cameras down to a level undetectable by our data. Secondly,
our assumption of the same color $V - R = 0.5$ mag for all the targets 
is clearly too simple and in any case the color correction derived
from stars is not an accurate model for blazars which have different
spectral energy distributions (SEDs) from the stars.
Thirdly, blazars display color variations, e.g. a "bluer when brighter" type
of behavior \citep[e.g.][]{2011PASJ...63..639I}, which produces
small brightness-dependent errors in our data. Given the
$zeta$-values in Table \ref{ccdcameras} and the range of (V-R)
color variations ($\sim$ 0.1) mag, this error is negligible compared
to the error bars.

We also checked if the error bars $\sigma_a$ obtained by
the above procedure could be underestimated.  We tested the control
star light curves for variability using the chi squared test with the
null hypothesis that the stars are intrinsically not variable.  The
chi squared statistic was computed from the formula
\begin{equation}
\label{chikaava}
\chi^2 = \sum_{i=1}^N \frac{(\langle F \rangle - F_{a}(i))^2}{\sigma_a^2(i)}\ ,
\end{equation}
where $\langle F \rangle$ is the average flux of the light curve.
We also computed the probability $p$ that the null hypothesis can 
be rejected and assigned a limit $p < 0.05$ for a target to be
considered variable.

\begin{figure}
\centering
\includegraphics[width=8cm]{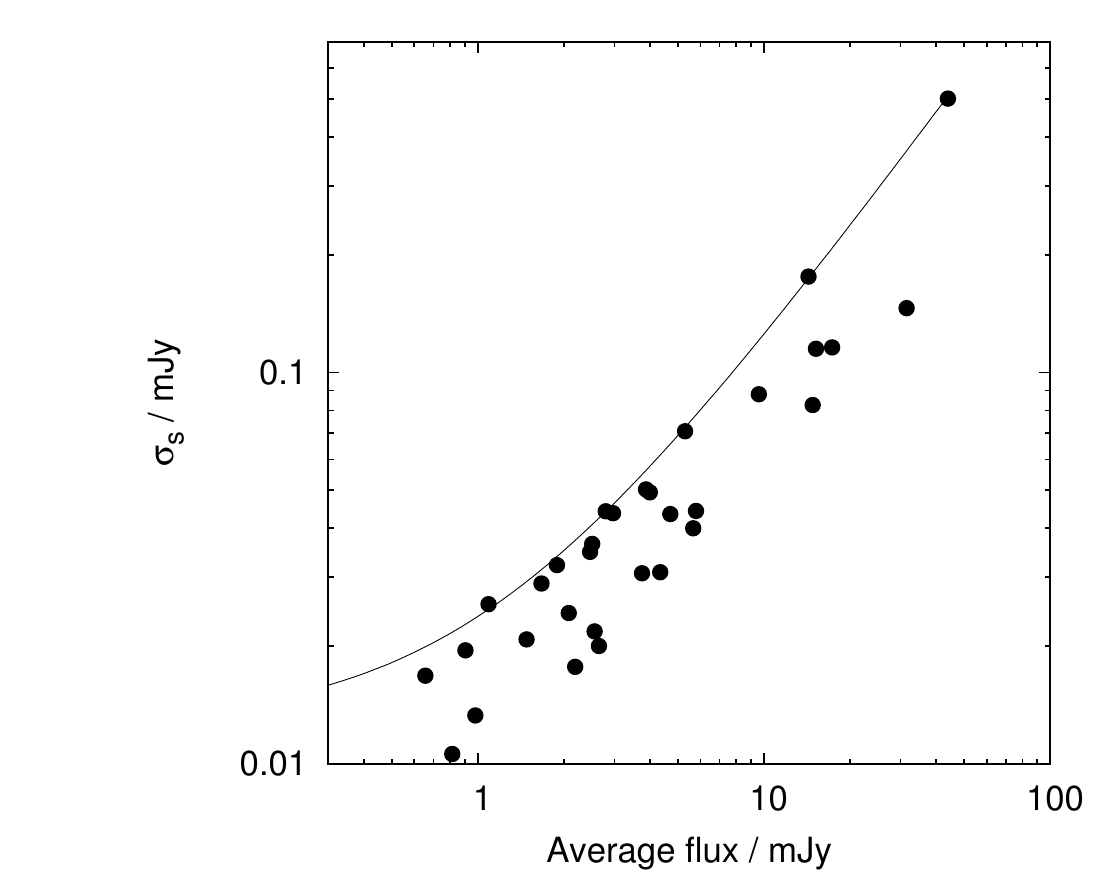}
\caption{\label{sigmaplot} 
The dependence of the additional error term $\sigma_s$ on the average
flux level. The solid line shows the relationship in Eq. \ref{extrakaava},
i.e. the relationship, which will make 95\% of the control stars nonvariable.
This relationship is applied to our data.
}
\end{figure}

Applying this procedure to the control stars, we found $p < 0.01$ for
every control star. Rather than classifying all control stars variable,
we assumed that the error bars derived from Eqs.  \ref{staterrkaava} -
\ref{weighterr} were too small. We thus added in quadrature an additional
error term $\sigma_s$ to the error bars $\sigma_a$ and determined for each star
the smallest $\sigma_s$ which made the star non-variable at the
5\% level. Plotting the smallest $\sigma_s$ against the average flux
$\langle F \rangle$ (Fig. \ref{sigmaplot}), we found $\sigma_s$ to
increase linearly with $\langle F \rangle$ with a slope of
0.0078$\pm$0.0014 and an intercept of (9$\pm$4) $\mu$Jy.  The linear
dependence indicates that $\sigma_s$ is always a constant fraction of
the total flux, leading us to attribute this linear behavior to
flat-fielding errors, which are multiplicative in nature. The
intercept is barely significant, but we nevertheless included it into
our noise model since such a noise limit is expected, and without this
term the noise of faint targets is systematically
underestimated. Since the relation above is an average dependence,
adding $\sigma_s$ from this relation makes $\sim$ 50\% of the control
stars non variable. To be consistent with the 5\% variability limit we
thus multiplied this relation until only 2 of the control stars (6\%)
remained variable, resulting in
\begin{equation}
\label{extrakaava}
\sigma_s = 13\,\mu {\rm Jy} + 0.011 \times F_a\ .
\end{equation}
The final error bars $\sigma$ for the binned average $F_a$ is then obtained
from
\begin{equation}
\label{finalerr}
\sigma = \sqrt{ \sigma_a^2 + \sigma_s^2 }\ .
\end{equation}

A small random error remains in the light curves of those blazars
where the host galaxy component is relatively strong. Variable seeing
causes different fractions of host galaxy and comparison star light to
be included inside the aperture due to the difference in the surface
brightness profiles \citep{1991AJ....101.1196C,2000AJ....119.1534C}.
However, for most of our targets this effect is very small. For instance,
Mkn~501 has one of the strongest host galaxies in our sample and the
effect of FWHM changing from 2 to 5 arcsec is $\sim$ 0.02 mag
\citep[see Fig. 3 in][]{2007A&A...475..199N}. Targets with a nearby
companion galaxy or a foreground star are most affected by the variable seeing.
These targets are discussed in more detail in Section \ref{analyysi}.

In the Tables in Appendix \ref{fluxtables} the errors have been converted into a magnitude
errors $\sigma_m$ via
\begin{equation}
\label{magekaava}
\sigma_{m} = 
\frac{2.5 \log{(F_a+\sigma)} - 2.5 \log{(F_a-\sigma)}}{2}\ ,
\end{equation}
i.e. the asymmetric magnitude errors have been made symmetric by taking the
average of the upward and downward magnitude errors. The flux
errors $\sigma$ can be recovered from magnitude errors $\sigma_m$ by marking
\begin{equation}
\label{palautus1}
k=10^{\sigma_m/(0.5*2.5)}
\end{equation}
and using
\begin{equation}
\label{palautus2}
\sigma = \frac{k-1}{k+1}\ F_a\ .
\end{equation}

To summarize our procedure: we first obtain the counts for the target,
comparison star and control star, $N$, $N_{comp}$ and $N_{ctrl}$,
respectively, via aperture photometry. Then we determine $c$ for each CCD
frame from Eq. \ref{scalekaava} and the target and control star fluxes
from Eqs. \ref{fluxkaava2} and \ref{fluxkaava3} and their errors from
Eqs. \ref{staterrkaava} and \ref{ferr}.  We compute one hour averages
using Eq. \ref{weightavg} and their errors from Eqs. \ref{weighterr},
\ref{extrakaava} and \ref{finalerr}.  Finally we convert fluxes to
magnitudes via Eqs. \ref{fkaava} and \ref{magekaava}.

\section{Analysis methods\label{analyysi}}

As a first step in the analysis we subtracted the host galaxy contribution
from the observed fluxes, corrected the light curves for the galactic
extinction and applied the K-correction.

As was mentioned above, the presence of a host galaxy makes the fluxes
to depend on both aperture and seeing. By using a constant aperture
per target we have eliminated the aperture dependence, but an
additional step was needed to account for the seeing effect.  The host
galaxy fluxes for different apertures and seeing conditions for the
topmost 24 targets in Table \ref{cstars} are tabulated in
\cite{2007A&A...475..199N}. This work used observed, high-resolution
(FWHM 0.5-1.0 arcsec) R-band images of our targets, convolved to a
range of seeing values and and measured with different of aperture
radii. We extracted from the tables in \cite{2007A&A...475..199N} the
host galaxy fluxes for each target using the corresponding measurement
aperture and a seeing value of 2.0 arcsec for the KVA data and 5.0
arcsec for the Tuorla data. These values represent average seeing
conditions at the two sites.  Using different seeing values for the
KVA and Tuorla data effectively reduces the shift between the two data
sets, especially for 1ES~0120+340, Mkn~180 and 1ES~1544+820, all of
which have a relatively strong nearby object leaking light into the
measurement aperture. These targets are also most strongly affected by
the varying seeing conditions, which increase their apparent
variability.

For the 7 targets not included in \cite{2007A&A...475..199N} we used
the analytical formulae in \cite{2005PASA...22..118G} and literature
data to integrate the host galaxy light inside the aperture.  These
formulae do not take into account the smoothing by seeing, whose
effect on the host galaxy fluxes is complicated due to the
differential mode used.  We thus applied the analytical formulae to
the topmost 24 targets in Table \ref{cstars} and checked the results
against the more rigorously obtained values given in
\cite{2007A&A...475..199N}. This comparison indicated that the
analytic expression overestimates the host galaxy fluxes by only
3\%. We thus divided the analytical host galaxy fluxes by 1.03 to be
consistent with the other targets.

The galactic extinction was corrected by extracting the R-band
extinction value $A_R$ from the NED\footnote{NASA/IPAC
  Extragalactic Database (NED) is operated by the Jet Propulsion
  Laboratory, California Institute of Technology, under contract with
  the National Aeronautics and Space Administration.}
and applying the correction.
These values based on the results in
\cite{1998ApJ...500..525S}. Finally, the light curves were corrected
for the cosmological expansion by dividing the time scales by $1+z$ and
applying the K-correction by multiplying the fluxes by 
$(1+z)^{3+\alpha}$ with $\alpha = 1.1$
$(F_{\nu} \propto \nu^{-\alpha})$. The spectral slope chosen here
corresponds to the mode of the $\alpha$ distribution of HBL
in \cite{2004A&A...419...25F}. The LBL have generally steeper spectra
($\alpha_{\rm mode}$ $\sim$ 1.5), so the transformed fluxes of LBL
are likely to be underestimated. We note that this transform does not
correct the light curves for the beaming effect caused by bulk relativistic
motion in the jet.

\subsection{Variability strength}

As a general indicator of how variable our targets are, we use the chi
squared obtained by fitting a constant flux model to the data
(Eq. \ref{chikaava}). This also provides us with the significance of
the variations. Only significantly variable targets are submitted to
further tests.  As discussed above, the error bars include a
noise-term scaled in such a way that the light curves of the control
stars are consistent with a non variable target.

\subsection{Synchrotron peak frequencies}

In order to determine the peak frequency $\nu_{peak}$ of the synchrotron
component, we extracted the archival broadband flux data for all 31 targets
from the Roma-BZCAT \citep[5th Edition,][]{2015Ap&SS.357...75M}
using the SED builder at the ASI Science Data Center
\footnote{http://www.asdc.asi.it/bzcat/}. In cases where there were
few data points in the optical, were augmented the data by our
host galaxy subtracted R-band monitoring data.

We fit simultaneously two log-parabolic spectra
\citep[e.g][]{2004A&A...413..489M}, one for the synchrotron hump and
another for the Inverse Compton (IC) hump, to the broadband spectral
energy distribution (SED) of the targets, including only data with
$\log \nu / ({\rm Hz}) > 8.5$.  Since the archival data are
non-simultaneous and $\nu_{peak}$ is known to change with the activity
state in blazars \citep[e.g.][]{2009ApJ...705.1624A}, we can expect
the fitted $\nu_{peak}$ to depend on the frequencies covered and on
the number of observing epochs.  To roughly estimate how much this
could affect $\nu_{peak}$ we binned the data starting from $\log \nu /
({\rm Hz}) = 8.5$.  The first bin had a width of 0.25 in log space,
followed by bins increasing by a factor of two in width.  We computed
the mean flux in each bin and assigned an error bar equal to the
standard error of the mean in each bin.

The two humps require 8 parameters \citep{2004A&A...413..489M}, two of
which, the pivot energies, were held constant and the remaining 6 were
free. The fit was made by applying a Bayesian approach, sampling the
posteriori distribution of the six free parameters with an Monte Carlo
Markov Chain (MCMC) sampler and with ensemble sampling and 30
walkers. At each iteration $i$, the synchrotron peak frequency
$\nu_{\rm peak}^i$ was computed from current parameters. Then the
distribution of $\nu_{\rm peak}^i$ was used to determine $\nu_{\rm
  peak}$ and its uncertainty by a Gaussian fit made to this
distribution.

The values of $\nu_{peak}$ are tabulated in Table \ref{betatulos} and all
the SEDs together with the best-fitting curve are show in in the
Appendix (Figure \ref{sed1}).  It is obvious that the radio part is
poorly fitted, but this does not not seem to introduce a large shift in
the fitted synchrotron component with respect to the data. However, it
may add a small systematic error not taken into account by our error
estimate. In some cases the IC peak fit can be considered questionable,
but given that, for the most part of the synchrotron spectrum, the
contribution of the IC peak is negligible, no large errors are expected
for $\nu_{peak}$.

\subsection{\label{slopesection}PSD Power-law slope}

Next we proceeded with estimating the slope of the intrinsic power
spectral density (PSD) $P(f)$ of the targets under the assumption that
the PSD has a power-law form, i.e.  $P(f) \propto f^{\beta}$ where
$f$ is the temporal frequency with in units of day$^{-1}$ and $\beta$ is the
power-law slope.  The PSD is equal to the square of the Fourier
transform of the underlying time series. In practice we can only
produce an estimate $p(f)$ of $P(f)$ by computing the discrete Fourier
transform (DFT) of the observed time series. Inferring $P(f)$ from
$p(f)$ is notoriously difficult due to the instrumental noise
and the sampling effects \citep[see e.g.][]{2002MNRAS.332..231U}.  The
observed Fourier transform is a convolution of the true underlying
Fourier transform and the window function $W(x)$. The latter can be a
very complicated function of $f$ resulting in a distorted PSD $p(f)$.
Furthermore, due to the limited length of the times series and
discrete sampling, the PSD can be estimated only within a limited
window between $f_{min}$ to $f_{max}$. If the true PSD contains
significant power outside this window, limited data length and
sampling cause power outside the window to leak into the window,
further distorting the $p(f)$.  Especially in the case of a power-law
PSD, power from frequencies below $f_{min}$, where the PSD is
strongest, leaks into the frequency window (the so called "red noise
leak").

Many different approaches have been developed over the years to
overcome the problems associated with time series dominated by
power-law noise
\citep[e.g.][]{2010MNRAS.404..931E,2014MNRAS.445..437M,2016MNRAS.461.3145V}.
The most recent methods use a "forward casting" approach: starting
from a model $P(f)$, a large number of time series are generated with
the same sampling and noise properties as in the observed data. The
simulated sets are then used to derive an estimate of the statistical
properties of $P(f)$, and the observed data are tested against these
distributions. By varying the model parameters, the best-fitting
parameters can then be found by a suitable statistic.  The distortions
of $P(f)$ are imprinted into the probability distributions and thus
naturally taken into account.

The sampling patterns of our light curves are highly irregular and
contain large gaps due to the target being close to the Sun. Thus we
decided to reject any method relying on binning or interpolating in
the time domain. We performed $P(f)$ estimation using the multiple fractions
variance function (MFVF) presented in \cite{2011A&A...531A.123K}.

The method \citep{2011A&A...531A.123K} studies the variance of the
time series as function of time window.  The algorithm works as
follows: first, it computes the variance $\sigma_0^2$ of the whole
time series and the corresponding "frequency" $1/\Delta t_0$, where
$\Delta t_0$ is the length of the data train. Next, it divides the
times series into two "fragments" in the middle and compute the
variances $\sigma_1^2$ and $\sigma_2^2$ of the two subsets together
with their corresponding "frequencies". This process of subsequent
halving is repeated until there are less than 10 data points in a
fragment. This process results in a set of variances $\sigma_i$ over a
number of frequencies $f_i = 1/\Delta t_i$ which can be analyzed with
the same tools as the Fourier spectra.

\begin{figure*}
\centering
\includegraphics[width=9.5cm]{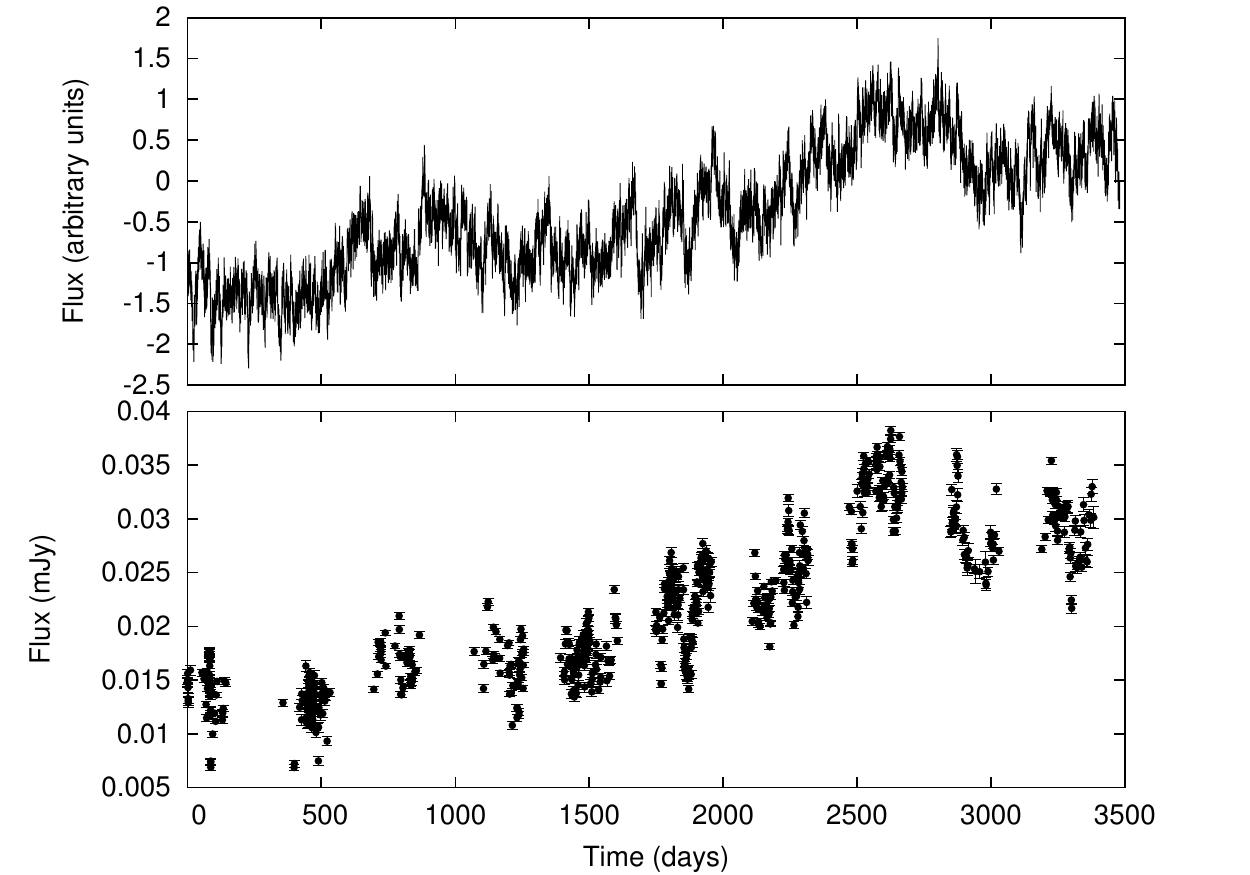}
\hspace*{-1.5cm}
\includegraphics[width=9.8cm]{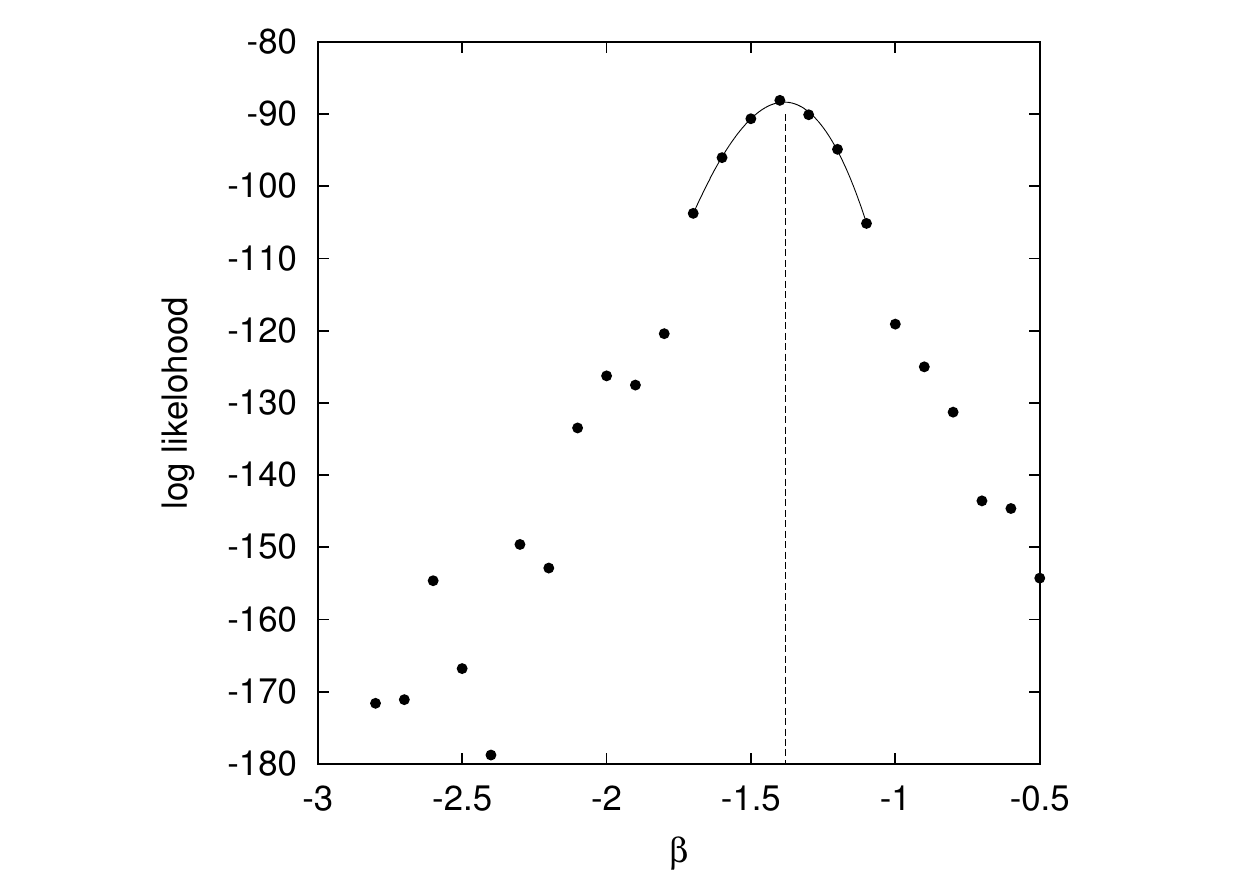}
\caption{\label{likekuva} 
Illustration of different phases of the analysis. {\em upper left}: An evenly-sampled
light curve generated with $\beta = -1.4$. This curve us cut from a longer set with
a length $\sim$ 100 longer than shown here. {\em lower left}: The simulated curve has
been resampled to the observing epochs of Mkn~421 and instrumental noise has been added.
{\em right}: Likelihood curve of Mkn~421 with the MFVF.
The maximum of the polynomial fit (solid line) corresponds to $\beta = -1.38$.
}
\end{figure*}

Our procedure to estimate the PSD slope $\beta$ is thus the following:

\noindent
1. Let $\beta$ vary from -2.8 to -1.0 with step 0.1. At each $\beta$ repeat
steps 2--8:

\noindent
2. Generate 5000 evenly sampled light curves with a length of $\sim$
100 times longer than the observed curve and a sampling of 10 samples
per day by inverse Fourier transform from the assumed model PSD
  \begin{equation}
    \label{psdeq}
 P(f) \propto f^{\beta}
\end{equation}
(Fig. \ref{likekuva}, upper left).  The dense sampling ensures that
  the high frequencies of the power-law noise are properly presented
  in the data and that the long simulation length incorporates the red
  noise leak into the simulation. In our case the number of data
  points was $2^{22} = 4~194~304$.  The time series are generated
  using the prescription of \cite{1995A&A...300..707T}. Note that our
  model includes no flattening of the spectrum at low frequencies and
  the probability density function (PDF) of the time series is assumed
  to be Gaussian. Furthermore, our model does not implicitly include a
  white noise component. These points are discussed in more detail
  below.

\noindent
3. Resample the simulated light curves into the observing epochs (Fig. \ref{likekuva},
lower left).

\noindent
4. Scale the light curve to have the same variance as the observed
data and add to each point a Gaussian random number with $\sigma$
equal to the observational error of that point to simulate
observational errors. The observed variance cannot be directly used to
scale the simulated curve, because the former contains instrumental
noise, which increases the variance. We use the normalized excess
variance \citep[NXV;][]{1997ApJ...476...70N} to estimate the intrinsic
variance $\sigma_I^2$, via the equation
\begin{equation}
\sigma_I^2 = \frac{1}{N} \sum_{k=1}^N \left[
(x(k) - \overline{x})^2 - \sigma_k^2
\right]\ ,
\end{equation}
where $\overline{x}$ is the average of the data series and $\sigma_k$ is the error
of the $k$th data point. We then scale the simulated and resampled  curve to have
a variance equal to $\sigma_I^2$ and add Gaussian random noise to each data point.

\noindent
5. Compute the MFVF of the simulated curves.

\noindent
6. Bin the MFVF data into frequency bins $f_i$ with roughly
a factor two increase in frequency per bin.

\noindent
7. At each frequency bin $f_i$, estimate the probability density
function (PDF) $p(f_i)$ of the MFVF variance from the 5000
  simulated values using Gaussian kernel density estimation.

\noindent
8. Compute the log likelihood of $\beta$ from
\begin{equation}
\ln{p} = \sum_{i=1}^{N_f} \ln{p(f_i)}\ ,
\end{equation}
where $p(f_i)$ is the value of the PDF at $f_i$ and the summation is
over all $N_f$ frequency bins
\citep{2005A&A...431..391V,2010MNRAS.402..307V}. The MFVF
  transform uses variances in time windows of various lengths, so each
  point in the MFVF ``spectrum'' is distributed like chi squared
  $\chi^2_{n-1}$, where $n$ is the number of points in each time
  window. However, due to the possible effects of uneven sampling and
  power-law nature of PSD, we don't use an analytical formula for
  $p(f_i)$.  As explained in step 7, $p(f_i)$ was derived from the
  simulated spectra using a Gaussian kernel smoothing of the simulated
  points. The resulting $p(f_i)$s do visually correspond to a chi
  squared distribution with then appropriate degree of freedom, giving
  us further confidence that the simulations are producing correct
  results.  

\noindent
9. After scanning through the whole range in $\beta$, find the $\beta$ corresponding
to the maximum likelihood. The maximum was found by fitting a 3rd degree polynomial
to the 7 points straddling the highest likelihood found, and by finding the maximum 
of this polynomial. Figure
\ref{likekuva} (right) shows a typical example of the likelihood curve and the fit.

We tested through Monte Carlo simulations the capability of the MFVF
in recovering the correct power-law slope $\beta$. We generated 200
light curves with $\beta_in$ between -1.0 and -2.3 and ran the MFVF
analysis on each of them. For the temporal sampling and instrumental
noise we used the light curve of 3C~66A with 644 data points.

\begin{figure}
\centering
\includegraphics[width=9cm]{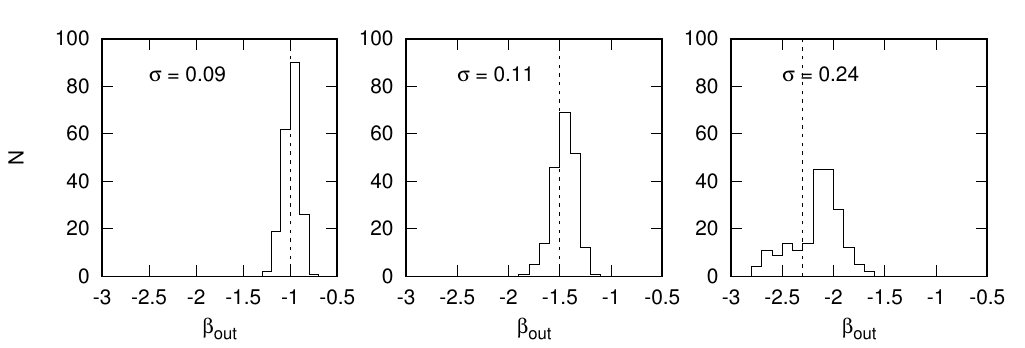}
\caption{\label{betatest} 
Distributions of power-low slopes $\beta_{out}$ for three different input slopes $\beta_{in}$,
-1.0 ({\em left}), -1,5 ({\em middle}) and -2.3 ({\em right})
using the MFVF function. The rms scatter of the distributions are also indicated.
}
\end{figure}

The results are summarized in Fig. \ref{betatest}.
Two things are readily apparent from this figure:
a) the capability to recover the correct power-law slope gets increasingly worse when the input slope becomes steeper, and b) there is a small bias to underestimate
the slope, which is statistically significant in some cases, but nevertheless at least
a factor of $\sim$ 2 smaller than the internal scatter.

We note that MFVF method is applied here in its simplest form,
i.e. the results are computed directly from the observed points
without binning or interpolating the data or applying any filtering
technique. The performance of MFVF could probably be improved for
steep power-law spectra with suitable filtering, but this is out of
the scope of this paper. In any case, all derived PSD slopes are
$>-1.9$, indicating that the most troublesome $\beta$ range is mostly
avoided in our study. We also note that although our PSD model
(Eq. \ref{psdeq}) does not specify a white noise component, it is
taken into account in step 4, where we add Gaussian noise to the
simulated data points. When the simulated light curves are then
transformed by the MFVM, this white noise gets imprinted into the
probability density distribution at each frequency.

Errors on the derived $\beta$ values were estimated by Monte Carlo simulations of
artificial light curves, generated in the same way as in points 2--4 above. We generated
100 such curves, computed their PSDs and MFVF data, ran the likelihood analysis
for each of the 100 curves and recorded the rms scatter of the obtained $\beta$ values.

\subsection{Search for periodicities}

The difficulty of reliably identifying a periodic signal in a red
noise background has been discussed in detail e.g. by
\cite{2005A&A...431..391V}. We estimated the PSD by computing the
periodogram, i.e. the amplitude of the discrete Fourier transform of
the light curve in the case of uneven sampling. Before computing the
periodogram, the data were binned into bins of 3.0 days in order to
avoid dependencies between different frequencies.  As in
\cite{2005A&A...431..391V} we denote the true periodogram at
frequencies $f_j = 1/\Delta t{_j}$ with $\mathcal{P}(f_j)$, the
observed periodogram with $I(f_j)$ and the true probability density
function (PDF) of $\mathcal{P}(f_j)$ with $p(f_j)$.

We created 35~000 simulated light curves per target, again with
similar mean, variance and sampling as in the observed data and with
the power-law slope derived in the previous step (Sect.
\ref{slopesection}). We then computed the periodogram $I(f_j)$ for
each simulation and an estimate of the PDF $p(f_j)$, denoted
here $\hat{p}(f_j)$, from the ensemble of 35~000 points at each frequency
$f_j$ via Gaussian kernel estimation. The high number of simulations
was needed to sample the high end $p(f_j) > 0.99$ properly.  The
probability that the power $x$ at single frequency $f_j$ exceeds the
observed value $I(f_j)$ was computed from
\begin{equation}
P = Pr \left\{ x > I(f_j) \right\} = \int_{I(f_j)}^{\infty} \hat{p}(f_j)\,d x\ .
\end{equation}
Since the possible periodic signal lies on top of a power-law
background, it does not necessarily appear as the highest peak in the
PSD.  For this reason we chose the frequency with the highest
significance (lowest $P = P_{\rm min}$) as a candidate for a periodic
signal and computed the probability $P_N$ of finding such a peak in
the absence of a periodic signal when $N$ frequencies are examined
from
\begin{equation}
P_N = 1 - (1 - P_{\rm min})^N\ .
\end{equation}
Finally, we set $P_N < 5$\% as a limit for significant detection. In
a sample of 31 targets we would then expect $\sim$ 2 targets to show
significant periodicity by chance only.

\section{Results\label{resu}}

Here we list shortly the main results and discuss them further in the
next section.  Table \ref{esimtaulu} gives a sample of the photometric
tables, available for all 31 targets through
Vizier\footnote{http://vizier.u-strasbg.fr/viz-bin/VizieR}.  A
conversion from magnitudes to fluxes can be made through
Eqs. \ref{fkaava}, \ref{palautus1}, and \ref{palautus2}.  Note that the
presented magnitudes have not been corrected for the galactic
extinction or the host galaxy component.

\begin{table}
\centering
\caption{\label{esimtaulu}Sample of the light curve data available
  electronically at Vizier. The target is 3C~66A. Only the first 10
  lines of the table are shown.}
\begin{tabular}{cccc}
Target & JD & R-mag & err\\
\hline
3C 66A & 2452528.40571 & 14.311 & 0.015\\
3C 66A & 2452529.43809 & 14.392 & 0.015\\
3C 66A & 2452550.38235 & 14.872 & 0.017\\
3C 66A & 2452556.31446 & 14.919 & 0.018\\
3C 66A & 2452567.38406 & 14.881 & 0.019\\
3C 66A & 2452577.38589 & 14.821 & 0.017\\
3C 66A & 2452590.41486 & 14.711 & 0.017\\
3C 66A & 2452613.45107 & 15.008 & 0.019\\
3C 66A & 2452613.51023 & 14.998 & 0.018\\
3C 66A & 2452615.45196 & 15.087 & 0.017\\
... & ... & ... & ...\\
\hline
\end{tabular}
\end{table}

Figures \ref{ekavalo}-\ref{vikavalo}, available only electronically,
show on the left the light curves after subtracting the host galaxy
and correcting for galactic extinction. The next panel shows the MVFV
spectrum and the rightmost panel the periodogram. Figures
\ref{sed1}-\ref{vikased} show the SEDs and their corresponding fits.

Table \ref{betatulos} summarizes the main results of our analysis.  We
show the reduced $\chi^2$ obtained by fitting a constant flux model to the
target light curve (Col. 2) , the synchrotron peak frequency
$\nu_{peak}$ from our fits (Col. 3).  The BL Lac subclass division
Col. (4) (LBL/IBL/HBL) in Table \ref{bllist} is based on the value in
Col. (3).  The PSD slope $\beta$ is shown (Col. 5) and the period with
the highest significance (Col. 6) with its probability $P_N$ (Col. 7).

\begin{figure}
\centering \includegraphics[width=9cm]{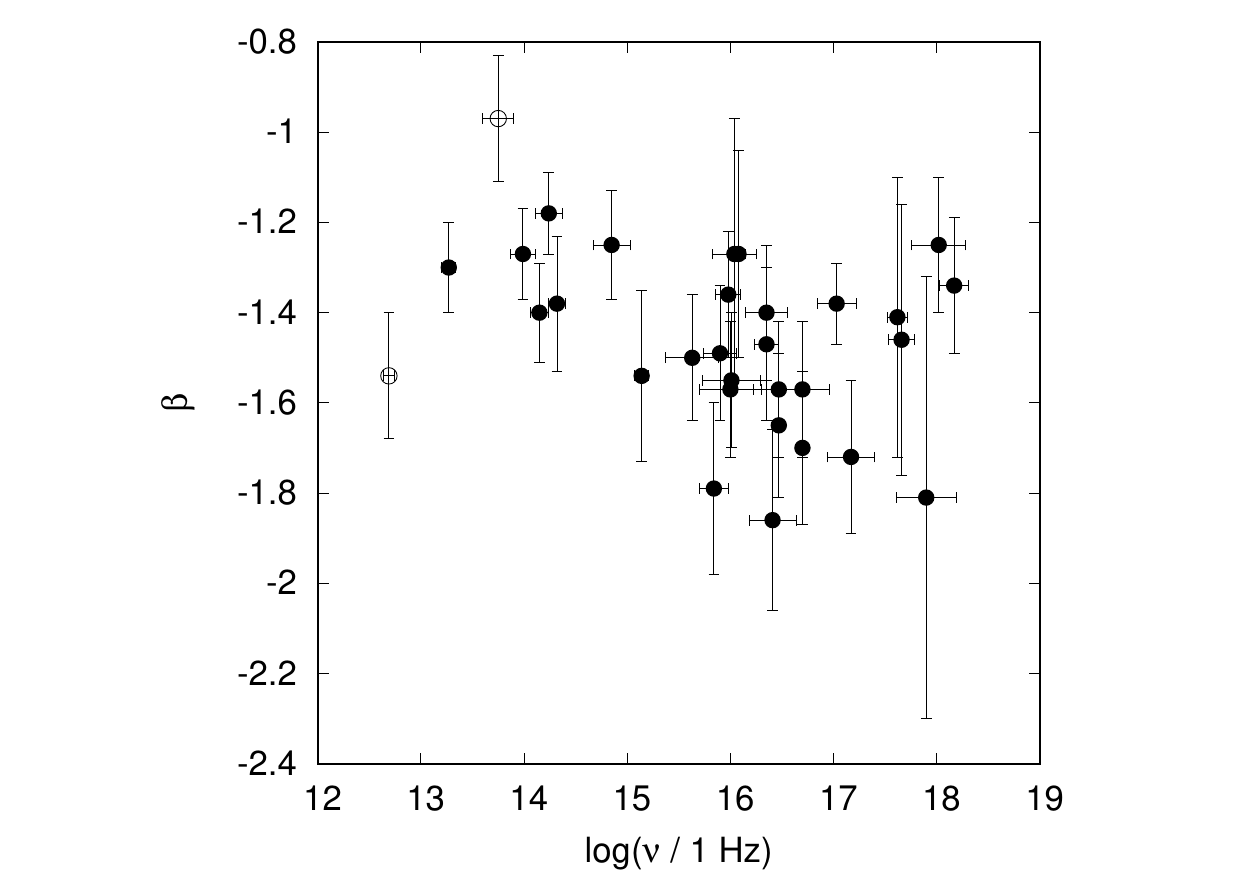}
\caption{\label{nuubeta} 
Best-fitting PSD slope against synchrotron peak frequency.
Filled symbols are BL Lacs, open symbols FSRQs.
}
\end{figure}

Using the chi squared test, we find that the null hypothesis that the
target flux does not vary with time can be rejected for all of our
targets with $p < 0.0001$. As discussed above, the control stars are
by design non-variable by the same test.  The 30 targets we analyzed
therefore exhibit significant variability, so we apply our variability
analysis to all of them, except to 1ES~1544+820, which has
significantly lower number of data points compared to the other
targets.

In Fig. \ref{nuubeta} we plot the power-law slope $\beta$
vs. $\nu_{peak}$. A weak correlation seems to be present, so we tested
the significance by a chi squared test with the null hypothesis that
the $\beta$ values are drawn from a distribution $\beta =
\beta_0$. PKS~1510-089 was excluded from this analysis since its light
is dominated by a single huge flare and our assumption of a powerlaw
PSD with Gaussian PDF is clearly not valid.  Fitting a constant
$\beta$ to the data we obtain $\beta_{avg} = -1.42$, which we use as a
surrogate for the population $\beta_0$. Applying the chi squared test
yields $\chi^2_{red} = 1.36$ with a probability of $p = 0.098$ that the
null hypothesis can be rejected, assuming our ``model'' of constant
$\beta$ is true. Thus we do not find any significant deviation from a
single PSD slope for our sample.

Our periodicity search finds a significant PSD peak in one target,
Mkn~421 with a rest frame period of 477 days. Finding one periodicity
in 31 targets is consistent with the expected false alarm rate. Our
result is thus consistent with no significant periodicities in any of
our targets, but see discussion below on Mkn~421.

\begin{table*}
\caption{\label{betatulos} Main results of the analysis. The columns are: (1) Target name,
(2) Reduced chi squared from the variability analysis, (3) Synchrotron peak frequency,
(4) Classification, (5) Power Spectral Density slope, (6) the most significant rest frame period
(days) and (7) the probability in percent of finding such period in case of pure power-law noise input spectrum. }
\begin{tabular}{lrcccrc}
(1)    &   (2)    &     (3)           & (4)  &  (5) & (6) & (7)\\
Target & $\chi^2$ & $\log[{\nu_{peak}({\rm Hz})]}$ & Class & $\beta$ & $f_P(d)$ & $P_N$\\
\hline
1ES 0033+595      &    5.9 &   18.17 $\pm$ 0.14  & HBL      &   -1.34 $\pm$ 0.15 &    7&  48.3\\
1ES 0120+340      &    2.4 &   17.66 $\pm$ 0.13  & HBL      &   -1.46 $\pm$ 0.30 &  61 &  44.0\\
RGB 0136+391      &   60.5 &   16.00 $\pm$ 0.30  & HBL      &   -1.57 $\pm$ 0.15 &  10 &   6.8\\
RGB 0214+517      &    2.4 &   16.08 $\pm$ 0.07  & HBL      &   -1.27 $\pm$ 0.23 & 130 &  99.1\\
3C 66A            &  899.0 &   14.15 $\pm$ 0.09  & IBL      &   -1.40 $\pm$ 0.11 &  15 &  77.5\\
1ES 0647+250      &   80.8 &   16.41 $\pm$ 0.23  & HBL      &   -1.86 $\pm$ 0.20 &  13 &  88.6\\
1ES 0806+524      &  185.2 &   15.84 $\pm$ 0.14  & HBL      &   -1.79 $\pm$ 0.19 &  38 &  54.0\\
OJ 287            & 1393.6 &   13.27 $\pm$ 0.07  & LBL      &   -1.30 $\pm$ 0.10 &  40 &  18.5\\
1ES 1011+496      &  155.0 &   15.63 $\pm$ 0.26  & HBL      &   -1.50 $\pm$ 0.14 &  16 &  30.6\\
1ES 1028+511      &   33.1 &   16.70 $\pm$ 0.26  & HBL      &   -1.57 $\pm$ 0.15 &   6 &  96.4\\
Mkn 421           &  355.2 &   17.03 $\pm$ 0.19  & HBL      &   -1.38 $\pm$ 0.09 & 477 &   0.1\\
RGB 1117+202      &   21.0 &   15.98 $\pm$ 0.12  & HBL      &   -1.36 $\pm$ 0.14 &  40 &  84.8\\
Mkn 180           &   48.2 &   16.47 $\pm$ 0.25  & HBL      &   -1.57 $\pm$ 0.15 &  29 &  99.9\\
RGB 1136+676      &    3.1 &   17.90 $\pm$ 0.29  & HBL      &   -1.81 $\pm$ 0.49 &  40 &  51.7\\
ON 325            &  137.2 &   14.85 $\pm$ 0.18  & IBL/HBL  &   -1.25 $\pm$ 0.12 &  28 &  99.5\\
1ES 1218+304      &   93.7 &   17.17 $\pm$ 0.23  & HBL      &   -1.72 $\pm$ 0.17 &  33 &  34.3\\
RGB 1417+257      &    2.7 &   17.62 $\pm$ 0.10  & HBL      &   -1.41 $\pm$ 0.31 &  20 &  88.8\\
1ES 1426+428      &    4.5 &   18.02 $\pm$ 0.26  & HBL      &   -1.25 $\pm$ 0.15 &  12 &  90.0\\
1ES 1544+820      &     -  &   16.04 $\pm$ 0.21  & HBL      &     -    &   -    &   -  \\
Mkn 501           &    8.3 &   16.47 $\pm$ 0.06  & HBL      &   -1.65 $\pm$ 0.16 &  15 &  97.9\\
OT 546            &   10.3 &   16.35 $\pm$ 0.20  & HBL      &   -1.40 $\pm$ 0.15 &  18 &  64.8\\
1ES 1959+650      &  249.1 &   16.70 $\pm$ 0.04  & HBL      &   -1.70 $\pm$ 0.17 & 1050 & 84.7\\
BL Lac            &  849.6 &   13.99 $\pm$ 0.12  & LBL      &   -1.27 $\pm$ 0.10 &  197 & 94.5\\
1ES 2344+514      &    5.7 &   16.35 $\pm$ 0.12  & HBL      &   -1.47 $\pm$ 0.17 &  14 &  10.0\\
\hline
S5 0716+714       & 2761.4 &   14.24 $\pm$ 0.13  & IBL      &   -1.18 $\pm$ 0.09 & 163 &  20.8\\
ON 231            &  354.6 &   14.32 $\pm$ 0.08  & IBL      &   -1.38 $\pm$ 0.15 &  18 &  98.8\\
3C 279            & 1597.3 &   12.69 $\pm$ 0.05  & FSRQ     &   -1.54 $\pm$ 0.14 & 202 &  80.0\\
PG 1424+240       &  162.5 &   15.14 $\pm$ 0.07  & IBL/HBL  &   -1.54 $\pm$ 0.19 &  17 &  88.8\\
PKS 1510-089      &  248.3 &   13.75 $\pm$ 0.15  & FSRQ     &   -0.97 $\pm$ 0.14 & 155 &  40.1\\
PG 1553+113       &  323.6 &   15.90 $\pm$ 0.16  & HBL      &   -1.49 $\pm$ 0.15 & 174 &  31.5\\
PKS 2155-304      & 1980.8 &   16.01 $\pm$ 0.28  & HBL      &   -1.55 $\pm$ 0.15 &  99 &  81.1\\
\hline
\end{tabular}
\end{table*}

\section{Discussion}

\subsection{PSD slopes}

\begin{figure}[htb]

\centering
\includegraphics[width=9cm]{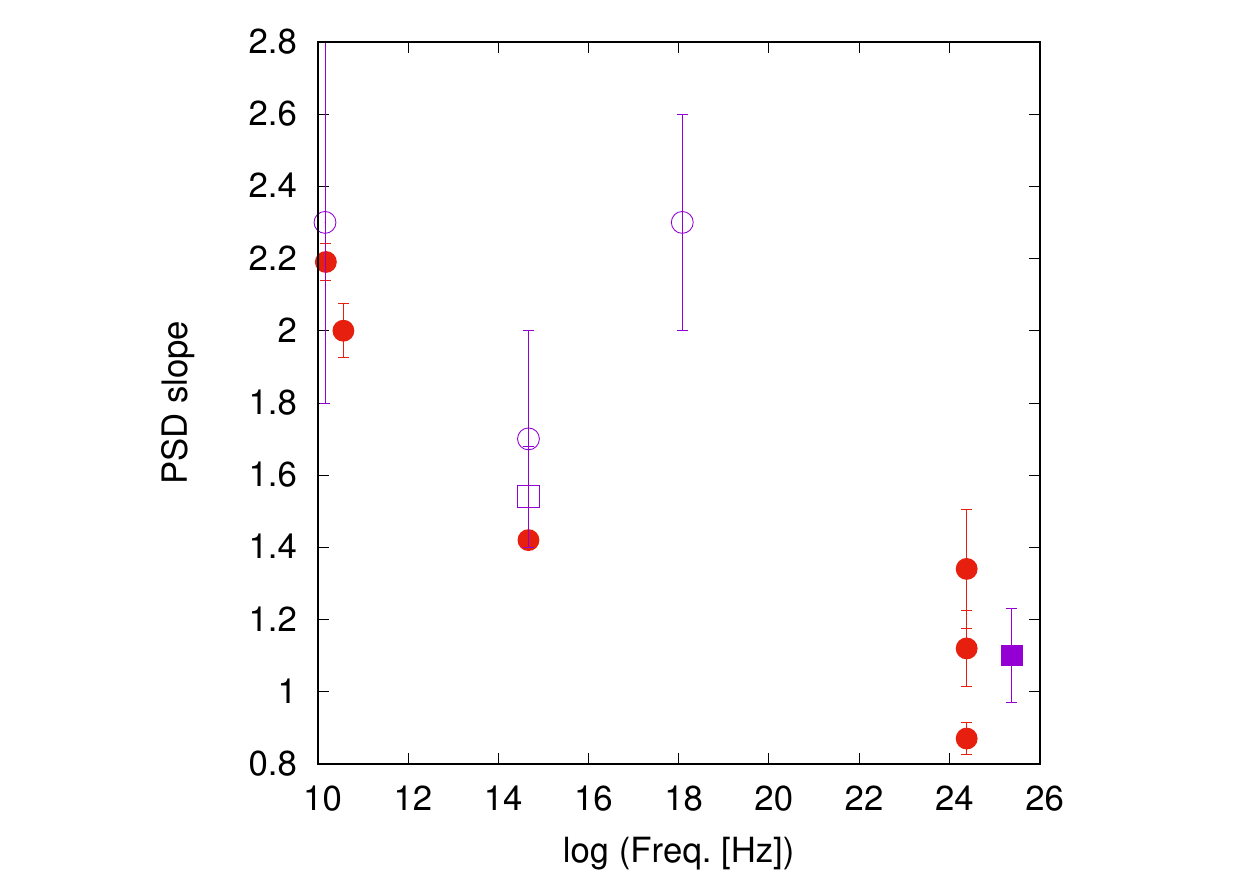}
\caption{\label{slopevsband} PSD slope vs. observing frequency for
  samples of BL Lacs and a few individual targets. 
Filled red symbols : Samples in Table \ref{betavertailu};
Open purple square: FSRQ 3C~279 (this work); 
Open purple: 3C~279 \citep[]{2008ApJ...689...79C};
Filled purple square: PKS~2155-304 \citep{2016arXiv161003311H}.
}
\end{figure}

In Table \ref{betavertailu} and Fig. \ref{slopevsband} we compare our
average PSD slope $-1.42 \pm 0.12$ to the values in the literature
obtained recently at radio, optical and gamma-rays for samples
comparable in size to ours and by using similar
methodology. Particularly, these studies considered carefully the
distortions caused by uneven sampling and noise. 

\begin{table}
\caption{\label{betavertailu} PSD slopes of BL Lacs in recent studies}
\centering
\begin{tabular}{lllll}
Band & $\log({\rm Freq.})$ & $\beta \pm {\rm err}$  & N  & ref.\\
\hline
R-band    & 14.67 & $1.46 \pm 0.18$ & 26 & 1\\ 
15 GHz    & 10.18 & $2.19 \pm 0.17$ & 11 & 2\\
Fermi LAT & 24.38 & $1.34 \pm 0.55$ & 11 & 2\\
37 GHz    & 10.57 & $2.00 \pm 0.27$ & 13 & 3\\
Fermi LAT & 24.38 & $1.12 \pm 0.36$ & 12 & 3\\
Fermi LAT & 24.38 & $0.87 \pm 0.16$ &  5 & 4\\
\hline
\end{tabular}
\tablebib{
(1) This work;
(2) \cite{2014MNRAS.445..428M};
(3) \cite{2015MNRAS.452.1280R};
(4) \cite{2014ApJ...786..143S}.
}
\end{table}

The number of results is still small and one cannot draw firm
conclusions, but a trend of decreasing beta with increasing frequency
is apparent, or at the very least the radio slopes appear
significantly different from the rest. The same trend seems to
continue in the FSRQ 3C~279, although the results are more noisy for a
single target compared to samples of targets. This result, if
confirmed, would mean that in the regions emitting at radio
frequencies variability preferentially occurs over long time scales,
rather than over short time scales, i.e. the radio emitting regions
have a longer ``memory'' of their previous state than the optical and
gamma-ray emitting regions. This could simply be due to larger
emitting volume in the radio than in the gamma-rays and in the
optical.

We emphasize, however, that although we fit a power-law PSD to the
data, this does not imply that the underlying process is indeed a
power-law process or even that the light curves at different wavebands
result from the same process. For instance, the 22 and 37 GHz light
curves of blazars can apparently be decomposed into a series of
exponential flares \citep[e.g][]{1999ApJS..120...95V} with some
regular features, like the decay times scale always being 1.3 times
the rising time scale. A visual inspection of our optical light curves
gives an impression that such a decomposition might be possible in
some cases (e.g. 1ES 1959+650), but in most cases not. The apparent
regularity in the radio suggests that the steeper PSD slope could
simply be a result of fitting a noise process to a light curve that is
not a result of such a process.

\begin{figure}
\centering
\hspace*{-5mm}
\includegraphics[width=9.5cm]{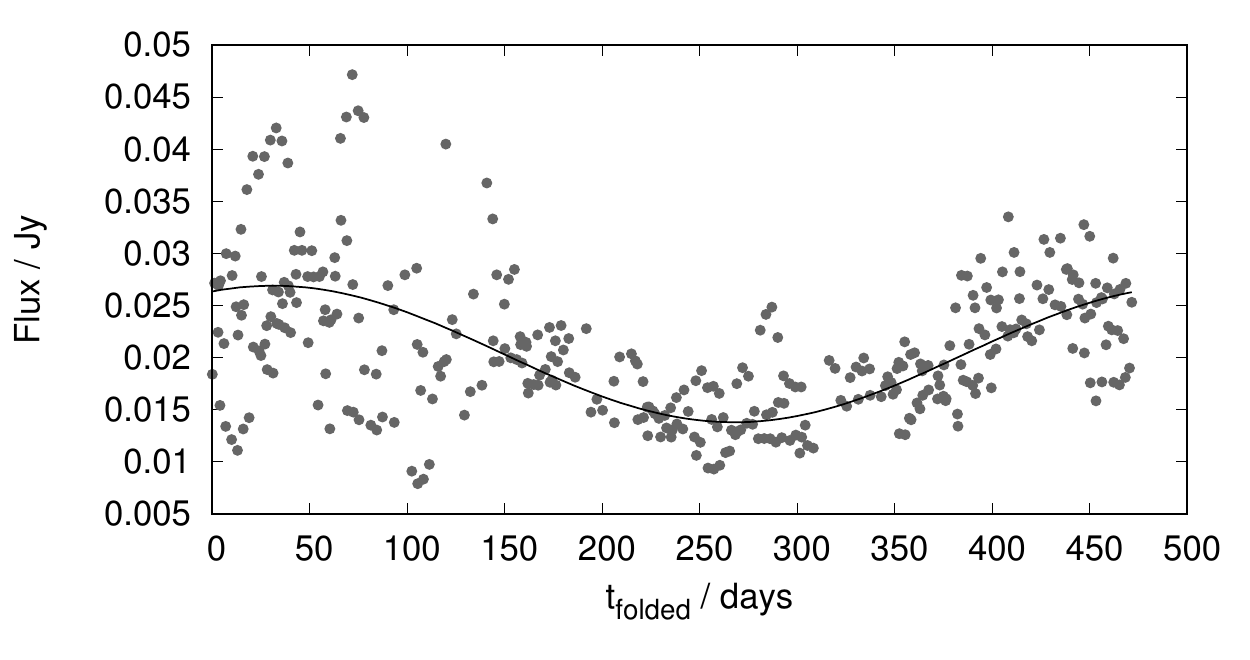}
 \caption{\label{mkn421period} The folded light curve of Mkn~421 using
   a rest-frame period of 477 days (grey symbols). The black line
   shows the harmonic function corresponding to the phase and
   amplitude in the periodogram.  }
\end{figure}

In the optical, we do not find a significant correlation between the
synchrotron peak frequency and the PSD slope. Such a correlation would
not be unexpected. In Low-Peaked BL Lacs (LBL) the synchrotron peak is
below the observation frequency and thus we are observing electrons in
the high-energy tail of the energy distribution.  In contrast, the
optical emission from High-Peaked BL Lacs (HBL) is originating from
electrons radiating below the peak energy. The cooling times scales of
the high- and low energy electrons are very different, and thus one
might expect differences in the variability characteristics of LBL and
HBL. However, our sample is not complete and contains only few
LBL. Therefore, such a correlation could be biased, if even found.

\subsection{Periodicities}

Looking at the sample as a whole, we did not find any evidence of
periodic variations over the 10-year time span studied here. Our
analysis takes into account the power-law background and is expected
to be less sensitive to spurious peaks in the periodogram than many
previous studies. We found significant periodicity ($P_N < 0.05$) in
one target only, a 477 days rest frame period in Mkn~421. Finding one
significant period among 31 targets is just what we would expect from
chance alone.

However, the PSD peak in Mkn~421 is very strong, which warrants
further consideration. Figure \ref{mkn421period} shows the folded
light curve of Mkn~421 over 7 cycles. The variations seem consistently
sinusoidal, except during the first $\sim$ 150 days of the
cycle. There is thus an intriguing possibility of periodic variations
in this source with extra activity triggered at certain phase of the
cycle. Considering that we have tested the periodicity at $>$100
frequencies over 31 targets, a chance coincidence cannot be completely
ruled out, however. \cite{2016PASP..128g4101L} found periods of
280-310 days in radio, x-ray and gamma-ray light curves of Mkn~421 in
data spanning 6 to 10 years. The period found here is longer, but
since we do not interpolate the spectrum in order to retain
independence between the frequencies, our frequency resolution is
quite low. Indeed, the adjacent frequencies in our PSD correspond to
periods of 830 and 330 days. The difference cannot be completely
explained by resolution only, but the actual difference cannot be well
determined considering the differences in the analyses.

\begin{figure}
\centering
\hspace*{-5mm}
\includegraphics[width=9.5cm]{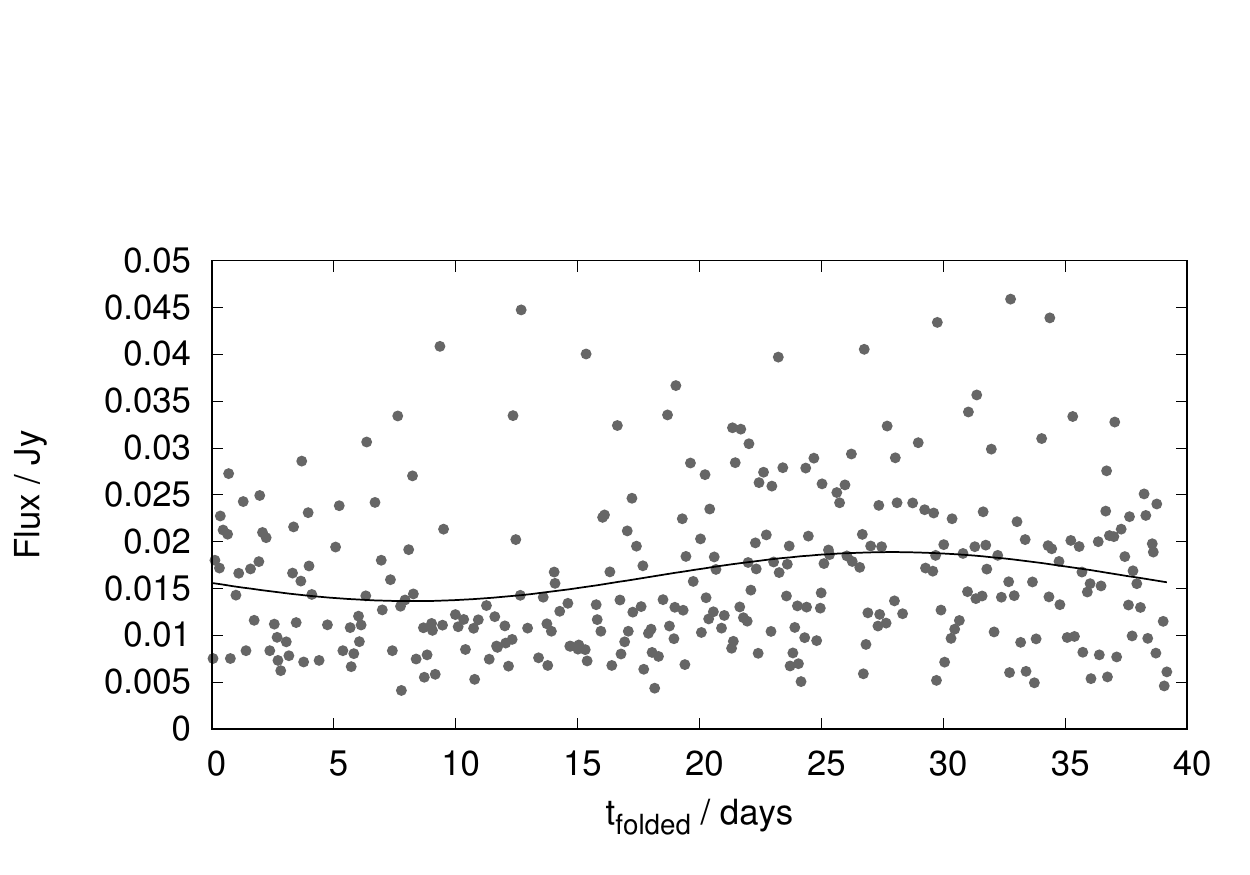}
 \caption{\label{oj287period} The folded light curve of OJ~287 using
   a rest-frame period of 477 days (grey symbols). The black line
   shows the harmonic function corresponding to the phase and
   amplitude in the periodogram.  }
\end{figure}

We finally comment on some periods claimed to have been found in our
sample objects.  Periodicities or quasiperiodicities have been claimed
for S5~0716+714, but with much shorter time scales, e.g. 25-73 minutes
\citep{2009ApJ...690..216G} or 15 min \citep{2010ApJ...719L.153R}.
Our sampling is too sparse to investigate this.  In
\cite{2013MNRAS.434.3122P} a 50 day period was found in OJ~287 from a
2-year densely sampled data set taken in 2004-06. This study used
partly the same data as here, but the number of common data points is
very small. Out of the 3991 data points in \cite{2013MNRAS.434.3122P},
only about 140 originate from the data presented here. Our data also
cover a time span longer than \cite{2013MNRAS.434.3122P} by a factor
of $\sim$ 5 and hence the two data sets are largely independent.  We
find a very similar period of 52 days in the observed frame, but the
significance is below our detection threshold. The folded light curve in
Fig. \ref{oj287period} gives an indication why the results could
differ between different authors. There seems to be a stable periodic
signal at low flux levels intermixed by a few high flux points at
random phases. These high points are due to the double flares that
occur in this source at $\sim$ 12 year intervals
\citep[e.g]{2006ApJ...646...36V}, which are very likely to be caused
by a process completely unrelated to the periodic
variations. \citep{2016ApJ...819L..37V}. The inclusion or exclusion of
these flares will certainly affect the Fourier analysis pushing the
result beyond the significance level in our case.

Another significant periodicity reported recently is the $798\pm30$
day period found in Fermi gamma-ray data of PG~1553+113 and further
supported by optical data with a period of $754\pm20$ days
\citep{2015ApJ...813L..41A}. The significance of the optical PSD peak
was reported to be $<$5\%. We do not detect this period in our data,
although our data set is almost entirely included in
\cite{2015ApJ...813L..41A}, forming about half of their sample.
However, our frequency resolution is again very poor at periods of
$\sim$ 800 days due to the relatively short time span with respect to
this period. The fact that similar time scale was found in gamma-rays
strengthens the case of significant periodic variations in
PG~1553+113.

There are many other reports of detected periodicities, which we did
not find here, like the optical 65-day period of 3C 66A
\citep{1999ApJ...521..561L}, or the $\sim$ 1 year optical periods
tentatively, but not conclusively detected in OJ~287, PKS~1510-089 and
PKS~2155-304 by \cite{2016AJ....151...54S}.  Our analysis, and these
examples, illustrate the difficulty of finding a weak periodic signal
in a red noise background using data suffering from unknown systematic
errors and sparse and uneven sampling \citep{2016MNRAS.461.3145V}.  If
persistent or recurrent periods were actually found, the time scale
could shed some light onto their origin. The optical emission in BL
Lacs is dominated by synchrotron emission from the jet, so periodic
variations could be a result of a precession of the jet.  This model
has been used to explain e.g. the trajectories of the parsec-scale
Very Long Baseline (VLBI) components in BL Lac
\citep{2013MNRAS.428..280C}, although in this particular case no
optical variations have been detected in the derived precession period
of 12.1 years. Other possibilities exist, like helical structure of
the jet, which can form as a consequence of current-driven
instabilities in the jet \citep{2004ApJ...617..123N}.

Regular changes in the accretion mechanism that feeds the jet could
also lead to periodic or quasiperiodic changes in the jet.
\cite{2013MNRAS.434.3122P} attributed the 50 day period found in
OJ~287 to a spiral density wave in the accretion disk and performed
particle N-body simulations to show that a spiral wave configuration
results in a periodic influx of material with approximately the same
period as observed in OJ~287. Spiral density waves seem to be
naturally generated around single \citep{2001ApJ...551..874L} and
binary \citep{2010ApJ...708..485H} black hole systems. In the former
study, high-pressure vortices formed in the accretion disk, providing
a natural source for increased accretion. Also in the latter study
the spiral waves exhibited oscillations, which could lead to episodes
of periodic variations in the matter influx.

\subsection{Caveats and future work}

Our results and conclusions have to be taken with some caveats:
firstly, we assume a Gaussian probability density function (PDF) when
doing the simulations and secondly, our simulated spectra have no low-
or high frequency cutoffs.  The assumption of Gaussian PDF is clearly
not always valid and a log normal distribution would in many cases
better represent the PDF, especially in targets whose light curve is
dominated by a single or a few strong flares with apparently
exponential growth and decay.  Recently, a method has been presented
to generate non-Gaussian light curves \citep{2013MNRAS.433..907E}, but
the application of this procedure was left to future studies.  At
least \cite{2015ApJ...798...27I} found their simulated x-ray PSDs of
Mrk~421 to depend only little on the assumption of the PDF.

Since a power-law spectrum extending all the way to $f = 0$ would
imply infinite power output, the PSDs of BL Lacs are expected to level
off at some long time scale $t_b$, which would reveal itself as a
break in the spectrum at $f_b = 1 / t_b$. Many of our PSDs show this
kind of break, but these could be a result of finite data length
rather than true breaks. Our simulations do not include this break as
an input, but this is necessarily not a problem since the break time
scale could be far longer than the 10-year interval studied here. In
order to look for true breaks in blazar PSDs, long-term historical
data needs to be collected and analyzed, like in
e.g. \cite{2007A&A...467..465C}.

\section{Conclusions}

We have presented R-band monitoring data of 31 blazars (29 BL Lacs and
2 FSRQs) observed over a time span of 10 years. In addition to
presenting the light curves and describing in detail the data
reduction process, we have analyzed the light curves by determining
their PSD slopes and by searching for periodic variations in the light
curves. These analyses were augmented by substantial number of
simulations to take into account the effects of uneven sampling and
detector noise and to calibrate the false alarm rate of the
periodicity search. Our results can be summarized as follows:
\begin{itemize}
\item [1)] We present for the first time all our R-band monitoring
  data in tabular form, altogether 11820 photometric data points.
\item[2)] By applying a chi squared test we find that all 32 targets
  show significant variability with respect to the comparison stars.
\item[3)] The average PSD slope of the 29 targets in our sample
  -1.42$\pm$0.12 (1$\sigma$ standard deviation). The PSD slope is not significantly
  ($p = 9.8$\%), correlated with the synchrotron peak frequency.
\item[4)] Our average PSD slope $\overline{\beta}$ is consistent with
  values found in the literature.  Comparing our average PSD slope to
  those in the literature, we find that in the radio the slope tends to be
  steeper than in the optical and gamma-ray bands.
\item[5)] The periodicity search returned one target, Mkn~421, with a
  significant ($p<5$\%) peak in the periodogram. This is consistent
  with the expected false alarm rate, but the signal is Mrk~421 is
  very strong ($p=0.1$\%) and warrants further study with longer time
  span.  The 52 day period found in OJ~287 is now confirmed by us, but
  we note that high flare states caused by an unrelated emission
  process may complicate the analysis.

\end{itemize}

\begin{acknowledgements} 

  This paper is dedicated to the memory of our colleague and dear
  friend Leo Takalo 1952-2018, who played a crucial role in starting
  this monitoring effort and contributed significantly to the data
  acquisition.  We would like to thank the Instituto de Astrofísica de
  Canarias for the excellent working conditions at the Observatorio
  del Roque de los Muchachos in La Palma. Part of this work is based
  on archival data, software, online services provided by the ASI
  Space Science Data Center (ASI-SSDC).

\end{acknowledgements} 

\bibliographystyle{aa}

\bibliography{monitoringv7.bib}

\begin{thebibliography}{127}
\expandafter\ifx\csname natexlab\endcsname\relax\def\natexlab#1{#1}\fi

\bibitem[{{Abdo} {et~al.}(2010{\natexlab{a}}){Abdo}, {Ackermann}, {Agudo},
  {Ajello}, {Aller}, {Aller}, {Angelakis}, {Arkharov}, {Axelsson}, {Bach}, \&
  et~al.}]{2010ApJ...716...30A}
{Abdo}, A.~A., {Ackermann}, M., {Agudo}, I., {et~al.} 2010{\natexlab{a}}, \apj,
  716, 30

\bibitem[{{Abdo} {et~al.}(2011{\natexlab{a}}){Abdo}, {Ackermann}, {Ajello},
  {Allafort}, {Baldini}, {Ballet}, {Barbiellini}, {Baring}, {Bastieri},
  {Bechtol}, \& et~al.}]{2011ApJ...727..129A}
{Abdo}, A.~A., {Ackermann}, M., {Ajello}, M., {et~al.} 2011{\natexlab{a}},
  \apj, 727, 129

\bibitem[{{Abdo} {et~al.}(2010{\natexlab{b}}){Abdo}, {Ackermann}, {Ajello},
  {Axelsson}, {Baldini}, {Ballet}, {Barbiellini}, {Bastieri}, {Baughman},
  {Bechtol}, \& et~al.}]{2010Natur.463..919A}
{Abdo}, A.~A., {Ackermann}, M., {Ajello}, M., {et~al.} 2010{\natexlab{b}},
  \nat, 463, 919

\bibitem[{{Abdo} {et~al.}(2011{\natexlab{b}}){Abdo}, {Ackermann}, {Ajello},
  {Baldini}, {Ballet}, {Barbiellini}, {Bastieri}, {Bechtol}, {Bellazzini},
  {Berenji}, \& et~al.}]{2011ApJ...736..131A}
{Abdo}, A.~A., {Ackermann}, M., {Ajello}, M., {et~al.} 2011{\natexlab{b}},
  \apj, 736, 131

\bibitem[{{Acciari} {et~al.}(2011{\natexlab{a}}){Acciari}, {Aliu}, {Arlen},
  {Aune}, {Beilicke}, {Benbow}, {Boltuch}, {Bradbury}, {Buckley}, {Bugaev},
  {Byrum}, {Cannon}, {Cesarini}, {Ciupik}, {Cui}, {Dickherber}, {Duke},
  {Falcone}, {Finley}, {Finnegan}, {Fortson}, {Furniss}, {Galante}, {Gall},
  {Gillanders}, {Godambe}, {Grube}, {Guenette}, {Gyuk}, {Hanna}, {Holder},
  {Hui}, {Humensky}, {Imran}, {Kaaret}, {Karlsson}, {Kertzman}, {Kieda},
  {Konopelko}, {Krawczynski}, {Krennrich}, {Lang}, {Maier}, {McArthur},
  {McCutcheon}, {Moriarty}, {Ong}, {Otte}, {Ouellette}, {Pandel}, {Perkins},
  {Pichel}, {Pohl}, {Quinn}, {Ragan}, {Reyes}, {Reynolds}, {Roache}, {Rose},
  {Rovero}, {Schroedter}, {Sembroski}, {Senturk}, {Steele}, {Swordy},
  {Theiling}, {Thibadeau}, {Varlotta}, {Vassiliev}, {Vincent}, {Wagner},
  {Wakely}, {Ward}, {Weekes}, {Weinstein}, {Weisgarber}, {Williams}, {Wissel},
  {Wood}, {Zitzer}, {Garson}, {Lee}, {Sadun}, {Carini}, {Barnaby}, {Cook},
  {Maune}, {Pease}, {Smith}, {Walters}, {Berdyugin}, {Lindfors}, {Nilsson},
  {Pasanen}, {Sainio}, {Sillanp\"a\"a}, {Takalo}, {Villforth}, {Montaruli},
  {Baker}, {Lahteenmaki}, {Tornikoski}, {Hovatta}, {Nieppola}, {Aller}, \&
  {Aller}}]{2011ApJ...738...25A}
{Acciari}, V.~A., {Aliu}, E., {Arlen}, T., {et~al.} 2011{\natexlab{a}}, \apj,
  738, 25

\bibitem[{{Acciari} {et~al.}(2009{\natexlab{a}}){Acciari}, {Aliu}, {Aune},
  {Beilicke}, {Benbow}, {B{\"o}ttcher}, {Boltuch}, {Buckley}, {Bradbury},
  {Bugaev}, {Byrum}, {Cannon}, {Cesarini}, {Ciupik}, {Cogan}, {Cui},
  {Dickherber}, {Duke}, {Falcone}, {Finley}, {Fortin}, {Fortson}, {Furniss},
  {Galante}, {Gall}, {Gibbs}, {Gillanders}, {Grube}, {Guenette}, {Gyuk},
  {Hanna}, {Holder}, {Hui}, {Humensky}, {Kaaret}, {Karlsson}, {Kertzman},
  {Kieda}, {Konopelko}, {Krawczynski}, {Krennrich}, {Lang}, {Le Bohec},
  {Maier}, {McArthur}, {McCann}, {McCutcheon}, {Millis}, {Moriarty}, {Ong},
  {Otte}, {Pandel}, {Perkins}, {Pichel}, {Pohl}, {Quinn}, {Ragan}, {Reyes},
  {Reynolds}, {Roache}, {Rose}, {Sembroski}, {Smith}, {Steele}, {Theiling},
  {Thibadeau}, {Varlotta}, {Vassiliev}, {Vincent}, {Wakely}, {Ward}, {Weekes},
  {Weinstein}, {Weisgarber}, {Williams}, {Wissel}, {Wood}, {Pian},
  {Vercellone}, {Donnarumma}, {D'Ammando}, {Bulgarelli}, {Chen}, {Giuliani},
  {Longo}, {Pacciani}, {Pucella}, {Vittorini}, {Tavani}, {Argan},
  {Barbiellini}, {Caraveo}, {Cattaneo}, {Cocco}, {Costa}, {Del Monte}, {De
  Paris}, {Di Cocco}, {Evangelista}, {Feroci}, {Fiorini}, {Froysland},
  {Frutti}, {Fuschino}, {Galli}, {Gianotti}, {Labanti}, {Lapshov},
  {Lazzarotto}, {Lipari}, {Marisaldi}, {Mastropietro}, {Mereghetti}, {Morelli},
  {Morselli}, {Pellizzoni}, {Perotti}, {Piano}, {Picozza}, {Pilia},
  {Porrovecchio}, {Prest}, {Rapisarda}, {Rappoldi}, {Rubini}, {Sabatini},
  {Soffitta}, {Trifoglio}, {Trois}, {Vallazza}, {Zambra}, {Zanello}, {Pittori},
  {Santolamazza}, {Verrecchia}, {Giommi}, {Colafrancesco}, {Salotti},
  {Villata}, {Raiteri}, {Aller}, {Aller}, {Arkharov}, {Efimova}, {Larionov},
  {Leto}, {Ligustri}, {Lindfors}, {Pasanen}, {Kurtanidze}, {Tetradze},
  {Lahteenmaki}, {Kotiranta}, {Cucchiara}, {Romano}, {Nesci}, {Pursimo},
  {Heidt}, {Benitez}, {Hiriart}, {Nilsson}, {Berdyugin}, {Mujica}, {Dultzin},
  {Lopez}, {Mommert}, {Sorcia}, \& {de la Calle Perez}}]{2009ApJ...707..612A}
{Acciari}, V.~A., {Aliu}, E., {Aune}, T., {et~al.} 2009{\natexlab{a}}, \apj,
  707, 612

\bibitem[{{Acciari} {et~al.}(2009{\natexlab{b}}){Acciari}, {Aliu}, {Aune},
  {Beilicke}, {Benbow}, {B{\"o}ttcher}, {Bradbury}, {Buckley}, {Bugaev},
  {Butt}, \& et~al.}]{2009ApJ...703..169A}
{Acciari}, V.~A., {Aliu}, E., {Aune}, T., {et~al.} 2009{\natexlab{b}}, \apj,
  703, 169

\bibitem[{{Acciari} {et~al.}(2011{\natexlab{b}}){Acciari}, {Arlen}, {Aune},
  {Beilicke}, {Benbow}, {B{\"o}ttcher}, {Boltuch}, {Bradbury}, {Buckley},
  {Bugaev}, \& et~al.}]{2011ApJ...729....2A}
{Acciari}, V.~A., {Arlen}, T., {Aune}, T., {et~al.} 2011{\natexlab{b}}, \apj,
  729, 2

\bibitem[{{Ackermann} {et~al.}(2015){Ackermann}, {Ajello}, {Albert}, {Atwood},
  {Baldini}, {Ballet}, {Barbiellini}, {Bastieri}, {Becerra Gonzalez},
  {Bellazzini}, {Bissaldi}, {Blandford}, {Bloom}, {Bonino}, {Bottacini},
  {Bregeon}, {Bruel}, {Buehler}, {Buson}, {Caliandro}, {Cameron}, {Caputo},
  {Caragiulo}, {Caraveo}, {Cavazzuti}, {Cecchi}, {Chekhtman}, {Chiang},
  {Chiaro}, {Ciprini}, {Cohen-Tanugi}, {Conrad}, {Cutini}, {D'Ammando}, {de
  Angelis}, {de Palma}, {Desiante}, {Di Venere}, {Dominguez}, {Drell},
  {Favuzzi}, {Fegan}, {Ferrara}, {Focke}, {Fuhrmann}, {Fukazawa}, {Fusco},
  {Gargano}, {Gasparrini}, {Giglietto}, {Giommi}, {Giordano}, {Giroletti},
  {Godfrey}, {Green}, {Grenier}, {Grove}, {Guiriec}, {Harding}, {Hays},
  {Hewitt}, {Hill}, {Horan}, {Jogler}, {J{\'o}hannesson}, {Johnson}, {Kamae},
  {Kuss}, {Larsson}, {Latronico}, {Li}, {Li}, {Longo}, {Loparco}, {Lott},
  {Lovellette}, {Lubrano}, {Magill}, {Maldera}, {Manfreda}, {Max-Moerbeck},
  {Mayer}, {Mazziotta}, {McEnery}, {Michelson}, {Mizuno}, {Monzani},
  {Morselli}, {Moskalenko}, {Murgia}, {Nuss}, {Ohno}, {Ohsugi}, {Ojha},
  {Omodei}, {Orlando}, {Ormes}, {Paneque}, {Pearson}, {Perkins}, {Perri},
  {Pesce-Rollins}, {Petrosian}, {Piron}, {Pivato}, {Porter}, {Rain{\`o}},
  {Rando}, {Razzano}, {Readhead}, {Reimer}, {Reimer}, {Schulz}, {Sgr{\`o}},
  {Siskind}, {Spada}, {Spandre}, {Spinelli}, {Suson}, {Takahashi}, {Thayer},
  {Thompson}, {Tibaldo}, {Torres}, {Tosti}, {Troja}, {Uchiyama}, {Vianello},
  {Wood}, {Wood}, {Zimmer}, {Berdyugin}, {Corbet}, {Hovatta}, {Lindfors},
  {Nilsson}, {Reinthal}, {Sillanp{\"a}{\"a}}, {Stamerra}, {Takalo}, \&
  {Valtonen}}]{2015ApJ...813L..41A}
{Ackermann}, M., {Ajello}, M., {Albert}, A., {et~al.} 2015, \apjl, 813, L41

\bibitem[{{Ackermann} {et~al.}(2014){Ackermann}, {Ajello}, {Allafort},
  {Antolini}, {Barbiellini}, {Bastieri}, {Bellazzini}, {Bissaldi}, {Bonamente},
  {Bregeon}, \& et~al.}]{2014ApJ...786..157A}
{Ackermann}, M., {Ajello}, M., {Allafort}, A., {et~al.} 2014, \apj, 786, 157

\bibitem[{{Ahnen} {et~al.}(2016{\natexlab{a}}){Ahnen}, {Ansoldi}, {Antonelli},
  {Antoranz}, {Babic}, {Banerjee}, {Bangale}, {Barres de Almeida}, {Barrio},
  {Becerra Gonz{\'a}lez}, {Bednarek}, {Bernardini}, {Biasuzzi}, {Biland},
  {Blanch}, {Bonnefoy}, {Bonnoli}, {Borracci}, {Bretz}, {Buson}, {Carosi},
  {Chatterjee}, {Clavero}, {Colin}, {Colombo}, {Contreras}, {Cortina},
  {Covino}, {Da Vela}, {Dazzi}, {De Angelis}, {De Lotto}, {de O{\~n}a
  Wilhelmi}, {Di Pierro}, {Dom{\'{\i}}nguez}, {Dominis Prester}, {Dorner},
  {Doro}, {Einecke}, {Eisenacher Glawion}, {Elsaesser}, {Fern{\'a}ndez-Barral},
  {Fidalgo}, {Fonseca}, {Font}, {Frantzen}, {Fruck}, {Galindo}, {Garc{\'{\i}}a
  L{\'o}pez}, {Garczarczyk}, {Garrido Terrats}, {Gaug}, {Giammaria},
  {Godinovi{\'c}}, {Gonz{\'a}lez Mu{\~n}oz}, {Gora}, {Guberman}, {Hadasch},
  {Hahn}, {Hanabata}, {Hayashida}, {Herrera}, {Hose}, {Hrupec}, {Hughes},
  {Idec}, {Kodani}, {Konno}, {Kubo}, {Kushida}, {La Barbera}, {Lelas},
  {Lindfors}, {Lombardi}, {Longo}, {L{\'o}pez}, {L{\'o}pez-Coto}, {Majumdar},
  {Makariev}, {Mallot}, {Maneva}, {Manganaro}, {Mannheim}, {Maraschi},
  {Marcote}, {Mariotti}, {Mart{\'{\i}}nez}, {Mazin}, {Menzel}, {Miranda},
  {Mirzoyan}, {Moralejo}, {Moretti}, {Nakajima}, {Neustroev}, {Niedzwiecki},
  {Nievas Rosillo}, {Nilsson}, {Nishijima}, {Noda}, {Nogu{\'e}s}, {Orito},
  {Overkemping}, {Paiano}, {Palacio}, {Palatiello}, {Paneque}, {Paoletti},
  {Paredes}, {Paredes-Fortuny}, {Pedaletti}, {Perri}, {Persic}, {Poutanen},
  {Prada Moroni}, {Prandini}, {Puljak}, {Rhode}, {Rib{\'o}}, {Rico}, {Rodriguez
  Garcia}, {Saito}, {Satalecka}, {Schultz}, {Schweizer}, {Shore},
  {Sillanp{\"a}{\"a}}, {Sitarek}, {Snidaric}, {Sobczynska}, {Stamerra},
  {Steinbring}, {Strzys}, {Takalo}, {Takami}, {Tavecchio}, {Temnikov},
  {Terzi{\'c}}, {Tescaro}, {Teshima}, {Thaele}, {Torres}, {Toyama}, {Treves},
  {Verguilov}, {Vovk}, {Ward}, {Will}, {Wu}, \& {Zanin}}]{2016A&A...593A..91A}
{Ahnen}, M.~L., {Ansoldi}, S., {Antonelli}, L.~A., {et~al.} 2016{\natexlab{a}},
  \aap, 593, A91

\bibitem[{{Ahnen} {et~al.}(2016{\natexlab{b}}){Ahnen}, {Ansoldi}, {Antonelli},
  {Antoranz}, {Babic}, {Banerjee}, {Bangale}, {de Almeida}, {Barrio},
  {Gonz{\'a}lez}, {Bednarek}, {Bernardini}, {Biasuzzi}, {Biland}, {Blanch},
  {Bonnefoy}, {Bonnoli}, {Borracci}, {Bretz}, {Carmona}, {Carosi},
  {Chatterjee}, {Clavero}, {Colin}, {Colombo}, {Contreras}, {Cortina},
  {Covino}, {Da Vela}, {Dazzi}, {De Angelis}, {De Caneva}, {De Lotto}, {de
  O{\~n}a Wilhelmi}, {Mendez}, {Pierro}, {Prester}, {Dorner}, {Doro},
  {Einecke}, {Elsaesser}, {Fern{\'a}ndez-Barral}, {Fidalgo}, {Fonseca}, {Font},
  {Frantzen}, {Fruck}, {Galindo}, {L{\'o}pez}, {Garczarczyk}, {Terrats},
  {Gaug}, {Giammaria}, {(Eisenacher)}, {Godinovi{\'c}}, {Mu{\~n}oz},
  {Guberman}, {Hanabata}, {Hayashida}, {Herrera}, {Hose}, {Hrupec}, {Hughes},
  {Idec}, {Kodani}, {Konno}, {Kubo}, {Kushida}, {Barbera}, {Lelas}, {Lindfors},
  {Lombardi}, {Longo}, {L{\'o}pez}, {L{\'o}pez-Coto}, {L{\'o}pez-Oramas},
  {Lorenz}, {Majumdar}, {Makariev}, {Mallot}, {Maneva}, {Manganaro},
  {Mannheim}, {Maraschi}, {Marcote}, {Mariotti}, {Mart{\'{\i}}nez}, {Mazin},
  {Menzel}, {Miranda}, {Mirzoyan}, {Moralejo}, {Nakajima}, {Neustroev},
  {Niedzwiecki}, {Rosillo}, {Nilsson}, {Nishijima}, {Noda}, {Orito},
  {Overkemping}, {Paiano}, {Palacio}, {Palatiello}, {Paneque}, {Paoletti},
  {Paredes}, {Paredes-Fortuny}, {Persic}, {Poutanen}, {Moroni}, {Prandini},
  {Puljak}, {Reinthal}, {Rhode}, {Rib{\'o}}, {Rico}, {Garcia}, {R{\"u}gamer},
  {Saito}, {Satalecka}, {Scapin}, {Schultz}, {Schweizer}, {Shore},
  {Sillanp{\"a}{\"a}}, {Sitarek}, {Snidaric}, {Sobczynska}, {Stamerra},
  {Steinbring}, {Strzys}, {Takalo}, {Takami}, {Tavecchio}, {Temnikov},
  {Terzi{\'c}}, {Tescaro}, {Teshima}, {Thaele}, {Torres}, {Toyama}, {Treves},
  {Verguilov}, {Vovk}, {Ward}, {Will}, {Wu}, {Zanin}, {Lucarelli}, {Pittori},
  {Vercellone}, {Berdyugin}, {Carini}, {L{\"a}hteenm{\"a}ki}, {Pasanen},
  {Pease}, {Sainio}, {Tornikoski}, \& {Walters}}]{2016MNRAS.459.2286A}
{Ahnen}, M.~L., {Ansoldi}, S., {Antonelli}, L.~A., {et~al.} 2016{\natexlab{b}},
  \mnras, 459, 2286

\bibitem[{{Albert} {et~al.}(2006{\natexlab{a}}){Albert}, {Aliu}, {Anderhub},
  {Antoranz}, {Armada}, {Asensio}, {Baixeras}, {Barrio}, {Bartko}, {Bastieri},
  {Becker}, {Bednarek}, {Berger}, {Bigongiari}, {Biland}, {Bisesi}, {Bock},
  {Bordas}, {Bosch-Ramon}, {Bretz}, {Britvitch}, {Camara}, {Carmona},
  {Chilingarian}, {Ciprini}, {Coarasa}, {Commichau}, {Contreras}, {Cortina},
  {Curtef}, {Danielyan}, {Dazzi}, {De Angelis}, {de los Reyes}, {De Lotto},
  {Domingo-Santamar{\'{\i}}a}, {Dorner}, {Doro}, {Errando}, {Fagiolini},
  {Ferenc}, {Fern{\'a}ndez}, {Firpo}, {Flix}, {Fonseca}, {Font}, {Fuchs},
  {Galante}, {Garczarczyk}, {Gaug}, {Giller}, {Goebel}, {Hakobyan},
  {Hayashida}, {Hengstebeck}, {H{\"o}hne}, {Hose}, {Hsu}, {Jacon}, {Kalekin},
  {Kosyra}, {Kranich}, {Laatiaoui}, {Laille}, {Lenisa}, {Liebing}, {Lindfors},
  {Lombardi}, {Longo}, {L{\'o}pez}, {L{\'o}pez}, {Lorenz}, {Majumdar},
  {Maneva}, {Mannheim}, {Mansutti}, {Mariotti}, {Mart{\'{\i}}nez}, {Mazin},
  {Merck}, {Meucci}, {Meyer}, {Miranda}, {Mirzoyan}, {Mizobuchi}, {Moralejo},
  {Nilsson}, {Ninkovic}, {O{\~n}a-Wilhelmi}, {Ordu{\~n}a}, {Otte}, {Oya},
  {Paneque}, {Paoletti}, {Paredes}, {Pasanen}, {Pascoli}, {Pauss}, {Pegna},
  {Persic}, {Peruzzo}, {Piccioli}, {Poller}, {Prandini}, {Raymers}, {Rhode},
  {Rib{\'o}}, {Rico}, {Riegel}, {Rissi}, {Robert}, {R{\"u}gamer}, {Saggion},
  {S{\'a}nchez}, {Sartori}, {Scalzotto}, {Scapin}, {Schmitt}, {Schweizer},
  {Shayduk}, {Shinozaki}, {Shore}, {Sidro}, {Sillanp{\"a}{\"a}}, {Sobczynska},
  {Stamerra}, {Stark}, {Takalo}, {Temnikov}, {Tescaro}, {Teshima}, {Tonello},
  {Torres}, {Torres}, {Turini}, {Vankov}, {Vitale}, {Wagner}, {Wibig},
  {Wittek}, {Zanin}, \& {Zapatero}}]{2006ApJ...648L.105A}
{Albert}, J., {Aliu}, E., {Anderhub}, H., {et~al.} 2006{\natexlab{a}}, \apjl,
  648, L105

\bibitem[{{Albert} {et~al.}(2006{\natexlab{b}}){Albert}, {Aliu}, {Anderhub},
  {Antoranz}, {Armada}, {Asensio}, {Baixeras}, {Barrio}, {Bartko}, {Bastieri},
  {Bednarek}, {Berger}, {Bigongiari}, {Biland}, {Bisesi}, {Bock}, {Bretz},
  {Britvitch}, {Camara}, {Chilingarian}, {Ciprini}, {Coarasa}, {Commichau},
  {Contreras}, {Cortina}, {Danielyan}, {Dazzi}, {De Angelis}, {de los Reyes},
  {De Lotto}, {Domingo-Santamar{\'{\i}}a}, {Dorner}, {Doro}, {Errando},
  {Ferenc}, {Fern{\'a}ndez}, {Firpo}, {Flix}, {Fonseca}, {Font}, {Galante},
  {Garczarczyk}, {Gaug}, {Gebauer}, {Giannitrapani}, {Giller}, {Goebel},
  {Hakobyan}, {Hayashida}, {Hengstebeck}, {H{\"o}hne}, {Hose}, {Jacon},
  {Kalekin}, {Kranich}, {Laille}, {Lenisa}, {Liebing}, {Lindfors}, {Longo},
  {L{\'o}pez}, {L{\'o}pez}, {Lorenz}, {Lucarelli}, {Majumdar}, {Maneva},
  {Mannheim}, {Mariotti}, {Mart{\'{\i}}nez}, {Mase}, {Mazin}, {Merck}, {Merck},
  {Meucci}, {Meyer}, {Miranda}, {Mirzoyan}, {Mizobuchi}, {Moralejo}, {Nilsson},
  {O{\~n}a-Wilhelmi}, {Ordu{\~n}a}, {Otte}, {Oya}, {Paneque}, {Paoletti},
  {Pasanen}, {Pascoli}, {Pauss}, {Pavel}, {Pegna}, {Peruzzo}, {Piccioli},
  {Pin}, {Prandini}, {Rico}, {Rhode}, {Riegel}, {Rissi}, {Robert}, {Rossato},
  {R{\"u}gamer}, {Saggion}, {S{\'a}nchez}, {Sartori}, {Scalzotto}, {Schmitt},
  {Schweizer}, {Shayduk}, {Shinozaki}, {Sidro}, {Sillanp{\"a}{\"a}},
  {Sobczynska}, {Stamerra}, {Stark}, {Takalo}, {Temnikov}, {Tescaro},
  {Teshima}, {Tonello}, {Torres}, {Torres}, {Turini}, {Vankov}, {Vitale},
  {Wagner}, {Wibig}, {Wittek}, \& {Zapatero}}]{2006ApJ...639..761A}
{Albert}, J., {Aliu}, E., {Anderhub}, H., {et~al.} 2006{\natexlab{b}}, \apj,
  639, 761

\bibitem[{{Albert} {et~al.}(2007{\natexlab{a}}){Albert}, {Aliu}, {Anderhub},
  {Antoranz}, {Armada}, {Baixeras}, {Barrio}, {Bartko}, {Bastieri}, {Becker},
  {Bednarek}, {Berger}, {Bigongiari}, {Biland}, {Bock}, {Bordas},
  {Bosch-Ramon}, {Bretz}, {Britvitch}, {Camara}, {Carmona}, {Chilingarian},
  {Ciprini}, {Coarasa}, {Commichau}, {Contreras}, {Cortina}, {Curtef},
  {Danielyan}, {Dazzi}, {De Angelis}, {de los Reyes}, {De Lotto},
  {Domingo-Santamar{\'{\i}}a}, {Dorner}, {Doro}, {Errando}, {Fagiolini},
  {Ferenc}, {Fern{\'a}ndez}, {Firpo}, {Flix}, {Fonseca}, {Font}, {Fuchs},
  {Galante}, {Garczarczyk}, {Gaug}, {Giller}, {Goebel}, {Hakobyan},
  {Hayashida}, {Hengstebeck}, {H{\"o}hne}, {Hose}, {Hsu}, {Jacon}, {Jogler},
  {Kalekin}, {Kosyra}, {Kranich}, {Kritzer}, {Laille}, {Liebing}, {Lindfors},
  {Lombardi}, {Longo}, {L{\'o}pez}, {L{\'o}pez}, {Lorenz}, {Majumdar},
  {Maneva}, {Mannheim}, {Mansutti}, {Mariotti}, {Mart{\'{\i}}nez}, {Mazin},
  {Merck}, {Meucci}, {Meyer}, {Miranda}, {Mirzoyan}, {Mizobuchi}, {Moralejo},
  {Nilsson}, {Ninkovic}, {O{\~n}a-Wilhelmi}, {Otte}, {Oya}, {Paneque},
  {Paoletti}, {Paredes}, {Pasanen}, {Pascoli}, {Pauss}, {Pegna}, {Persic},
  {Peruzzo}, {Piccioli}, {Poller}, {Puchades}, {Prandini}, {Raymers}, {Rhode},
  {Rib{\'o}}, {Rico}, {Rissi}, {Robert}, {R{\"u}gamer}, {Saggion},
  {S{\'a}nchez}, {Sartori}, {Scalzotto}, {Scapin}, {Schmitt}, {Schweizer},
  {Shayduk}, {Shinozaki}, {Sidro}, {Sillanp{\"a}{\"a}}, {Sobczynska},
  {Stamerra}, {Stark}, {Takalo}, {Temnikov}, {Tescaro}, {Teshima}, {Tonello},
  {Torres}, {Turini}, {Vankov}, {Vitale}, {Wagner}, {Wibig}, {Wittek}, {Zanin},
  \& {Zapatero}}]{2007ApJ...654L.119A}
{Albert}, J., {Aliu}, E., {Anderhub}, H., {et~al.} 2007{\natexlab{a}}, \apjl,
  654, L119

\bibitem[{{Albert} {et~al.}(2007{\natexlab{b}}){Albert}, {Aliu}, {Anderhub},
  {Antoranz}, {Armada}, {Baixeras}, {Barrio}, {Bartko}, {Bastieri}, {Becker},
  {Bednarek}, {Berger}, {Bigongiari}, {Biland}, {Bock}, {Bordas},
  {Bosch-Ramon}, {Bretz}, {Britvitch}, {Camara}, {Carmona}, {Chilingarian},
  {Coarasa}, {Commichau}, {Contreras}, {Cortina}, {Costado}, {Curtef},
  {Danielyan}, {Dazzi}, {De Angelis}, {Delgado}, {de los Reyes}, {De Lotto},
  {Domingo-Santamar{\'{\i}}a}, {Dorner}, {Doro}, {Errando}, {Fagiolini},
  {Ferenc}, {Fern{\'a}ndez}, {Firpo}, {Flix}, {Fonseca}, {Font}, {Fuchs},
  {Galante}, {Garc{\'{\i}}a-L{\'o}pez}, {Garczarczyk}, {Gaug}, {Giller},
  {Goebel}, {Hakobyan}, {Hayashida}, {Hengstebeck}, {Herrero}, {H{\"o}hne},
  {Hose}, {Hsu}, {Jacon}, {Jogler}, {Kosyra}, {Kranich}, {Kritzer}, {Laille},
  {Lindfors}, {Lombardi}, {Longo}, {L{\'o}pez}, {L{\'o}pez}, {Lorenz},
  {Majumdar}, {Maneva}, {Mannheim}, {Mansutti}, {Mariotti}, {Mart{\'{\i}}nez},
  {Mazin}, {Merck}, {Meucci}, {Meyer}, {Miranda}, {Mirzoyan}, {Mizobuchi},
  {Moralejo}, {Nieto}, {Nilsson}, {Ninkovic}, {O{\~n}a-Wilhelmi}, {Otte},
  {Oya}, {Paneque}, {Panniello}, {Paoletti}, {Paredes}, {Pasanen}, {Pascoli},
  {Pauss}, {Pegna}, {Perlman}, {Persic}, {Peruzzo}, {Piccioli}, {Prandini},
  {Puchades}, {Raymers}, {Rhode}, {Rib{\'o}}, {Rico}, {Rissi}, {Robert},
  {R{\"u}gamer}, {Saggion}, {Saito}, {S{\'a}nchez}, {Sartori}, {Scalzotto},
  {Scapin}, {Schmitt}, {Schweizer}, {Shayduk}, {Shinozaki}, {Shore}, {Sidro},
  {Sillanp{\"a}{\"a}}, {Sobczynska}, {Stamerra}, {Stark}, {Takalo},
  {Tavecchio}, {Temnikov}, {Tescaro}, {Teshima}, {Torres}, {Turini}, {Vankov},
  {Vitale}, {Wagner}, {Wibig}, {Wittek}, {Zandanel}, {Zanin}, \&
  {Zapatero}}]{2007ApJ...667L..21A}
{Albert}, J., {Aliu}, E., {Anderhub}, H., {et~al.} 2007{\natexlab{b}}, \apjl,
  667, L21

\bibitem[{{Albert} {et~al.}(2007{\natexlab{c}}){Albert}, {Aliu}, {Anderhub},
  {Antoranz}, {Armada}, {Baixeras}, {Barrio}, {Bartko}, {Bastieri}, {Becker},
  {Bednarek}, {Berger}, {Bigongiari}, {Biland}, {Bock}, {Bordas},
  {Bosch-Ramon}, {Bretz}, {Britvitch}, {Camara}, {Carmona}, {Chilingarian},
  {Coarasa}, {Commichau}, {Contreras}, {Cortina}, {Costado}, {Curtef},
  {Danielyan}, {Dazzi}, {De Angelis}, {Delgado}, {de los Reyes}, {De Lotto},
  {Domingo-Santamar{\'{\i}}a}, {Dorner}, {Doro}, {Errando}, {Fagiolini},
  {Ferenc}, {Fern{\'a}ndez}, {Firpo}, {Flix}, {Fonseca}, {Font}, {Fuchs},
  {Galante}, {Garc{\'{\i}}a-L{\'o}pez}, {Garczarczyk}, {Gaug}, {Giller},
  {Goebel}, {Hakobyan}, {Hayashida}, {Hengstebeck}, {Herrero}, {H{\"o}hne},
  {Hose}, {Hrupec}, {Hsu}, {Jacon}, {Jogler}, {Kosyra}, {Kranich}, {Kritzer},
  {Laille}, {Lindfors}, {Lombardi}, {Longo}, {L{\'o}pez}, {L{\'o}pez},
  {Lorenz}, {Majumdar}, {Maneva}, {Mannheim}, {Mansutti}, {Mariotti},
  {Mart{\'{\i}}nez}, {Mazin}, {Merck}, {Meucci}, {Meyer}, {Miranda},
  {Mirzoyan}, {Mizobuchi}, {Moralejo}, {Nieto}, {Nilsson}, {Ninkovic},
  {O{\~n}a-Wilhelmi}, {Otte}, {Oya}, {Paneque}, {Panniello}, {Paoletti},
  {Paredes}, {Pasanen}, {Pascoli}, {Pauss}, {Pegna}, {Persic}, {Peruzzo},
  {Piccioli}, {Prandini}, {Puchades}, {Raymers}, {Rhode}, {Rib{\'o}}, {Rico},
  {Rissi}, {Robert}, {R{\"u}gamer}, {Saggion}, {Saito}, {S{\'a}nchez},
  {Sartori}, {Scalzotto}, {Scapin}, {Schmitt}, {Schweizer}, {Shayduk},
  {Shinozaki}, {Shore}, {Sidro}, {Sillanp{\"a}{\"a}}, {Sobczynska}, {Stamerra},
  {Stark}, {Takalo}, {Tavecchio}, {Temnikov}, {Tescaro}, {Teshima}, {Torres},
  {Turini}, {Vankov}, {Vitale}, {Wagner}, {Wibig}, {Wittek}, {Zandanel},
  {Zanin}, \& {Zapatero}}]{2007ApJ...669..862A}
{Albert}, J., {Aliu}, E., {Anderhub}, H., {et~al.} 2007{\natexlab{c}}, \apj,
  669, 862

\bibitem[{{Albert} {et~al.}(2007{\natexlab{d}}){Albert}, {Aliu}, {Anderhub},
  {Antoranz}, {Armada}, {Baixeras}, {Barrio}, {Bartko}, {Bastieri}, {Becker},
  {Bednarek}, {Berger}, {Bigongiari}, {Biland}, {Bock}, {Bordas},
  {Bosch-Ramon}, {Bretz}, {Britvitch}, {Camara}, {Carmona}, {Chilingarian},
  {Coarasa}, {Commichau}, {Contreras}, {Cortina}, {Costado}, {Curtef},
  {Danielyan}, {Dazzi}, {De Angelis}, {Delgado}, {de los Reyes}, {De Lotto},
  {Domingo-Santamar{\'{\i}}a}, {Dorner}, {Doro}, {Errando}, {Fagiolini},
  {Ferenc}, {Fern{\'a}ndez}, {Firpo}, {Flix}, {Fonseca}, {Font}, {Fuchs},
  {Galante}, {Garc{\'{\i}}a-L{\'o}pez}, {Garczarczyk}, {Gaug}, {Giller},
  {Goebel}, {Hakobyan}, {Hayashida}, {Hengstebeck}, {Herrero}, {H{\"o}hne},
  {Hose}, {Hsu}, {Jacon}, {Jogler}, {Kosyra}, {Kranich}, {Kritzer}, {Laille},
  {Lindfors}, {Lombardi}, {Longo}, {L{\'o}pez}, {L{\'o}pez}, {Lorenz},
  {Majumdar}, {Maneva}, {Mannheim}, {Mansutti}, {Mariotti}, {Mart{\'{\i}}nez},
  {Mazin}, {Merck}, {Meucci}, {Meyer}, {Miranda}, {Mirzoyan}, {Mizobuchi},
  {Moralejo}, {Nilsson}, {Ninkovic}, {O{\~n}a-Wilhelmi}, {Otte}, {Oya},
  {Paneque}, {Panniello}, {Paoletti}, {Paredes}, {Pasanen}, {Pascoli}, {Pauss},
  {Pegna}, {Persic}, {Peruzzo}, {Piccioli}, {Poller}, {Prandini}, {Puchades},
  {Raymers}, {Rhode}, {Rib{\'o}}, {Rico}, {Rissi}, {Robert}, {R{\"u}gamer},
  {Saggion}, {S{\'a}nchez}, {Sartori}, {Scalzotto}, {Scapin}, {Schmitt},
  {Schweizer}, {Shayduk}, {Shinozaki}, {Shore}, {Sidro}, {Sillanp{\"a}{\"a}},
  {Sobczynska}, {Stamerra}, {Stark}, {Takalo}, {Temnikov}, {Tescaro},
  {Teshima}, {Tonello}, {Torres}, {Turini}, {Vankov}, {Vitale}, {Wagner},
  {Wibig}, {Wittek}, {Zandanel}, {Zanin}, \& {Zapatero}}]{2007ApJ...666L..17A}
{Albert}, J., {Aliu}, E., {Anderhub}, H., {et~al.} 2007{\natexlab{d}}, \apjl,
  666, L17

\bibitem[{{Albert} {et~al.}(2007{\natexlab{e}}){Albert}, {Aliu}, {Anderhub},
  {Antoranz}, {Armada}, {Baixeras}, {Barrio}, {Bartko}, {Bastieri}, {Becker},
  {Bednarek}, {Berger}, {Bigongiari}, {Biland}, {Bock}, {Bordas},
  {Bosch-Ramon}, {Bretz}, {Britvitch}, {Camara}, {Carmona}, {Chilingarian},
  {Ciprini}, {Coarasa}, {Commichau}, {Contreras}, {Cortina}, {Costado},
  {Curtef}, {Danielyan}, {Dazzi}, {De Angelis}, {Delgado}, {de los Reyes}, {De
  Lotto}, {Domingo-Santamar{\'{\i}}a}, {Dorner}, {Doro}, {Errando},
  {Fagiolini}, {De Angelis}, {Ferenc}, {Fern{\'a}ndez}, {Firpo}, {Flix},
  {Fonseca}, {Font}, {Fuchs}, {Galante}, {Garc{\'{\i}}a-L{\'o}pez},
  {Garczarczyk}, {Gaug}, {Giller}, {Goebel}, {Hakobyan}, {Hayashida},
  {Hengstebeck}, {Herrero}, {H{\"o}hne}, {Hose}, {Hsu}, {Jacon}, {Jogler},
  {Kalekin}, {Kosyra}, {Kranich}, {Kritzer}, {Laille}, {Liebing}, {Lindfors},
  {Lombardi}, {Longo}, {L{\'o}pez}, {L{\'o}pez}, {Lorenz}, {Majumdar},
  {Maneva}, {Mannheim}, {Mansutti}, {Mariotti}, {Mart{\'{\i}}nez}, {Mazin},
  {Merck}, {Meucci}, {Meyer}, {Miranda}, {Mirzoyan}, {Mizobuchi}, {Moralejo},
  {Nilsson}, {Ninkovic}, {On{\~n}a-Wilhelmi}, {Otte}, {Oya}, {Paneque},
  {Panniello}, {Paoletti}, {Paredes}, {Pasanen}, {Pascoli}, {Pauss}, {Pegna},
  {Persic}, {Peruzzo}, {Piccioli}, {Poller}, {Prandini}, {Puchades}, {Raymers},
  {Rhode}, {Rib{\'o}}, {Rico}, {Rissi}, {Robert}, {R{\"u}gamer}, {Saggion},
  {S{\'a}nchez}, {Sartori}, {Scalzotto}, {Scapin}, {Schmitt}, {Schweizer},
  {Shayduk}, {Shinozaki}, {Shore}, {Sidro}, {Sillanp{\"a}{\"a}}, {Sobczynska},
  {Stamerra}, {Stark}, {Takalo}, {Temnikov}, {Tescaro}, {Teshima}, {Tonello},
  {Torres}, {Turini}, {Vankov}, {Vitale}, {Wagner}, {Wibig}, {Wittek},
  {Zandanel}, {Zanin}, \& {Zapatero}}]{2007ApJ...662..892A}
{Albert}, J., {Aliu}, E., {Anderhub}, H., {et~al.} 2007{\natexlab{e}}, \apj,
  662, 892

\bibitem[{{Albert} {et~al.}(2009){Albert}, {Aliu}, {Anderhub}, {Antoranz},
  {Baixeras}, {Barrio}, {Bartko}, {Bastieri}, {Becker}, {Bednarek},
  {Berdyugin}, {Berger}, {Bigongiari}, {Biland}, {Bock}, {Bordas},
  {Bosch-Ramon}, {Bretz}, {Britvitch}, {Camara}, {Carmona}, {Chilingarian},
  {Commichau}, {Contreras}, {Cortina}, {Costado}, {Curtef}, {Danielyan},
  {Dazzi}, {de Angelis}, {Delgado}, {de Los Reyes}, {de Lotto}, {Dorner},
  {Doro}, {Errando}, {Fagiolini}, {Ferenc}, {Fern{\'a}ndez}, {Firpo},
  {Fonseca}, {Font}, {Fuchs}, {Galante}, {Garc{\'{\i}}a-L{\'o}pez},
  {Garczarczyk}, {Gaug}, {Goebel}, {Hakobyan}, {Hayashida}, {Hengstebeck},
  {Herrero}, {H{\"o}hne}, {Hose}, {Hsu}, {Huber}, {Jacon}, {Jogler}, {Kosyra},
  {Kranich}, {Kritzer}, {Laille}, {Lindfors}, {Lombardi}, {Longo}, {L{\'o}pez},
  {Lorenz}, {Majumdar}, {Maneva}, {Mannheim}, {Mariotti}, {Mart{\'{\i}}nez},
  {Mazin}, {Merck}, {Meucci}, {Meyer}, {Miranda}, {Mirzoyan}, {Mizobuchi},
  {Moralejo}, {Nieto}, {Nilsson}, {Ninkovic}, {O{\~n}a-Wilhelmi}, {Otte},
  {Oya}, {Panniello}, {Paoletti}, {Pasanen}, {Pascoli}, {Pauss}, {Pegna},
  {Persic}, {Peruzzo}, {Piccioli}, {Prandini}, {Puchades}, {Raymers}, {Rico},
  {Rhode}, {Rico}, {Rissi}, {Robert}, {R{\"u}gamer}, {Saggion}, {Saito},
  {S{\'a}nchez}, {Sartori}, {Scalzotto}, {Scapin}, {Schmitt}, {Schweizer},
  {Shayduk}, {Shinozaki}, {Shore}, {Sidro}, {Sillanp{\"a}{\"a}}, {Sobczynska},
  {Spanier}, {Stamerra}, {Stark}, {Takalo}, {Temnikov}, {Tescaro}, {Teshima},
  {Torres}, {Turini}, {Vankov}, {Venturini}, {Vitale}, {Wagner}, {Wibig},
  {Wittek}, {Zandanel}, {Zanin}, \& {Zapatero}}]{2009A&A...493..467A}
{Albert}, J., {Aliu}, E., {Anderhub}, H., {et~al.} 2009, \aap, 493, 467

\bibitem[{{Aleksi{\'c}} {et~al.}(2012{\natexlab{a}}){Aleksi{\'c}}, {Alvarez},
  {Antonelli}, {Antoranz}, {Ansoldi}, {Asensio}, {Backes}, {Barres de Almeida},
  {Barrio}, {Bastieri}, {Becerra Gonz{\'a}lez}, {Bednarek}, {Berger},
  {Bernardini}, {Biland}, {Blanch}, {Bock}, {Boller}, {Bonnoli}, {Borla
  Tridon}, {Bretz}, {Ca{\~n}ellas}, {Carmona}, {Carosi}, {Colin}, {Colombo},
  {Contreras}, {Cortina}, {Cossio}, {Covino}, {Da Vela}, {Dazzi}, {De Angelis},
  {De Caneva}, {De Cea del Pozo}, {De Lotto}, {Delgado Mendez}, {Diago Ortega},
  {Doert}, {Dom{\'{\i}}nguez}, {Dominis Prester}, {Dorner}, {Doro},
  {Eisenacher}, {Elsaesser}, {Ferenc}, {Fonseca}, {Font}, {Fruck},
  {Garc{\'{\i}}a L{\'o}pez}, {Garczarczyk}, {Garrido Terrats}, {Gaug},
  {Giavitto}, {Godinovi{\'c}}, {Gonz{\'a}lez Mu{\~n}oz}, {Gozzini}, {Hadasch},
  {H{\"a}fner}, {Herrero}, {Hildebrand}, {Hose}, {Hrupec}, {Huber},
  {Jankowski}, {Jogler}, {Kadenius}, {Kellermann}, {Klepser},
  {Kr{\"a}henb{\"u}hl}, {Krause}, {La Barbera}, {Lelas}, {Leonardo},
  {Lewandowska}, {Lindfors}, {Lombardi}, {L{\'o}pez}, {L{\'o}pez-Coto},
  {L{\'o}pez-Oramas}, {Lorenz}, {Makariev}, {Maneva}, {Mankuzhiyil},
  {Mannheim}, {Maraschi}, {Mariotti}, {Mart{\'{\i}}nez}, {Mazin}, {Meucci},
  {Miranda}, {Mirzoyan}, {Mold{\'o}n}, {Moralejo}, {Munar-Adrover},
  {Niedzwiecki}, {Nieto}, {Nilsson}, {Nowak}, {Orito}, {Paiano}, {Paneque},
  {Paoletti}, {Pardo}, {Paredes}, {Partini}, {Perez-Torres}, {Persic}, {Pilia},
  {Pochon}, {Prada}, {Prada Moroni}, {Prandini}, {Puerto Gimenez}, {Puljak},
  {Reichardt}, {Reinthal}, {Rhode}, {Rib{\'o}}, {Rico}, {R{\"u}gamer},
  {Saggion}, {Saito}, {Saito}, {Salvati}, {Satalecka}, {Scalzotto}, {Scapin},
  {Schultz}, {Schweizer}, {Shore}, {Sillanp{\"a}{\"a}}, {Sitarek}, {Snidaric},
  {Sobczynska}, {Spanier}, {Spiro}, {Stamatescu}, {Stamerra}, {Steinke},
  {Storz}, {Strah}, {Sun}, {Suri{\'c}}, {Takalo}, {Takami}, {Tavecchio},
  {Temnikov}, {Terzi{\'c}}, {Tescaro}, {Teshima}, {Tibolla}, {Torres},
  {Treves}, {Uellenbeck}, {Vogler}, {Wagner}, {Weitzel}, {Zabalza}, {Zandanel},
  {Zanin}, {Berdyugin}, {Buson}, {J{\"a}rvel{\"a}}, {Larsson},
  {L{\"a}hteenm{\"a}ki}, \& {Tammi}}]{2012A&A...544A.142A}
{Aleksi{\'c}}, J., {Alvarez}, E.~A., {Antonelli}, L.~A., {et~al.}
  2012{\natexlab{a}}, \aap, 544, A142

\bibitem[{{Aleksi{\'c}} {et~al.}(2012{\natexlab{b}}){Aleksi{\'c}}, {Alvarez},
  {Antonelli}, {Antoranz}, {Asensio}, {Backes}, {Barrio}, {Bastieri}, {Becerra
  Gonz{\'a}lez}, {Bednarek}, {Berdyugin}, {Berger}, {Bernardini}, {Biland},
  {Blanch}, {Bock}, {Boller}, {Bonnoli}, {Borla Tridon}, {Braun}, {Bretz},
  {Ca{\~n}ellas}, {Carmona}, {Carosi}, {Colin}, {Colombo}, {Contreras},
  {Cortina}, {Cossio}, {Covino}, {Dazzi}, {De Angelis}, {De Caneva}, {De Cea
  del Pozo}, {De Lotto}, {Delgado Mendez}, {Diago Ortega}, {Doert},
  {Dom{\'{\i}}nguez}, {Dominis Prester}, {Dorner}, {Doro}, {Elsaesser},
  {Ferenc}, {Fonseca}, {Font}, {Fruck}, {Garc{\'{\i}}a L{\'o}pez},
  {Garczarczyk}, {Garrido}, {Giavitto}, {Godinovi{\'c}}, {Hadasch},
  {H{\"a}fner}, {Herrero}, {Hildebrand}, {H{\"o}hne-M{\"o}nch}, {Hose},
  {Hrupec}, {Huber}, {Jogler}, {Kellermann}, {Klepser}, {Kr{\"a}henb{\"u}hl},
  {Krause}, {La Barbera}, {Lelas}, {Leonardo}, {Lindfors}, {Lombardi},
  {L{\'o}pez}, {L{\'o}pez}, {Lorenz}, {Makariev}, {Maneva}, {Mankuzhiyil},
  {Mannheim}, {Maraschi}, {Mariotti}, {Mart{\'{\i}}nez}, {Mazin}, {Meucci},
  {Miranda}, {Mirzoyan}, {Miyamoto}, {Mold{\'o}n}, {Moralejo}, {Munar-Adrover},
  {Nieto}, {Nilsson}, {Orito}, {Oya}, {Paneque}, {Paoletti}, {Pardo},
  {Paredes}, {Partini}, {Pasanen}, {Pauss}, {Perez-Torres}, {Persic},
  {Peruzzo}, {Pilia}, {Pochon}, {Prada}, {Prada Moroni}, {Prandini}, {Puljak},
  {Reichardt}, {Reinthal}, {Rhode}, {Rib{\'o}}, {Rico}, {R{\"u}gamer},
  {Saggion}, {Saito}, {Saito}, {Salvati}, {Satalecka}, {Scalzotto}, {Scapin},
  {Schultz}, {Schweizer}, {Shayduk}, {Shore}, {Sillanp{\"a}{\"a}}, {Sitarek},
  {Sobczynska}, {Spanier}, {Spiro}, {Stamerra}, {Steinke}, {Storz}, {Strah},
  {Suri{\'c}}, {Takalo}, {Takami}, {Tavecchio}, {Temnikov}, {Terzi{\'c}},
  {Tescaro}, {Teshima}, {Tibolla}, {Torres}, {Treves}, {Uellenbeck}, {Vankov},
  {Vogler}, {Wagner}, {Weitzel}, {Zabalza}, {Zandanel}, \&
  {Zanin}}]{2012A&A...542A.100A}
{Aleksi{\'c}}, J., {Alvarez}, E.~A., {Antonelli}, L.~A., {et~al.}
  2012{\natexlab{b}}, \aap, 542, A100

\bibitem[{{Aleksi{\'c}} {et~al.}(2012{\natexlab{c}}){Aleksi{\'c}}, {Alvarez},
  {Antonelli}, {Antoranz}, {Asensio}, {Backes}, {Barrio}, {Bastieri}, {Becerra
  Gonz{\'a}lez}, {Bednarek}, {Berdyugin}, {Berger}, {Bernardini}, {Biland},
  {Blanch}, {Bock}, {Boller}, {Bonnoli}, {Borla Tridon}, {Braun}, {Bretz},
  {Ca{\~n}ellas}, {Carmona}, {Carosi}, {Colin}, {Colombo}, {Contreras},
  {Cortina}, {Cossio}, {Covino}, {Dazzi}, {De Angelis}, {De Caneva}, {De Cea
  del Pozo}, {De Lotto}, {Delgado Mendez}, {Diago Ortega}, {Doert},
  {Dom{\'{\i}}nguez}, {Dominis Prester}, {Dorner}, {Doro}, {Elsaesser},
  {Ferenc}, {Fonseca}, {Font}, {Fruck}, {Garc{\'{\i}}a L{\'o}pez},
  {Garczarczyk}, {Garrido}, {Giavitto}, {Godinovi{\'c}}, {Hadasch},
  {H{\"a}fner}, {Herrero}, {Hildebrand}, {H{\"o}hne-M{\"o}nch}, {Hose},
  {Hrupec}, {Huber}, {Jogler}, {Kellermann}, {Klepser}, {Kr{\"a}henb{\"u}hl},
  {Krause}, {La Barbera}, {Lelas}, {Leonardo}, {Lindfors}, {Lombardi},
  {L{\'o}pez}, {L{\'o}pez-Oramas}, {Lorenz}, {Makariev}, {Maneva},
  {Mankuzhiyil}, {Mannheim}, {Maraschi}, {Mariotti}, {Mart{\'{\i}}nez},
  {Mazin}, {Meucci}, {Miranda}, {Mirzoyan}, {Miyamoto}, {Mold{\'o}n},
  {Moralejo}, {Munar-Adrover}, {Nieto}, {Nilsson}, {Orito}, {Oya}, {Paneque},
  {Paoletti}, {Pardo}, {Paredes}, {Partini}, {Pasanen}, {Pauss},
  {Perez-Torres}, {Persic}, {Peruzzo}, {Pilia}, {Pochon}, {Prada}, {Prada
  Moroni}, {Prandini}, {Puljak}, {Reichardt}, {Reinthal}, {Rhode}, {Rib{\'o}},
  {Rico}, {R{\"u}gamer}, {Saggion}, {Saito}, {Saito}, {Salvati}, {Satalecka},
  {Scalzotto}, {Scapin}, {Schultz}, {Schweizer}, {Shayduk}, {Shore},
  {Sillanp{\"a}{\"a}}, {Sitarek}, {Sobczynska}, {Spanier}, {Spiro},
  {Stamatescu}, {Stamerra}, {Steinke}, {Storz}, {Strah}, {Suri{\'c}}, {Takalo},
  {Takami}, {Tavecchio}, {Temnikov}, {Terzi{\'c}}, {Tescaro}, {Teshima},
  {Tibolla}, {Torres}, {Treves}, {Uellenbeck}, {Vankov}, {Vogler}, {Wagner},
  {Weitzel}, {Zabalza}, {Zandanel}, {Zanin}, {Buson}, {Horan}, {Larsson}, \&
  {D'Ammando}}]{2012ApJ...748...46A}
{Aleksi{\'c}}, J., {Alvarez}, E.~A., {Antonelli}, L.~A., {et~al.}
  2012{\natexlab{c}}, \apj, 748, 46

\bibitem[{{Aleksi{\'c}} {et~al.}(2010{\natexlab{a}}){Aleksi{\'c}}, {Anderhub},
  {Antonelli}, {Antoranz}, {Backes}, {Baixeras}, {Balestra}, {Barrio},
  {Bastieri}, {Becerra Gonz{\'a}lez}, {Becker}, {Bednarek}, {Berdyugin},
  {Berger}, {Bernardini}, {Biland}, {Bock}, {Bonnoli}, {Bordas}, {Borla
  Tridon}, {Bosch-Ramon}, {Bose}, {Braun}, {Bretz}, {Britzger}, {Camara},
  {Carmona}, {Carosi}, {Colin}, {Commichau}, {Contreras}, {Cortina}, {Costado},
  {Covino}, {Dazzi}, {de Angelis}, {de Cea Del Pozo}, {de Los Reyes}, {de
  Lotto}, {de Maria}, {de Sabata}, {Delgado Mendez}, {Doert},
  {Dom{\'{\i}}nguez}, {Dominis Prester}, {Dorner}, {Doro}, {Elsaesser},
  {Errando}, {Ferenc}, {Fonseca}, {Font}, {Garc{\'{\i}}a L{\'o}pez},
  {Garczarczyk}, {Gaug}, {Godinovic}, {Hadasch}, {Herrero}, {Hildebrand},
  {H{\"o}hne-M{\"o}nch}, {Hose}, {Hrupec}, {Hsu}, {Jogler}, {Klepser},
  {Kr{\"a}henb{\"u}hl}, {Kranich}, {La Barbera}, {Laille}, {Leonardo},
  {Lindfors}, {Lombardi}, {Longo}, {L{\'o}pez}, {Lorenz}, {Majumdar}, {Maneva},
  {Mankuzhiyil}, {Mannheim}, {Maraschi}, {Mariotti}, {Mart{\'{\i}}nez},
  {Mazin}, {Meucci}, {Miranda}, {Mirzoyan}, {Miyamoto}, {Mold{\'o}n}, {Moles},
  {Moralejo}, {Nieto}, {Nilsson}, {Ninkovic}, {Orito}, {Oya}, {Paoletti},
  {Paredes}, {Partini}, {Pasanen}, {Pascoli}, {Pauss}, {Pegna}, {Perez-Torres},
  {Persic}, {Peruzzo}, {Prada}, {Prandini}, {Puchades}, {Puljak}, {Reichardt},
  {Rhode}, {Rib{\'o}}, {Rico}, {Rissi}, {R{\"u}gamer}, {Saggion}, {Saito},
  {Salvati}, {S{\'a}nchez-Conde}, {Satalecka}, {Scalzotto}, {Scapin},
  {Schweizer}, {Shayduk}, {Shore}, {Sierpowska-Bartosik}, {Sillanp{\"a}{\"a}},
  {Sitarek}, {Sobczynska}, {Spanier}, {Spiro}, {Stamerra}, {Steinke}, {Strah},
  {Struebig}, {Suric}, {Takalo}, {Tavecchio}, {Temnikov}, {Tescaro}, {Teshima},
  {Torres}, {Vankov}, {Wagner}, {Zabalza}, {Zandanel}, {Zanin}, \& {MAGIC
  Collaboration}}]{2010A&A...519A..32A}
{Aleksi{\'c}}, J., {Anderhub}, H., {Antonelli}, L.~A., {et~al.}
  2010{\natexlab{a}}, \aap, 519, A32

\bibitem[{{Aleksi{\'c}} {et~al.}(2010{\natexlab{b}}){Aleksi{\'c}}, {Anderhub},
  {Antonelli}, {Antoranz}, {Backes}, {Baixeras}, {Balestra}, {Barrio},
  {Bastieri}, {Becerra Gonz{\'a}lez}, {Becker}, {Bednarek}, {Berdyugin},
  {Berger}, {Bernardini}, {Biland}, {Bock}, {Bonnoli}, {Bordas}, {Borla
  Tridon}, {Bosch-Ramon}, {Bose}, {Braun}, {Bretz}, {Britzger}, {Camara},
  {Carmona}, {Carosi}, {Colin}, {Commichau}, {Contreras}, {Cortina}, {Costado},
  {Covino}, {Dazzi}, {De Angelis}, {de Cea Del Pozo}, {De Los Reyes}, {De
  Lotto}, {De Maria}, {De Sabata}, {Delgado Mendez}, {Dom{\'{\i}}nguez},
  {Dominis Prester}, {Dorner}, {Doro}, {Elsaesser}, {Errando}, {Ferenc},
  {Fern{\'a}ndez}, {Firpo}, {Fonseca}, {Font}, {Galante}, {Garc{\'{\i}}a
  L{\'o}pez}, {Garczarczyk}, {Gaug}, {Godinovic}, {Goebel}, {Hadasch},
  {Herrero}, {Hildebrand}, {H{\"o}hne-M{\"o}nch}, {Hose}, {Hrupec}, {Hsu},
  {Jogler}, {Klepser}, {Kr{\"a}henb{\"u}hl}, {Kranich}, {La Barbera}, {Laille},
  {Leonardo}, {Lindfors}, {Lombardi}, {Longo}, {L{\'o}pez}, {Lorenz},
  {Majumdar}, {Maneva}, {Mankuzhiyil}, {Mannheim}, {Maraschi}, {Mariotti},
  {Mart{\'{\i}}nez}, {Mazin}, {Meucci}, {Miranda}, {Mirzoyan}, {Miyamoto},
  {Mold{\'o}n}, {Moles}, {Moralejo}, {Nieto}, {Nilsson}, {Ninkovic}, {Orito},
  {Oya}, {Paoletti}, {Paredes}, {Pasanen}, {Pascoli}, {Pauss}, {Pegna},
  {Perez-Torres}, {Persic}, {Peruzzo}, {Prada}, {Prandini}, {Puchades},
  {Puljak}, {Reichardt}, {Rhode}, {Rib{\'o}}, {Rico}, {Rissi}, {R{\"u}gamer},
  {Saggion}, {Saito}, {Salvati}, {S{\'a}nchez-Conde}, {Satalecka}, {Scalzotto},
  {Scapin}, {Schweizer}, {Shayduk}, {Shore}, {Sierpowska-Bartosik},
  {Sillanp{\"a}{\"a}}, {Sitarek}, {Sobczynska}, {Spanier}, {Spiro}, {Stamerra},
  {Steinke}, {Strah}, {Struebig}, {Suric}, {Takalo}, {Tavecchio}, {Temnikov},
  {Tescaro}, {Teshima}, {Torres}, {Turini}, {Vankov}, {Wagner}, {Zabalza},
  {Zandanel}, {Zanin}, {Zapatero}, {Pian}, {Bianchin}, {D'Ammando}, {Di Cocco},
  {Fugazza}, {Ghisellini}, {Kurtanidze}, {Raiteri}, {Tosti}, {Treves},
  {Vercellone}, {Villata}, \& {MAGIC Collaboration}}]{2010A&A...515A..76A}
{Aleksi{\'c}}, J., {Anderhub}, H., {Antonelli}, L.~A., {et~al.}
  2010{\natexlab{b}}, \aap, 515, A76

\bibitem[{{Aleksi{\'c}} {et~al.}(2016){Aleksi{\'c}}, {Ansoldi}, {Antonelli},
  {Antoranz}, {Arcaro}, {Babic}, {Bangale}, {Barres de Almeida}, {Barrio},
  {Becerra Gonz{\'a}lez}, {Bednarek}, {Bernardini}, {Biasuzzi}, {Biland},
  {Blanch}, {Bonnefoy}, {Bonnoli}, {Borracci}, {Bretz}, {Carmona}, {Carosi},
  {Colin}, {Colombo}, {Contreras}, {Cortina}, {Covino}, {Da Vela}, {Dazzi}, {De
  Angelis}, {De Caneva}, {De Lotto}, {de O{\~n}a Wilhelmi}, {Delgado Mendez},
  {Di Pierro}, {Dominis Prester}, {Dorner}, {Doro}, {Einecke}, {Eisenacher},
  {Elsaesser}, {Fern{\'a}ndez-Barral}, {Fidalgo}, {Fonseca}, {Font},
  {Frantzen}, {Fruck}, {Galindo}, {Garc{\'{\i}}a L{\'o}pez}, {Garczarczyk},
  {Garrido Terrats}, {Gaug}, {Godinovi{\'c}}, {Gonz{\'a}lez Mu{\~n}oz},
  {Gozzini}, {Hadasch}, {Hanabata}, {Hayashida}, {Herrera}, {Hose}, {Hrupec},
  {Idec}, {Kadenius}, {Kellermann}, {Knoetig}, {Kodani}, {Konno}, {Krause},
  {Kubo}, {Kushida}, {La Barbera}, {Lelas}, {Lewandowska}, {Lindfors},
  {Lombardi}, {Longo}, {L{\'o}pez}, {L{\'o}pez-Coto}, {L{\'o}pez-Oramas},
  {Lorenz}, {Lozano}, {Makariev}, {Mallot}, {Maneva}, {Mannheim}, {Maraschi},
  {Marcote}, {Mariotti}, {Mart{\'{\i}}nez}, {Mazin}, {Menzel}, {Miranda},
  {Mirzoyan}, {Moralejo}, {Munar-Adrover}, {Nakajima}, {Neustroev},
  {Niedzwiecki}, {Nievas Rosillo}, {Nilsson}, {Nishijima}, {Noda}, {Orito},
  {Overkemping}, {Paiano}, {Palatiello}, {Paneque}, {Paoletti}, {Paredes},
  {Paredes-Fortuny}, {Persic}, {Poutanen}, {Prada Moroni}, {Prandini},
  {Puljak}, {Reinthal}, {Rhode}, {Rib{\'o}}, {Rico}, {Rodriguez Garcia},
  {Saito}, {Saito}, {Satalecka}, {Scalzotto}, {Scapin}, {Schweizer}, {Shore},
  {Sillanp{\"a}{\"a}}, {Sitarek}, {Snidaric}, {Sobczynska}, {Stamerra},
  {Steinbring}, {Strzys}, {Takalo}, {Takami}, {Tavecchio}, {Temnikov},
  {Terzi{\'c}}, {Tescaro}, {Teshima}, {Thaele}, {Torres}, {Toyama}, {Treves},
  {Vogler}, {Will}, {Zanin}, {Buson}, {D'Ammando}, {L{\"a}hteenm{\"a}ki},
  {Hovatta}, {Kovalev}, {Lister}, {Max-Moerbeck}, {Mundell}, {Pushkarev},
  {Rastorgueva-Foi}, {Readhead}, {Richards}, {Tammi}, {Sanchez}, {Tornikoski},
  {Savolainen}, \& {Steele}}]{2016A&A...591A..10A}
{Aleksi{\'c}}, J., {Ansoldi}, S., {Antonelli}, L.~A., {et~al.} 2016, \aap, 591,
  A10

\bibitem[{{Aleksi{\'c}} {et~al.}(2014{\natexlab{a}}){Aleksi{\'c}}, {Ansoldi},
  {Antonelli}, {Antoranz}, {Babic}, {Bangale}, {Barres de Almeida}, {Barrio},
  {Becerra Gonz{\'a}lez}, {Bednarek}, {Berger}, {Bernardini}, {Biland},
  {Blanch}, {Bock}, {Bonnefoy}, {Bonnoli}, {Borracci}, {Bretz}, {Carmona},
  {Carosi}, {Carreto Fidalgo}, {Colin}, {Colombo}, {Contreras}, {Cortina},
  {Covino}, {Da Vela}, {Dazzi}, {De Angelis}, {De Caneva}, {De Lotto}, {Delgado
  Mendez}, {Doert}, {Dom{\'{\i}}nguez}, {Dominis Prester}, {Dorner}, {Doro},
  {Einecke}, {Eisenacher}, {Elsaesser}, {Farina}, {Ferenc}, {Fonseca}, {Font},
  {Frantzen}, {Fruck}, {Garc{\'{\i}}a L{\'o}pez}, {Garczarczyk}, {Garrido
  Terrats}, {Gaug}, {Giavitto}, {Godinovi{\'c}}, {Gonz{\'a}lez Mu{\~n}oz},
  {Gozzini}, {Hadasch}, {Herrero}, {Hildebrand}, {Hose}, {Hrupec}, {Idec},
  {Kadenius}, {Kellermann}, {Knoetig}, {Kodani}, {Konno}, {Krause}, {Kubo},
  {Kushida}, {La Barbera}, {Lelas}, {Lewandowska}, {Lindfors}, {Lombardi},
  {L{\'o}pez}, {L{\'o}pez-Coto}, {L{\'o}pez-Oramas}, {Lorenz}, {Lozano},
  {Makariev}, {Mallot}, {Maneva}, {Mankuzhiyil}, {Mannheim}, {Maraschi},
  {Marcote}, {Mariotti}, {Mart{\'{\i}}nez}, {Mazin}, {Menzel}, {Meucci},
  {Miranda}, {Mirzoyan}, {Moralejo}, {Munar-Adrover}, {Nakajima},
  {Niedzwiecki}, {Nilsson}, {Nishijima}, {Nowak}, {Orito}, {Overkemping},
  {Paiano}, {Palatiello}, {Paneque}, {Paoletti}, {Paredes}, {Paredes-Fortuny},
  {Partini}, {Persic}, {Prada}, {Prada Moroni}, {Prandini}, {Preziuso},
  {Puljak}, {Reinthal}, {Rhode}, {Rib{\'o}}, {Rico}, {Rodriguez Garcia},
  {R{\"u}gamer}, {Saggion}, {Saito}, {Saito}, {Salvati}, {Satalecka},
  {Scalzotto}, {Scapin}, {Schultz}, {Schweizer}, {Shore}, {Sillanp{\"a}{\"a}},
  {Sitarek}, {Snidaric}, {Sobczynska}, {Spanier}, {Stamatescu}, {Stamerra},
  {Steinbring}, {Storz}, {Sun}, {Suri{\'c}}, {Takalo}, {Takami}, {Tavecchio},
  {Temnikov}, {Terzi{\'c}}, {Tescaro}, {Teshima}, {Thaele}, {Tibolla},
  {Torres}, {Toyama}, {Treves}, {Vogler}, {Wagner}, {Zandanel}, \&
  {Zanin}}]{2014A&A...567A..41A}
{Aleksi{\'c}}, J., {Ansoldi}, S., {Antonelli}, L.~A., {et~al.}
  2014{\natexlab{a}}, \aap, 567, A41

\bibitem[{{Aleksi{\'c}} {et~al.}(2014{\natexlab{b}}){Aleksi{\'c}}, {Ansoldi},
  {Antonelli}, {Antoranz}, {Babic}, {Bangale}, {Barres de Almeida}, {Barrio},
  {Becerra Gonz{\'a}lez}, {Bednarek}, {Berger}, {Bernardini}, {Biland},
  {Blanch}, {Bock}, {Bonnefoy}, {Bonnoli}, {Borracci}, {Bretz}, {Carmona},
  {Carosi}, {Carreto Fidalgo}, {Colin}, {Colombo}, {Contreras}, {Cortina},
  {Covino}, {Da Vela}, {Dazzi}, {De Angelis}, {De Caneva}, {De Lotto}, {Delgado
  Mendez}, {Doert}, {Dom{\'{\i}}nguez}, {Dominis Prester}, {Dorner}, {Doro},
  {Einecke}, {Eisenacher}, {Elsaesser}, {Farina}, {Ferenc}, {Fonseca}, {Font},
  {Frantzen}, {Fruck}, {Garc{\'{\i}}a L{\'o}pez}, {Garczarczyk}, {Garrido
  Terrats}, {Gaug}, {Giavitto}, {Godinovi{\'c}}, {Gonz{\'a}lez Mu{\~n}oz},
  {Gozzini}, {Hadasch}, {Hayashida}, {Herrero}, {Hildebrand}, {Hose}, {Hrupec},
  {Idec}, {Kadenius}, {Kellermann}, {Kodani}, {Konno}, {Krause}, {Kubo},
  {Kushida}, {La Barbera}, {Lelas}, {Lewandowska}, {Lindfors}, {Lombardi},
  {L{\'o}pez}, {L{\'o}pez-Coto}, {L{\'o}pez-Oramas}, {Lorenz}, {Lozano},
  {Makariev}, {Mallot}, {Maneva}, {Mankuzhiyil}, {Mannheim}, {Maraschi},
  {Marcote}, {Mariotti}, {Mart{\'{\i}}nez}, {Mazin}, {Menzel}, {Meucci},
  {Miranda}, {Mirzoyan}, {Moralejo}, {Munar-Adrover}, {Nakajima},
  {Niedzwiecki}, {Nilsson}, {Nishijima}, {Nowak}, {Orito}, {Overkemping},
  {Paiano}, {Palatiello}, {Paneque}, {Paoletti}, {Paredes}, {Paredes-Fortuny},
  {Partini}, {Persic}, {Prada}, {Prada Moroni}, {Prandini}, {Preziuso},
  {Puljak}, {Reinthal}, {Rhode}, {Rib{\'o}}, {Rico}, {Rodriguez Garcia},
  {R{\"u}gamer}, {Saggion}, {Saito}, {Saito}, {Salvati}, {Satalecka},
  {Scalzotto}, {Scapin}, {Schultz}, {Schweizer}, {Shore}, {Sillanp{\"a}{\"a}},
  {Sitarek}, {Snidaric}, {Sobczynska}, {Spanier}, {Stamatescu}, {Stamerra},
  {Steinbring}, {Storz}, {Sun}, {Suri{\'c}}, {Takalo}, {Takami}, {Tavecchio},
  {Temnikov}, {Terzi{\'c}}, {Tescaro}, {Teshima}, {Thaele}, {Tibolla},
  {Torres}, {Toyama}, {Treves}, {Uellenbeck}, {Vogler}, {Wagner}, {Zandanel},
  {Zanin}, {MAGIC Collaboration}, {Cutini}, {Gasparrini}, {Furniss}, {Hovatta},
  {Kangas}, {Kankare}, {Kotilainen}, {Lister}, {L{\"a}hteenm{\"a}ki},
  {Max-Moerbeck}, {Pavlidou}, {Readhead}, \& {Richards}}]{2014A&A...567A.135A}
{Aleksi{\'c}}, J., {Ansoldi}, S., {Antonelli}, L.~A., {et~al.}
  2014{\natexlab{b}}, \aap, 567, A135

\bibitem[{{Aleksi{\'c}} {et~al.}(2015{\natexlab{a}}){Aleksi{\'c}}, {Ansoldi},
  {Antonelli}, {Antoranz}, {Babic}, {Bangale}, {Barres de Almeida}, {Barrio},
  {Becerra Gonz{\'a}lez}, {Bednarek}, {Berger}, {Bernardini}, {Biland},
  {Blanch}, {Bonnefoy}, {Bonnoli}, {Borracci}, {Bretz}, {Carmona}, {Carosi},
  {Carreto Fidalgo}, {Colin}, {Colombo}, {Contreras}, {Cortina}, {Covino}, {da
  Vela}, {Dazzi}, {de Angelis}, {de Caneva}, {de Lotto}, {Delgado Mendez},
  {Doert}, {Dom{\'{\i}}nguez}, {Dominis Prester}, {Dorner}, {Doro}, {Einecke},
  {Eisenacher}, {Elsaesser}, {Farina}, {Ferenc}, {Fonseca}, {Font}, {Frantzen},
  {Fruck}, {Garc{\'{\i}}a L{\'o}pez}, {Garczarczyk}, {Garrido Terrats}, {Gaug},
  {Godinovi{\'c}}, {Gonz{\'a}lez Mu{\~n}oz}, {Gozzini}, {Hadasch}, {Hayashida},
  {Herrera}, {Herrero}, {Hildebrand}, {Hose}, {Hrupec}, {Idec}, {Kadenius},
  {Kellermann}, {Kodani}, {Konno}, {Krause}, {Kubo}, {Kushida}, {La Barbera},
  {Lelas}, {Lewandowska}, {Lindfors}, {Lombardi}, {L{\'o}pez},
  {L{\'o}pez-Coto}, {L{\'o}pez-Oramas}, {Lorenz}, {Lozano}, {Makariev},
  {Mallot}, {Maneva}, {Mankuzhiyil}, {Mannheim}, {Maraschi}, {Marcote},
  {Mariotti}, {Mart{\'{\i}}nez}, {Mazin}, {Menzel}, {Meucci}, {Miranda},
  {Mirzoyan}, {Moralejo}, {Munar-Adrover}, {Nakajima}, {Niedzwiecki},
  {Nilsson}, {Nishijima}, {Noda}, {Nowak}, {Orito}, {Overkemping}, {Paiano},
  {Palatiello}, {Paneque}, {Paoletti}, {Paredes}, {Paredes-Fortuny}, {Partini},
  {Persic}, {Prada}, {Moroni}, {Prandini}, {Preziuso}, {Puljak}, {Reinthal},
  {Rhode}, {Rib{\'o}}, {Rico}, {Rodriguez Garcia}, {R{\"u}gamer}, {Saggion},
  {Saito}, {Saito}, {Satalecka}, {Scalzotto}, {Scapin}, {Schultz}, {Schweizer},
  {Shore}, {Sillanp{\"a}{\"a}}, {Sitarek}, {Snidaric}, {Sobczynska}, {Spanier},
  {Stamatescu}, {Stamerra}, {Steinbring}, {Storz}, {Sun}, {Suri{\'c}},
  {Takalo}, {Takami}, {Tavecchio}, {Temnikov}, {Terzi{\'c}}, {Tescaro},
  {Teshima}, {Thaele}, {Tibolla}, {Torres}, {Toyama}, {Treves}, {Uellenbeck},
  {Vogler}, {Wagner}, {Zandanel}, {Zanin}, {MAGIC Collaboration}, {Tronconi},
  {Buson}, \& {Borghese}}]{2015MNRAS.446..217A}
{Aleksi{\'c}}, J., {Ansoldi}, S., {Antonelli}, L.~A., {et~al.}
  2015{\natexlab{a}}, \mnras, 446, 217

\bibitem[{{Aleksi{\'c}} {et~al.}(2014{\natexlab{c}}){Aleksi{\'c}}, {Ansoldi},
  {Antonelli}, {Antoranz}, {Babic}, {Bangale}, {Barres de Almeida}, {Barrio},
  {Becerra Gonz{\'a}lez}, {Bednarek}, \& et~al.}]{2014A&A...569A..46A}
{Aleksi{\'c}}, J., {Ansoldi}, S., {Antonelli}, L.~A., {et~al.}
  2014{\natexlab{c}}, \aap, 569, A46

\bibitem[{{Aleksi{\'c}} {et~al.}(2015{\natexlab{b}}){Aleksi{\'c}}, {Ansoldi},
  {Antonelli}, {Antoranz}, {Babic}, {Bangale}, {Barres de Almeida}, {Barrio},
  {Becerra Gonz{\'a}lez}, {Bednarek}, \& et~al.}]{2015A&A...573A..50A}
{Aleksi{\'c}}, J., {Ansoldi}, S., {Antonelli}, L.~A., {et~al.}
  2015{\natexlab{b}}, \aap, 573, A50

\bibitem[{{Aleksi{\'c}} {et~al.}(2015{\natexlab{c}}){Aleksi{\'c}}, {Ansoldi},
  {Antonelli}, {Antoranz}, {Babic}, {Bangale}, {Barres de Almeida}, {Barrio},
  {Becerra Gonz{\'a}lez}, {Bednarek}, \& et~al.}]{2015A&A...576A.126A}
{Aleksi{\'c}}, J., {Ansoldi}, S., {Antonelli}, L.~A., {et~al.}
  2015{\natexlab{c}}, \aap, 576, A126

\bibitem[{{Aleksi{\'c}} {et~al.}(2015{\natexlab{d}}){Aleksi{\'c}}, {Ansoldi},
  {Antonelli}, {Antoranz}, {Babic}, {Bangale}, {Barres de Almeida}, {Barrio},
  {Becerra Gonz{\'a}lez}, {Bednarek}, \& et~al.}]{2015A&A...578A..22A}
{Aleksi{\'c}}, J., {Ansoldi}, S., {Antonelli}, L.~A., {et~al.}
  2015{\natexlab{d}}, \aap, 578, A22

\bibitem[{{Aleksi{\'c}} {et~al.}(2015{\natexlab{e}}){Aleksi{\'c}}, {Ansoldi},
  {Antonelli}, {Antoranz}, {Babic}, {Bangale}, {Barrio}, {Becerra
  Gonz{\'a}lez}, {Bednarek}, {Bernardini}, {Biasuzzi}, {Biland}, {Blanch},
  {Bonnefoy}, {Bonnoli}, {Borracci}, {Bretz}, {Carmona}, {Carosi}, {Colin},
  {Colombo}, {Contreras}, {Cortina}, {Covino}, {Da Vela}, {Dazzi}, {De
  Angelis}, {De Caneva}, {De Lotto}, {de O{\~n}a Wilhelmi}, {Delgado Mendez},
  {Di Pierro}, {Dominis Prester}, {Dorner}, {Doro}, {Einecke}, {Eisenacher},
  {Elsaesser}, {Fern{\'a}ndez-Barral}, {Fidalgo}, {Fonseca}, {Font},
  {Frantzen}, {Fruck}, {Galindo}, {Garc{\'{\i}}a L{\'o}pez}, {Garczarczyk},
  {Garrido Terrats}, {Gaug}, {Godinovi{\'c}}, {Gonz{\'a}lez Mu{\~n}oz},
  {Gozzini}, {Hadasch}, {Hanabata}, {Hayashida}, {Herrera}, {Hose}, {Hrupec},
  {Idec}, {Kadenius}, {Kellermann}, {Knoetig}, {Kodani}, {Konno}, {Krause},
  {Kubo}, {Kushida}, {La Barbera}, {Lelas}, {Lewandowska}, {Lindfors},
  {Lombardi}, {Longo}, {L{\'o}pez}, {L{\'o}pez-Coto}, {L{\'o}pez-Oramas},
  {Lorenz}, {Lozano}, {Makariev}, {Mallot}, {Maneva}, {Mannheim}, {Maraschi},
  {Marcote}, {Mariotti}, {Mart{\'{\i}}nez}, {Mazin}, {Menzel}, {Miranda},
  {Mirzoyan}, {Moralejo}, {Munar-Adrover}, {Nakajima}, {Neustroev},
  {Niedzwiecki}, {Nievas Rosillo}, {Nilsson}, {Nishijima}, {Noda}, {Orito},
  {Overkemping}, {Paiano}, {Palatiello}, {Paneque}, {Paoletti}, {Paredes},
  {Paredes-Fortuny}, {Persic}, {Poutanen}, {Prada Moroni}, {Prandini},
  {Puljak}, {Reinthal}, {Rhode}, {Rib{\'o}}, {Rico}, {Rodriguez Garcia},
  {Saito}, {Saito}, {Satalecka}, {Scalzotto}, {Scapin}, {Schultz}, {Schweizer},
  {Shore}, {Sillanp{\"a}{\"a}}, {Sitarek}, {Snidaric}, {Sobczynska},
  {Stamerra}, {Steinbring}, {Strzys}, {Takalo}, {Takami}, {Tavecchio},
  {Temnikov}, {Terzi{\'c}}, {Tescaro}, {Teshima}, {Thaele}, {Torres}, {Toyama},
  {Treves}, {Vogler}, {Will}, {Zanin}, {Berger}, {Buson}, {D'Ammando},
  {Gasparrini}, {Hovatta}, {Max-Moerbeck}, {Readhead}, \&
  {Richards}}]{2015MNRAS.451..739A}
{Aleksi{\'c}}, J., {Ansoldi}, S., {Antonelli}, L.~A., {et~al.}
  2015{\natexlab{e}}, \mnras, 451, 739

\bibitem[{{Aleksi{\'c}} {et~al.}(2015{\natexlab{f}}){Aleksi{\'c}}, {Ansoldi},
  {Antonelli}, {Antoranz}, {Babic}, {Bangale}, {Barrio}, {Becerra
  Gonz{\'a}lez}, {Bednarek}, {Bernardini}, {Biasuzzi}, {Biland}, {Blanch},
  {Bonnefoy}, {Bonnoli}, {Borracci}, {Bretz}, {Carmona}, {Carosi}, {Colin},
  {Colombo}, {Contreras}, {Cortina}, {Covino}, {da Vela}, {Dazzi}, {de
  Angelis}, {de Caneva}, {de Lotto}, {de O{\~n}a Wilhelmi}, {Delgado Mendez},
  {Dominis Prester}, {Dorner}, {Doro}, {Einecke}, {Eisenacher}, {Elsaesser},
  {Fidalgo}, {Fonseca}, {Font}, {Frantzen}, {Fruck}, {Galindo}, {Garc{\'{\i}}a
  L{\'o}pez}, {Garczarczyk}, {Garrido Terrats}, {Gaug}, {Godinovi{\'c}},
  {Gonz{\'a}lez Mu{\~n}oz}, {Gozzini}, {Hadasch}, {Hanabata}, {Hayashida},
  {Herrera}, {Hildebrand}, {Hose}, {Hrupec}, {Idec}, {Kadenius}, {Kellermann},
  {Knoetig}, {Kodani}, {Konno}, {Krause}, {Kubo}, {Kushida}, {La Barbera},
  {Lelas}, {Lewandowska}, {Lindfors}, {Lombardi}, {Longo}, {L{\'o}pez},
  {L{\'o}pez-Coto}, {L{\'o}pez-Oramas}, {Lorenz}, {Lozano}, {Makariev},
  {Mallot}, {Maneva}, {Mannheim}, {Maraschi}, {Marcote}, {Mariotti},
  {Mart{\'{\i}}nez}, {Mazin}, {Menzel}, {Miranda}, {Mirzoyan}, {Moralejo},
  {Munar-Adrover}, {Nakajima}, {Neustroev}, {Niedzwiecki}, {Nilsson},
  {Nishijima}, {Noda}, {Orito}, {Overkemping}, {Paiano}, {Palatiello},
  {Paneque}, {Paoletti}, {Paredes}, {Paredes-Fortuny}, {Persic}, {Poutanen},
  {Prada Moroni}, {Prandini}, {Puljak}, {Reinthal}, {Rhode}, {Rib{\'o}},
  {Rico}, {Rodriguez Garcia}, {R{\"u}gamer}, {Saito}, {Saito}, {Satalecka},
  {Scalzotto}, {Scapin}, {Schultz}, {Schweizer}, {Sillanp{\"a}{\"a}},
  {Sitarek}, {Snidaric}, {Sobczynska}, {Spanier}, {Stamerra}, {Steinbring},
  {Storz}, {Strzys}, {Takalo}, {Takami}, {Tavecchio}, {Temnikov}, {Terzi{\'c}},
  {Tescaro}, {Teshima}, {Thaele}, {Tibolla}, {Torres}, {Toyama}, {Treves},
  {Vogler}, {Will}, {Zanin}, {MAGIC Collaboration}, {D'Ammando}, {Buson},
  {L{\"a}hteenm{\"a}ki}, {Tornikoski}, {Hovatta}, {Readhead}, {Max-Moerbeck},
  \& {Richards}}]{2015MNRAS.450.4399A}
{Aleksi{\'c}}, J., {Ansoldi}, S., {Antonelli}, L.~A., {et~al.}
  2015{\natexlab{f}}, \mnras, 450, 4399

\bibitem[{{Aleksi{\'c}} {et~al.}(2014{\natexlab{d}}){Aleksi{\'c}}, {Antonelli},
  {Antoranz}, {Asensio}, {Backes}, {Barres de Almeida}, {Barrio}, {Becerra
  Gonz{\'a}lez}, {Bednarek}, {Berger}, {Bernardini}, {Biland}, {Blanch},
  {Bock}, {Boller}, {Bonnefoy}, {Bonnoli}, {Borla Tridon}, {Borracci}, {Bretz},
  {Carmona}, {Carosi}, {Carreto Fidalgo}, {Colin}, {Colombo}, {Contreras},
  {Cortina}, {Cossio}, {Covino}, {da Vela}, {Dazzi}, {de Angelis}, {de Caneva},
  {de Lotto}, {Delgado Mendez}, {Doert}, {Dom{\'{\i}}nguez}, {Dominis Prester},
  {Dorner}, {Doro}, {Eisenacher}, {Elsaesser}, {Farina}, {Ferenc}, {Fonseca},
  {Font}, {Fruck}, {Garc{\'{\i}}a L{\'o}pez}, {Garczarczyk}, {Garrido Terrats},
  {Gaug}, {Giavitto}, {Godinovi{\'c}}, {Gonz{\'a}lez Mu{\~n}oz}, {Gozzini},
  {Hadamek}, {Hadasch}, {H{\"a}fner}, {Herrero}, {Hose}, {Hrupec}, {Idec},
  {Jankowski}, {Kadenius}, {Klepser}, {Knoetig}, {Kr{\"a}henb{\"u}hl},
  {Krause}, {Kushida}, {La Barbera}, {Lelas}, {Lewandowska}, {Lindfors},
  {Lombardi}, {L{\'o}pez}, {L{\'o}pez-Coto}, {L{\'o}pez-Oramas}, {Lorenz},
  {Lozano}, {Makariev}, {Mallot}, {Maneva}, {Mankuzhiyil}, {Mannheim},
  {Maraschi}, {Marcote}, {Mariotti}, {Mart{\'{\i}}nez}, {Masbou}, {Mazin},
  {Meucci}, {Miranda}, {Mirzoyan}, {Mold{\'o}n}, {Moralejo}, {Munar-Adrover},
  {Nakajima}, {Niedzwiecki}, {Nilsson}, {Nowak}, {Orito}, {Paiano},
  {Palatiello}, {Paneque}, {Paoletti}, {Paredes}, {Partini}, {Persic}, {Prada},
  {Prada Moroni}, {Prandini}, {Puljak}, {Reichardt}, {Reinthal}, {Rhode},
  {Rib{\'o}}, {Rico}, {R{\"u}gamer}, {Saggion}, {Saito}, {Saito}, {Salvati},
  {Satalecka}, {Scalzotto}, {Scapin}, {Schultz}, {Schweizer}, {Shore},
  {Sillanp{\"a}{\"a}}, {Sitarek}, {Snidaric}, {Sobczynska}, {Spanier}, {Spiro},
  {Stamatescu}, {Stamerra}, {Steinke}, {Storz}, {Sun}, {Suri{\'c}}, {Takalo},
  {Takami}, {Tavecchio}, {Temnikov}, {Terzi{\'c}}, {Tescaro}, {Teshima},
  {Tibolla}, {Torres}, {Toyama}, {Treves}, {Uellenbeck}, {Vogler}, {Wagner},
  {Weitzel}, {Zandanel}, {Zanin}, \& {MAGIC
  Collaboration}}]{2014A&A...563A..90A}
{Aleksi{\'c}}, J., {Antonelli}, L.~A., {Antoranz}, P., {et~al.}
  2014{\natexlab{d}}, \aap, 563, A90

\bibitem[{{Aleksi{\'c}} {et~al.}(2013){Aleksi{\'c}}, {Antonelli}, {Antoranz},
  {Asensio}, {Backes}, {Barres de Almeida}, {Barrio}, {Bednarek}, {Berger},
  {Bernardini}, {Biland}, {Blanch}, {Bock}, {Boller}, {Bonnefoy}, {Bonnoli},
  {Borla Tridon}, {Bretz}, {Carmona}, {Carosi}, {Carreto Fidalgo}, {Colin},
  {Colombo}, {Contreras}, {Cortina}, {Cossio}, {Covino}, {Da Vela}, {Dazzi},
  {De Angelis}, {De Caneva}, {De Lotto}, {Delgado Mendez}, {Doert},
  {Dom{\'{\i}}nguez}, {Dominis Prester}, {Dorner}, {Doro}, {Eisenacher},
  {Elsaesser}, {Ferenc}, {Fonseca}, {Font}, {Fruck}, {Garc{\'{\i}}a L{\'o}pez},
  {Garczarczyk}, {Garrido Terrats}, {Gaug}, {Giavitto}, {Godinovi{\'c}},
  {Gonz{\'a}lez Mu{\~n}oz}, {Gozzini}, {Hadamek}, {Hadasch}, {Herrero}, {Hose},
  {Hrupec}, {Jankowski}, {Kadenius}, {Klepser}, {Knoetig},
  {Kr{\"a}henb{\"u}hl}, {Krause}, {Kushida}, {La Barbera}, {Lelas}, {Leonardo},
  {Lewandowska}, {Lindfors}, {Lombardi}, {L{\'o}pez}, {L{\'o}pez-Coto},
  {L{\'o}pez-Oramas}, {Lorenz}, {Lozano}, {Makariev}, {Mallot}, {Maneva},
  {Mankuzhiyil}, {Mannheim}, {Maraschi}, {Marcote}, {Mariotti},
  {Mart{\'{\i}}nez}, {Masbou}, {Mazin}, {Meucci}, {Miranda}, {Mirzoyan},
  {Mold{\'o}n}, {Moralejo}, {Munar-Adrover}, {Nakajima}, {Niedzwiecki},
  {Nieto}, {Nilsson}, {Nowak}, {Orito}, {Paiano}, {Palatiello}, {Paneque},
  {Paoletti}, {Paredes}, {Partini}, {Persic}, {Pilia}, {Prada}, {Prada Moroni},
  {Prandini}, {Puljak}, {Reichardt}, {Reinthal}, {Rhode}, {Rib{\'o}}, {Rico},
  {R{\"u}gamer}, {Saggion}, {Saito}, {Saito}, {Salvati}, {Satalecka},
  {Scalzotto}, {Scapin}, {Schultz}, {Schweizer}, {Shore}, {Sillanp{\"a}{\"a}},
  {Sitarek}, {Snidaric}, {Sobczynska}, {Spanier}, {Spiro}, {Stamatescu},
  {Stamerra}, {Steinke}, {Storz}, {Sun}, {Suri{\'c}}, {Takalo}, {Takami},
  {Tavecchio}, {Temnikov}, {Terzi{\'c}}, {Tescaro}, {Teshima}, {Tibolla},
  {Torres}, {Toyama}, {Treves}, {Uellenbeck}, {Vogler}, {Wagner}, {Weitzel},
  {Zandanel}, {Zanin}, {MAGIC Collaboration}, {Longo}, {Lucarelli}, {Pittori},
  {Vercellone}, {AGILE Team}, {Bastieri}, {Sbarra}, {Fermi-LAT Collaboration},
  {Angelakis}, {Fuhrmann}, {Nestoras}, {Krichbaum}, {Sievers}, {Zensus},
  {F-GAMMA program}, {Antonyuk}, {Baumgartner}, {Berduygin}, {Carini}, {Cook},
  {Gehrels}, {Kadler}, {Kovalev}, {Kovalev}, {Krauss}, {Krimm},
  {L{\"a}hteenm{\"a}ki}, {Lister}, {Max-Moerbeck}, {Pasanen}, {Pushkarev},
  {Readhead}, {Richards}, {Sainio}, {Shakhovskoy}, {Sokolovsky}, {Tornikoski},
  {Tueller}, {Weidinger}, \& {Wilms}}]{2013A&A...556A..67A}
{Aleksi{\'c}}, J., {Antonelli}, L.~A., {Antoranz}, P., {et~al.} 2013, \aap,
  556, A67

\bibitem[{{Aleksi{\'c}} {et~al.}(2010{\natexlab{c}}){Aleksi{\'c}}, {Antonelli},
  {Antoranz}, {Backes}, {Baixeras}, {Barrio}, {Bastieri}, {Becerra
  Gonz{\'a}lez}, {Bednarek}, {Berdyugin}, {Berger}, {Bernardini}, {Biland},
  {Blanch}, {Bock}, {Bonnoli}, {Bordas}, {Borla Tridon}, {Bosch-Ramon}, {Bose},
  {Braun}, {Bretz}, {Britzger}, {Camara}, {Carmona}, {Carosi}, {Colin},
  {Commichau}, {Contreras}, {Cortina}, {Costado}, {Covino}, {Dazzi}, {de
  Angelis}, {de Cea Del Pozo}, {de Los Reyes}, {de Lotto}, {de Maria}, {de
  Sabata}, {Delgado Mendez}, {Doert}, {Dom{\'{\i}}nguez}, {Dominis Prester},
  {Dorner}, {Doro}, {Elsaesser}, {Errando}, {Ferenc}, {Fonseca}, {Font},
  {Garc{\'{\i}}a L{\'o}pez}, {Garczarczyk}, {Gaug}, {Godinovic}, {Hadasch},
  {Herrero}, {Hildebrand}, {H{\"o}hne-M{\"o}nch}, {Hose}, {Hrupec}, {Hsu},
  {Jogler}, {Klepser}, {Kr{\"a}henb{\"u}hl}, {Kranich}, {La Barbera}, {Laille},
  {Leonardo}, {Lindfors}, {Lombardi}, {Longo}, {L{\'o}pez}, {Lorenz},
  {Majumdar}, {Maneva}, {Mankuzhiyil}, {Mannheim}, {Maraschi}, {Mariotti},
  {Mart{\'{\i}}nez}, {Mazin}, {Meucci}, {Miranda}, {Mirzoyan}, {Miyamoto},
  {Mold{\'o}n}, {Moles}, {Moralejo}, {Nieto}, {Nilsson}, {Ninkovic}, {Orito},
  {Oya}, {Paiano}, {Paoletti}, {Paredes}, {Partini}, {Pasanen}, {Pascoli},
  {Pauss}, {Pegna}, {Perez-Torres}, {Persic}, {Peruzzo}, {Prada}, {Prandini},
  {Puchades}, {Puljak}, {Reichardt}, {Rhode}, {Rib{\'o}}, {Rico}, {Rissi},
  {R{\"u}gamer}, {Saggion}, {Saito}, {Salvati}, {S{\'a}nchez-Conde},
  {Satalecka}, {Scalzotto}, {Scapin}, {Schultz}, {Schweizer}, {Shayduk},
  {Shore}, {Sierpowska-Bartosik}, {Sillanp{\"a}{\"a}}, {Sitarek}, {Sobczynska},
  {Spanier}, {Spiro}, {Stamerra}, {Steinke}, {Struebig}, {Suric}, {Takalo},
  {Tavecchio}, {Temnikov}, {Terzic}, {Tescaro}, {Teshima}, {Torres}, {Vankov},
  {Wagner}, {Weitzel}, {Zabalza}, {Zandanel}, {Zanin}, {Neronov}, \&
  {Semikoz}}]{2010A&A...524A..77A}
{Aleksi{\'c}}, J., {Antonelli}, L.~A., {Antoranz}, P., {et~al.}
  2010{\natexlab{c}}, \aap, 524, A77

\bibitem[{{Aleksi{\'c}} {et~al.}(2011){Aleksi{\'c}}, {Antonelli}, {Antoranz},
  {Backes}, {Barrio}, {Bastieri}, {Becerra Gonz{\'a}lez}, {Bednarek},
  {Berdyugin}, {Berger}, {Bernardini}, {Biland}, {Blanch}, {Bock}, {Boller},
  {Bonnoli}, {Borla Tridon}, {Braun}, {Bretz}, {Ca{\~n}ellas}, {Carmona},
  {Carosi}, {Colin}, {Colombo}, {Contreras}, {Cortina}, {Cossio}, {Covino},
  {Dazzi}, {de Angelis}, {de Cea Del Pozo}, {de Lotto}, {Delgado Mendez},
  {Diago Ortega}, {Doert}, {Dom{\'{\i}}nguez}, {Dominis Prester}, {Dorner},
  {Doro}, {Elsaesser}, {Ferenc}, {Fonseca}, {Font}, {Fruck}, {Garc{\'{\i}}a
  L{\'o}pez}, {Garczarczyk}, {Garrido}, {Giavitto}, {Godinovi{\'c}}, {Hadasch},
  {H{\"a}fner}, {Herrero}, {Hildebrand}, {Hose}, {Hrupec}, {Huber}, {Jogler},
  {Klepser}, {Kr{\"a}henb{\"u}hl}, {Krause}, {La Barbera}, {Lelas}, {Leonardo},
  {Lindfors}, {Lombardi}, {L{\'o}pez}, {Lorenz}, {Majumdar}, {Makariev},
  {Maneva}, {Mankuzhiyil}, {Mannheim}, {Maraschi}, {Mariotti},
  {Mart{\'{\i}}nez}, {Mazin}, {Meucci}, {Miranda}, {Mirzoyan}, {Miyamoto},
  {Mold{\'o}n}, {Moralejo}, {Nieto}, {Nilsson}, {Orito}, {Oya}, {Paoletti},
  {Pardo}, {Paredes}, {Partini}, {Pasanen}, {Pauss}, {Perez-Torres}, {Persic},
  {Peruzzo}, {Pilia}, {Pochon}, {Prada}, {Prada Moroni}, {Prandini}, {Puljak},
  {Reichardt}, {Reinthal}, {Rhode}, {Rib{\'o}}, {Rico}, {R{\"u}gamer},
  {R{\"u}ger}, {Saggion}, {Saito}, {Saito}, {Salvati}, {Satalecka},
  {Scalzotto}, {Scapin}, {Schultz}, {Schweizer}, {Shayduk}, {Shore},
  {Sillanp{\"a}{\"a}}, {Sitarek}, {Sobczynska}, {Spanier}, {Spiro}, {Stamerra},
  {Steinke}, {Storz}, {Strah}, {Suri{\'c}}, {Takalo}, {Tavecchio}, {Temnikov},
  {Terzi{\'c}}, {Tescaro}, {Teshima}, {Thom}, {Tibolla}, {Torres}, {Treves},
  {Vankov}, {Vogler}, {Wagner}, {Weitzel}, {Zabalza}, {Zandanel}, \&
  {Zanin}}]{2011A&A...530A...4A}
{Aleksi{\'c}}, J., {Antonelli}, L.~A., {Antoranz}, P., {et~al.} 2011, \aap,
  530, A4

\bibitem[{{Aliu} {et~al.}(2009){Aliu}, {Anderhub}, {Antonelli}, {Antoranz},
  {Backes}, {Baixeras}, {Balestra}, {Barrio}, {Bartko}, {Bastieri}, {Becerra
  Gonz{\'a}lez}, {Becker}, {Bednarek}, {Berger}, {Bernardini}, {Biland},
  {Bock}, {Bonnoli}, {Bordas}, {Borla Tridon}, {Bosch-Ramon}, {Bretz},
  {Britvitch}, {Camara}, {Carmona}, {Chilingarian}, {Commichau}, {Contreras},
  {Cortina}, {Costado}, {Covino}, {Curtef}, {Dazzi}, {DeAngelis}, {DeCea del
  Pozo}, {de los Reyes}, {DeLotto}, {DeMaria}, {DeSabata}, {Delgado Mendez},
  {Dominguez}, {Dorner}, {Doro}, {Elsaesser}, {Errando}, {Ferenc},
  {Fern{\'a}ndez}, {Firpo}, {Fonseca}, {Font}, {Galante}, {Garc{\'{\i}}a
  L{\'o}pez}, {Garczarczyk}, {Gaug}, {Goebel}, {Hadasch}, {Hayashida},
  {Herrero}, {H{\"o}hne-M{\"o}nch}, {Hose}, {Hsu}, {Huber}, {Jogler},
  {Kranich}, {La Barbera}, {Laille}, {Leonardo}, {Lindfors}, {Lombardi},
  {Longo}, {L{\'o}pez}, {Lorenz}, {Majumdar}, {Maneva}, {Mankuzhiyil},
  {Mannheim}, {Maraschi}, {Mariotti}, {Mart{\'{\i}}nez}, {Mazin}, {Meucci},
  {Meyer}, {Miranda}, {Mirzoyan}, {Mold{\'o}n}, {Moles}, {Moralejo}, {Nieto},
  {Nilsson}, {Ninkovic}, {Otte}, {Oya}, {Paoletti}, {Paredes}, {Pasanen},
  {Pascoli}, {Pauss}, {Pegna}, {Perez-Torres}, {Persic}, {Peruzzo}, {Prada},
  {Prandini}, {Puchades}, {Raymers}, {Rhode}, {Rib{\'o}}, {Rico}, {Rissi},
  {Robert}, {R{\"u}gamer}, {Saggion}, {Saito}, {Salvati}, {Sanchez-Conde},
  {Sartori}, {Satalecka}, {Scalzotto}, {Scapin}, {Schweizer}, {Shayduk},
  {Shinozaki}, {Shore}, {Sidro}, {Sierpowska-Bartosik}, {Sillanp{\"a}{\"a}},
  {Sitarek}, {Sobczynska}, {Spanier}, {Stamerra}, {Stark}, {Takalo},
  {Tavecchio}, {Temnikov}, {Tescaro}, {Teshima}, {Tluczykont}, {Torres},
  {Turini}, {Vankov}, {Venturini}, {Vitale}, {Wagner}, {Wittek}, {Zabalza},
  {Zandanel}, {Zanin}, \& {Zapatero}}]{2009ApJ...692L..29A}
{Aliu}, E., {Anderhub}, H., {Antonelli}, L.~A., {et~al.} 2009, \apjl, 692, L29

\bibitem[{{Aller} {et~al.}(2003){Aller}, {Aller}, \&
  {Hughes}}]{2003ApJ...586...33A}
{Aller}, M.~F., {Aller}, H.~D., \& {Hughes}, P.~A. 2003, \apj, 586, 33

\bibitem[{{Anderhub} {et~al.}(2009{\natexlab{a}}){Anderhub}, {Antonelli},
  {Antoranz}, {Backes}, {Baixeras}, {Balestra}, {Barrio}, {Bastieri}, {Becerra
  Gonz{\'a}lez}, {Becker}, {Bednarek}, {Berdyugin}, {Berger}, {Bernardini},
  {Biland}, {Bock}, {Bonnoli}, {Bordas}, {Borla Tridon}, {Bosch-Ramon}, {Bose},
  {Braun}, {Bretz}, {Britzger}, {Camara}, {Carmona}, {Carosi}, {Colin},
  {Commichau}, {Contreras}, {Cortina}, {Costado}, {Covino}, {Dazzi}, {De
  Angelis}, {de Cea del Pozo}, {De los Reyes}, {De Lotto}, {De Maria}, {De
  Sabata}, {Delgado Mendez}, {Dom{\'{\i}}nguez}, {Dominis Prester}, {Dorner},
  {Doro}, {Elsaesser}, {Errando}, {Ferenc}, {Fern{\'a}ndez}, {Firpo},
  {Fonseca}, {Font}, {Galante}, {Garc{\'{\i}}a L{\'o}pez}, {Garczarczyk},
  {Gaug}, {Godinovic}, {Goebel}, {Hadasch}, {Herrero}, {Hildebrand},
  {H{\"o}hne-M{\"o}nch}, {Hose}, {Hrupec}, {Hsu}, {Jogler}, {Klepser},
  {Kranich}, {La Barbera}, {Laille}, {Leonardo}, {Lindfors}, {Lombardi},
  {Longo}, {L{\'o}pez}, {Lorenz}, {Majumdar}, {Maneva}, {Mankuzhiyil},
  {Mannheim}, {Maraschi}, {Mariotti}, {Mart{\'{\i}}nez}, {Mazin}, {Meucci},
  {Miranda}, {Mirzoyan}, {Miyamoto}, {Mold{\'o}n}, {Moles}, {Moralejo},
  {Nieto}, {Nilsson}, {Ninkovic}, {Orito}, {Oya}, {Paoletti}, {Paredes},
  {Pasanen}, {Pascoli}, {Pauss}, {Pegna}, {Perez-Torres}, {Persic}, {Peruzzo},
  {Prada}, {Prandini}, {Puchades}, {Puljak}, {Reichardt}, {Rhode}, {Rib{\'o}},
  {Rico}, {Rissi}, {Robert}, {R{\"u}gamer}, {Saggion}, {Sainio}, {Saito},
  {Salvati}, {S{\'a}nchez-Conde}, {Satalecka}, {Scalzotto}, {Scapin},
  {Schweizer}, {Shayduk}, {Shore}, {Sierpowska-Bartosik}, {Sillanp{\"a}{\"a}},
  {Sitarek}, {Sobczynska}, {Spanier}, {Spiro}, {Stamerra}, {Stark}, {Suric},
  {Takalo}, {Tavecchio}, {Temnikov}, {Tescaro}, {Teshima}, {Torres}, {Turini},
  {Vankov}, {Wagner}, {Villforth}, {Zabalza}, {Zandanel}, {Zanin}, \&
  {Zapatero}}]{2009ApJ...704L.129A}
{Anderhub}, H., {Antonelli}, L.~A., {Antoranz}, P., {et~al.}
  2009{\natexlab{a}}, \apjl, 704, L129

\bibitem[{{Anderhub} {et~al.}(2009{\natexlab{b}}){Anderhub}, {Antonelli},
  {Antoranz}, {Backes}, {Baixeras}, {Balestra}, {Barrio}, {Bastieri}, {Becerra
  Gonz{\'a}lez}, {Becker}, {Bednarek}, {Berger}, {Bernardini}, {Biland},
  {Bock}, {Bonnoli}, {Bordas}, {Borla Tridon}, {Bosch-Ramon}, {Bose}, {Braun},
  {Bretz}, {Britvitch}, {Camara}, {Carmona}, {Commichau}, {Contreras},
  {Cortina}, {Costado}, {Covino}, {Curtef}, {Dazzi}, {De Angelis}, {De Cea del
  Pozo}, {de los Reyes}, {De Lotto}, {De Maria}, {De Sabata}, {Delgado Mendez},
  {Dominguez}, {Dorner}, {Doro}, {Elsaesser}, {Errando}, {Ferenc},
  {Fern{\'a}ndez}, {Firpo}, {Fonseca}, {Font}, {Galante}, {Garc{\'{\i}}a
  L{\'o}pez}, {Garczarczyk}, {Gaug}, {Goebel}, {Hadasch}, {Hayashida},
  {Herrero}, {Hildebrand}, {H{\"o}hne-M{\"o}nch}, {Hose}, {Hsu}, {Jogler},
  {Kranich}, {La Barbera}, {Laille}, {Leonardo}, {Lindfors}, {Lombardi},
  {Longo}, {L{\'o}pez}, {Lorenz}, {Majumdar}, {Maneva}, {Mankuzhiyil},
  {Mannheim}, {Maraschi}, {Mariotti}, {Mart{\'{\i}}nez}, {Mazin}, {Meucci},
  {Meyer}, {Miranda}, {Mirzoyan}, {Miyamoto}, {Mold{\'o}n}, {Moles},
  {Moralejo}, {Nieto}, {Nilsson}, {Ninkovic}, {Otte}, {Oya}, {Paoletti},
  {Paredes}, {Pasanen}, {Pascoli}, {Pauss}, {Pegna}, {Perez-Torres}, {Persic},
  {Peruzzo}, {Prada}, {Prandini}, {Puchades}, {Reichardt}, {Rhode}, {Rib{\'o}},
  {Rico}, {Rissi}, {Robert}, {R{\"u}gamer}, {Saggion}, {Saito}, {Salvati},
  {Sanchez-Conde}, {Satalecka}, {Scalzotto}, {Scapin}, {Schweizer}, {Shayduk},
  {Shore}, {Sidro}, {Sierpowska-Bartosik}, {Sillanp{\"a}{\"a}}, {Sitarek},
  {Sobczynska}, {Spanier}, {Stamerra}, {Stark}, {Takalo}, {Tavecchio},
  {Temnikov}, {Tescaro}, {Teshima}, {Tluczykont}, {Torres}, {Turini}, {Vankov},
  {Wagner}, {Wittek}, {Zabalza}, {Zandanel}, {Zanin}, {Zapatero}, {MAGIC
  Collaboration}, {Sato}, {Ushio}, {Kataoka}, {Madejski}, \&
  {Takahashi}}]{2009ApJ...705.1624A}
{Anderhub}, H., {Antonelli}, L.~A., {Antoranz}, P., {et~al.}
  2009{\natexlab{b}}, \apj, 705, 1624

\bibitem[{{Bhatta} {et~al.}(2013){Bhatta}, {Webb}, {Hollingsworth}, {Dhalla},
  {Khanuja}, {Bachev}, {Blinov}, {B{\"o}ttcher}, {Bravo Calle}, {Calcidese},
  {Capezzali}, {Carosati}, {Chigladze}, {Collins}, {Coloma}, {Efimov}, {Gupta},
  {Hu}, {Kurtanidze}, {Lamerato}, {Larionov}, {Lee}, {Lindfors}, {Murphy},
  {Nilsson}, {Ohlert}, {Oksanen}, {P{\"a}{\"a}kk{\"o}nen}, {Pollock}, {Rani},
  {Reinthal}, {Rodriguez}, {Ros}, {Roustazadeh}, {Sagar}, {Sanchez}, {Shastri},
  {Sillanp{\"a}{\"a}}, {Strigachev}, {Takalo}, {Vennes}, {Villata},
  {Villforth}, {Wu}, \& {Zhou}}]{2013A&A...558A..92B}
{Bhatta}, G., {Webb}, J.~R., {Hollingsworth}, H., {et~al.} 2013, \aap, 558, A92

\bibitem[{{B{\"o}ttcher} {et~al.}(2007){B{\"o}ttcher}, {Basu}, {Joshi},
  {Villata}, {Arai}, {Aryan}, {Asfandiyarov}, {Bach}, {Bachev}, {Berduygin},
  {Blaek}, {Buemi}, {Castro-Tirado}, {De Ugarte Postigo}, {Frasca}, {Fuhrmann},
  {Hagen-Thorn}, {Henson}, {Hovatta}, {Hudec}, {Ibrahimov}, {Ishii},
  {Ivanidze}, {Jel{\'{\i}}nek}, {Kamada}, {Kapanadze}, {Katsuura}, {Kotaka},
  {Kovalev}, {Kovalev}, {Kub{\'a}nek}, {Kurosaki}, {Kurtanidze},
  {L{\"a}hteenm{\"a}ki}, {Lanteri}, {Larionov}, {Larionova}, {Lee}, {Leto},
  {Lindfors}, {Marilli}, {Marshall}, {Miller}, {Mingaliev}, {Mirabal},
  {Mizoguchi}, {Nakamura}, {Nieppola}, {Nikolashvili}, {Nilsson}, {Nishiyama},
  {Ohlert}, {Osterman}, {Pak}, {Pasanen}, {Peters}, {Pursimo}, {Raiteri},
  {Robertson}, {Robertson}, {Ryle}, {Sadakane}, {Sadun}, {Sigua}, {Sohn},
  {Strigachev}, {Sumitomo}, {Takalo}, {Tamesue}, {Tanaka}, {Thorstensen},
  {Tosti}, {Trigilio}, {Umana}, {Vennes}, {Vitek}, {Volvach}, {Webb},
  {Yamanaka}, \& {Yim}}]{2007ApJ...670..968B}
{B{\"o}ttcher}, M., {Basu}, S., {Joshi}, M., {et~al.} 2007, \apj, 670, 968

\bibitem[{{B{\"o}ttcher} {et~al.}(2009){B{\"o}ttcher}, {Fultz}, {Aller},
  {Aller}, {Apodaca}, {Arkharov}, {Bach}, {Bachev}, {Berdyugin}, {Buemi},
  {Calcidese}, {Carosati}, {Charlot}, {Ciprini}, {Paola}, {Dolci}, {Efimova},
  {Scurrats}, {Frasca}, {Gupta}, {Hagen-Thorn}, {Heidt}, {Hiriart},
  {Konstantinova}, {Kopatskaya}, {L{\"a}hteenm{\"a}ki}, {Lanteri}, {Larionov},
  {LeCampion}, {Leto}, {Lindfors}, {Mihov}, {Nieppola}, {Nilsson}, {Ovcharov},
  {P{\"a}{\"a}kk{\"o}nen}, {Pasanen}, {Ragozzine}, {Raiteri}, {Ros}, {Sadun},
  {Sanchez}, {Semkov}, {Sorcia}, {Strigachev}, {Takalo}, {Tornikoski},
  {Trigilio}, {Umana}, {Valcheva}, {Villata}, {Volvach}, {Wu}, \&
  {Zhou}}]{2009ApJ...694..174B}
{B{\"o}ttcher}, M., {Fultz}, K., {Aller}, H.~D., {et~al.} 2009, \apj, 694, 174

\bibitem[{{Caproni} {et~al.}(2013){Caproni}, {Abraham}, \&
  {Monteiro}}]{2013MNRAS.428..280C}
{Caproni}, A., {Abraham}, Z., \& {Monteiro}, H. 2013, \mnras, 428, 280

\bibitem[{{Carini} {et~al.}(1991){Carini}, {Miller}, {Noble}, \&
  {Sadun}}]{1991AJ....101.1196C}
{Carini}, M.~T., {Miller}, H.~R., {Noble}, J.~C., \& {Sadun}, A.~C. 1991, \aj,
  101, 1196

\bibitem[{{Cellone} {et~al.}(2000){Cellone}, {Romero}, \&
  {Combi}}]{2000AJ....119.1534C}
{Cellone}, S.~A., {Romero}, G.~E., \& {Combi}, J.~A. 2000, \aj, 119, 1534

\bibitem[{{Chatterjee} {et~al.}(2008){Chatterjee}, {Jorstad}, {Marscher}, {Oh},
  {McHardy}, {Aller}, {Aller}, {Balonek}, {Miller}, {Ryle}, {Tosti},
  {Kurtanidze}, {Nikolashvili}, {Larionov}, \&
  {Hagen-Thorn}}]{2008ApJ...689...79C}
{Chatterjee}, R., {Jorstad}, S.~G., {Marscher}, A.~P., {et~al.} 2008, \apj,
  689, 79

\bibitem[{{Ciprini} {et~al.}(2007){Ciprini}, {Takalo}, {Tosti}, {Raiteri},
  {Fiorucci}, {Villata}, {Nucciarelli}, {Lanteri}, {Nilsson}, \&
  {Ros}}]{2007A&A...467..465C}
{Ciprini}, S., {Takalo}, L.~O., {Tosti}, G., {et~al.} 2007, \aap, 467, 465

\bibitem[{{Costamante} \& {Ghisellini}(2002)}]{2002A&A...384...56C}
{Costamante}, L. \& {Ghisellini}, G. 2002, \aap, 384, 56

\bibitem[{{D'Ammando} {et~al.}(2009){D'Ammando}, {Pucella}, {Raiteri},
  {Villata}, {Vittorini}, {Vercellone}, {Donnarumma}, {Longo}, {Tavani},
  {Argan}, {Barbiellini}, {Boffelli}, {Bulgarelli}, {Caraveo}, {Cattaneo},
  {Chen}, {Cocco}, {Costa}, {Del Monte}, {de Paris}, {Di Cocco}, {Evangelista},
  {Feroci}, {Ferrari}, {Fiorini}, {Froysland}, {Fuschino}, {Galli}, {Gianotti},
  {Giuliani}, {Labanti}, {Lapshov}, {Lazzarotto}, {Lipari}, {Marisaldi},
  {Mereghetti}, {Morselli}, {Pacciani}, {Pellizzoni}, {Perotti}, {Piano},
  {Picozza}, {Pilia}, {Prest}, {Rapisarda}, {Rappoldi}, {Sabatini}, {Soffitta},
  {Trifoglio}, {Trois}, {Vallazza}, {Zambra}, {Zanello}, {Agudo}, {Aller},
  {Aller}, {Arkharov}, {Bach}, {Benitez}, {Berdyugin}, {Blinov}, {Buemi},
  {Chen}, {di Paola}, {di Rico}, {Dultzin}, {Fuhrmann}, {G{\'o}mez}, {Gurwell},
  {Jorstad}, {Heidt}, {Hiriart}, {Hsiao}, {Kimeridze}, {Konstantinova},
  {Kopatskaya}, {Koptelova}, {Kurtanidze}, {Larionov}, {Leto}, {Lindfors},
  {Lopez}, {Marscher}, {McHardy}, {Melnichuk}, {Mommert}, {Mujica}, {Nilsson},
  {Pasanen}, {Roca-Sogorb}, {Sorcia}, {Takalo}, {Taylor}, {Trigilio},
  {Troitsky}, {Umana}, {Antonelli}, {Colafrancesco}, {Cutini}, {Gasparrini},
  {Pittori}, {Preger}, {Santolamazza}, {Verrecchia}, {Giommi}, \&
  {Salotti}}]{2009A&A...508..181D}
{D'Ammando}, F., {Pucella}, G., {Raiteri}, C.~M., {et~al.} 2009, \aap, 508, 181

\bibitem[{{D'Ammando} {et~al.}(2011){D'Ammando}, {Raiteri}, {Villata},
  {Romano}, {Pucella}, {Krimm}, {Covino}, {Orienti}, {Giovannini},
  {Vercellone}, {Pian}, {Donnarumma}, {Vittorini}, {Tavani}, {Argan},
  {Barbiellini}, {Boffelli}, {Bulgarelli}, {Caraveo}, {Cattaneo}, {Chen},
  {Cocco}, {Costa}, {Del Monte}, {de Paris}, {Di Cocco}, {Evangelista},
  {Feroci}, {Ferrari}, {Fiorini}, {Froysland}, {Frutti}, {Fuschino}, {Galli},
  {Gianotti}, {Giuliani}, {Labanti}, {Lapshov}, {Lazzarotto}, {Lipari},
  {Longo}, {Marisaldi}, {Mereghetti}, {Morselli}, {Pacciani}, {Pellizzoni},
  {Perotti}, {Piano}, {Picozza}, {Pilia}, {Porrovecchio}, {Prest}, {Rapisarda},
  {Rappoldi}, {Rubini}, {Sabatini}, {Soffitta}, {Striani}, {Trifoglio},
  {Trois}, {Vallazza}, {Zambra}, {Zanello}, {Agudo}, {Aller}, {Aller},
  {Arkharov}, {Bach}, {Benitez}, {Berdyugin}, {Blinov}, {Buemi}, {Chen}, {di
  Paola}, {Dolci}, {Forn{\'e}}, {Fuhrmann}, {G{\'o}mez}, {Gurwell}, {Jordan},
  {Jorstad}, {Heidt}, {Hiriart}, {Hovatta}, {Hsiao}, {Kimeridze},
  {Konstantinova}, {Kopatskaya}, {Koptelova}, {Kurtanidze}, {Kurtanidze},
  {Larionov}, {L{\"a}hteenm{\"a}ki}, {Leto}, {Lindfors}, {Marscher}, {McBreen},
  {McHardy}, {Morozova}, {Nilsson}, {Pasanen}, {Roca-Sogorb},
  {Sillanp{\"a}{\"a}}, {Takalo}, {Tornikoski}, {Trigilio}, {Troitsky}, {Umana},
  {Antonelli}, {Colafrancesco}, {Pittori}, {Santolamazza}, {Verrecchia},
  {Giommi}, \& {Salotti}}]{2011A&A...529A.145D}
{D'Ammando}, F., {Raiteri}, C.~M., {Villata}, M., {et~al.} 2011, \aap, 529,
  A145

\bibitem[{{Donnarumma} {et~al.}(2009){Donnarumma}, {Vittorini}, {Vercellone},
  {del Monte}, {Feroci}, {D'Ammando}, {Pacciani}, {Chen}, {Tavani},
  {Bulgarelli}, \& et~al.}]{2009ApJ...691L..13D}
{Donnarumma}, I., {Vittorini}, V., {Vercellone}, S., {et~al.} 2009, \apjl, 691,
  L13

\bibitem[{{Emmanoulopoulos} {et~al.}(2013){Emmanoulopoulos}, {McHardy}, \&
  {Papadakis}}]{2013MNRAS.433..907E}
{Emmanoulopoulos}, D., {McHardy}, I.~M., \& {Papadakis}, I.~E. 2013, \mnras,
  433, 907

\bibitem[{{Emmanoulopoulos} {et~al.}(2010){Emmanoulopoulos}, {McHardy}, \&
  {Uttley}}]{2010MNRAS.404..931E}
{Emmanoulopoulos}, D., {McHardy}, I.~M., \& {Uttley}, P. 2010, \mnras, 404, 931

\bibitem[{{Fiorucci} {et~al.}(2004){Fiorucci}, {Ciprini}, \&
  {Tosti}}]{2004A&A...419...25F}
{Fiorucci}, M., {Ciprini}, S., \& {Tosti}, G. 2004, \aap, 419, 25

\bibitem[{{Fiorucci} \& {Tosti}(1996)}]{1996A&AS..116..403F}
{Fiorucci}, M. \& {Tosti}, G. 1996, \aaps, 116, 403

\bibitem[{{Fiorucci} {et~al.}(1998){Fiorucci}, {Tosti}, \&
  {Rizzi}}]{1998PASP..110..105F}
{Fiorucci}, M., {Tosti}, G., \& {Rizzi}, N. 1998, \pasp, 110, 105

\bibitem[{{Graham} \& {Driver}(2005)}]{2005PASA...22..118G}
{Graham}, A.~W. \& {Driver}, S.~P. 2005, \pasa, 22, 118

\bibitem[{{Gupta} {et~al.}(2009){Gupta}, {Srivastava}, \&
  {Wiita}}]{2009ApJ...690..216G}
{Gupta}, A.~C., {Srivastava}, A.~K., \& {Wiita}, P.~J. 2009, \apj, 690, 216

\bibitem[{{H.~E.~S.~S.~Collaboration}
  {et~al.}(2016){H.~E.~S.~S.~Collaboration}, {Abdalla}, {Abramowski},
  {Aharonian}, {Ait Benkhali}, {Andersson}, {Arrieta}, {Aubert}, {Backes},
  {Balzer}, \& et~al.}]{2016arXiv161003311H}
{H.~E.~S.~S.~Collaboration}, {Abdalla}, H., {Abramowski}, A., {et~al.} 2016,
  ArXiv e-prints [\eprint[arXiv]{1610.03311}]

\bibitem[{{Hamuy} \& {Maza}(1989)}]{1989AJ.....97..720H}
{Hamuy}, M. \& {Maza}, J. 1989, \aj, 97, 720

\bibitem[{{Hanawa} {et~al.}(2010){Hanawa}, {Ochi}, \&
  {Ando}}]{2010ApJ...708..485H}
{Hanawa}, T., {Ochi}, Y., \& {Ando}, K. 2010, \apj, 708, 485

\bibitem[{{Hayashida} {et~al.}(2012){Hayashida}, {Madejski}, {Nalewajko},
  {Sikora}, {Wehrle}, {Ogle}, {Collmar}, {Larsson}, {Fukazawa}, {Itoh},
  {Chiang}, {Stawarz}, {Blandford}, {Richards}, {Max-Moerbeck}, {Readhead},
  {Buehler}, {Cavazzuti}, {Ciprini}, {Gehrels}, {Reimer}, {Szostek}, {Tanaka},
  {Tosti}, {Uchiyama}, {Kawabata}, {Kino}, {Sakimoto}, {Sasada}, {Sato},
  {Uemura}, {Yamanaka}, {Greiner}, {Kruehler}, {Rossi}, {Macquart}, {Bock},
  {Villata}, {Raiteri}, {Agudo}, {Aller}, {Aller}, {Arkharov}, {Bach},
  {Ben{\'{\i}}tez}, {Berdyugin}, {Blinov}, {Blumenthal}, {B{\"o}ttcher},
  {Buemi}, {Carosati}, {Chen}, {Di Paola}, {Dolci}, {Efimova}, {Forn{\'e}},
  {G{\'o}mez}, {Gurwell}, {Heidt}, {Hiriart}, {Jordan}, {Jorstad}, {Joshi},
  {Kimeridze}, {Konstantinova}, {Kopatskaya}, {Koptelova}, {Kurtanidze},
  {L{\"a}hteenm{\"a}ki}, {Lamerato}, {Larionov}, {Larionova}, {Larionova},
  {Leto}, {Lindfors}, {Marscher}, {McHardy}, {Molina}, {Morozova},
  {Nikolashvili}, {Nilsson}, {Reinthal}, {Roustazadeh}, {Sakamoto}, {Sigua},
  {Sillanp{\"a}{\"a}}, {Takalo}, {Tammi}, {Taylor}, {Tornikoski}, {Trigilio},
  {Troitsky}, \& {Umana}}]{2012ApJ...754..114H}
{Hayashida}, M., {Madejski}, G.~M., {Nalewajko}, K., {et~al.} 2012, \apj, 754,
  114

\bibitem[{{Hovatta} {et~al.}(2008){Hovatta}, {Lehto}, \&
  {Tornikoski}}]{2008A&A...488..897H}
{Hovatta}, T., {Lehto}, H.~J., \& {Tornikoski}, M. 2008, \aap, 488, 897

\bibitem[{{Hovatta} {et~al.}(2007){Hovatta}, {Tornikoski}, {Lainela}, {Lehto},
  {Valtaoja}, {Torniainen}, {Aller}, \& {Aller}}]{2007A&A...469..899H}
{Hovatta}, T., {Tornikoski}, M., {Lainela}, M., {et~al.} 2007, \aap, 469, 899

\bibitem[{{Howell}(1989)}]{1989PASP..101..616H}
{Howell}, S.~B. 1989, \pasp, 101, 616

\bibitem[{{Ikejiri} {et~al.}(2011){Ikejiri}, {Uemura}, {Sasada}, {Ito},
  {Yamanaka}, {Sakimoto}, {Arai}, {Fukazawa}, {Ohsugi}, {Kawabata}, {Yoshida},
  {Sato}, \& {Kino}}]{2011PASJ...63..639I}
{Ikejiri}, Y., {Uemura}, M., {Sasada}, M., {et~al.} 2011, \pasj, 63, 639

\bibitem[{{Isobe} {et~al.}(2015){Isobe}, {Sato}, {Ueda}, {Hayashida},
  {Shidatsu}, {Kawamuro}, {Ueno}, {Sugizaki}, {Sugimoto}, {Mihara}, {Matsuoka},
  \& {Negoro}}]{2015ApJ...798...27I}
{Isobe}, N., {Sato}, R., {Ueda}, Y., {et~al.} 2015, \apj, 798, 27

\bibitem[{{Jermak} {et~al.}(2016){Jermak}, {Steele}, {Lindfors}, {Hovatta},
  {Nilsson}, {Lamb}, {Mundell}, {Barres de Almeida}, {Berdyugin}, {Kadenius},
  {Reinthal}, \& {Takalo}}]{2016MNRAS.462.4267J}
{Jermak}, H., {Steele}, I.~A., {Lindfors}, E., {et~al.} 2016, \mnras, 462, 4267

\bibitem[{{Kastendieck} {et~al.}(2011){Kastendieck}, {Ashley}, \&
  {Horns}}]{2011A&A...531A.123K}
{Kastendieck}, M.~A., {Ashley}, M.~C.~B., \& {Horns}, D. 2011, \aap, 531, A123

\bibitem[{{Kiehlmann} {et~al.}(2016){Kiehlmann}, {Savolainen}, {Jorstad},
  {Sokolovsky}, {Schinzel}, {Marscher}, {Larionov}, {Agudo}, {Akitaya},
  {Ben{\'{\i}}tez}, {Berdyugin}, {Blinov}, {Bochkarev}, {Borman}, {Burenkov},
  {Casadio}, {Doroshenko}, {Efimova}, {Fukazawa}, {G{\'o}mez}, {Grishina},
  {Hagen-Thorn}, {Heidt}, {Hiriart}, {Itoh}, {Joshi}, {Kawabata}, {Kimeridze},
  {Kopatskaya}, {Korobtsev}, {Krajci}, {Kurtanidze}, {Kurtanidze}, {Larionova},
  {Larionova}, {Lindfors}, {L{\'o}pez}, {McHardy}, {Molina}, {Moritani},
  {Morozova}, {Nazarov}, {Nikolashvili}, {Nilsson}, {Pulatova}, {Reinthal},
  {Sadun}, {Sasada}, {Savchenko}, {Sergeev}, {Sigua}, {Smith}, {Sorcia},
  {Spiridonova}, {Takaki}, {Takalo}, {Taylor}, {Troitsky}, {Uemura},
  {Ugolkova}, {Ui}, {Yoshida}, {Zensus}, \& {Zhdanova}}]{2016A&A...590A..10K}
{Kiehlmann}, S., {Savolainen}, T., {Jorstad}, S.~G., {et~al.} 2016, \aap, 590,
  A10

\bibitem[{{King} {et~al.}(2013){King}, {Hovatta}, {Max-Moerbeck}, {Meier},
  {Pearson}, {Readhead}, {Reeves}, {Richards}, \&
  {Shepherd}}]{2013MNRAS.436L.114K}
{King}, O.~G., {Hovatta}, T., {Max-Moerbeck}, W., {et~al.} 2013, \mnras, 436,
  L114

\bibitem[{{Lainela} {et~al.}(1999){Lainela}, {Takalo}, {Sillanp{\"a}{\"a}},
  {Pursimo}, {Nilsson}, {Katajainen}, {Tosti}, {Fiorucci}, {Luciani},
  {Villata}, {Raiteri}, {De Francesco}, {Sobrito}, {Ben{\'{\i}}tez},
  {Dultzin-Hacyan}, {de Diego}, {Turner}, {Robertson}, \&
  {Honeycutt}}]{1999ApJ...521..561L}
{Lainela}, M., {Takalo}, L.~O., {Sillanp{\"a}{\"a}}, A., {et~al.} 1999, \apj,
  521, 561

\bibitem[{{Larionov} {et~al.}(2008){Larionov}, {Jorstad}, {Marscher},
  {Raiteri}, {Villata}, {Agudo}, {Aller}, {Arkharov}, {Asfandiyarov}, {Bach},
  {Bachev}, {Berdyugin}, {B{\"o}ttcher}, {Buemi}, {Calcidese}, {Carosati},
  {Charlot}, {Chen}, {di Paola}, {Dolci}, {Dogru}, {Doroshenko}, {Efimov},
  {Erdem}, {Frasca}, {Fuhrmann}, {Giommi}, {Glowienka}, {Gupta}, {Gurwell},
  {Hagen-Thorn}, {Hsiao}, {Ibrahimov}, {Jordan}, {Kamada}, {Konstantinova},
  {Kopatskaya}, {Kovalev}, {Kovalev}, {Kurtanidze}, {L{\"a}hteenm{\"a}ki},
  {Lanteri}, {Larionova}, {Leto}, {Le Campion}, {Lee}, {Lindfors}, {Marilli},
  {McHardy}, {Mingaliev}, {Nazarov}, {Nieppola}, {Nilsson}, {Ohlert},
  {Pasanen}, {Porter}, {Pursimo}, {Ros}, {Sadakane}, {Sadun}, {Sergeev},
  {Smith}, {Strigachev}, {Sumitomo}, {Takalo}, {Tanaka}, {Trigilio}, {Umana},
  {Ungerechts}, {Volvach}, \& {Yuan}}]{2008A&A...492..389L}
{Larionov}, V.~M., {Jorstad}, S.~G., {Marscher}, A.~P., {et~al.} 2008, \aap,
  492, 389

\bibitem[{{Li} {et~al.}(2001){Li}, {Colgate}, {Wendroff}, \&
  {Liska}}]{2001ApJ...551..874L}
{Li}, H., {Colgate}, S.~A., {Wendroff}, B., \& {Liska}, R. 2001, \apj, 551, 874

\bibitem[{{Li} {et~al.}(2016){Li}, {Jiang}, {Guo}, {Chen}, \&
  {Yi}}]{2016PASP..128g4101L}
{Li}, H.~Z., {Jiang}, Y.~G., {Guo}, D.~F., {Chen}, X., \& {Yi}, T.~F. 2016,
  \pasp, 128, 074101

\bibitem[{{Lichti} {et~al.}(2008){Lichti}, {Bottacini}, {Ajello}, {Charlot},
  {Collmar}, {Falcone}, {Horan}, {Huber}, {von Kienlin}, {L{\"a}hteenm{\"a}ki},
  {Lindfors}, {Morris}, {Nilsson}, {Petry}, {R{\"u}ger}, {Sillanp{\"a}{\"a}},
  {Spanier}, \& {Tornikoski}}]{2008A&A...486..721L}
{Lichti}, G.~G., {Bottacini}, E., {Ajello}, M., {et~al.} 2008, \aap, 486, 721

\bibitem[{{Lindfors} {et~al.}(2016){Lindfors}, {Hovatta}, {Nilsson},
  {Reinthal}, {Fallah Ramazani}, {Pavlidou}, {Max-Moerbeck}, {Richards},
  {Berdyugin}, {Takalo}, {Sillanp{\"a}{\"a}}, \&
  {Readhead}}]{2016A&A...593A..98L}
{Lindfors}, E.~J., {Hovatta}, T., {Nilsson}, K., {et~al.} 2016, \aap, 593, A98

\bibitem[{{MAGIC Collaboration} {et~al.}(2008){MAGIC Collaboration}, {Albert},
  {Aliu}, {Anderhub}, {Antonelli}, {Antoranz}, {Backes}, {Baixeras}, {Barrio},
  {Bartko}, {Bastieri}, {Becker}, {Bednarek}, {Berger}, {Bernardini},
  {Bigongiari}, {Biland}, {Bock}, {Bonnoli}, {Bordas}, {Bosch-Ramon}, {Bretz},
  {Britvitch}, {Camara}, {Carmona}, {Chilingarian}, {Commichau}, {Contreras},
  {Cortina}, {Costado}, {Covino}, {Curtef}, {Dazzi}, {De Angelis}, {Cea del
  Pozo}, {de los Reyes}, {De Lotto}, {De Maria}, {De Sabata}, {Mendez},
  {Dominguez}, {Dorner}, {Doro}, {Errando}, {Fagiolini}, {Ferenc},
  {Fern{\'a}ndez}, {Firpo}, {Fonseca}, {Font}, {Galante}, {Garc{\'{\i}}a
  L{\'o}pez}, {Garczarczyk}, {Gaug}, {Goebel}, {Hayashida}, {Herrero},
  {H{\"o}hne}, {Hose}, {Hsu}, {Huber}, {Jogler}, {Kneiske}, {Kranich}, {La
  Barbera}, {Laille}, {Leonardo}, {Lindfors}, {Lombardi}, {Longo}, {L{\'o}pez},
  {Lorenz}, {Majumdar}, {Maneva}, {Mankuzhiyil}, {Mannheim}, {Maraschi},
  {Mariotti}, {Mart{\'{\i}}nez}, {Mazin}, {Meucci}, {Meyer}, {Miranda},
  {Mirzoyan}, {Mizobuchi}, {Moles}, {Moralejo}, {Nieto}, {Nilsson}, {Ninkovic},
  {Otte}, {Oya}, {Panniello}, {Paoletti}, {Paredes}, {Pasanen}, {Pascoli},
  {Pauss}, {Pegna}, {Perez-Torres}, {Persic}, {Peruzzo}, {Piccioli}, {Prada},
  {Prandini}, {Puchades}, {Raymers}, {Rhode}, {Rib{\'o}}, {Rico}, {Rissi},
  {Robert}, {R{\"u}gamer}, {Saggion}, {Saito}, {Salvati}, {Sanchez-Conde},
  {Sartori}, {Satalecka}, {Scalzotto}, {Scapin}, {Schmitt}, {Schweizer},
  {Shayduk}, {Shinozaki}, {Shore}, {Sidro}, {Sierpowska-Bartosik},
  {Sillanp{\"a}{\"a}}, {Sobczynska}, {Spanier}, {Stamerra}, {Stark}, {Takalo},
  {Tavecchio}, {Temnikov}, {Tescaro}, {Teshima}, {Tluczykont}, {Torres},
  {Turini}, {Vankov}, {Venturini}, {Vitale}, {Wagner}, {Wittek}, {Zabalza},
  {Zandanel}, {Zanin}, \& {Zapatero}}]{2008Sci...320.1752M}
{MAGIC Collaboration}, {Albert}, J., {Aliu}, E., {et~al.} 2008, Science, 320,
  1752

\bibitem[{{Massaro} {et~al.}(2015){Massaro}, {Maselli}, {Leto}, {Marchegiani},
  {Perri}, {Giommi}, \& {Piranomonte}}]{2015Ap&SS.357...75M}
{Massaro}, E., {Maselli}, A., {Leto}, C., {et~al.} 2015, \apss, 357, 75

\bibitem[{{Massaro} {et~al.}(2004){Massaro}, {Perri}, {Giommi}, \&
  {Nesci}}]{2004A&A...413..489M}
{Massaro}, E., {Perri}, M., {Giommi}, P., \& {Nesci}, R. 2004, \aap, 413, 489

\bibitem[{{Max-Moerbeck} {et~al.}(2014{\natexlab{a}}){Max-Moerbeck}, {Hovatta},
  {Richards}, {King}, {Pearson}, {Readhead}, {Reeves}, {Shepherd}, {Stevenson},
  {Angelakis}, {Fuhrmann}, {Grainge}, {Pavlidou}, {Romani}, \&
  {Zensus}}]{2014MNRAS.445..428M}
{Max-Moerbeck}, W., {Hovatta}, T., {Richards}, J.~L., {et~al.}
  2014{\natexlab{a}}, \mnras, 445, 428

\bibitem[{{Max-Moerbeck} {et~al.}(2014{\natexlab{b}}){Max-Moerbeck},
  {Richards}, {Hovatta}, {Pavlidou}, {Pearson}, \&
  {Readhead}}]{2014MNRAS.445..437M}
{Max-Moerbeck}, W., {Richards}, J.~L., {Hovatta}, T., {et~al.}
  2014{\natexlab{b}}, \mnras, 445, 437

\bibitem[{{Nakamura} \& {Meier}(2004)}]{2004ApJ...617..123N}
{Nakamura}, M. \& {Meier}, D.~L. 2004, \apj, 617, 123

\bibitem[{{Nandra} {et~al.}(1997){Nandra}, {George}, {Mushotzky}, {Turner}, \&
  {Yaqoob}}]{1997ApJ...476...70N}
{Nandra}, K., {George}, I.~M., {Mushotzky}, R.~F., {Turner}, T.~J., \&
  {Yaqoob}, T. 1997, \apj, 476, 70

\bibitem[{{Nesci}(2010)}]{2010AJ....139.2425N}
{Nesci}, R. 2010, \aj, 139, 2425

\bibitem[{{Nilsson} {et~al.}(2007){Nilsson}, {Pasanen}, {Takalo}, {Lindfors},
  {Berdyugin}, {Ciprini}, \& {Pforr}}]{2007A&A...475..199N}
{Nilsson}, K., {Pasanen}, M., {Takalo}, L.~O., {et~al.} 2007, \aap, 475, 199

\bibitem[{{Ostorero} {et~al.}(2006){Ostorero}, {Wagner}, {Gracia}, {Ferrero},
  {Krichbaum}, {Britzen}, {Witzel}, {Nilsson}, {Villata}, {Bach}, {Barnaby},
  {Bernhart}, {Carini}, {Chen}, {Chen}, {Ciprini}, {Crapanzano}, {Doroshenko},
  {Efimova}, {Emmanoulopoulos}, {Fuhrmann}, {Gabanyi}, {Giltinan},
  {Hagen-Thorn}, {Hauser}, {Heidt}, {Hojaev}, {Hovatta}, {Hroch}, {Ibrahimov},
  {Impellizzeri}, {Ivanidze}, {Kachel}, {Kraus}, {Kurtanidze},
  {L{\"a}hteenm{\"a}ki}, {Lanteri}, {Larionov}, {Lin}, {Lindfors}, {Munz},
  {Nikolashvili}, {Nucciarelli}, {O'Connor}, {Ohlert}, {Pasanen}, {Pullen},
  {Raiteri}, {Rector}, {Robb}, {Sigua}, {Sillanp{\"a}{\"a}}, {Sixtova},
  {Smith}, {Strub}, {Takahashi}, {Takalo}, {Tapken}, {Tartar}, {Tornikoski},
  {Tosti}, {Tr{\"o}ller}, {Walters}, {Wilking}, {Wills}, {Agudo}, {Aller},
  {Aller}, {Angelakis}, {Klare}, {K{\"o}rding}, {Strom}, {Ter{\"a}sranta},
  {Ungerechts}, \& {Vila-Vilar{\'o}}}]{2006A&A...451..797O}
{Ostorero}, L., {Wagner}, S.~J., {Gracia}, J., {et~al.} 2006, \aap, 451, 797

\bibitem[{{Pian} {et~al.}(2005){Pian}, {Foschini}, {Beckmann},
  {Sillanp{\"a}{\"a}}, {Soldi}, {Tagliaferri}, {Takalo}, {Barr}, {Ghisellini},
  {Malaguti}, {Maraschi}, {Palumbo}, {Treves}, {Courvoisier}, {Di Cocco},
  {Gehrels}, {Giommi}, {Hudec}, {Lindfors}, {Marcowith}, {Nilsson}, {Pasanen},
  {Pursimo}, {Raiteri}, {Savolainen}, {Sikora}, {Tornikoski}, {Tosti},
  {T{\"u}rler}, {Valtaoja}, {Villata}, \& {Walter}}]{2005A&A...429..427P}
{Pian}, E., {Foschini}, L., {Beckmann}, V., {et~al.} 2005, \aap, 429, 427

\bibitem[{{Pihajoki} {et~al.}(2013{\natexlab{a}}){Pihajoki}, {Valtonen}, \&
  {Ciprini}}]{2013MNRAS.434.3122P}
{Pihajoki}, P., {Valtonen}, M., \& {Ciprini}, S. 2013{\natexlab{a}}, \mnras,
  434, 3122

\bibitem[{{Pihajoki} {et~al.}(2013{\natexlab{b}}){Pihajoki}, {Valtonen},
  {Zola}, {Liakos}, {Drozdz}, {Winiarski}, {Ogloza}, {Koziel-Wierzbowska},
  {Provencal}, {Nilsson}, {Berdyugin}, {Lindfors}, {Reinthal},
  {Sillanp{\"a}{\"a}}, {Takalo}, {Santangelo}, {Salo}, {Chandra}, {Ganesh},
  {Baliyan}, {Coggins-Hill}, \& {Gopakumar}}]{2013ApJ...764....5P}
{Pihajoki}, P., {Valtonen}, M., {Zola}, S., {et~al.} 2013{\natexlab{b}}, \apj,
  764, 5

\bibitem[{{Raiteri} {et~al.}(2001){Raiteri}, {Villata}, {Aller}, {Aller},
  {Heidt}, {Kurtanidze}, {Lanteri}, {Maesano}, {Massaro}, {Montagni}, {Nesci},
  {Nilsson}, {Nikolashvili}, {Nurmi}, {Ostorero}, {Pursimo}, {Rekola},
  {Sillanp{\"a}{\"a}}, {Takalo}, {Ter{\"a}sranta}, {Tosti}, {Balonek}, {Feldt},
  {Heines}, {Heisler}, {Hu}, {Kidger}, {Mattox}, {McGrath}, {Pati}, {Robb},
  {Sadun}, {Shastri}, {Wagner}, {Wei}, \& {Wu}}]{2001A&A...377..396R}
{Raiteri}, C.~M., {Villata}, M., {Aller}, H.~D., {et~al.} 2001, \aap, 377, 396

\bibitem[{{Raiteri} {et~al.}(2013){Raiteri}, {Villata}, {D'Ammando},
  {Larionov}, {Gurwell}, {Mirzaqulov}, {Smith}, {Acosta-Pulido}, {Agudo},
  {Ar{\'e}valo}, {Bachev}, {Ben{\'{\i}}tez}, {Berdyugin}, {Blinov}, {Borman},
  {B{\"o}ttcher}, {Bozhilov}, {Carnerero}, {Carosati}, {Casadio}, {Chen},
  {Doroshenko}, {Efimov}, {Efimova}, {Ehgamberdiev}, {G{\'o}mez},
  {Gonz{\'a}lez-Morales}, {Hiriart}, {Ibryamov}, {Jadhav}, {Jorstad}, {Joshi},
  {Kadenius}, {Klimanov}, {Kohli}, {Konstantinova}, {Kopatskaya}, {Koptelova},
  {Kimeridze}, {Kurtanidze}, {Larionova}, {Larionova}, {Ligustri}, {Lindfors},
  {Marscher}, {McBreen}, {McHardy}, {Metodieva}, {Molina}, {Morozova},
  {Nazarov}, {Nikolashvili}, {Nilsson}, {Okhmat}, {Ovcharov}, {Panwar},
  {Pasanen}, {Peneva}, {Phipps}, {Pulatova}, {Reinthal}, {Ros}, {Sadun},
  {Schwartz}, {Semkov}, {Sergeev}, {Sigua}, {Sillanp{\"a}{\"a}}, {Smith},
  {Stoyanov}, {Strigachev}, {Takalo}, {Taylor}, {Thum}, {Troitsky}, {Valcheva},
  {Wehrle}, \& {Wiesemeyer}}]{2013MNRAS.436.1530R}
{Raiteri}, C.~M., {Villata}, M., {D'Ammando}, F., {et~al.} 2013, \mnras, 436,
  1530

\bibitem[{{Raiteri} {et~al.}(1998){Raiteri}, {Villata}, {Lanteri}, {Cavallone},
  \& {Sobrito}}]{1998A&AS..130..495R}
{Raiteri}, C.~M., {Villata}, M., {Lanteri}, L., {Cavallone}, M., \& {Sobrito},
  G. 1998, \aaps, 130, 495

\bibitem[{{Ramakrishnan} {et~al.}(2015){Ramakrishnan}, {Hovatta}, {Nieppola},
  {Tornikoski}, {L{\"a}hteenm{\"a}ki}, \& {Valtaoja}}]{2015MNRAS.452.1280R}
{Ramakrishnan}, V., {Hovatta}, T., {Nieppola}, E., {et~al.} 2015, \mnras, 452,
  1280

\bibitem[{{Ramakrishnan} {et~al.}(2016){Ramakrishnan}, {Hovatta}, {Tornikoski},
  {Nilsson}, {Lindfors}, {Balokovi{\'c}}, {L{\"a}hteenm{\"a}ki}, {Reinthal}, \&
  {Takalo}}]{2016MNRAS.456..171R}
{Ramakrishnan}, V., {Hovatta}, T., {Tornikoski}, M., {et~al.} 2016, \mnras,
  456, 171

\bibitem[{{Rani} {et~al.}(2010){Rani}, {Gupta}, {Joshi}, {Ganesh}, \&
  {Wiita}}]{2010ApJ...719L.153R}
{Rani}, B., {Gupta}, A.~C., {Joshi}, U.~C., {Ganesh}, S., \& {Wiita}, P.~J.
  2010, \apjl, 719, L153

\bibitem[{{Sandrinelli} {et~al.}(2016{\natexlab{a}}){Sandrinelli}, {Covino},
  {Dotti}, \& {Treves}}]{2016AJ....151...54S}
{Sandrinelli}, A., {Covino}, S., {Dotti}, M., \& {Treves}, A.
  2016{\natexlab{a}}, \aj, 151, 54

\bibitem[{{Sandrinelli} {et~al.}(2014){Sandrinelli}, {Covino}, \&
  {Treves}}]{2014ApJ...793L...1S}
{Sandrinelli}, A., {Covino}, S., \& {Treves}, A. 2014, \apjl, 793, L1

\bibitem[{{Sandrinelli} {et~al.}(2016{\natexlab{b}}){Sandrinelli}, {Covino}, \&
  {Treves}}]{2016ApJ...820...20S}
{Sandrinelli}, A., {Covino}, S., \& {Treves}, A. 2016{\natexlab{b}}, \apj, 820,
  20

\bibitem[{{Scargle}(1982)}]{1982ApJ...263..835S}
{Scargle}, J.~D. 1982, \apj, 263, 835

\bibitem[{{Schlegel} {et~al.}(1998){Schlegel}, {Finkbeiner}, \&
  {Davis}}]{1998ApJ...500..525S}
{Schlegel}, D.~J., {Finkbeiner}, D.~P., \& {Davis}, M. 1998, \apj, 500, 525

\bibitem[{{Sillanp\"a\"a} {et~al.}(1988){Sillanp\"a\"a}, {Haarala}, {Valtonen},
  {Sundelius}, \& {Byrd}}]{1988ApJ...325..628S}
{Sillanp\"a\"a}, A., {Haarala}, S., {Valtonen}, M.~J., {Sundelius}, B., \&
  {Byrd}, G.~G. 1988, \apj, 325, 628

\bibitem[{{Smith} {et~al.}(1991){Smith}, {Jannuzi}, \&
  {Elston}}]{1991ApJS...77...67S}
{Smith}, P.~S., {Jannuzi}, B.~T., \& {Elston}, R. 1991, \apjs, 77, 67

\bibitem[{{Sobolewska} {et~al.}(2014){Sobolewska}, {Siemiginowska}, {Kelly}, \&
  {Nalewajko}}]{2014ApJ...786..143S}
{Sobolewska}, M.~A., {Siemiginowska}, A., {Kelly}, B.~C., \& {Nalewajko}, K.
  2014, \apj, 786, 143

\bibitem[{{Tagliaferri} {et~al.}(2008){Tagliaferri}, {Foschini}, {Ghisellini},
  {Maraschi}, {Tosti}, {Albert}, {Aliu}, {Anderhub}, {Antoranz}, {Baixeras},
  {Barrio}, {Bartko}, {Bastieri}, {Becker}, {Bednarek}, {Bedyugin}, {Berger},
  {Bigongiari}, {Biland}, {Bock}, {Bordas}, {Bosch-Ramon}, {Bretz},
  {Britvitch}, {Camara}, {Carmona}, {Chilingarian}, {Coarasa}, {Commichau},
  {Contreras}, {Cortina}, {Costado}, {Curtef}, {Danielyan}, {Dazzi}, {De
  Angelis}, {Delgado}, {de los Reyes}, {De Lotto}, {Dorner}, {Doro}, {Errando},
  {Fagiolini}, {Ferenc}, {Fern{\'a}ndez}, {Firpo}, {Fonseca}, {Font}, {Fuchs},
  {Galante}, {Garc{\'{\i}}a-L{\'o}pez}, {Garczarczyk}, {Gaug}, {Giller},
  {Goebel}, {Hakobyan}, {Hayashida}, {Hengstebeck}, {Herrero}, {H{\"o}hne},
  {Hose}, {Huber}, {Hsu}, {Jacon}, {Jogler}, {Kosyra}, {Kranich}, {Kritzer},
  {Laille}, {Lindfors}, {Lombardi}, {Longo}, {L{\'o}pez}, {Lorenz}, {Majumdar},
  {Maneva}, {Mannheim}, {Mariotti}, {Mart{\'{\i}}nez}, {Mazin}, {Merck},
  {Meucci}, {Meyer}, {Miranda}, {Mirzoyan}, {Mizobuchi}, {Moralejo}, {Nieto},
  {Nilsson}, {Ninkovic}, {O{\~n}a-Wilhelmi}, {Otte}, {Oya}, {Panniello},
  {Paoletti}, {Paredes}, {Pasanen}, {Pascoli}, {Pauss}, {Pegna}, {Persic},
  {Peruzzo}, {Piccioli}, {Prandini}, {Puchades}, {Raymers}, {Rhode},
  {Rib{\'o}}, {Rico}, {Rissi}, {Robert}, {R{\"u}gamer}, {Saggion}, {Saito},
  {S{\'a}nchez}, {Sartori}, {Scalzotto}, {Scapin}, {Schmitt}, {Schweizer},
  {Shayduk}, {Shinozaki}, {Shore}, {Sidro}, {Sillanp{\"a}{\"a}}, {Sobczynska},
  {Spanier}, {Stamerra}, {Stark}, {Takalo}, {Tavecchio}, {Temnikov}, {Tescaro},
  {Teshima}, {Torres}, {Turini}, {Vankov}, {Venturini}, {Vitale}, {Wagner},
  {Wibig}, {Wittek}, {Zandanel}, {Zanin}, {Zapatero}, \& {MAGIC
  Collaboration''}}]{2008ApJ...679.1029T}
{Tagliaferri}, G., {Foschini}, L., {Ghisellini}, G., {et~al.} 2008, \apj, 679,
  1029

\bibitem[{{Takalo} {et~al.}(2010){Takalo}, {Hagen-Thorn}, {Hagen-Thorn},
  {Ciprini}, \& {Sillanp{\"a}{\"a}}}]{2010A&A...517A..63T}
{Takalo}, L.~O., {Hagen-Thorn}, V.~A., {Hagen-Thorn}, E.~I., {Ciprini}, S., \&
  {Sillanp{\"a}{\"a}}, A. 2010, \aap, 517, A63

\bibitem[{{Takalo} {et~al.}(2008){Takalo}, {Nilsson}, {Lindfors},
  {Sillanp{\"a}{\"a}}, {Berdyugin}, \& {Pasanen}}]{2008AIPC.1085..705T}
{Takalo}, L.~O., {Nilsson}, K., {Lindfors}, E., {et~al.} 2008, in American
  Institute of Physics Conference Series, Vol. 1085, American Institute of
  Physics Conference Series, ed. F.~A. {Aharonian}, W.~{Hofmann}, \&
  F.~{Rieger}, 705--707

\bibitem[{{Timmer} \& {Koenig}(1995)}]{1995A&A...300..707T}
{Timmer}, J. \& {Koenig}, M. 1995, \aap, 300, 707

\bibitem[{{Uttley} {et~al.}(2002){Uttley}, {McHardy}, \&
  {Papadakis}}]{2002MNRAS.332..231U}
{Uttley}, P., {McHardy}, I.~M., \& {Papadakis}, I.~E. 2002, \mnras, 332, 231

\bibitem[{{Valtaoja} {et~al.}(1999){Valtaoja}, {L{\"a}hteenm{\"a}ki},
  {Ter{\"a}sranta}, \& {Lainela}}]{1999ApJS..120...95V}
{Valtaoja}, E., {L{\"a}hteenm{\"a}ki}, A., {Ter{\"a}sranta}, H., \& {Lainela},
  M. 1999, \apjs, 120, 95

\bibitem[{{Valtonen} {et~al.}(2008){Valtonen}, {Lehto}, {Nilsson}, {Heidt},
  {Takalo}, {Sillanp{\"a}{\"a}}, {Villforth}, {Kidger}, {Poyner}, {Pursimo},
  {Zola}, {Wu}, {Zhou}, {Sadakane}, {Drozdz}, {Koziel}, {Marchev}, {Ogloza},
  {Porowski}, {Siwak}, {Stachowski}, {Winiarski}, {Hentunen}, {Nissinen},
  {Liakos}, \& {Dogru}}]{2008Natur.452..851V}
{Valtonen}, M.~J., {Lehto}, H.~J., {Nilsson}, K., {et~al.} 2008, \nat, 452, 851

\bibitem[{{Valtonen} {et~al.}(2006){Valtonen}, {Lehto}, {Sillanp{\"a}{\"a}},
  {Nilsson}, {Mikkola}, {Hudec}, {Basta}, {Ter{\"a}sranta}, {Haque}, \&
  {Rampadarath}}]{2006ApJ...646...36V}
{Valtonen}, M.~J., {Lehto}, H.~J., {Sillanp{\"a}{\"a}}, A., {et~al.} 2006,
  \apj, 646, 36

\bibitem[{{Valtonen} {et~al.}(2009){Valtonen}, {Nilsson}, {Villforth}, {Lehto},
  {Takalo}, {Lindfors}, {Sillanp{\"a}{\"a}}, {Hentunen}, {Mikkola}, {Zola},
  {Drozdz}, {Koziel}, {Ogloza}, {Kurpinska-Winiarska}, {Siwak}, {Winiarski},
  {Heidt}, {Kidger}, {Pursimo}, {Wu}, {Zhou}, {Sadakane}, {Marchev},
  {Nissinen}, {Niarchos}, {Liakos}, {Gazeas}, {Dogru}, {Poyner}, {Dietrich},
  {Assef}, {Atlee}, {Bird}, {DePoy}, {Eastman}, {Peeples}, {Prieto}, {Watson},
  {Yee}, {Mattingly}, \& {Ohlert}}]{2009ApJ...698..781V}
{Valtonen}, M.~J., {Nilsson}, K., {Villforth}, C., {et~al.} 2009, \apj, 698,
  781

\bibitem[{{Valtonen} {et~al.}(2016){Valtonen}, {Zola}, {Ciprini}, {Gopakumar},
  {Matsumoto}, {Sadakane}, {Kidger}, {Gazeas}, {Nilsson}, {Berdyugin},
  {Piirola}, {Jermak}, {Baliyan}, {Alicavus}, {Boyd}, {Campas Torrent},
  {Campos}, {Carrillo G{\'o}mez}, {Caton}, {Chavushyan}, {Dalessio}, {Debski},
  {Dimitrov}, {Drozdz}, {Er}, {Erdem}, {Escartin P{\'e}rez}, {Fallah Ramazani},
  {Filippenko}, {Ganesh}, Query Results from~the ADS Database Retrieved
  1~abstracts, {G{\'o}mez Pinilla}, {Gopinathan}, {Haislip}, {Hudec}, {Hurst},
  {Ivarsen}, {Jelinek}, {Joshi}, {Kagitani}, {Kaur}, {Keel}, {LaCluyze}, {Lee},
  {Lindfors}, {Lozano de Haro}, {Moore}, {Mugrauer}, {Naves Nogues}, {Neely},
  {Nelson}, {Ogloza}, {Okano}, {Pandey}, {Perri}, {Pihajoki}, {Poyner},
  {Provencal}, {Pursimo}, {Raj}, {Reichart}, {Reinthal}, {Sadegi}, {Sakanoi},
  {Salto Gonz{\'a}lez}, {Sameer}, {Schweyer}, {Siwak}, {Sold{\'a}n Alfaro},
  {Sonbas}, {Steele}, {Stocke}, {Strobl}, {Takalo}, {Tomov}, {Tremosa Espasa},
  {Valdes}, {Valero P{\'e}rez}, {Verrecchia}, {Webb}, {Yoneda}, {Zejmo},
  {Zheng}, {Telting}, {Saario}, {Reynolds}, {Kvammen}, {Gafton}, {Karjalainen},
  {Harmanen}, \& {Blay}}]{2016ApJ...819L..37V}
{Valtonen}, M.~J., {Zola}, S., {Ciprini}, S., {et~al.} 2016, \apjl, 819, L37

\bibitem[{{Vaughan}(2005)}]{2005A&A...431..391V}
{Vaughan}, S. 2005, \aap, 431, 391

\bibitem[{{Vaughan}(2010)}]{2010MNRAS.402..307V}
{Vaughan}, S. 2010, \mnras, 402, 307

\bibitem[{{Vaughan} {et~al.}(2016){Vaughan}, {Uttley}, {Markowitz},
  {Huppenkothen}, {Middleton}, {Alston}, {Scargle}, \&
  {Farr}}]{2016MNRAS.461.3145V}
{Vaughan}, S., {Uttley}, P., {Markowitz}, A.~G., {et~al.} 2016, \mnras, 461,
  3145

\bibitem[{{Villata} {et~al.}(2004){Villata}, {Raiteri}, {Aller}, {Aller},
  {Ter{\"a}sranta}, {Koivula}, {Wiren}, {Kurtanidze}, {Nikolashvili},
  {Ibrahimov}, {Papadakis}, {Tosti}, {Hroch}, {Takalo}, {Sillanp{\"a}{\"a}},
  {Hagen-Thorn}, {Larionov}, {Schwartz}, {Basler}, {Brown}, \&
  {Balonek}}]{2004A&A...424..497V}
{Villata}, M., {Raiteri}, C.~M., {Aller}, H.~D., {et~al.} 2004, \aap, 424, 497

\bibitem[{{Villata} {et~al.}(1997){Villata}, {Raiteri}, {Ghisellini}, {de
  Francesco}, {Bosio}, {Latini}, {Bucciarelli}, {Chiaberge}, {Chiumiento},
  {Cora}, {Lanteri}, {Lattanzi}, {Massone}, {Peila}, {Racioppi}, {Smart}, \&
  {Scaltriti}}]{1997A&AS..121..119V}
{Villata}, M., {Raiteri}, C.~M., {Ghisellini}, G., {et~al.} 1997, \aaps, 121,
  119

\bibitem[{{Villata} {et~al.}(1998){Villata}, {Raiteri}, {Lanteri}, {Sobrito},
  \& {Cavallone}}]{1998A&AS..130..305V}
{Villata}, M., {Raiteri}, C.~M., {Lanteri}, L., {Sobrito}, G., \& {Cavallone},
  M. 1998, \aaps, 130, 305

\bibitem[{{Villata} {et~al.}(2008){Villata}, {Raiteri}, {Larionov},
  {Kurtanidze}, {Nilsson}, {Aller}, {Tornikoski}, {Volvach}, {Aller},
  {Arkharov}, {Bach}, {Beltrame}, {Bhatta}, {Buemi}, {B{\"o}ttcher},
  {Calcidese}, {Carosati}, {Castro-Tirado}, {da Rio}, {di Paola}, {Dolci},
  {Forn{\'e}}, {Frasca}, {Hagen-Thorn}, {Heidt}, {Hiriart}, {Jel{\'{\i}}nek},
  {Kimeridze}, {Konstantinova}, {Kopatskaya}, {Lanteri}, {Leto}, {Ligustri},
  {Lindfors}, {L{\"a}hteenm{\"a}ki}, {Marilli}, {Nieppola}, {Nikolashvili},
  {Pasanen}, {Ragozzine}, {Ros}, {Sigua}, {Smart}, {Sorcia}, {Takalo},
  {Tavani}, {Trigilio}, {Turchetti}, {Uckert}, {Umana}, {Vercellone}, \&
  {Webb}}]{2008A&A...481L..79V}
{Villata}, M., {Raiteri}, C.~M., {Larionov}, V.~M., {et~al.} 2008, \aap, 481,
  L79

\bibitem[{{Villata} {et~al.}(2009){Villata}, {Raiteri}, {Larionov},
  {Nikolashvili}, {Aller}, {Bach}, {Carosati}, {Hroch}, {Ibrahimov}, {Jorstad},
  {Kovalev}, {L{\"a}hteenm{\"a}ki}, {Nilsson}, {Ter{\"a}sranta}, {Tosti},
  {Aller}, {Arkharov}, {Berdyugin}, {Boltwood}, {Buemi}, {Casas}, {Charlot},
  {Coloma}, {di Paola}, {di Rico}, {Kimeridze}, {Konstantinova}, {Kopatskaya},
  {Kovalev}, {Kurtanidze}, {Lanteri}, {Larionova}, {Larionova}, {Le Campion},
  {Leto}, {Lindfors}, {Marscher}, {Marshall}, {McFarland}, {McHardy}, {Miller},
  {Nucciarelli}, {Osterman}, {Pasanen}, {Pursimo}, {Ros}, {Sadun}, {Sigua},
  {Sixtova}, {Takalo}, {Tornikoski}, {Trigilio}, {Umana}, {Xie}, {Zhang}, \&
  {Zhou}}]{2009A&A...501..455V}
{Villata}, M., {Raiteri}, C.~M., {Larionov}, V.~M., {et~al.} 2009, \aap, 501,
  455

\bibitem[{{Villforth} {et~al.}(2010){Villforth}, {Nilsson}, {Heidt}, {Takalo},
  {Pursimo}, {Berdyugin}, {Lindfors}, {Pasanen}, {Winiarski}, {Drozdz},
  {Ogloza}, {Kurpinska-Winiarska}, {Siwak}, {Koziel-Wierzbowska}, {Porowski},
  {Kuzmicz}, {Krzesinski}, {Kundera}, {Wu}, {Zhou}, {Efimov}, {Sadakane},
  {Kamada}, {Ohlert}, {Hentunen}, {Nissinen}, {Dietrich}, {Assef}, {Atlee},
  {Bird}, {Depoy}, {Eastman}, {Peeples}, {Prieto}, {Watson}, {Yee}, {Liakos},
  {Niarchos}, {Gazeas}, {Dogru}, {Donmez}, {Marchev}, {Coggins-Hill},
  {Mattingly}, {Keel}, {Haque}, {Aungwerojwit}, \&
  {Bergvall}}]{2010MNRAS.402.2087V}
{Villforth}, C., {Nilsson}, K., {Heidt}, J., {et~al.} 2010, \mnras, 402, 2087

\end{thebibliography}

\begin{appendix}

\section{Light curves}

\begin{figure*}
\includegraphics[width=\textwidth]{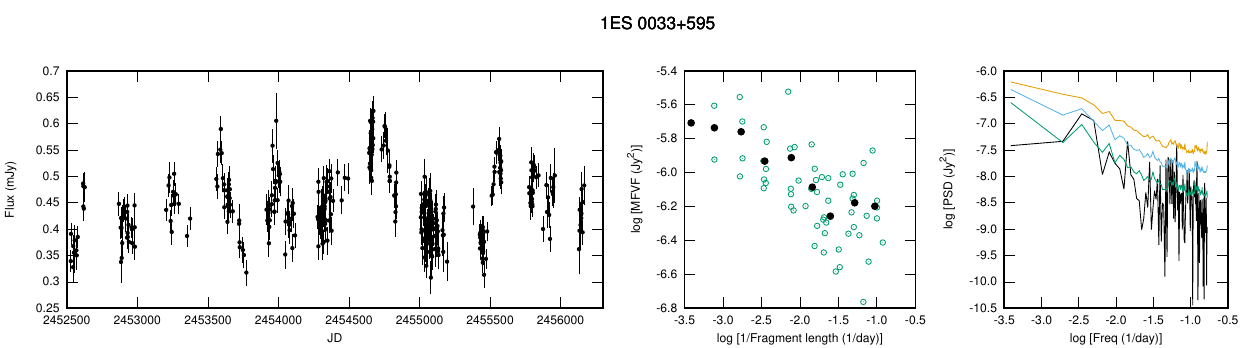}
\caption{\label{ekavalo} {\em Left}: Observed light curve. {\em
    Middle}: multiple fragments variance function (MFVF). Grey dots
  show the unbinned values and black dots the binned values. {\em
    Right}: Power spectral density (PSD) together with 67, 95 and 99.9
  percent limits for a single frequency, taking into account the
  number of frequencies covered.  The two rightmost panels were
  computed from data transformed to the rest frame.  }
\end{figure*}

\begin{figure*}
\includegraphics[width=\textwidth]{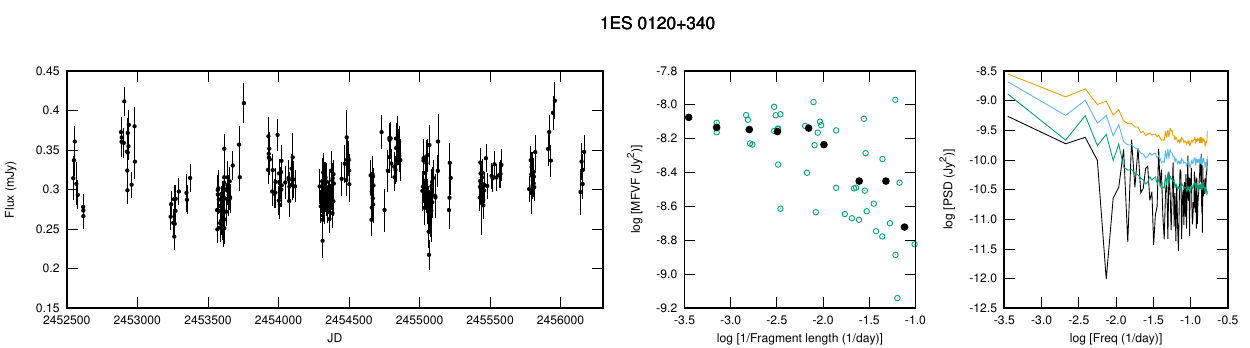}
\caption{See the caption of Fig. \ref{ekavalo}}
\end{figure*}

\begin{figure*}
\includegraphics[width=\textwidth]{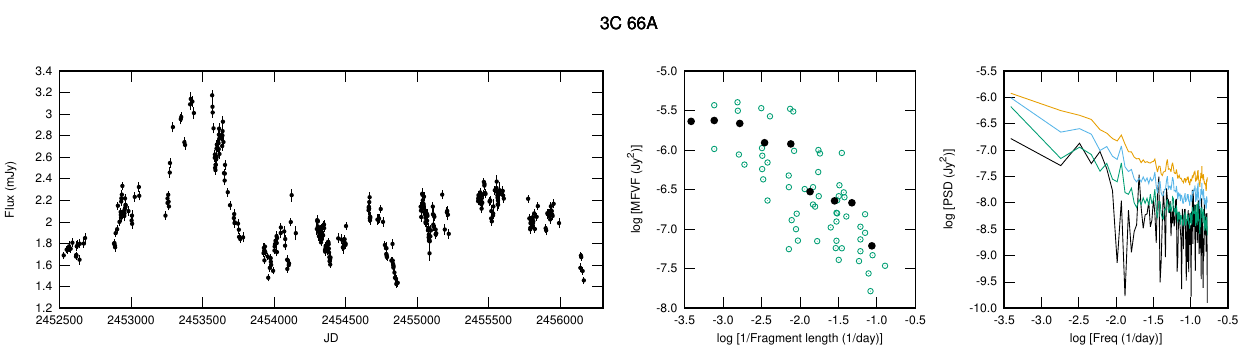}
\caption{See the caption of Fig. \ref{ekavalo}}
\end{figure*}

\begin{figure*}
\includegraphics[width=\textwidth]{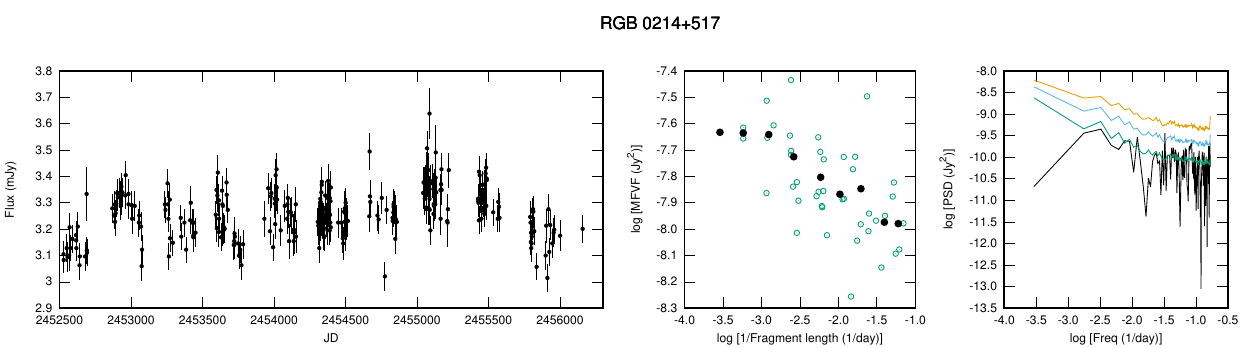}
\caption{See the caption of Fig. \ref{ekavalo}}
\end{figure*}

\begin{figure*}
\includegraphics[width=\textwidth]{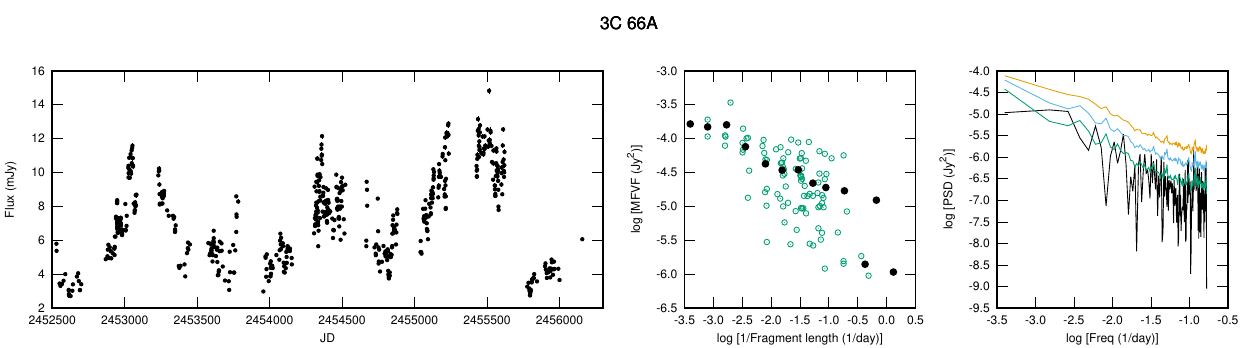}
\caption{See the caption of Fig. \ref{ekavalo}}
\end{figure*}

\begin{figure*}
\includegraphics[width=\textwidth]{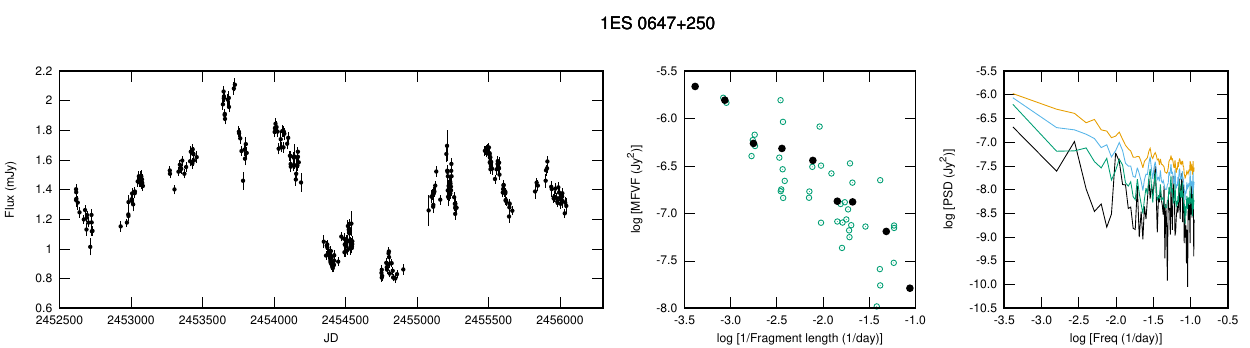}
\caption{See the caption of Fig. \ref{ekavalo}}
\end{figure*}

\begin{figure*}
\includegraphics[width=\textwidth]{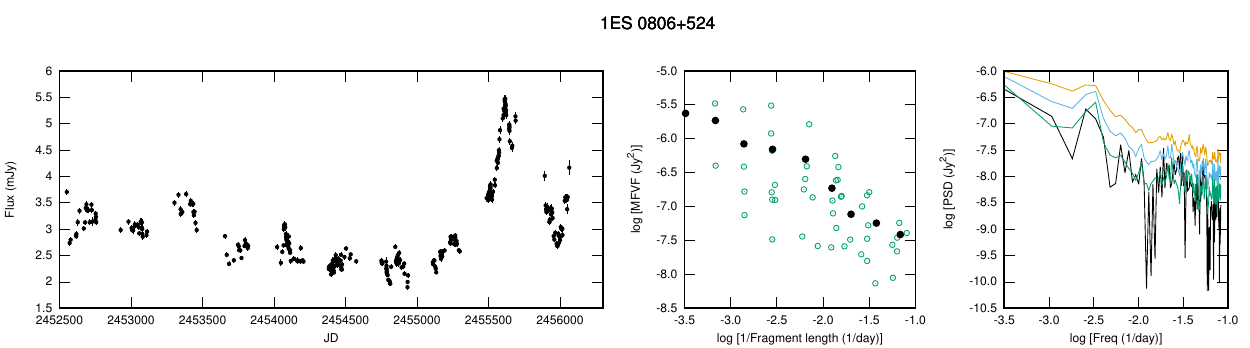}
\caption{See the caption of Fig. \ref{ekavalo}}
\end{figure*}

\begin{figure*}
\includegraphics[width=\textwidth]{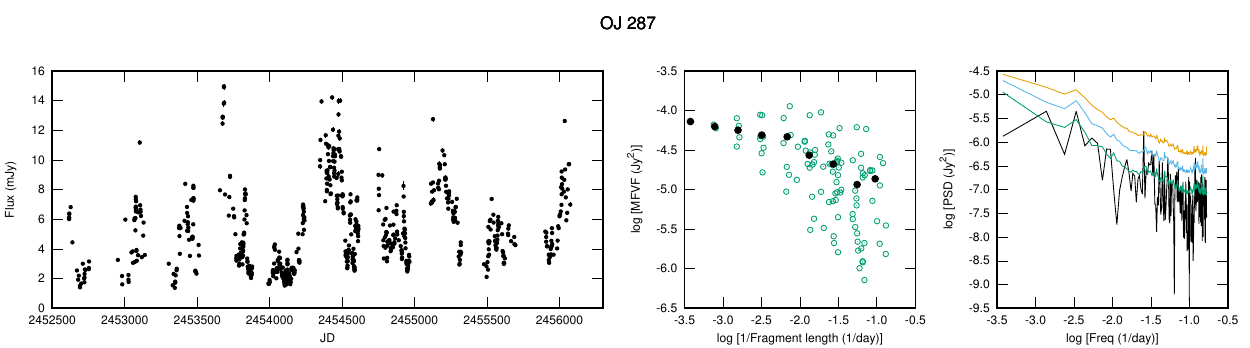}
\caption{See the caption of Fig. \ref{ekavalo}}
\end{figure*}

\begin{figure*}
\includegraphics[width=\textwidth]{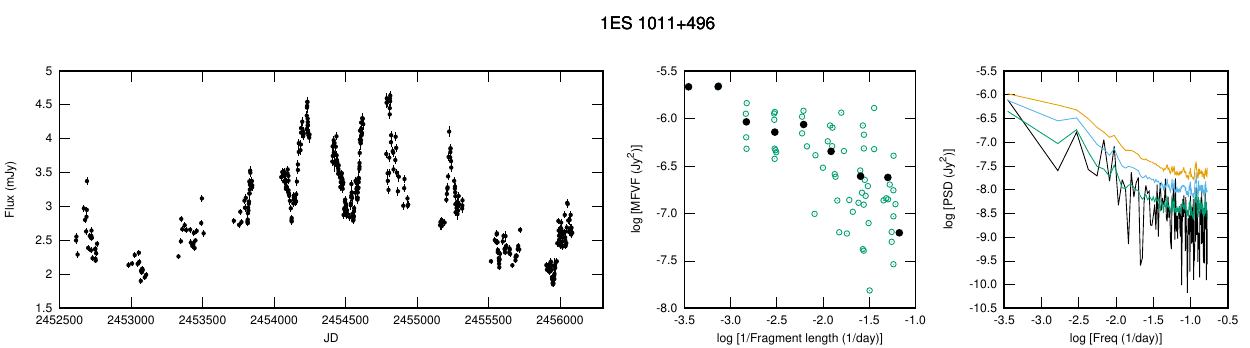}
\caption{See the caption of Fig. \ref{ekavalo}}
\end{figure*}

\begin{figure*}
\includegraphics[width=\textwidth]{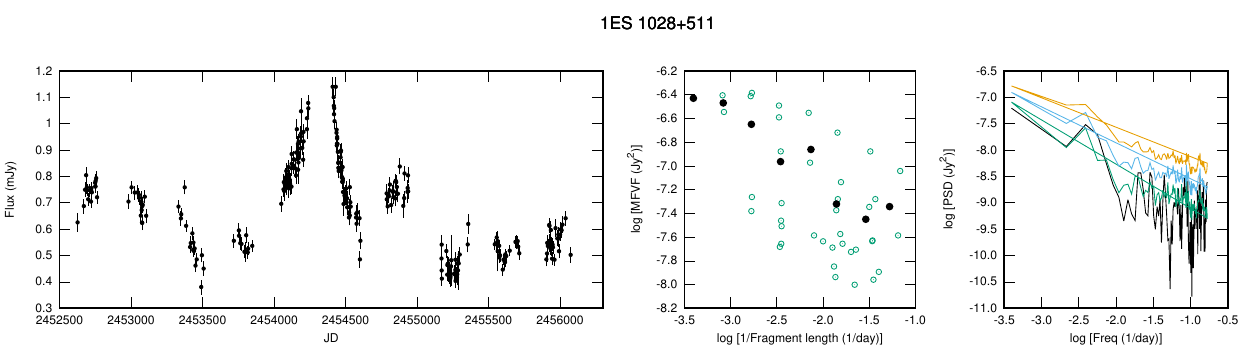}
\caption{See the caption of Fig. \ref{ekavalo}}
\end{figure*}

\begin{figure*}
\includegraphics[width=\textwidth]{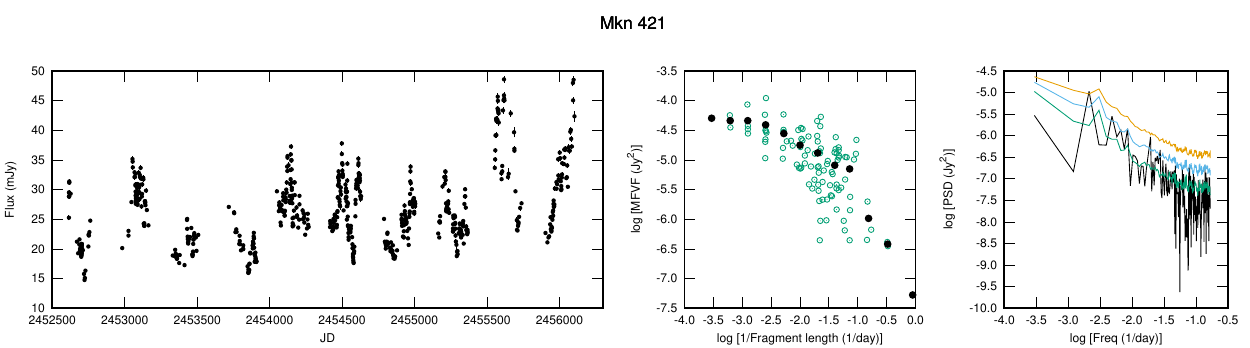}
\caption{See the caption of Fig. \ref{ekavalo}}
\end{figure*}

\begin{figure*}
\includegraphics[width=\textwidth]{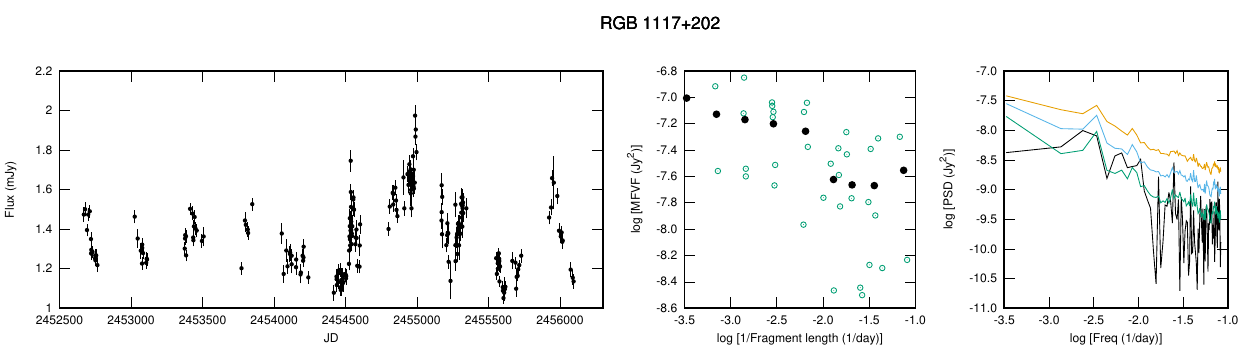}
\caption{See the caption of Fig. \ref{ekavalo}}
\end{figure*}

\begin{figure*}
\includegraphics[width=\textwidth]{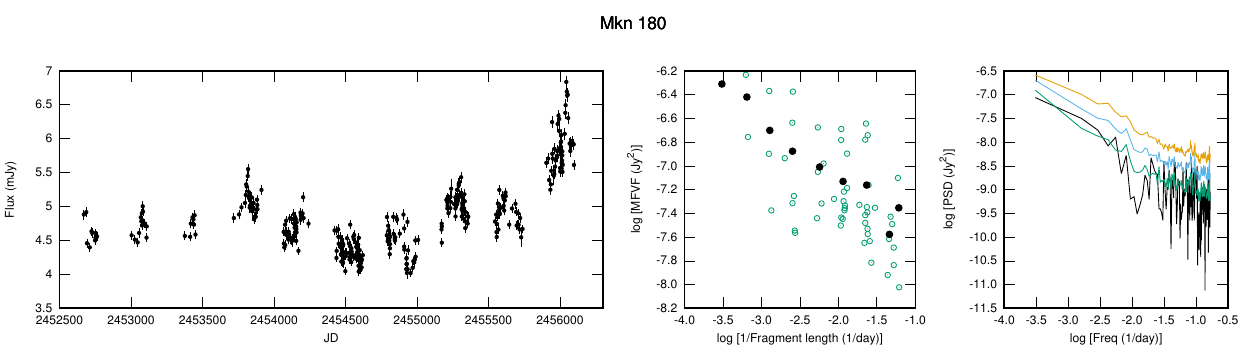}
\caption{See the caption of Fig. \ref{ekavalo}}
\end{figure*}

\begin{figure*}
\includegraphics[width=\textwidth]{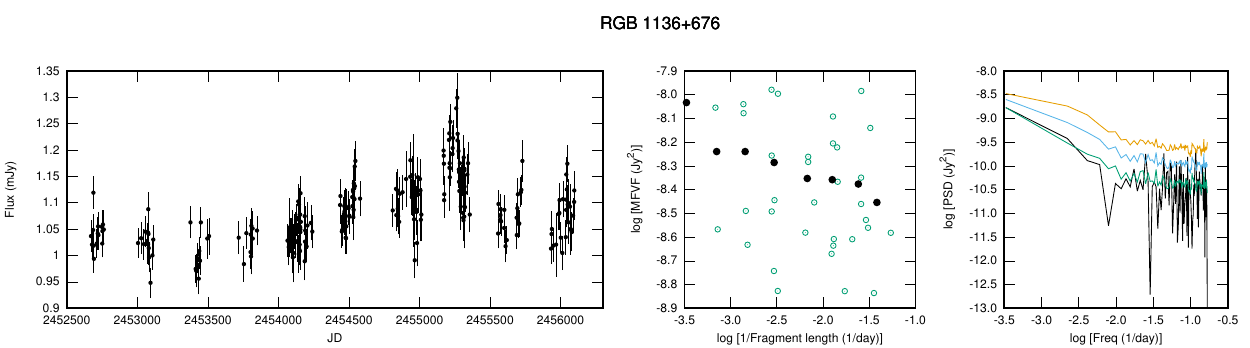}
\caption{See the caption of Fig. \ref{ekavalo}}
\end{figure*}

\begin{figure*}
\includegraphics[width=\textwidth]{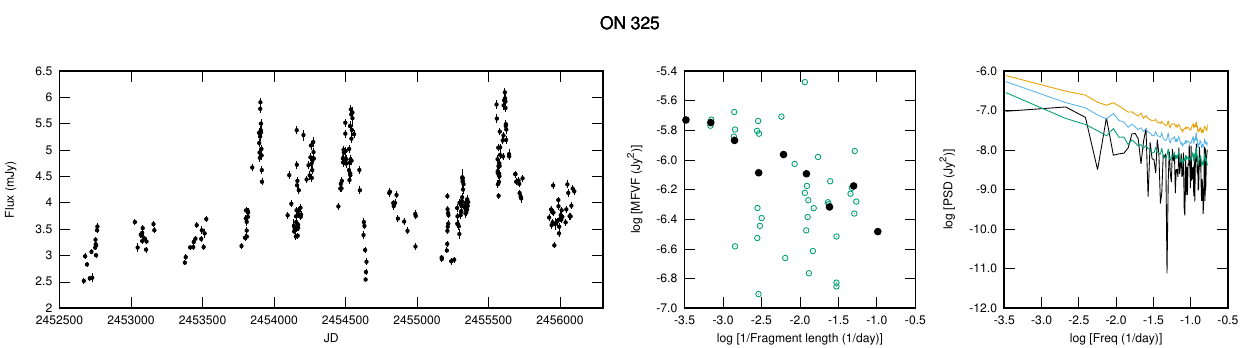}
\caption{See the caption of Fig. \ref{ekavalo}}
\end{figure*}

\begin{figure*}
\includegraphics[width=\textwidth]{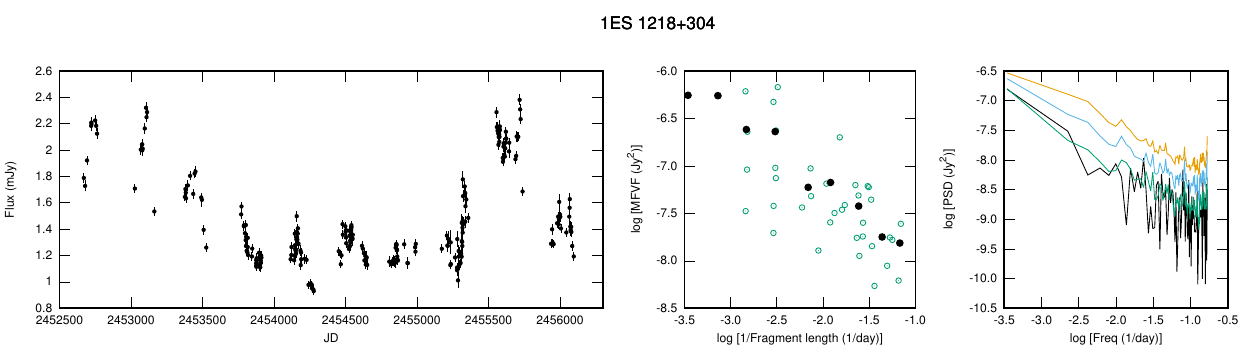}
\caption{See the caption of Fig. \ref{ekavalo}}
\end{figure*}

\begin{figure*}
\includegraphics[width=\textwidth]{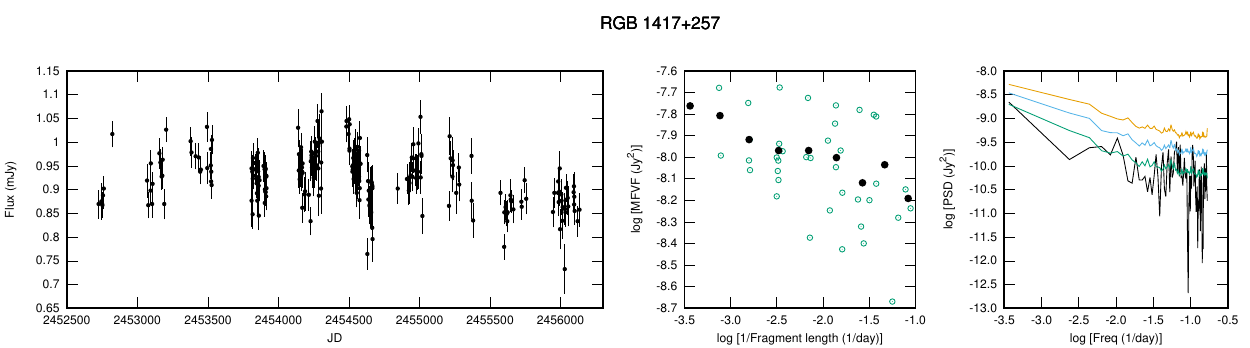}
\caption{See the caption of Fig. \ref{ekavalo}}
\end{figure*}

\begin{figure*}
\includegraphics[width=\textwidth]{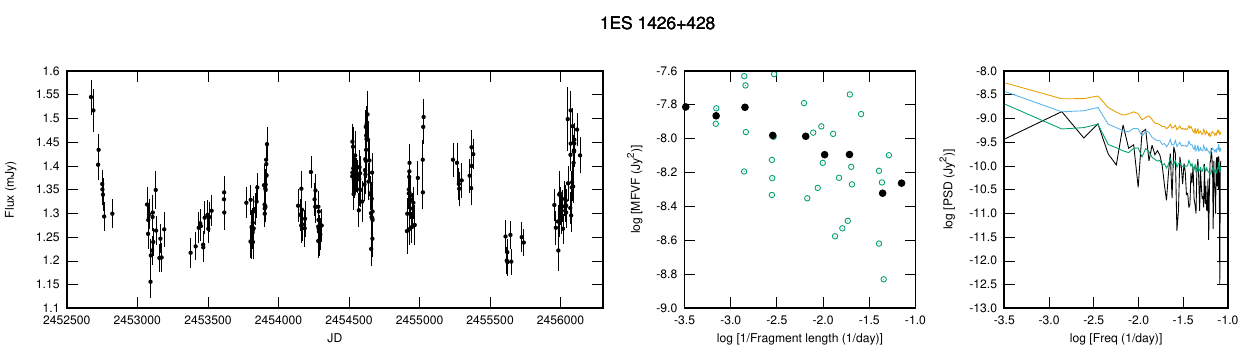}
\caption{See the caption of Fig. \ref{ekavalo}}
\end{figure*}

\begin{figure*}
\includegraphics[width=\textwidth]{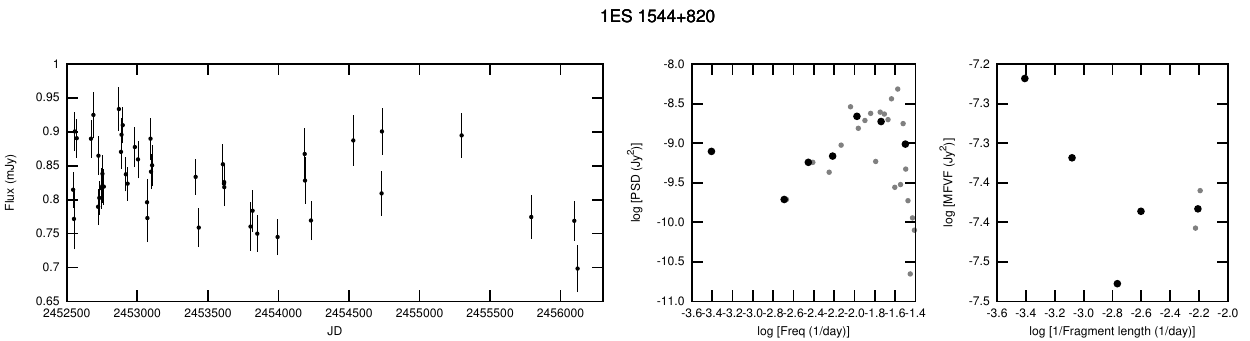}
\caption{See the caption of Fig. \ref{ekavalo}}
\end{figure*}

\begin{figure*}
\includegraphics[width=\textwidth]{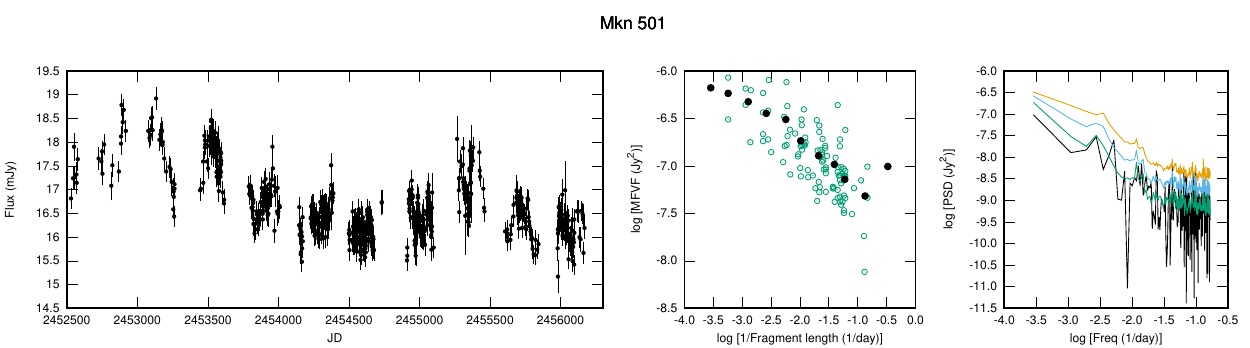}
\caption{See the caption of Fig. \ref{ekavalo}}
\end{figure*}

\begin{figure*}
\includegraphics[width=\textwidth]{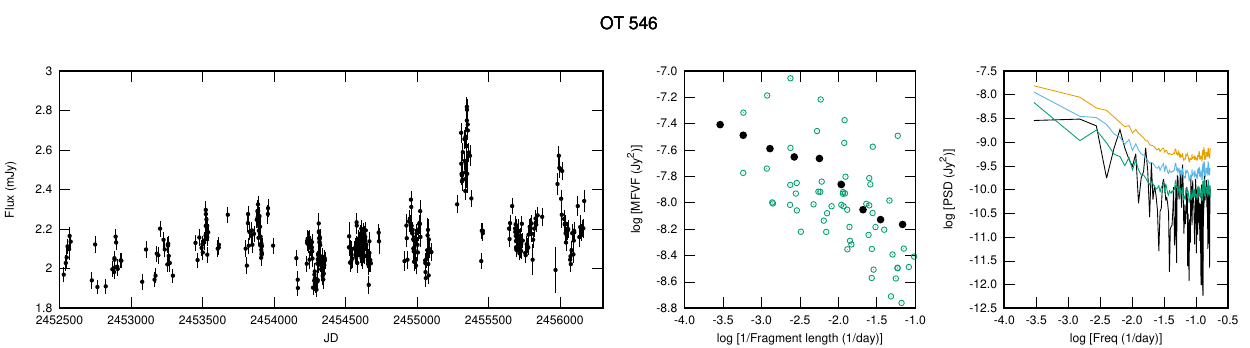}
\caption{See the caption of Fig. \ref{ekavalo}}
\end{figure*}

\begin{figure*}
\includegraphics[width=\textwidth]{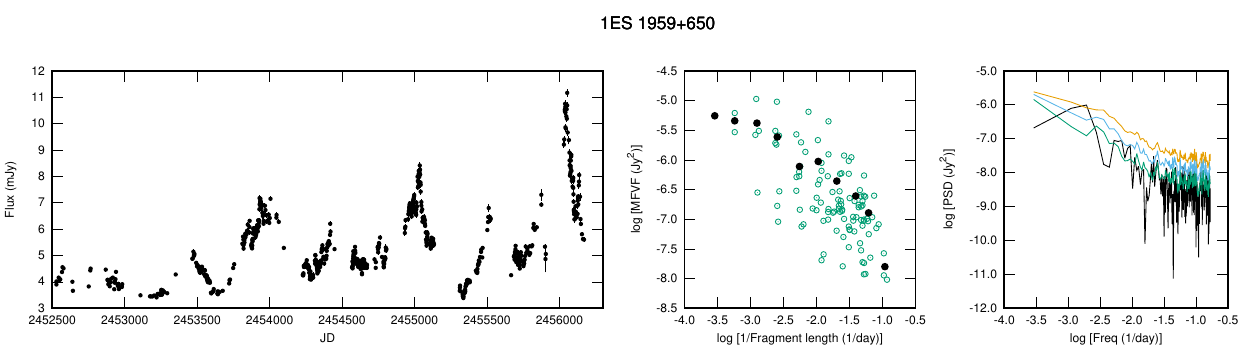}
\caption{See the caption of Fig. \ref{ekavalo}}
\end{figure*}

\begin{figure*}
\includegraphics[width=\textwidth]{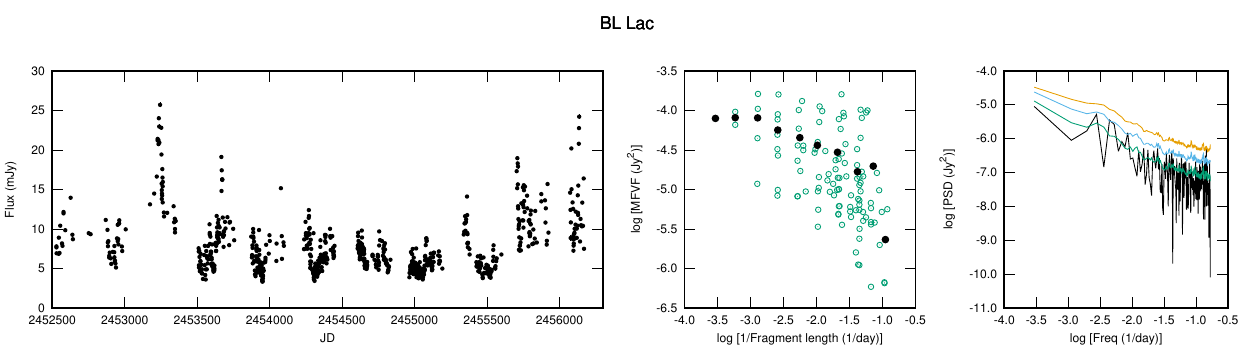}
\caption{See the caption of Fig. \ref{ekavalo}}
\end{figure*}

\begin{figure*}
\includegraphics[width=\textwidth]{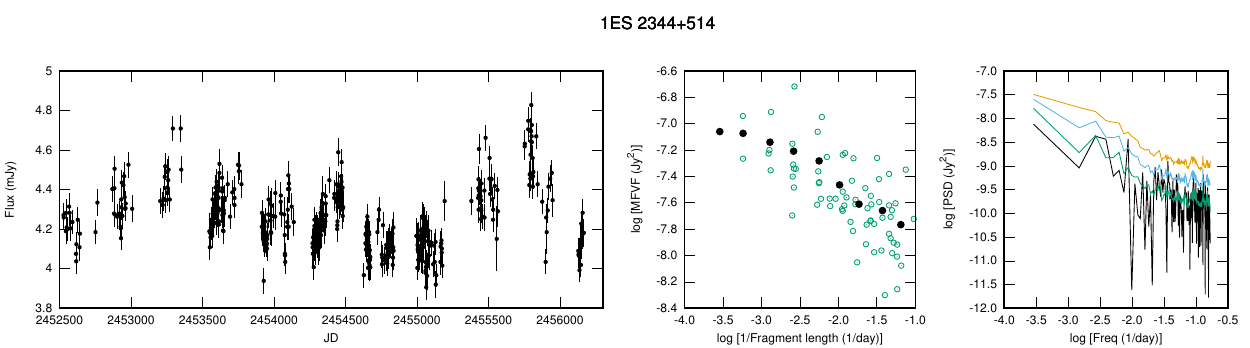}
\caption{See the caption of Fig. \ref{ekavalo}}
\end{figure*}

\begin{figure*}
\includegraphics[width=\textwidth]{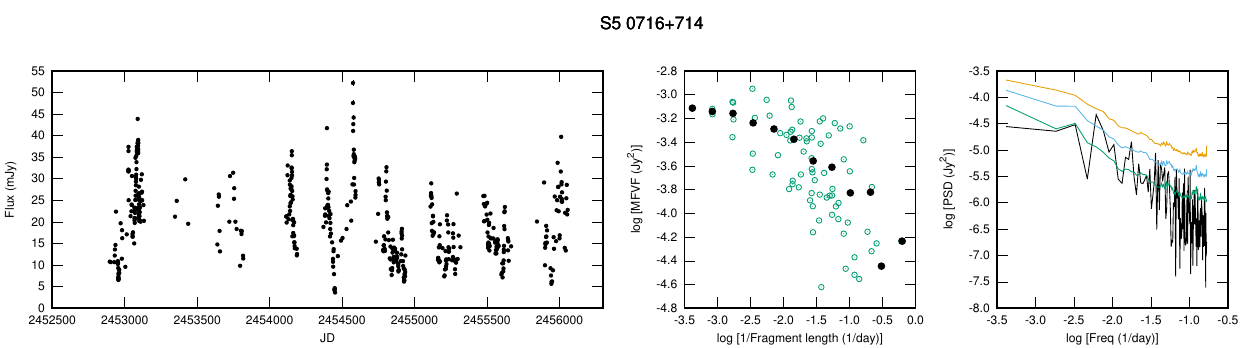}
\caption{See the caption of Fig. \ref{ekavalo}}
\end{figure*}

\begin{figure*}
\includegraphics[width=\textwidth]{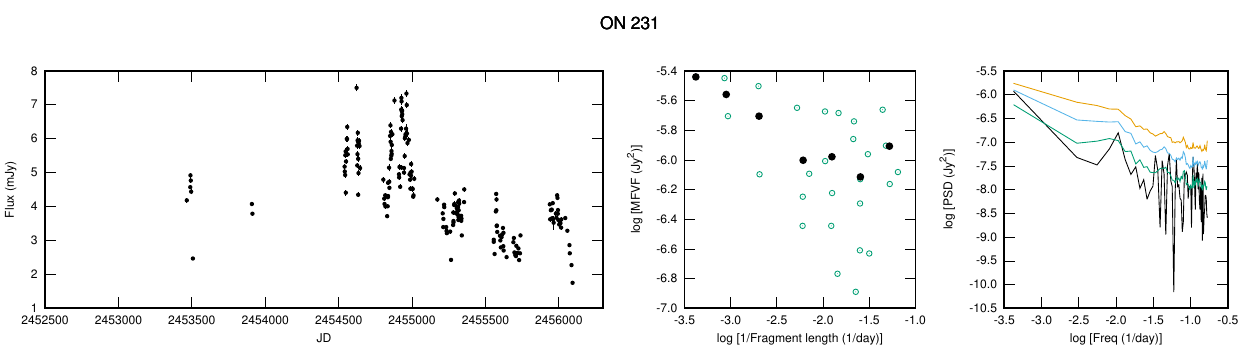}
\caption{See the caption of Fig. \ref{ekavalo}}
\end{figure*}

\begin{figure*}
\includegraphics[width=\textwidth]{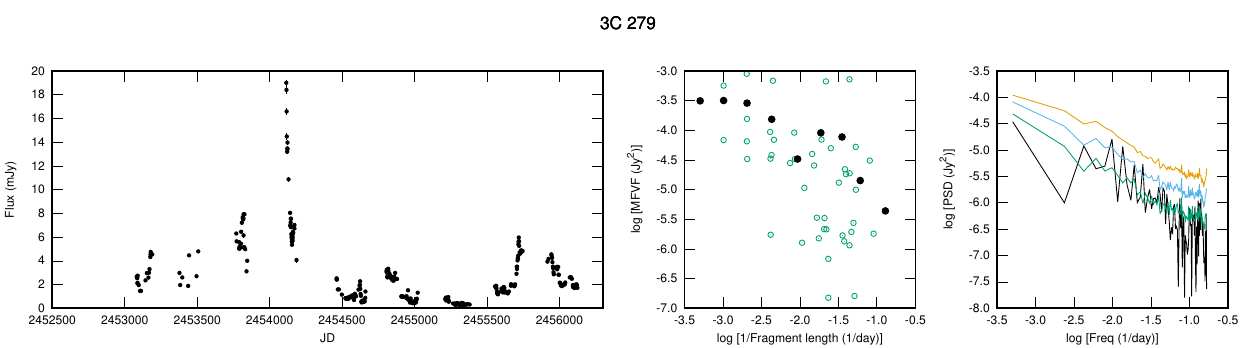}
\caption{See the caption of Fig. \ref{ekavalo}}
\end{figure*}

\begin{figure*}
\includegraphics[width=\textwidth]{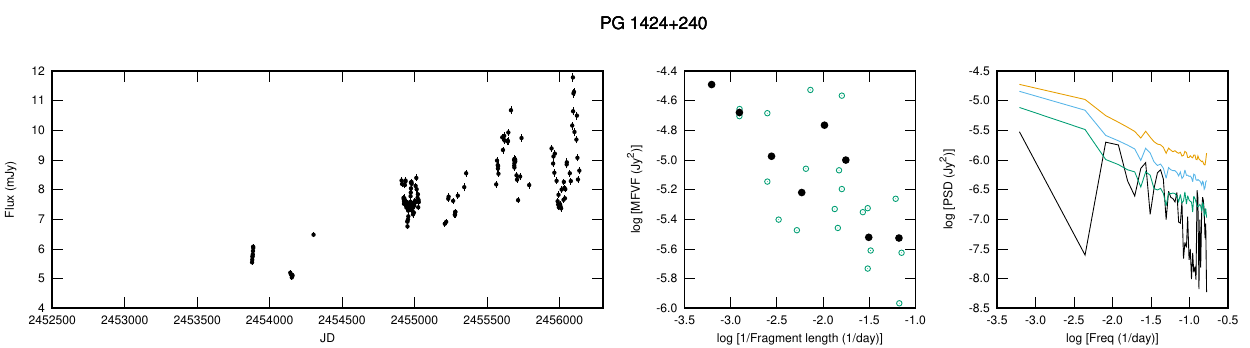}
\caption{See the caption of Fig. \ref{ekavalo}}
\end{figure*}

\begin{figure*}
\includegraphics[width=\textwidth]{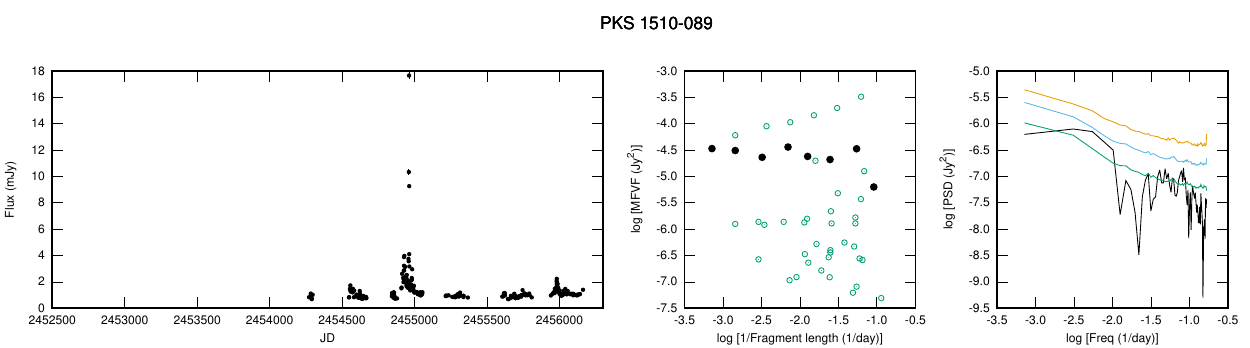}
\caption{See the caption of Fig. \ref{ekavalo}}
\end{figure*}

\begin{figure*}
\includegraphics[width=\textwidth]{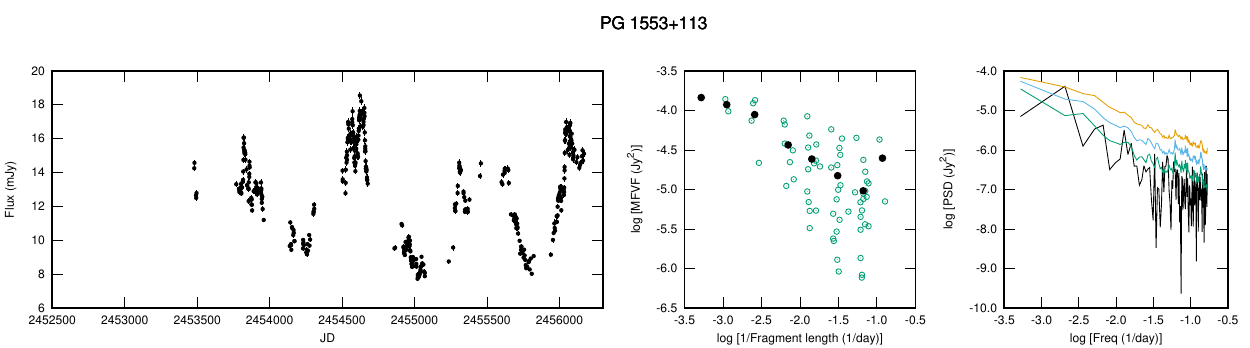}
\caption{See the caption of Fig. \ref{ekavalo}}
\end{figure*}

\begin{figure*}
\includegraphics[width=\textwidth]{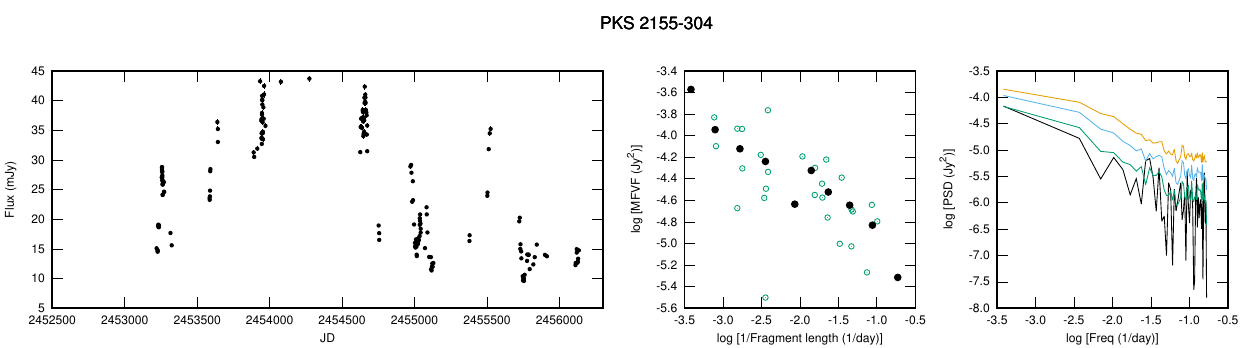}
\caption{\label{vikavalo}See the caption of Fig. \ref{ekavalo}}
\end{figure*}

\section{\label{seds}Spectral energy distributions}

\begin{figure*}
\includegraphics[width=0.45\textwidth]{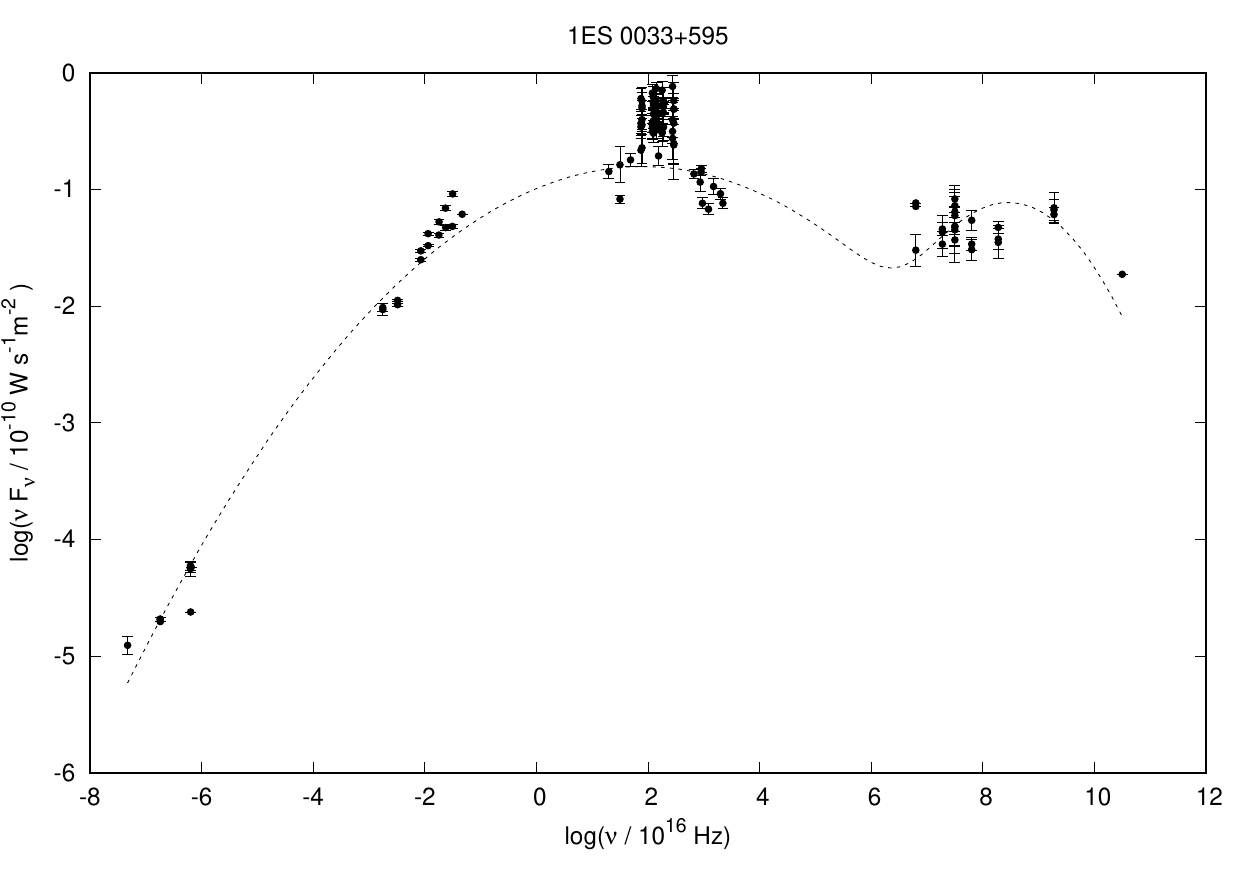}
\includegraphics[width=0.45\textwidth]{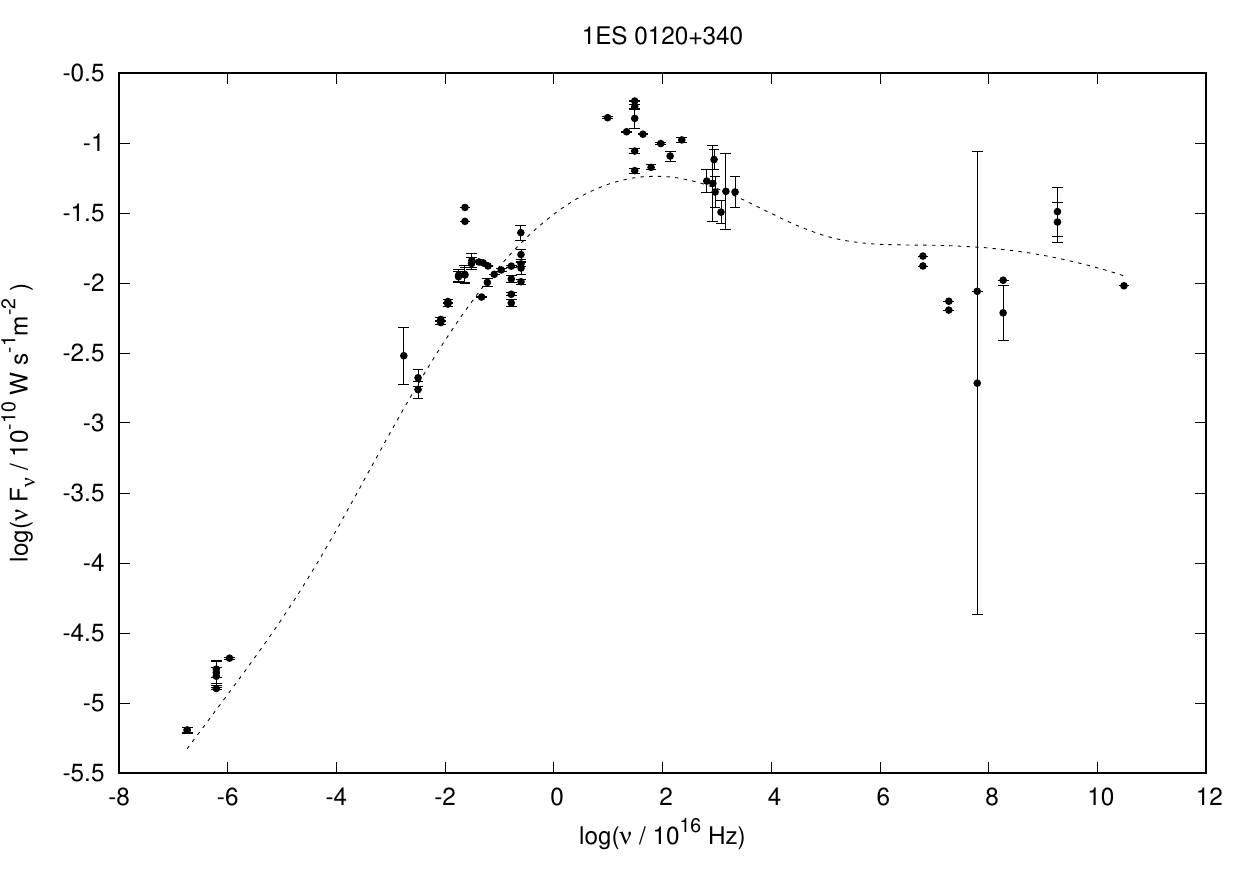}
\caption{\label{sed1}
The archive spectral energy distribution data used to determine
the synchrotron peak frequency. The dotted line shows the best-fit model.}
\end{figure*}

\begin{figure*}
\includegraphics[width=0.45\textwidth]{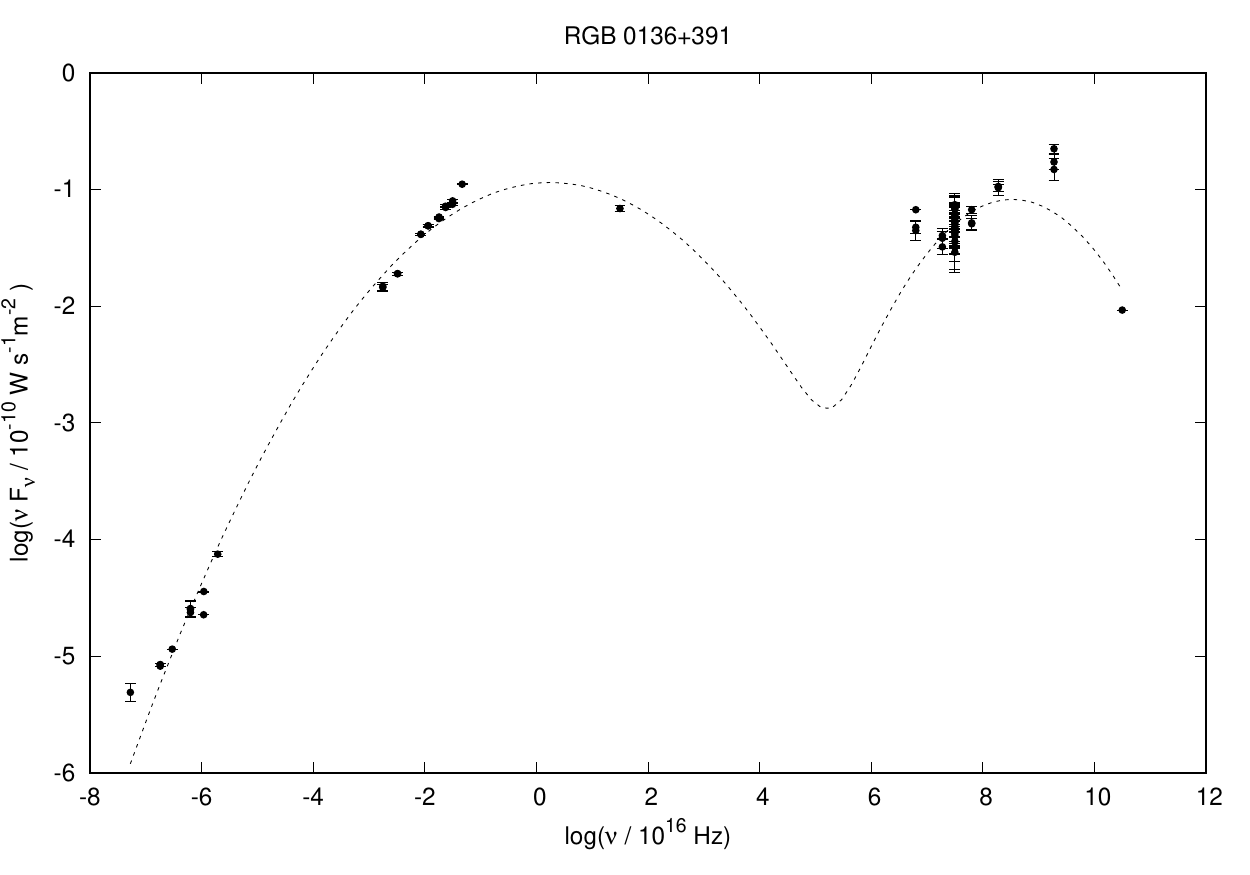}
\includegraphics[width=0.45\textwidth]{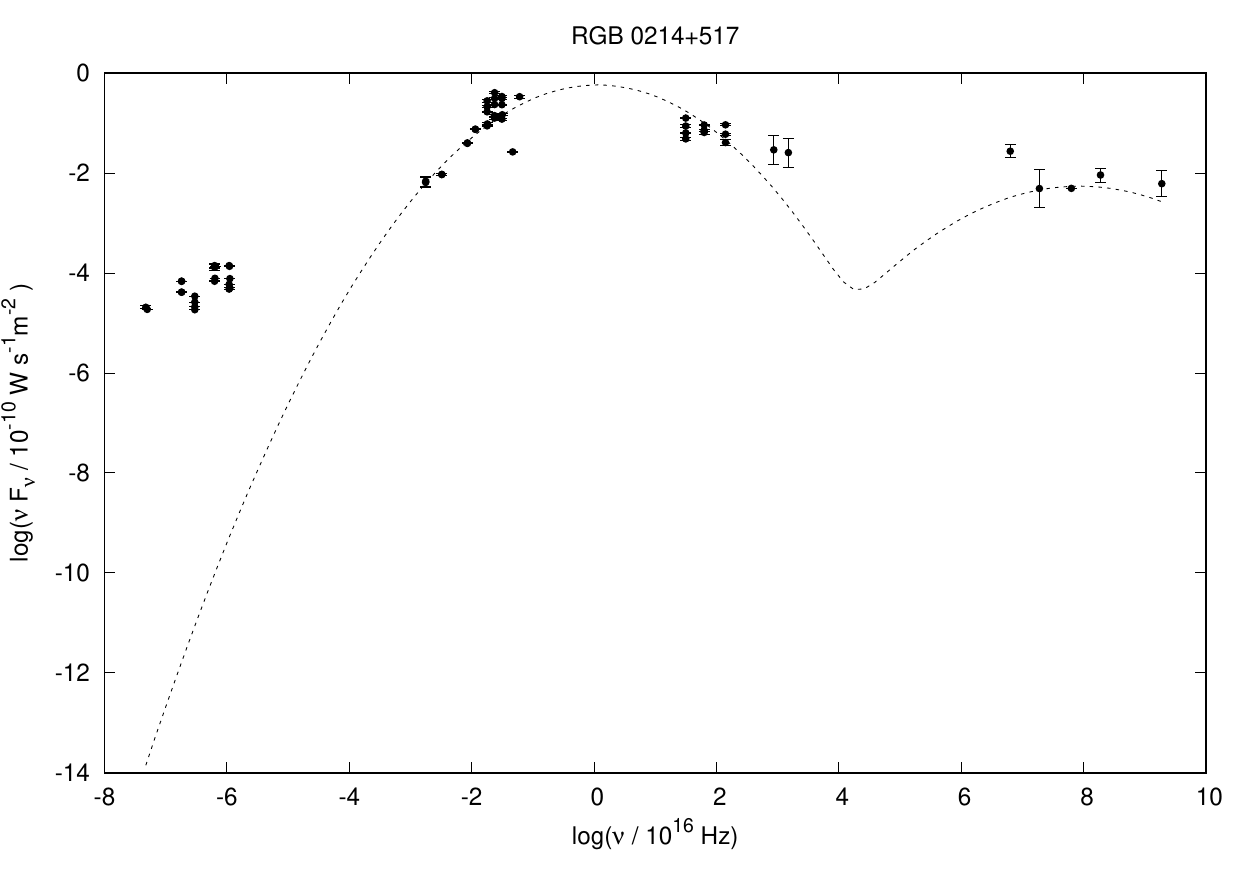}
\caption{See the caption of Fig. \ref{sed1}}
\label{Fig:sed3}
\end{figure*}

\begin{figure*}
\includegraphics[width=0.45\textwidth]{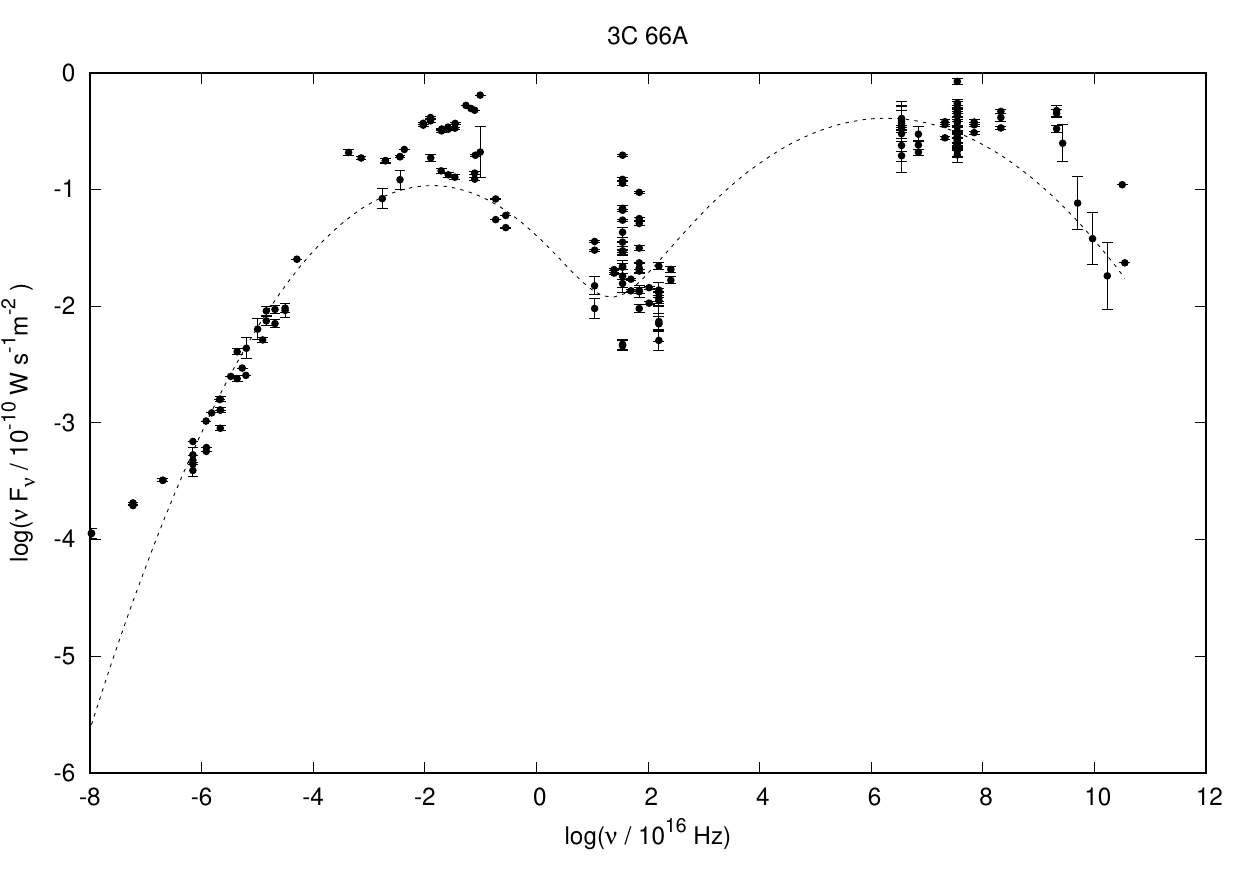}
\includegraphics[width=0.45\textwidth]{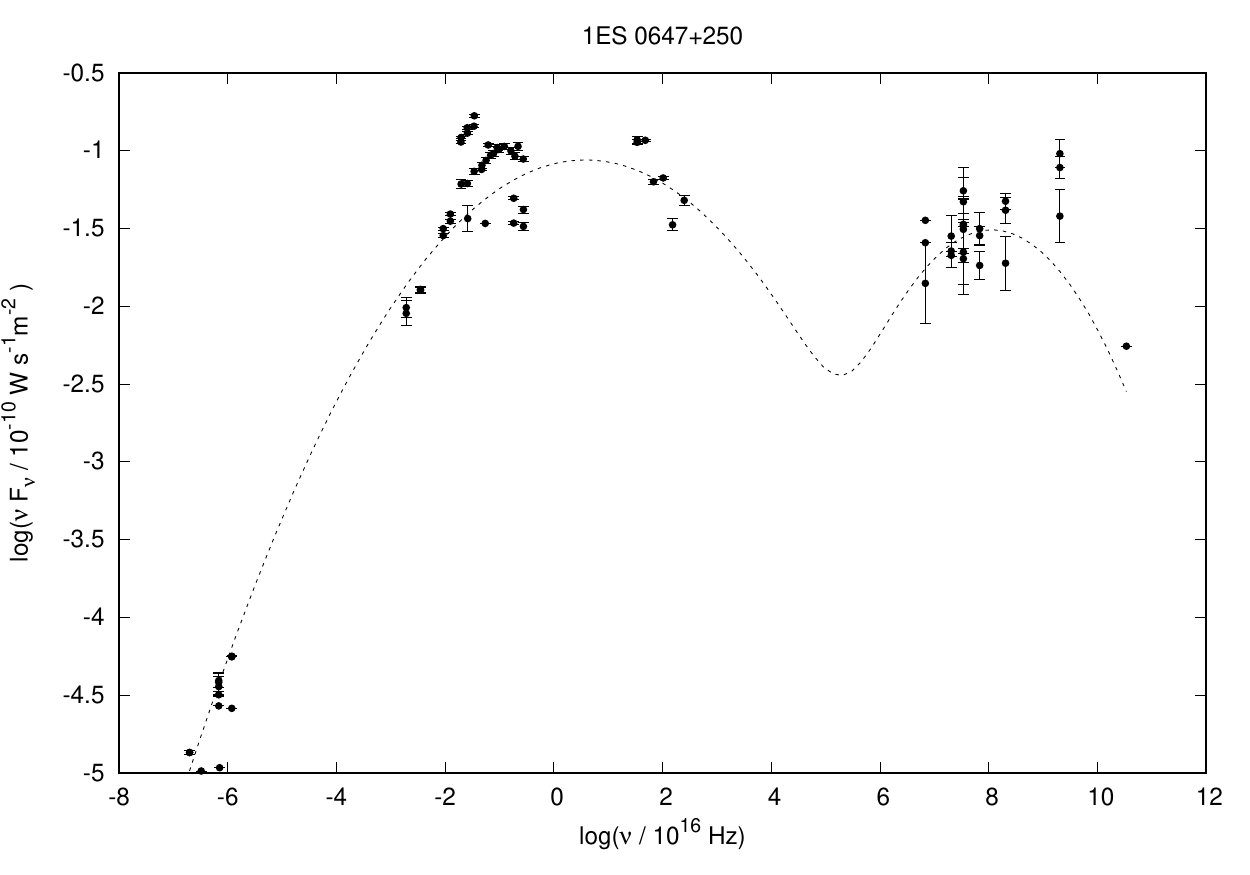}
\caption{See the caption of Fig. \ref{sed1}}
\end{figure*}

\begin{figure*}
\includegraphics[width=0.45\textwidth]{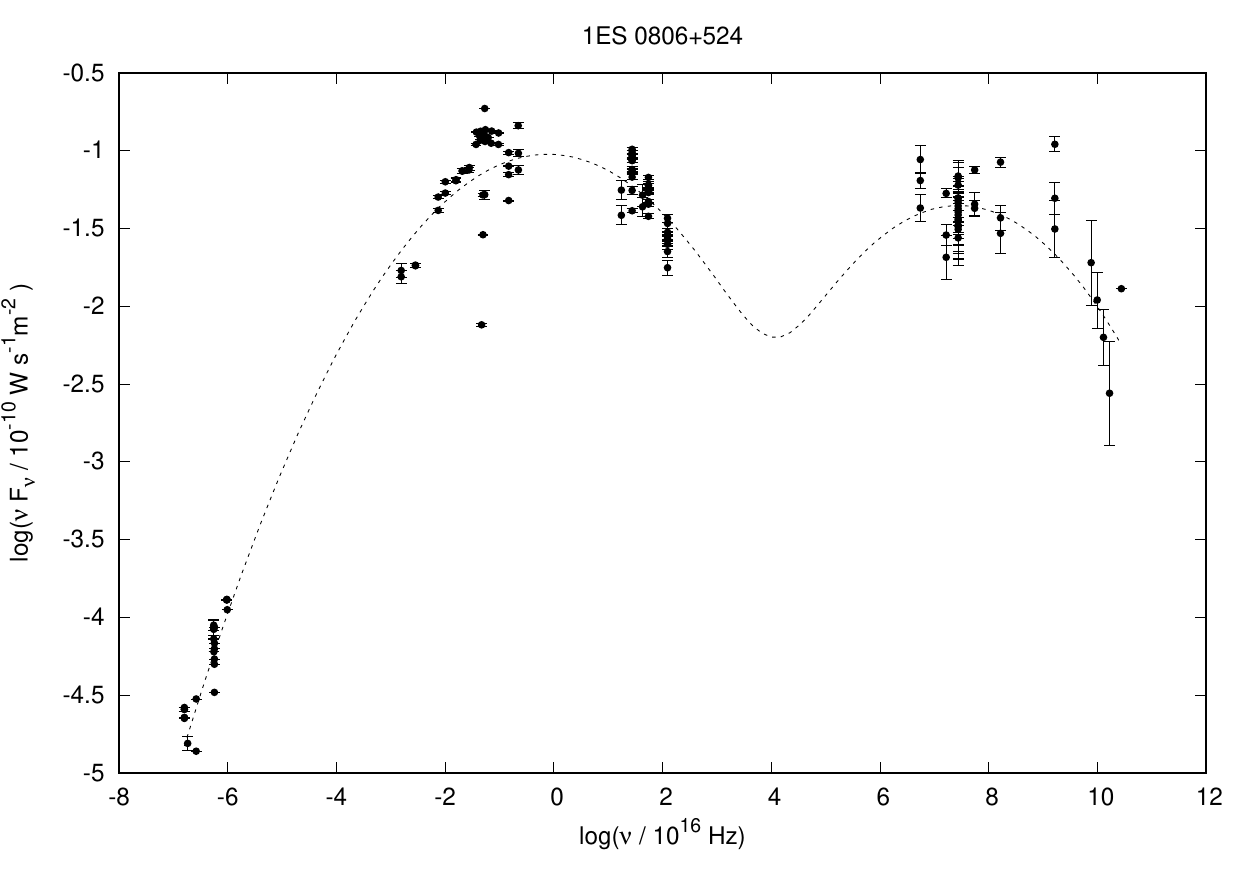}
\includegraphics[width=0.45\textwidth]{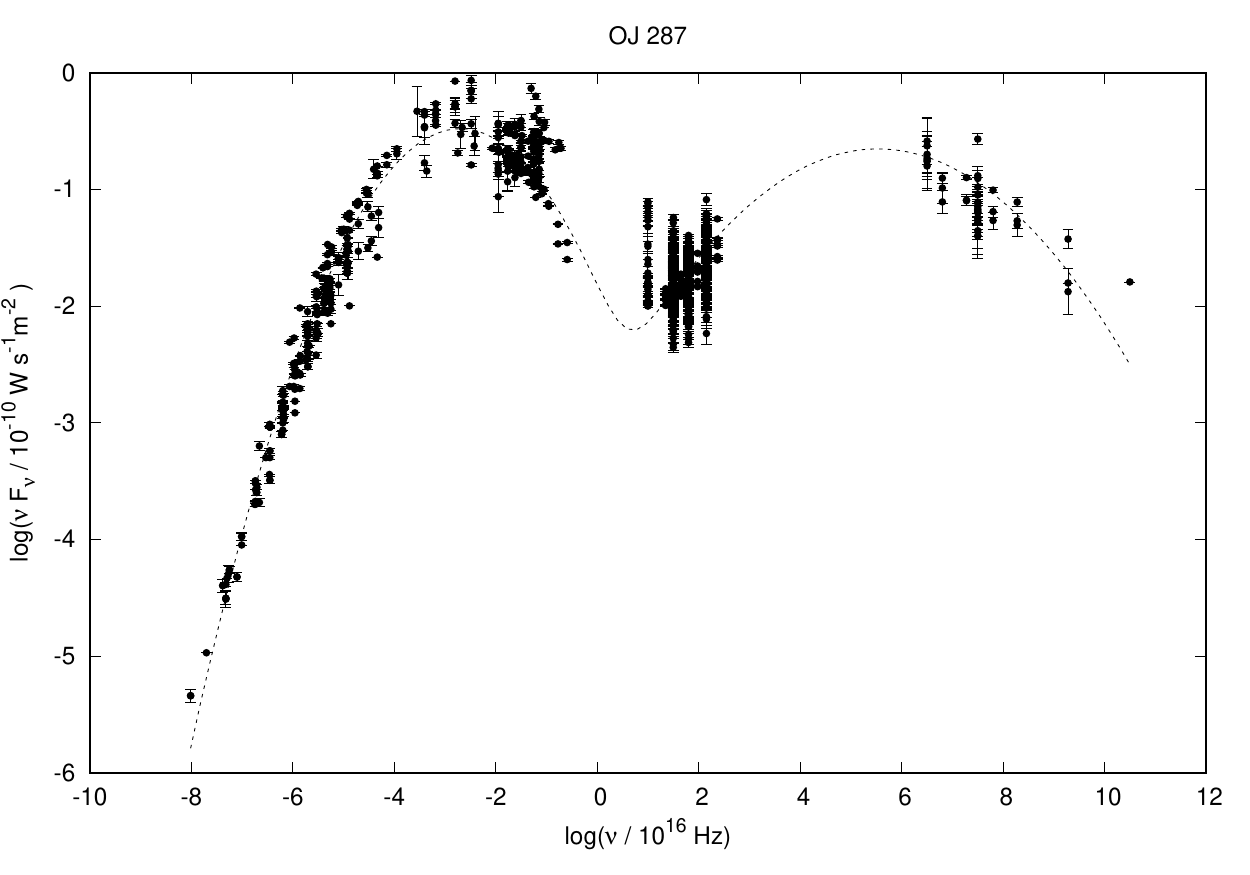}
\caption{See the caption of Fig. \ref{sed1}}
\label{Fig:sed6}
\end{figure*}

\begin{figure*}
\includegraphics[width=0.45\textwidth]{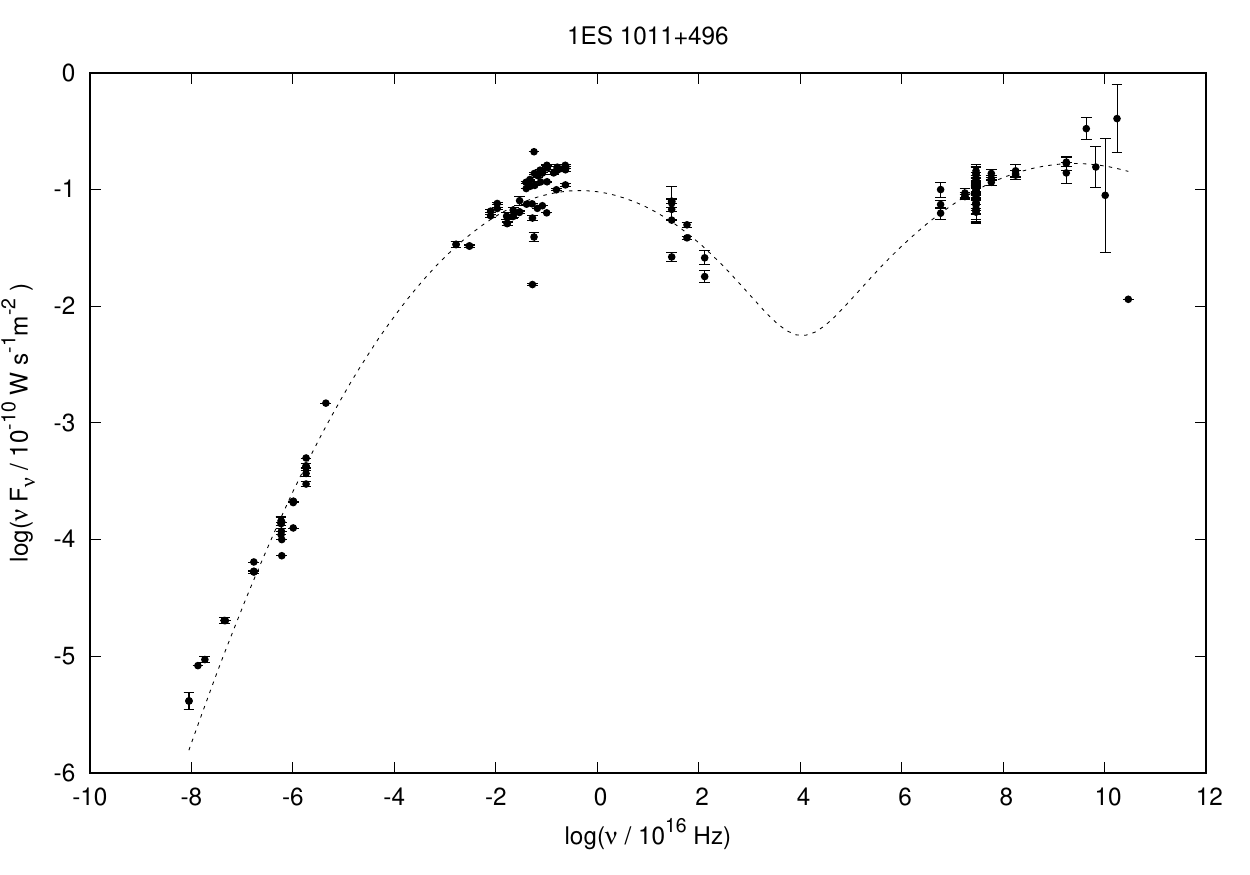}
\includegraphics[width=0.45\textwidth]{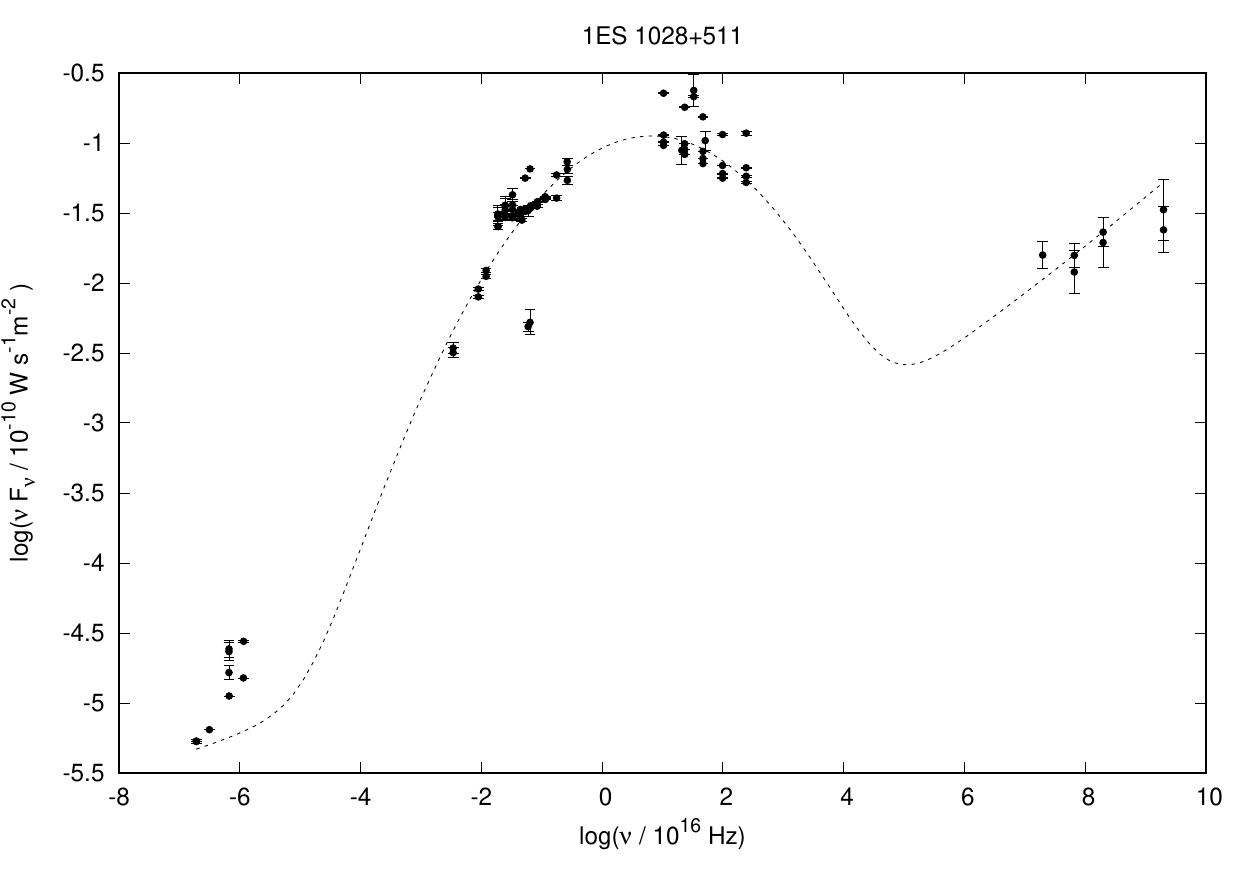}
\caption{See the caption of Fig. \ref{sed1}}
\label{Fig:sed9}
\end{figure*}

\begin{figure*}
\includegraphics[width=0.45\textwidth]{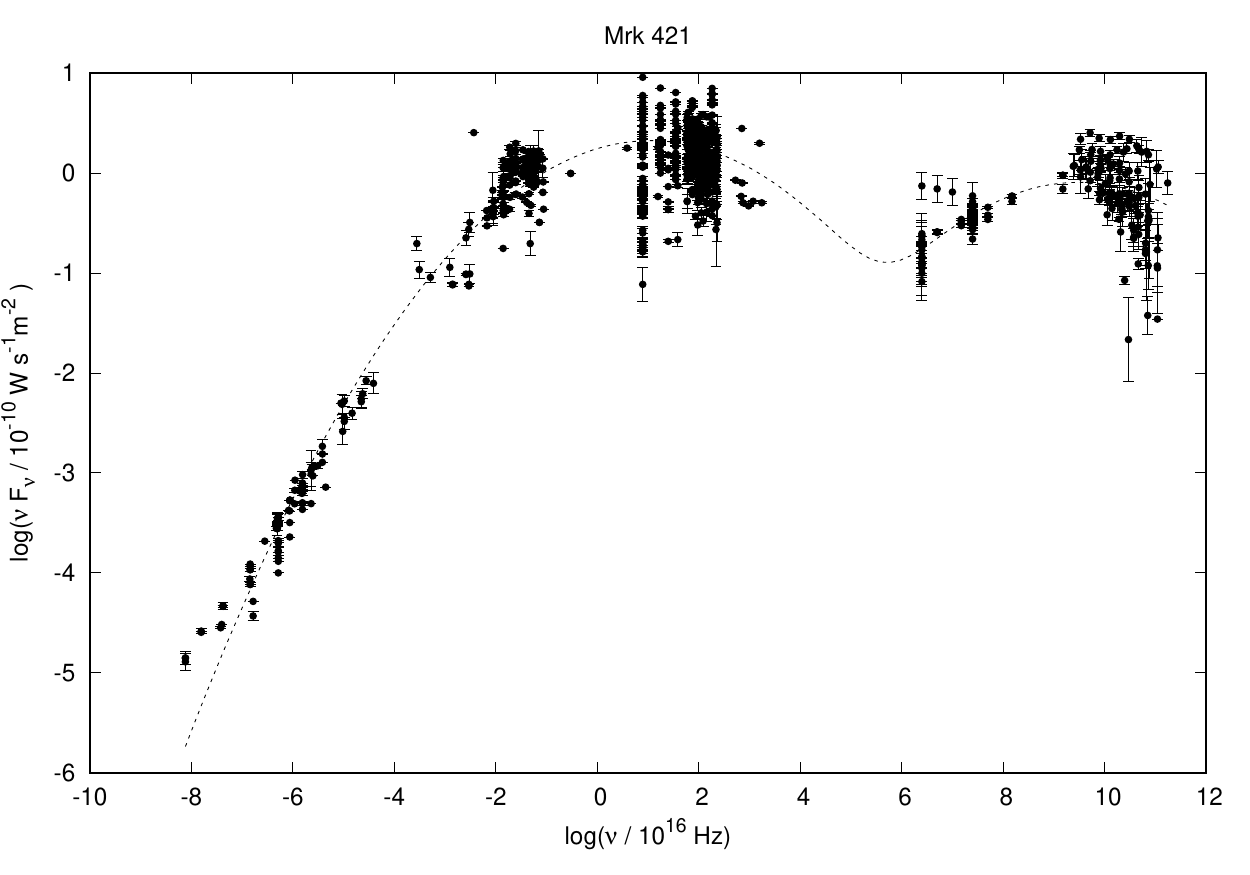}
\includegraphics[width=0.45\textwidth]{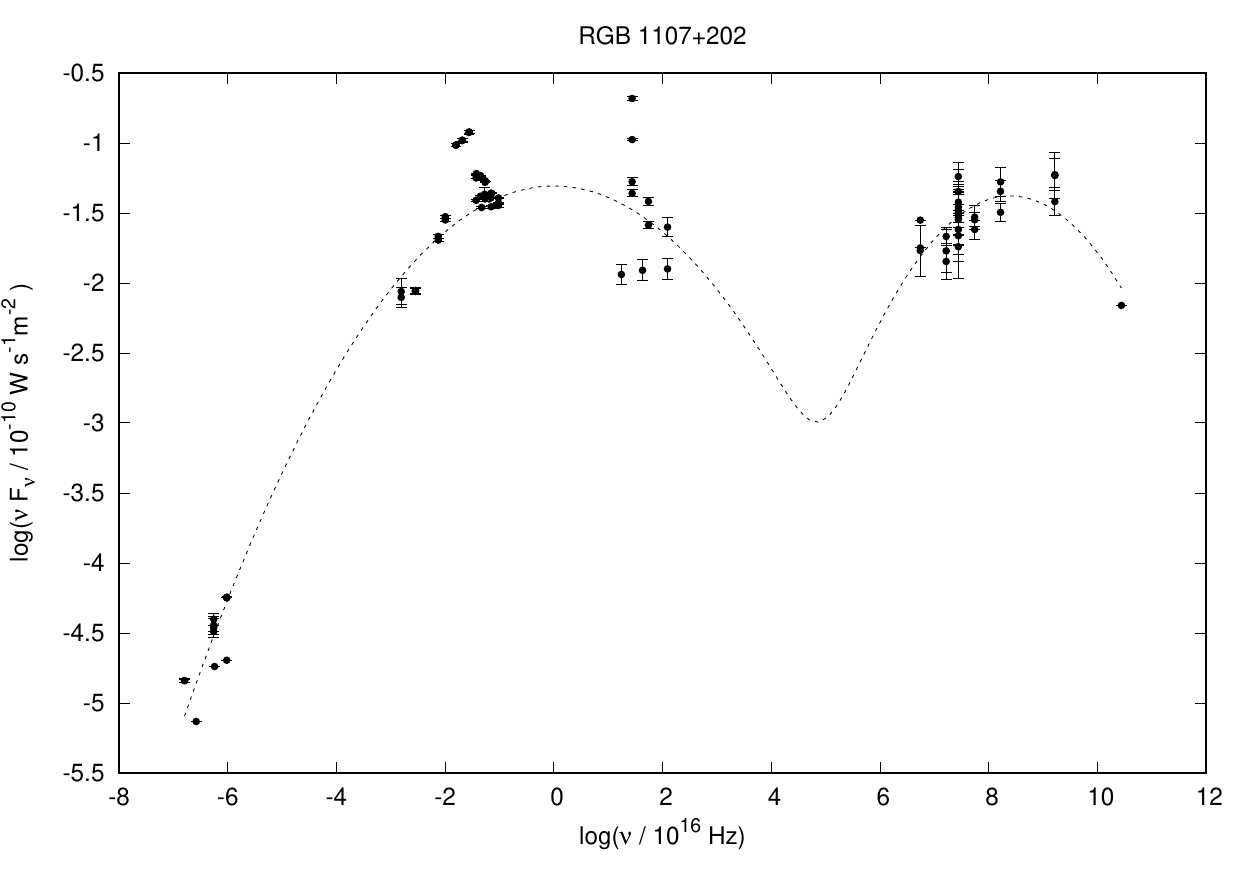}
\caption{See the caption of Fig. \ref{sed1}}
\label{Fig:sed12}
\end{figure*}

\begin{figure*}
\includegraphics[width=0.45\textwidth]{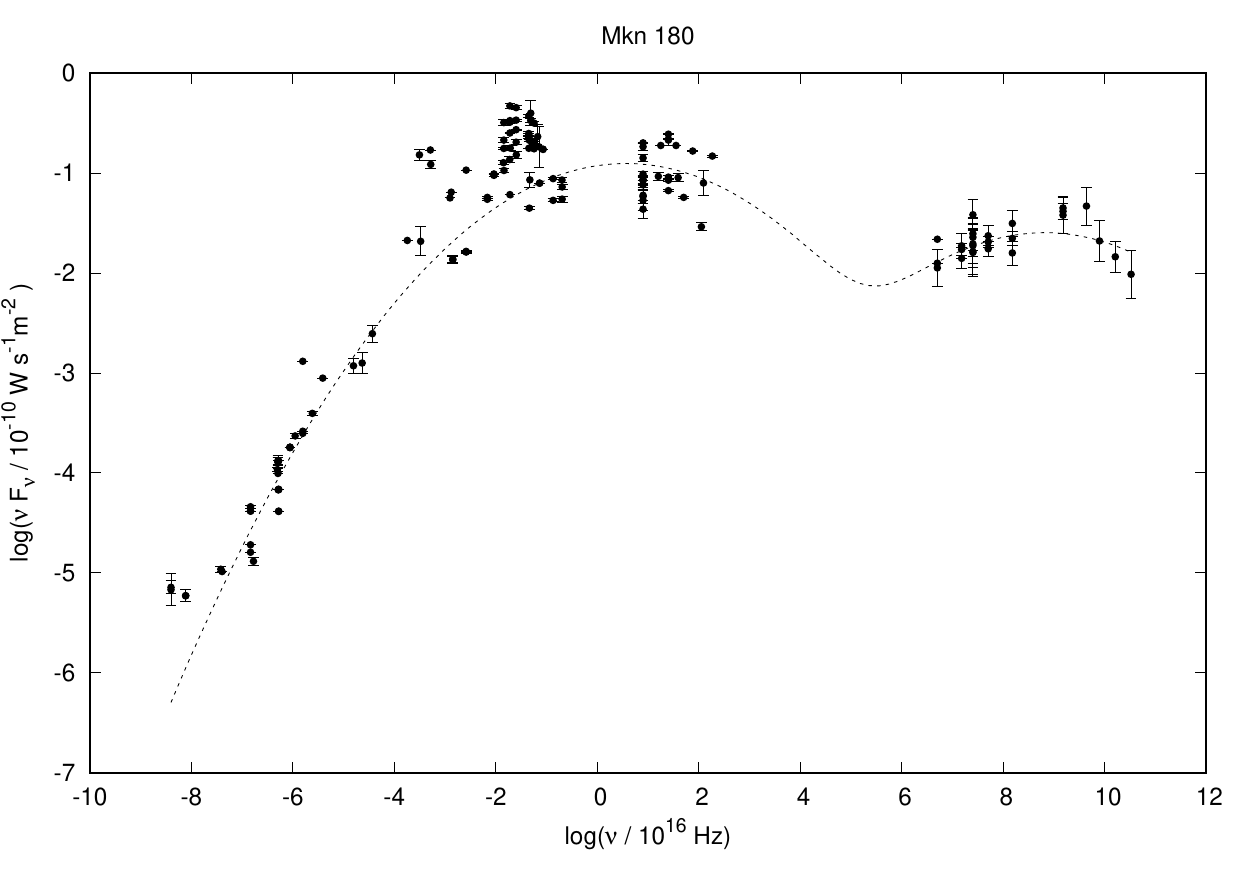}
\includegraphics[width=0.45\textwidth]{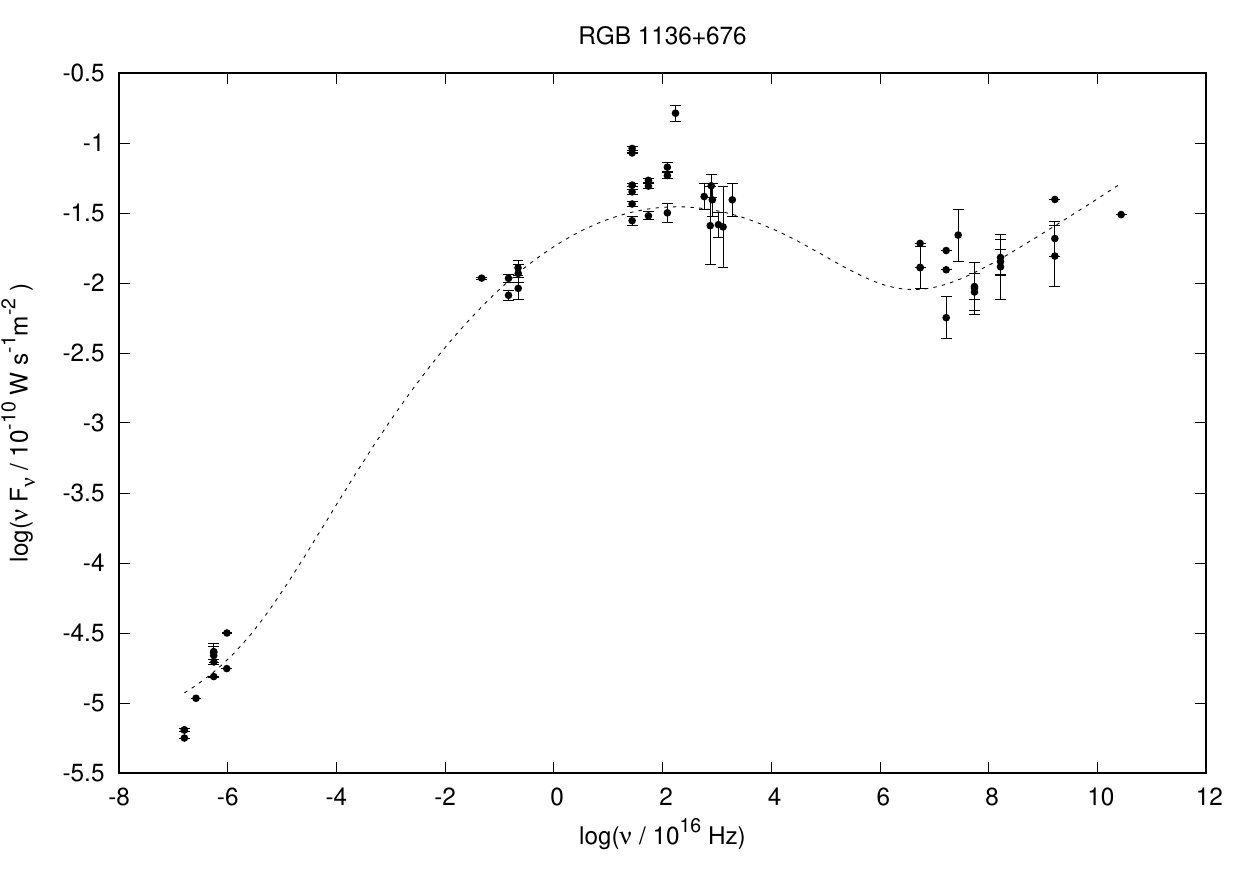}
\caption{See the caption of Fig. \ref{sed1}}
\label{Fig:sed14}
\end{figure*}

\begin{figure*}
\includegraphics[width=0.45\textwidth]{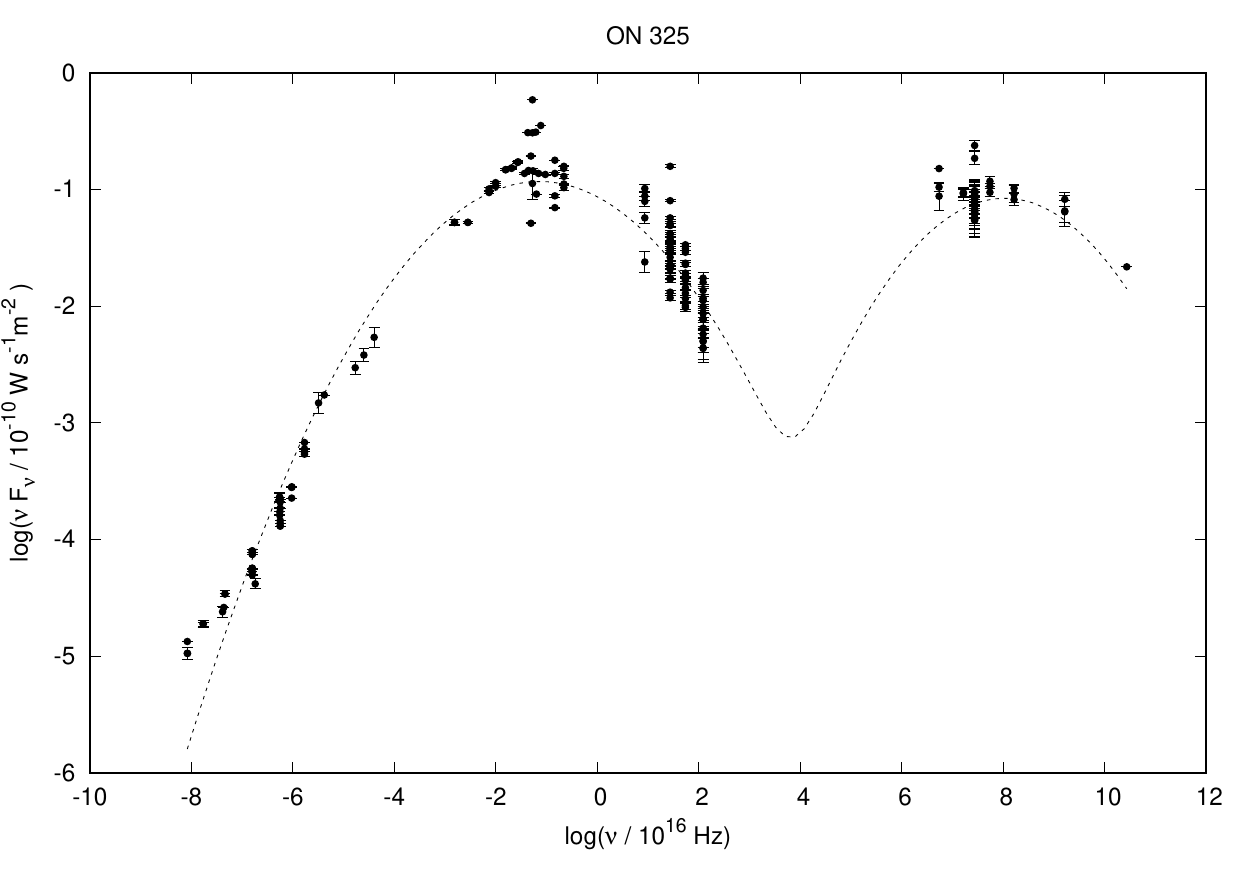}
\includegraphics[width=0.45\textwidth]{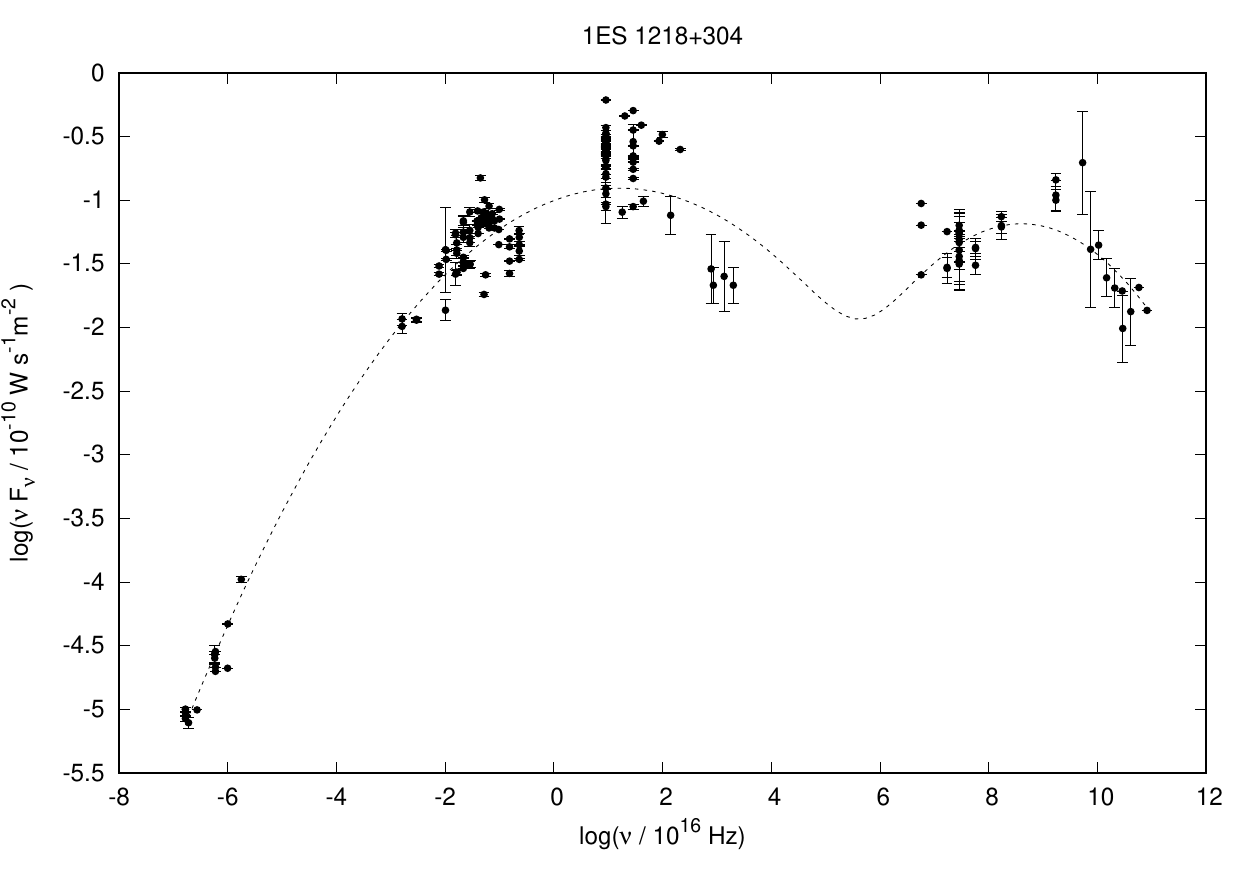}  
\caption{See the caption of Fig. \ref{sed1}}
\label{Fig:sed15}
\end{figure*}

\begin{figure*}
\includegraphics[width=0.45\textwidth]{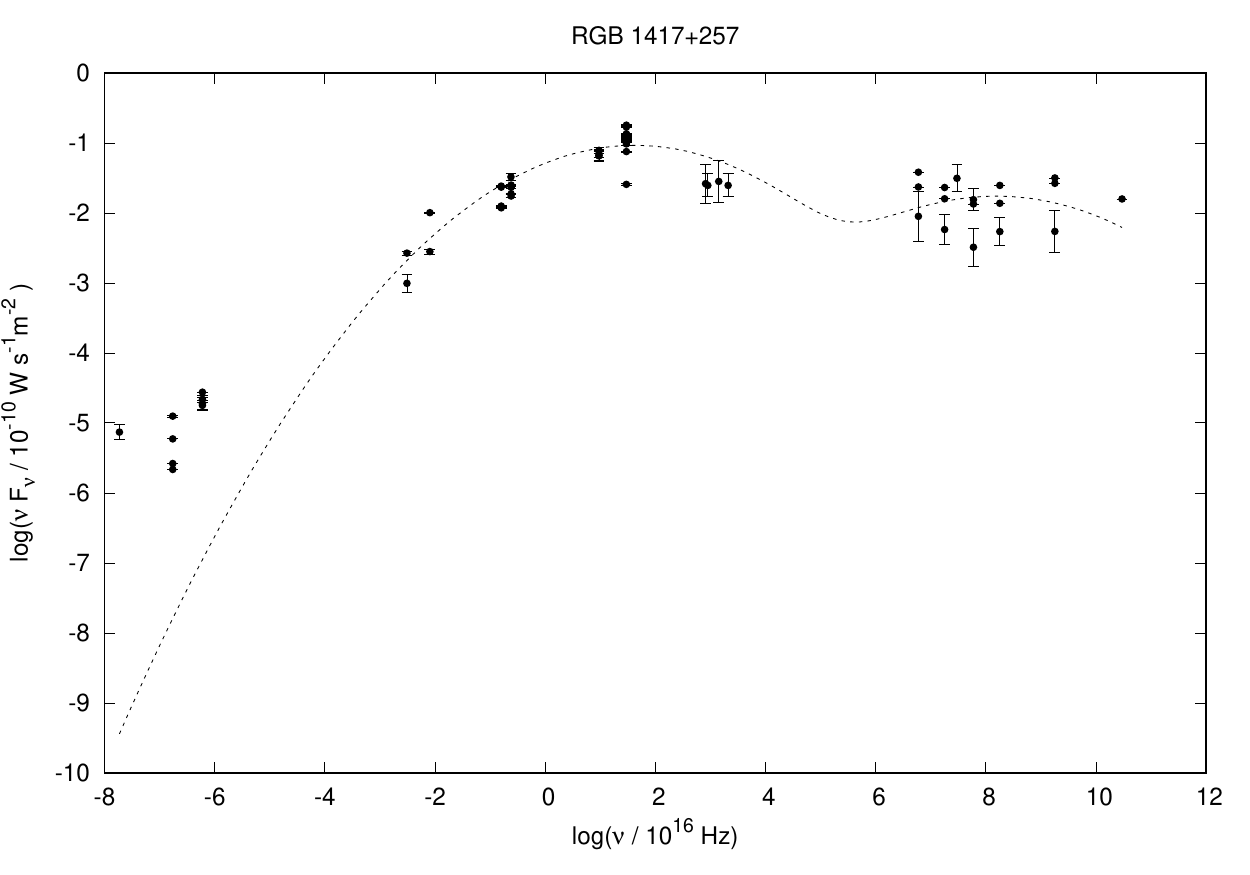}
\includegraphics[width=0.45\textwidth]{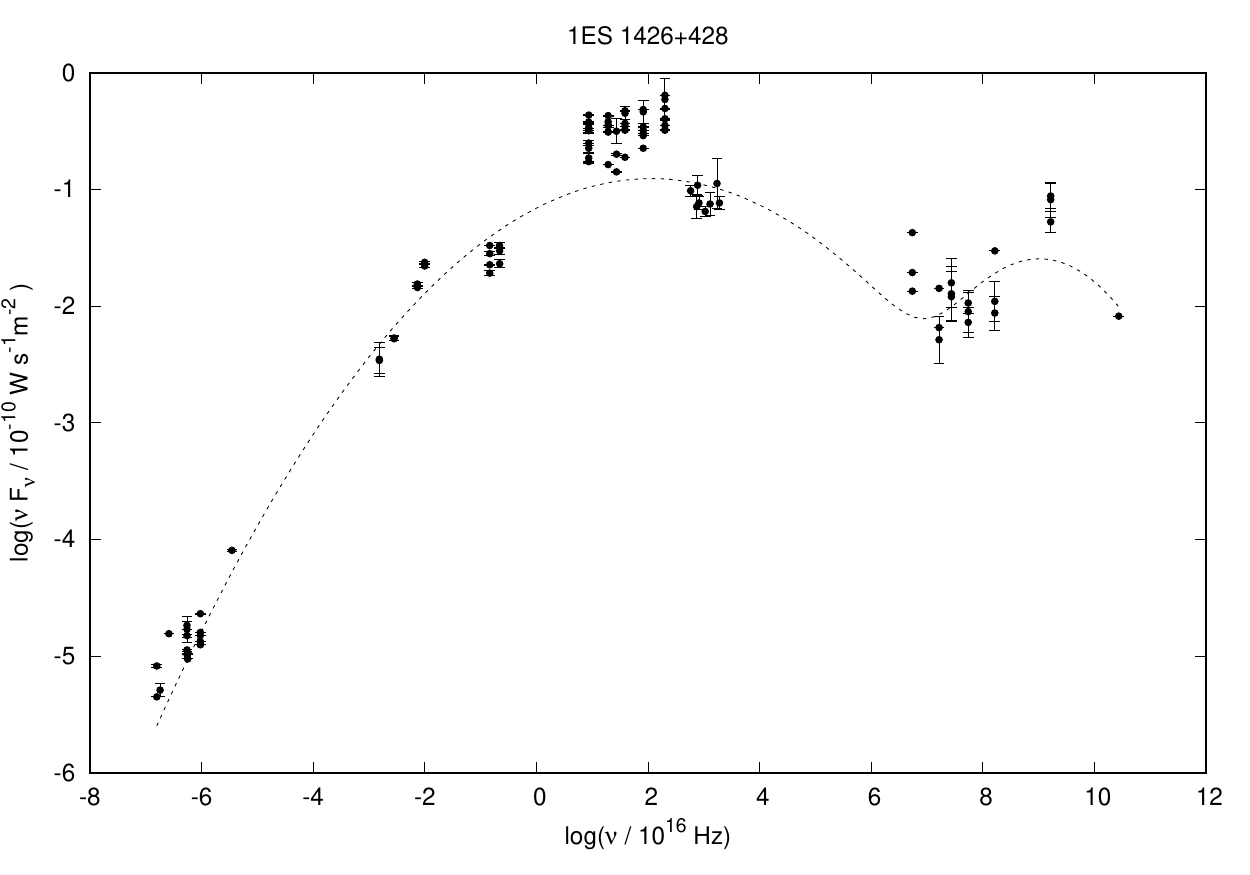}  
\caption{See the caption of Fig. \ref{sed1}}
\label{Fig:sed17}
\end{figure*}

\begin{figure*}
\includegraphics[width=0.45\textwidth]{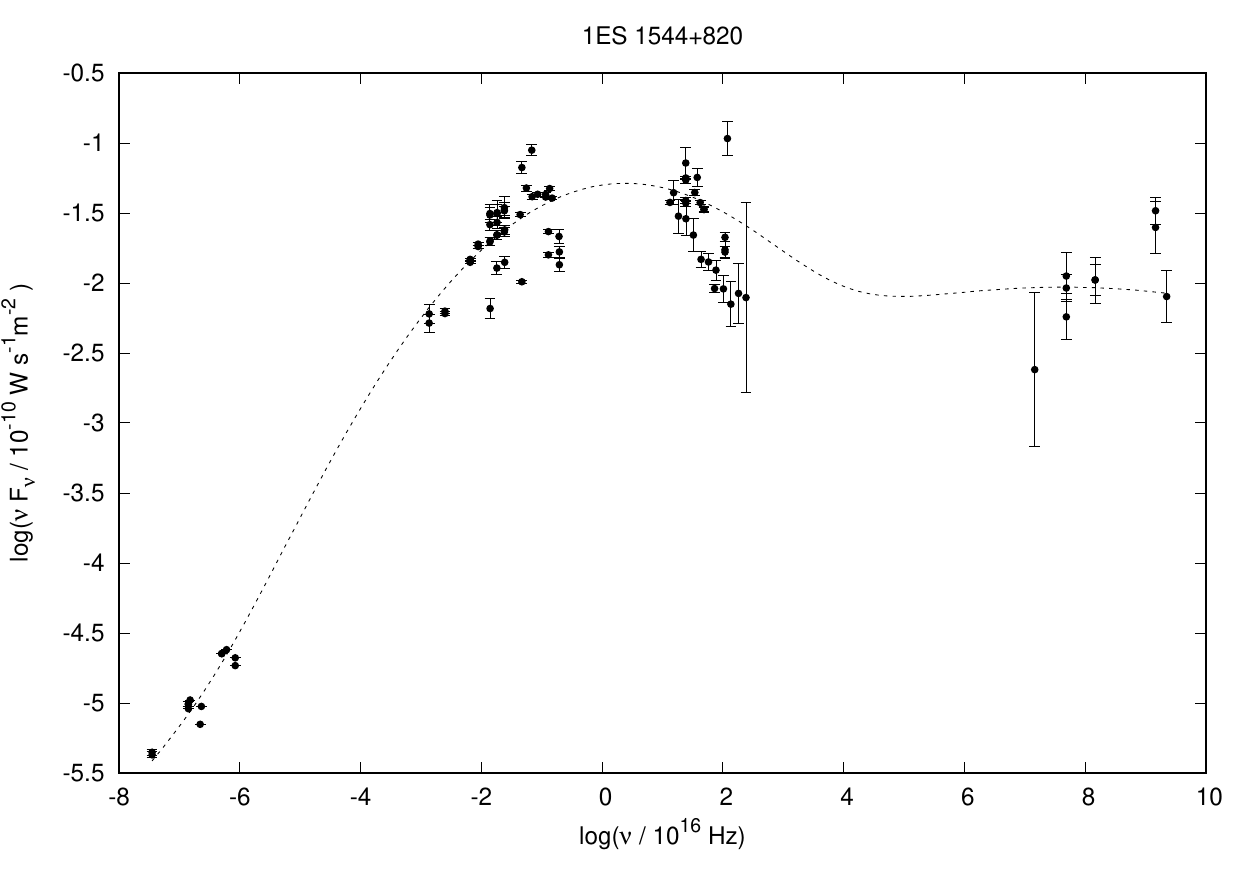}
\includegraphics[width=0.45\textwidth]{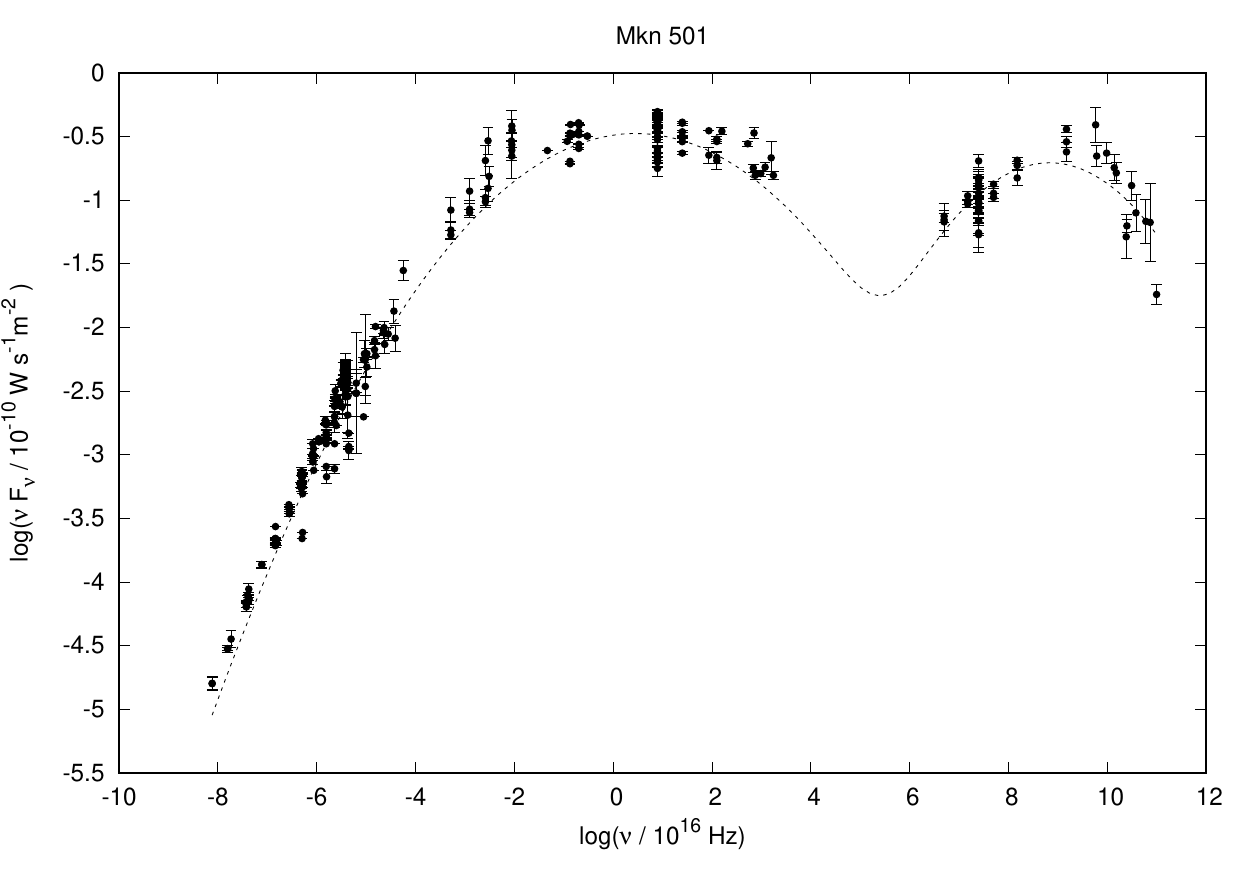}  
\caption{See the caption of Fig. \ref{sed1}}
\label{Fig:sed18b}
\end{figure*}

\begin{figure*}
\includegraphics[width=0.45\textwidth]{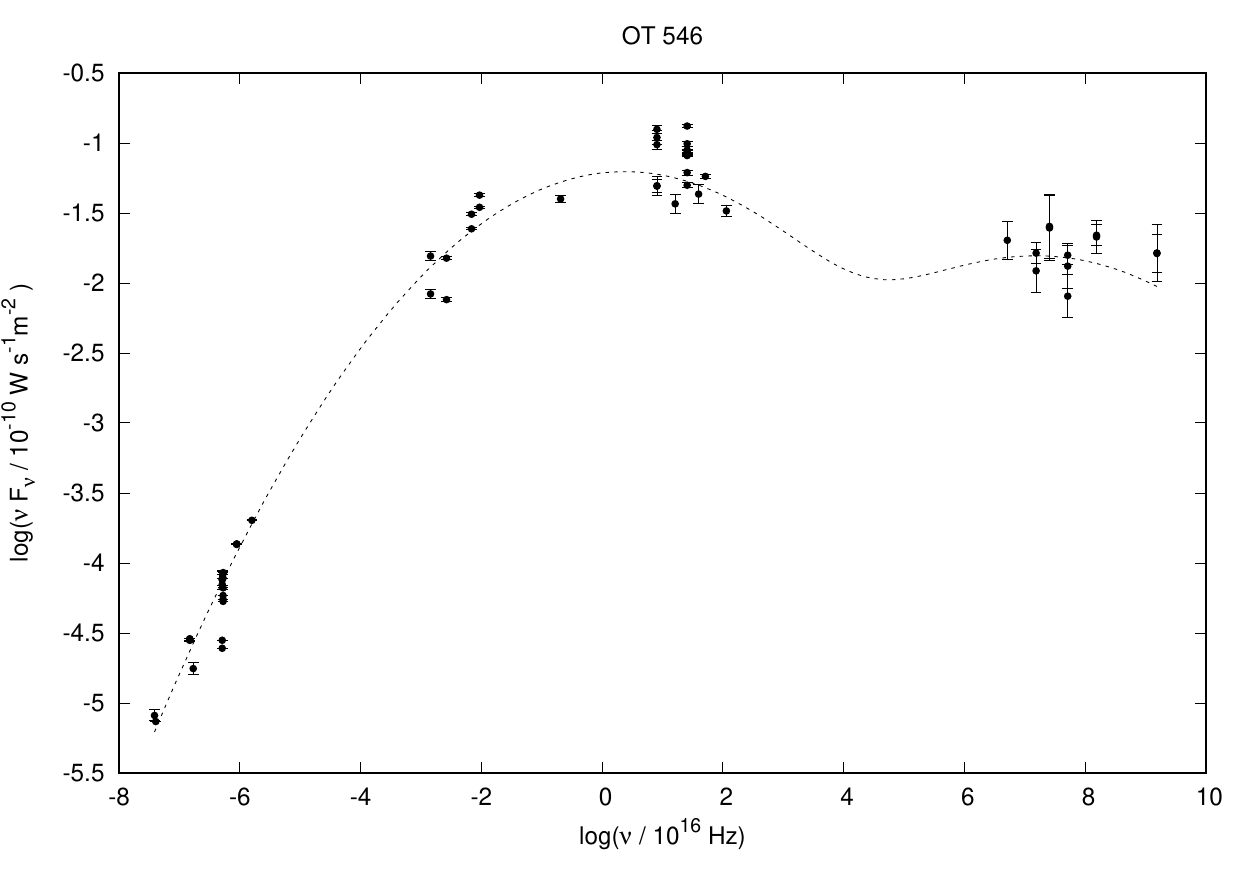}
\includegraphics[width=0.45\textwidth]{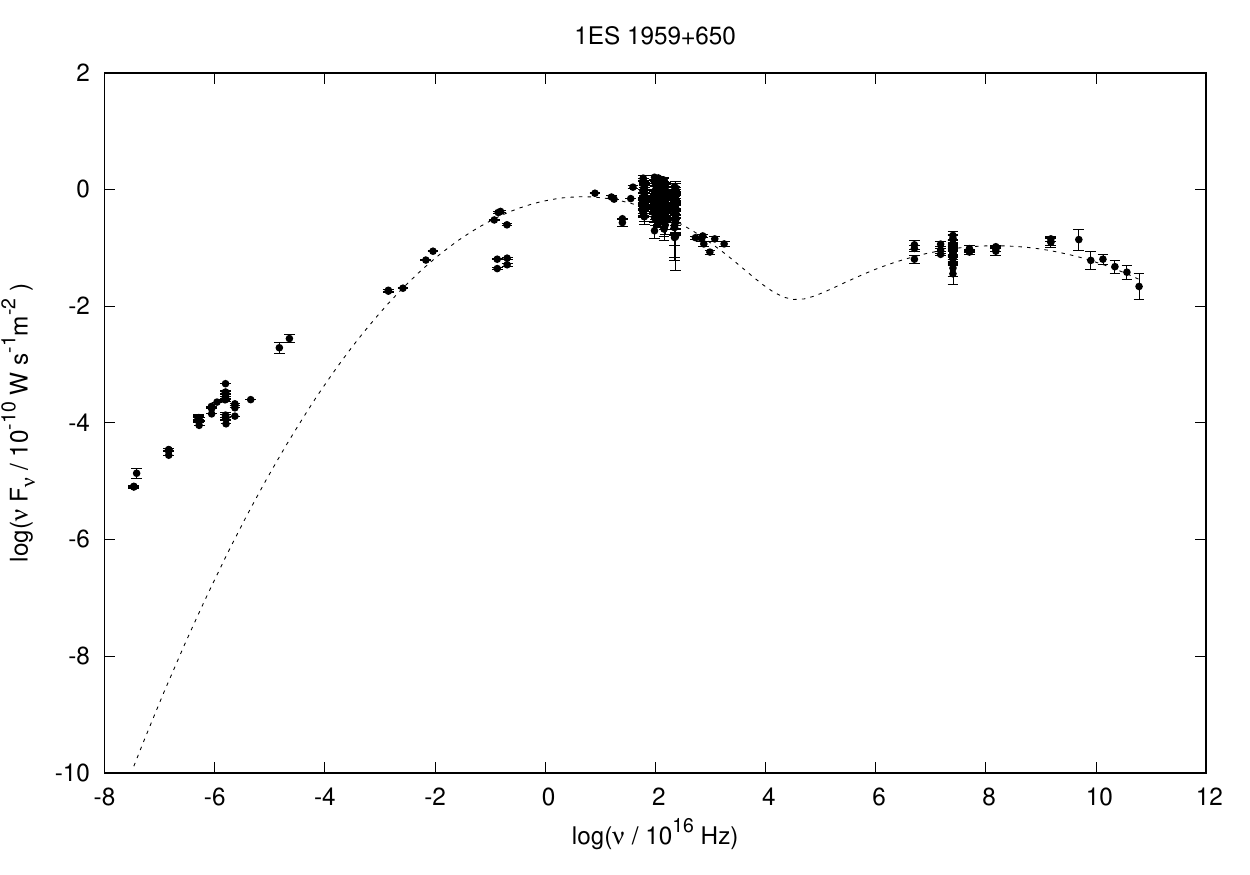}  
\caption{See the caption of Fig. \ref{sed1}}
\label{Fig:sed20}
\end{figure*}

\begin{figure*}
\includegraphics[width=0.45\textwidth]{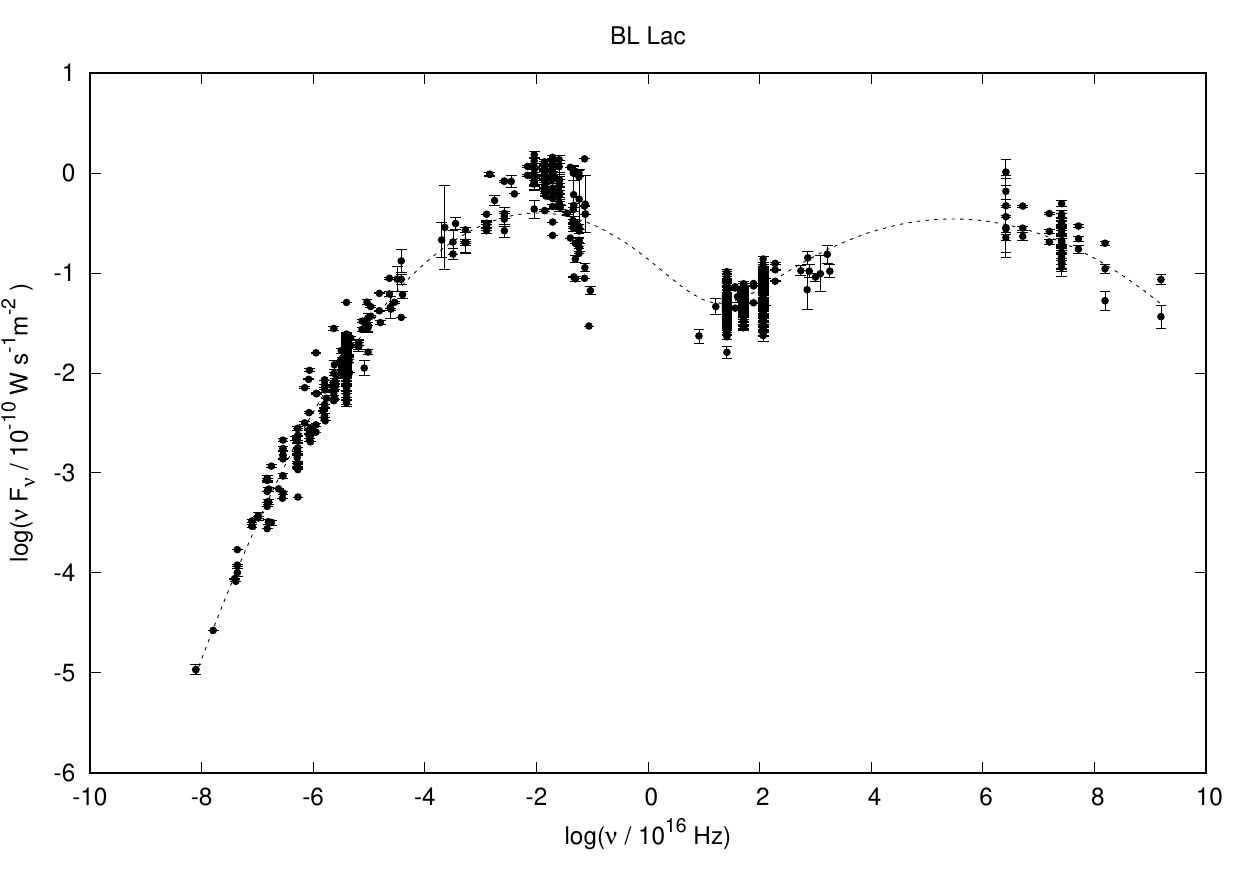}
\includegraphics[width=0.45\textwidth]{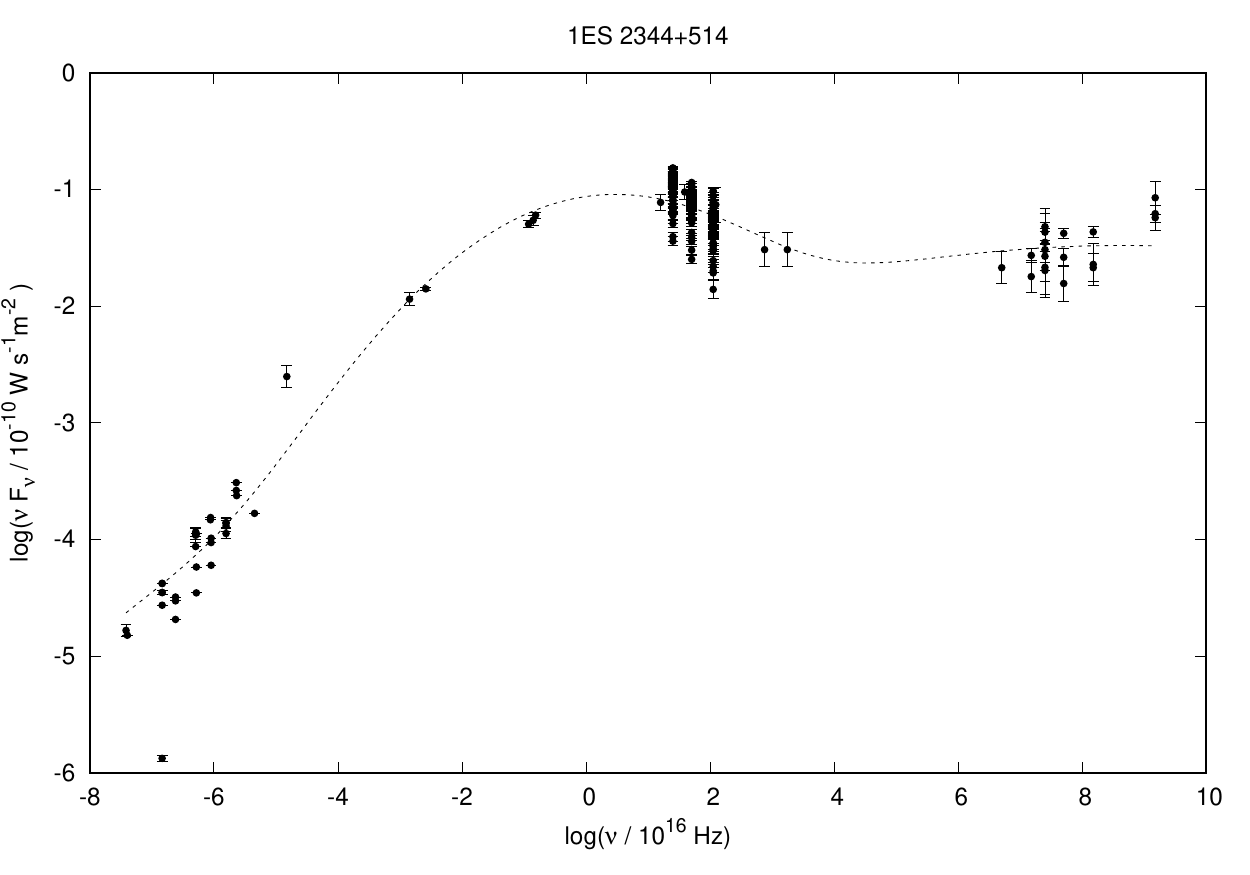}  
\caption{See the caption of Fig. \ref{sed1}}
\label{Fig:sed22}
\end{figure*}

\begin{figure*}
\includegraphics[width=0.45\textwidth]{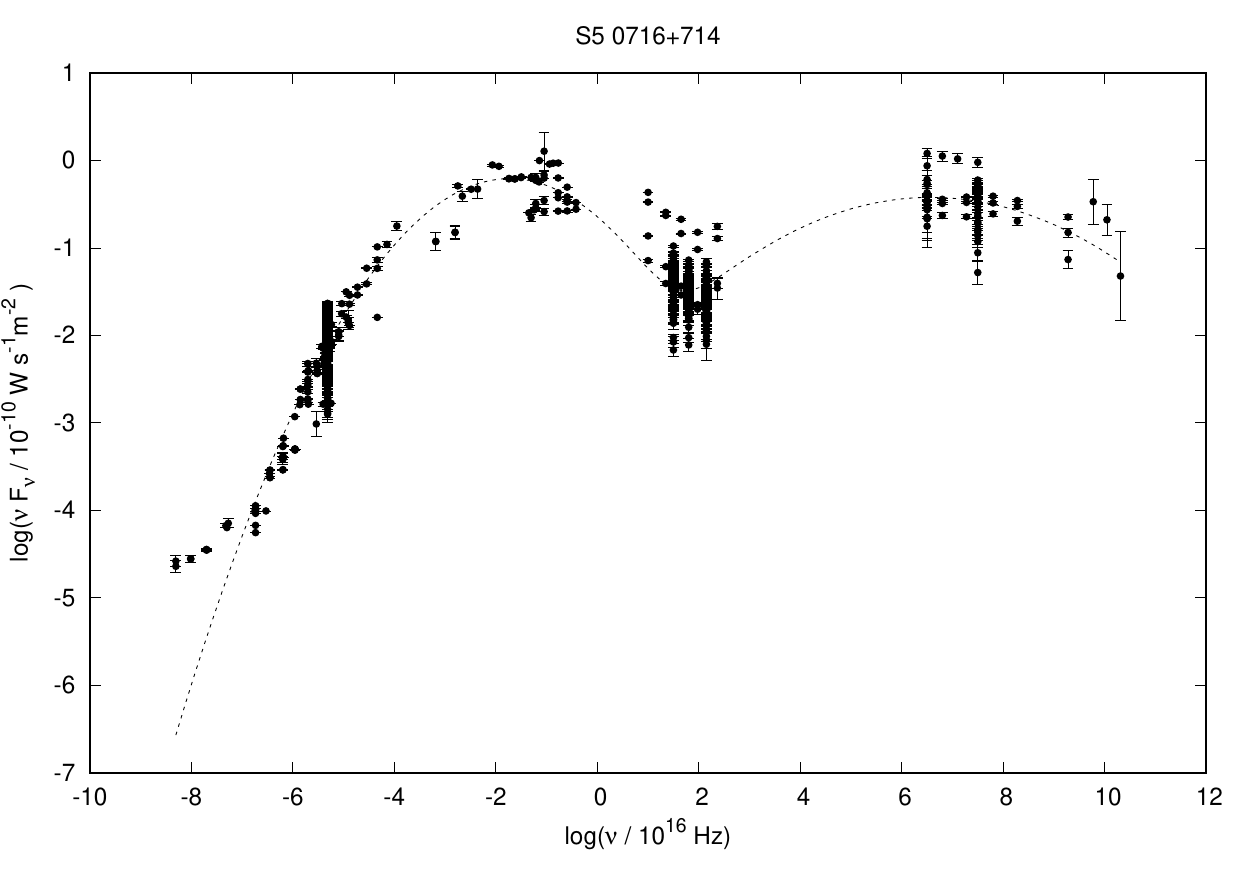}
\includegraphics[width=0.45\textwidth]{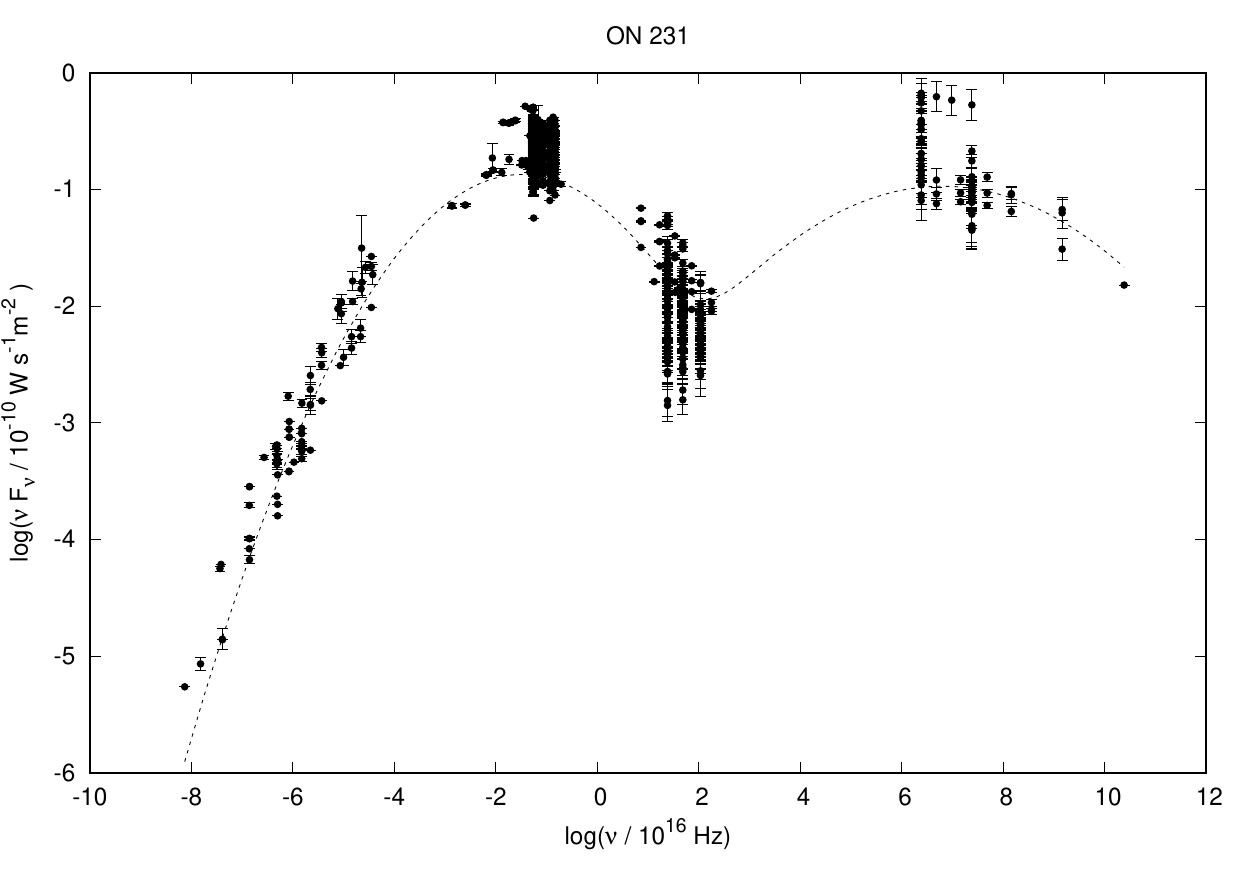}  
\caption{See the caption of Fig. \ref{sed1}}
\label{Fig:sed24}
\end{figure*}

\begin{figure*}
\includegraphics[width=0.45\textwidth]{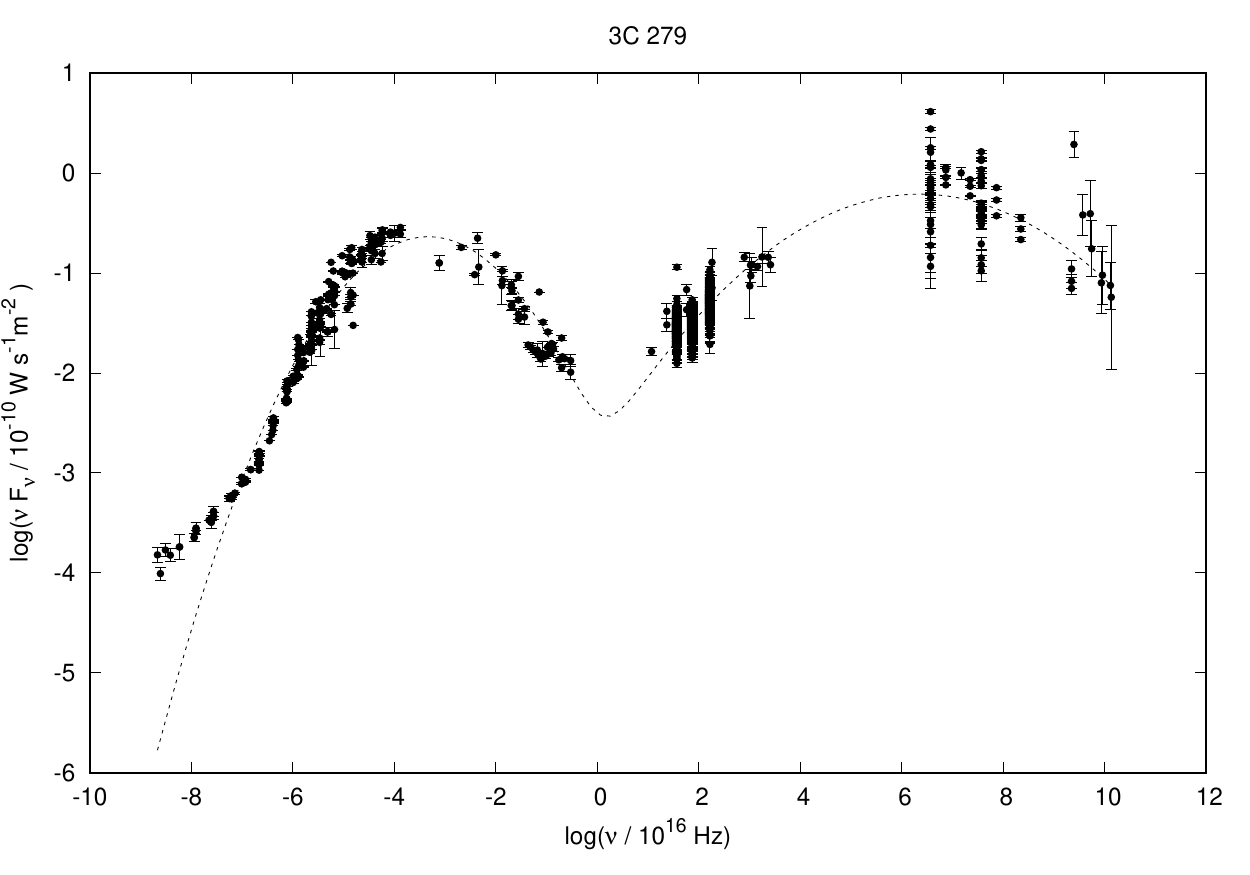}
\includegraphics[width=0.45\textwidth]{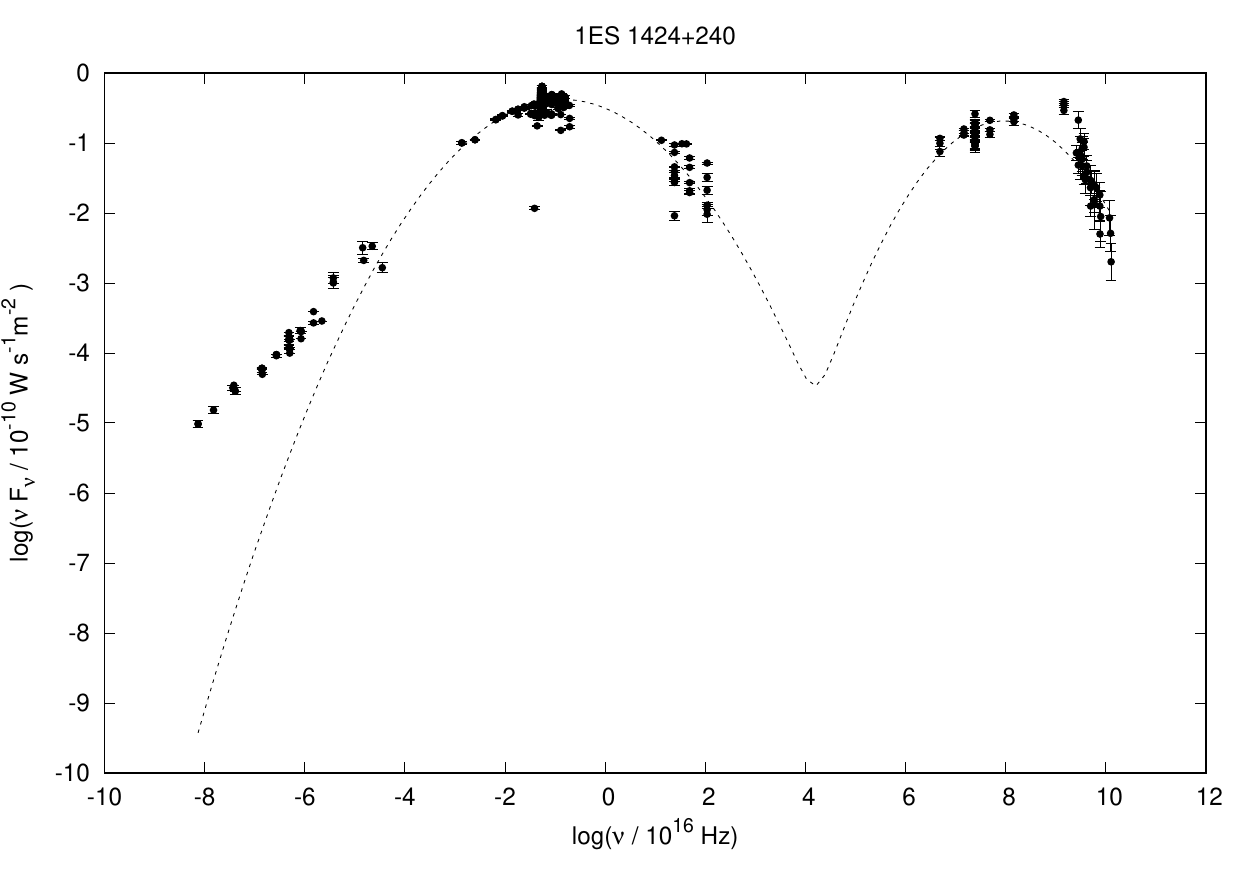}  
\caption{See the caption of Fig. \ref{sed1}}
\label{Fig:sed26}
\end{figure*}

\begin{figure*}
\includegraphics[width=0.45\textwidth]{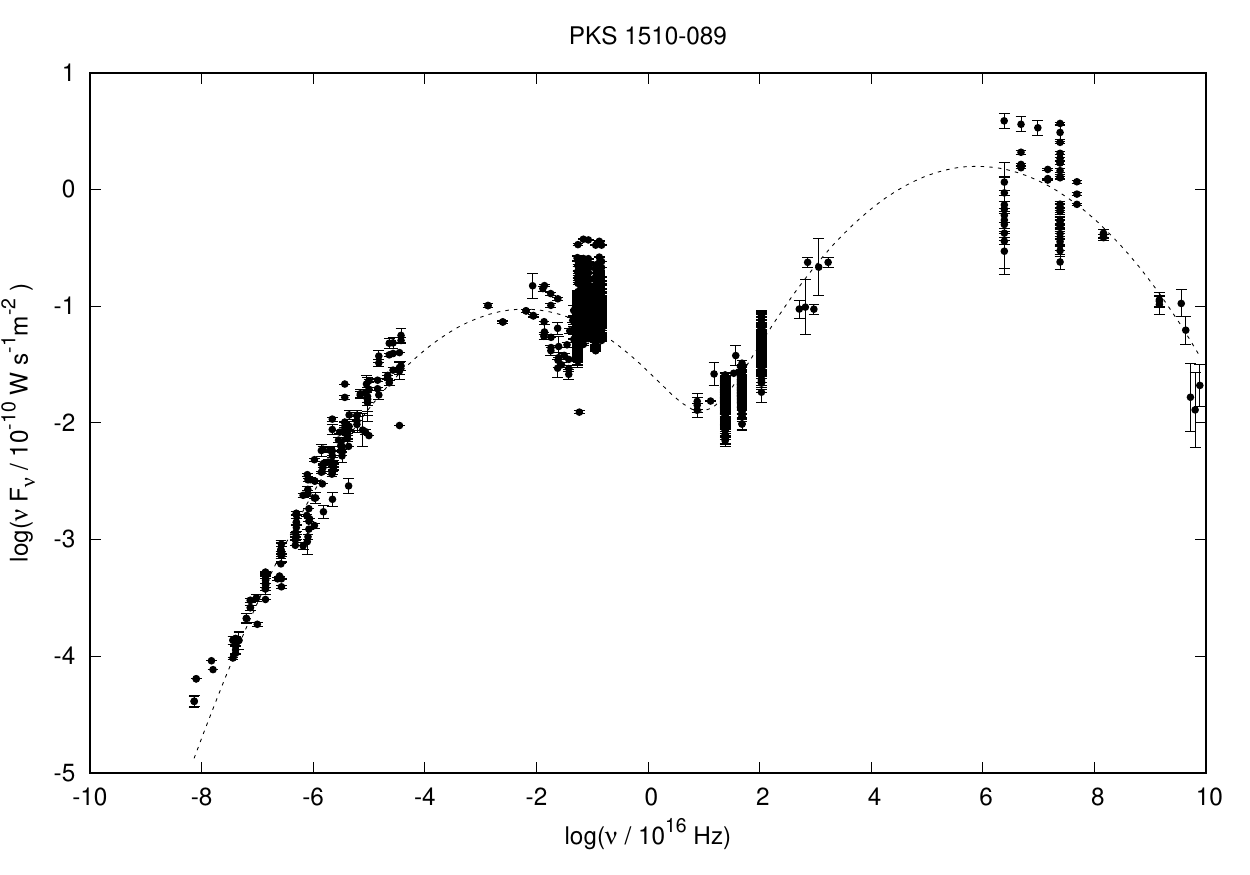}
\includegraphics[width=0.45\textwidth]{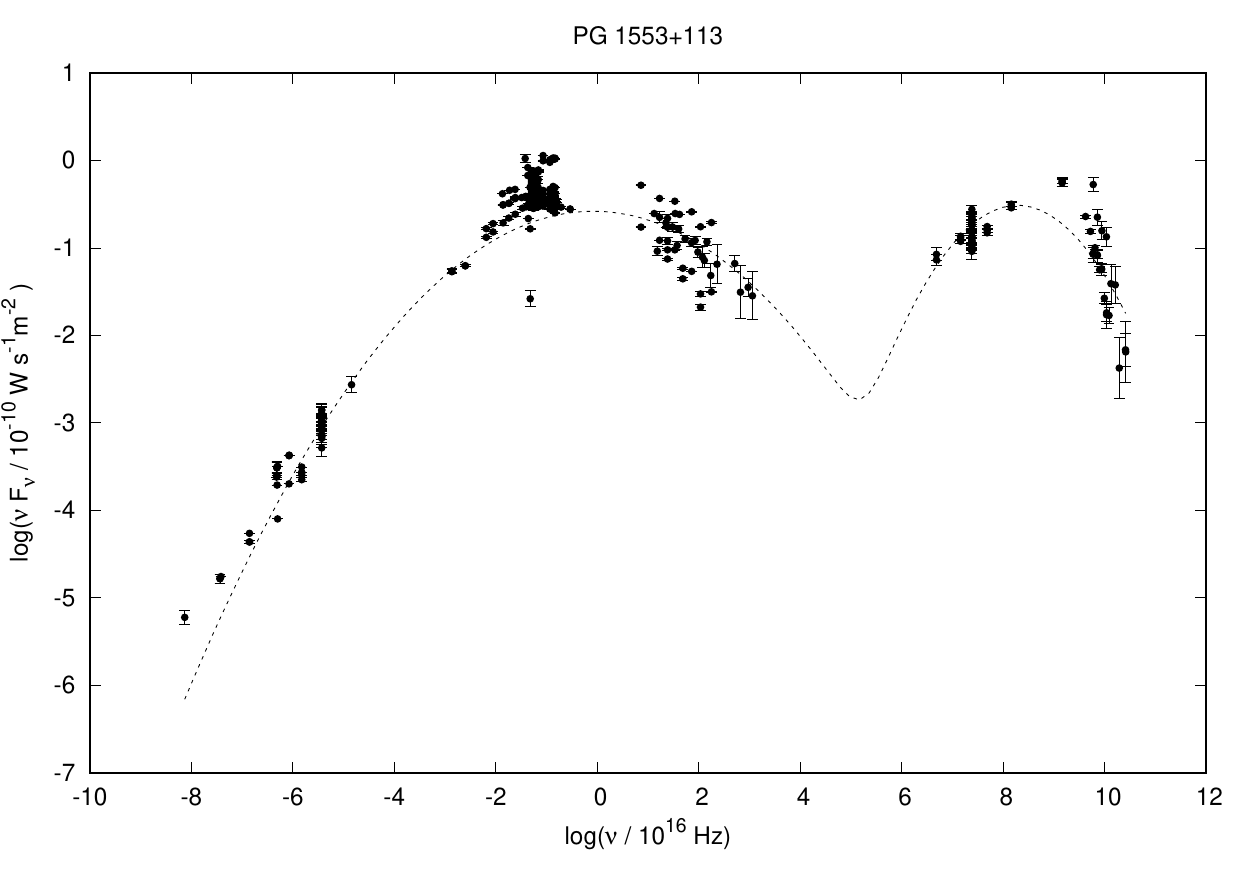}  
\caption{See the caption of Fig. \ref{sed1}}
\label{Fig:sed28}
\end{figure*}

\begin{figure*}
\includegraphics[width=0.45\textwidth]{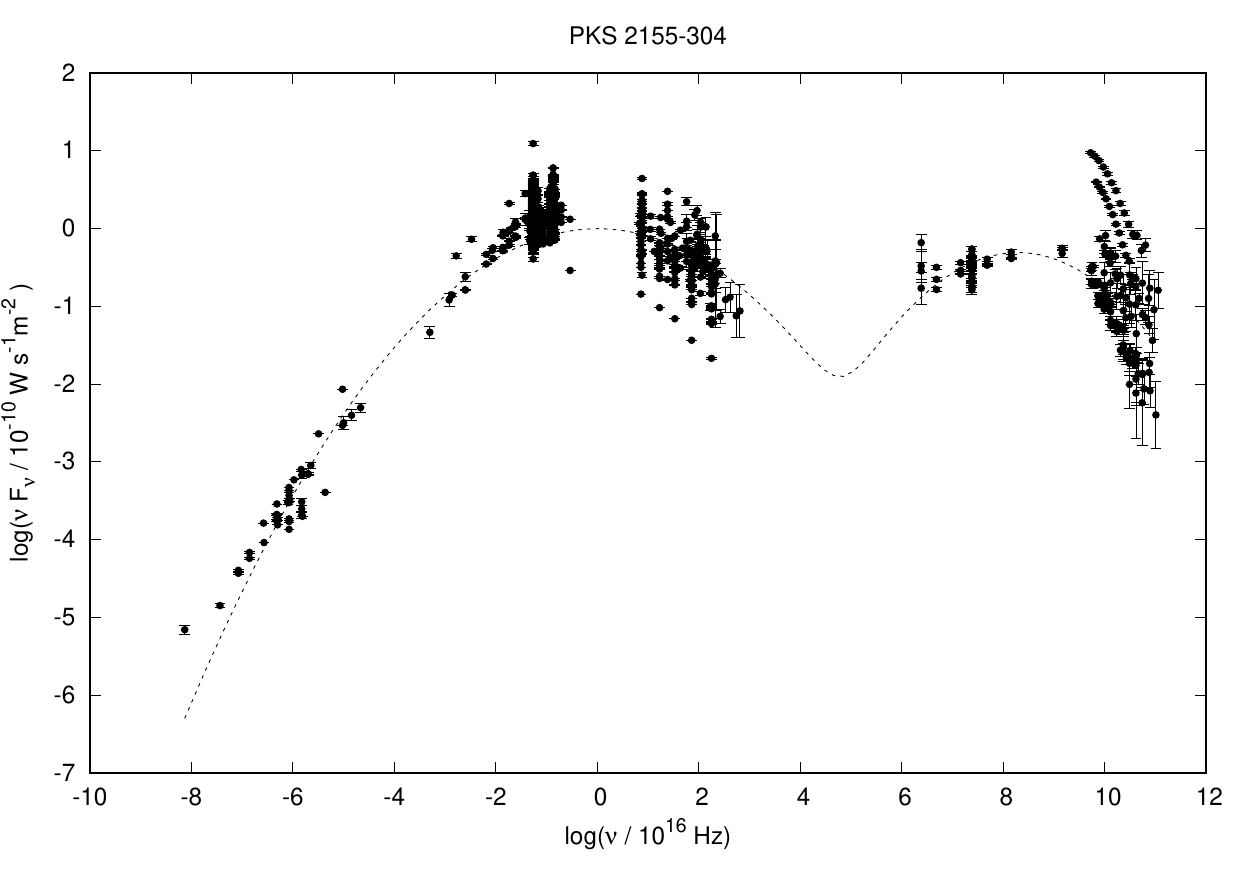}
\caption{See the caption of Fig. \ref{sed1}}
\label{vikased}
\end{figure*}

\section{\label{fluxtables}Tables}

In electronic form only.

\end{appendix}

\end{document}